 \documentclass[iop,apj,numberedappendix,twocolappendix]{emulateapj}

 \usepackage{apjfonts}

\slugcomment{to appear in the Astrophysical Journal, Supplement Series}

\shorttitle{Light Curves of Neon Novae}
\shortauthors{Hachisu et al.}

\begin{document}

\title{Light Curve Analysis of Neon Novae}


\author{Izumi Hachisu}
\affil{Department of Earth Science and Astronomy, 
College of Arts and Sciences, The University of Tokyo,
3-8-1 Komaba, Meguro-ku, Tokyo 153-8902, Japan} 
\email{hachisu@ea.c.u-tokyo.ac.jp}

\and

\author{Mariko Kato}
\affil{Department of Astronomy, Keio University, 
Hiyoshi, Kouhoku-ku, Yokohama 223-8521, Japan} 





\begin{abstract}
We analyzed light curves of five neon novae, QU~Vul, V351~Pup, V382~Vel,
V693~CrA, and V1974~Cyg, and determined their white dwarf (WD) masses and
distance moduli on the basis of theoretical light curves composed of
free-free and photospheric emission.  For QU~Vul, we obtained
a distance of $d\sim2.4$~kpc, reddening of $E(B-V)\sim0.55$,
and WD mass of $M_{\rm WD}=0.82$ -- $0.96~M_\sun$.  This suggests that an 
oxygen-neon WD lost a mass of more than $\sim0.1~M_\sun$ since its birth.
For V351~Pup, we obtained $d\sim5.5$~kpc, $E(B-V)\sim0.45$, and  
$M_{\rm WD}=0.98$ -- $1.1~M_\sun$.
For V382~Vel, we obtained $d\sim1.6$~kpc, $E(B-V)\sim0.15$, and
$M_{\rm WD}=1.13$ -- $1.28~M_\sun$.
For V693~CrA, we obtained $d\sim7.1$~kpc, $E(B-V)\sim0.05$, and
$M_{\rm WD}=1.15$ -- $1.25~M_\sun$.
For V1974~Cyg, we obtained $d\sim1.8$~kpc, $E(B-V)\sim0.30$, and
$M_{\rm WD}=0.95$ -- $1.1~M_\sun$.
For comparison, we added the carbon-oxygen nova V1668~Cyg
to our analysis and obtained $d\sim5.4$~kpc, $E(B-V)\sim0.30$,
and $M_{\rm WD}=0.98$--$1.1~M_\sun$.
In QU~Vul, photospheric emission contributes 0.4 -- 0.8 mag at most
to the optical light curve compared with free-free emission only.
In V351~Pup and V1974~Cyg, photospheric emission contributes very little
(0.2 -- 0.4 mag at most) to the optical light curve.
In V382~Vel and V693~CrA,
free-free emission dominates the continuum spectra, and
photospheric emission does not contribute to the optical magnitudes.
We also discuss the Maximum Magnitude versus Rate of Decline 
(MMRD) relation for these novae based on the universal decline law.
\end{abstract}


\keywords{novae, cataclysmic variables --- stars: individual 
(QU~Vul, V351~Pup, V382~Vel, V693~CrA, V1974~Cyg)}


\section{Introduction}
\label{introduction}
A classical nova is a thermonuclear runaway event on a mass-accreting
white dwarf (WD) in a binary system, in which mass is transferred from
the companion to the WD via Roche lobe overflow or winds.  
When the hydrogen-rich envelope mass reaches a critical
value, hydrogen ignites to trigger a shell flash at the bottom of the 
hydrogen-rich envelope.  The photosphere of the envelope expands
to a giant size of $\sim100R_\sun$, 
and the binary becomes a bright classical nova.

It is widely known that nova ejecta are enriched with heavy elements
such as C, O, and Ne \citep[e.g.,][]{geh98}.  These elements 
are thought to originate from the WD cores, because such amounts of
heavy elements cannot be synthesized in hydrogen burning on the WDs.
Thus, a nova having ejecta enriched by carbon and oxygen occurs on a
carbon--oxygen (CO) WD.  A nova is enriched by neon if it outbursts
on an oxygen-neon (ONe) WD.

Neon novae are a subclass of classical novae that show neon emission lines
stronger than the permitted lines in the nebular phase.    
If the neon in the ejecta comes from the core material of an ONe WD
\citep[e.g.,][]{geh98}, its natal WD mass is likely greater
than $\gtrsim1.07~M_\sun$ \citep[e.g.,][]{ume99}, or 
$\gtrsim1.0~M_\sun$ \citep[e.g.,][]{wei00}.
Evolution calculations suggest that a natal WD has a thin helium-rich layer
above a carbon--oxygen-rich mantle, e.g., a $0.035~M_\sun$ CO mantle for a 
$1.09~M_\sun$ ONe core \citep{gil01}.  Such a WD may undergo 
a number of nova explosions before the thin helium-rich layer is blown off.
Subsequently, the WD further undergoes a number of nova explosions before
the CO mantle on the ONe core is blown off.  
If the mass accretion rate is $\sim 10^{-9} M_\sun$~yr$^{-1}$,
the ignition mass is $\sim 10^{-5} M_\sun$, and the same amount of
WD material is dredged up in every nova outburst, we can expect that 3,000
to 4,000 outbursts (in a total time of at least 30 -- 40 Myr)
must occur before the WD is deprived of its $0.035~M_\sun$
carbon--oxygen-rich mantle and significant neon is detected in the ejecta. 
This could be a lower estimate because a carbon--oxygen-rich mantle
is more massive for the lower mass limit of the ONe core 
\citep[as massive as $\sim0.1~M_\sun$,][]{gil03}.  Therefore, 
the minimum masses could be slightly smaller than $\sim 1.0~M_\sun$.
In this paper, we determine the WD masses and other parameters
of various neon novae on the basis of our model light curve fitting.
It should be noted that, however, a modest enrichment in neon could
be originated even from a CO core \citep[][see discussion in Section 
\ref{wd_mass_neon_novae}]{liv94}.

\begin{figure}
\epsscale{1.15}
\plotone{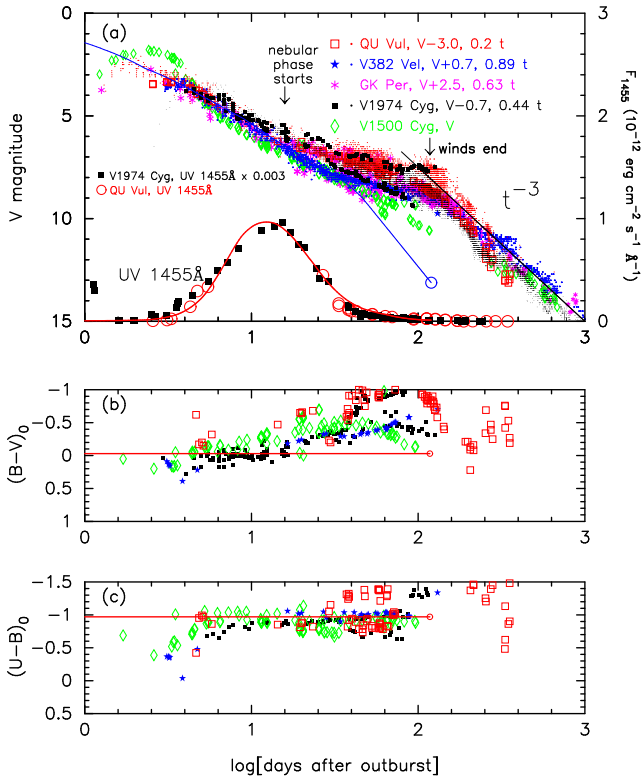}
\caption{
Five neon novae follow a universal decline law (blue solid line)
in the decay phase.
(a) $V$ band and UV 1455~\AA\  band light curves,
(b) $(B-V)_0$, and (c) $(U-B)_0$ color curves, for QU~Vul
(red open squares), V382~Vel (blue filled star marks), 
GK~Per (magenta asterisks), V1974~Cyg (black filled squares),
and V1500~Cyg (green open diamonds).
To make them overlap in the early decline phase,
we shift horizontally the logarithmic time of QU~Vul, V382~Vel, GK~Per,
and V1974~Cyg, by $-0.70=\log 0.2$, $-0.05=\log 0.89$,
$-0.20=\log 0.63$, and $-0.36=\log 0.63$, and vertically their magnitudes
by $-3.0$, $+0.7$, $+2.5$, and $-0.7$ mag, respectively, against V1500~Cyg.
UV~1455~\AA\  fluxes of V1974~Cyg are also rescaled against that of
QU~Vul as indicated in the figure.
Model light curves (blue solid line for optical, red solid line for
UV~1455~\AA) are also added for QU~Vul (a $0.96~M_\sun$ WD 
with the envelope chemical composition of Ne Nova 3 in Table 
\ref{chemical_composition_model}).
Colors of optically thick free-free emission are also added by
red solid lines, i.e., $(B-V)_0=-0.03$ in panel (b) and $(U-B)_0=-0.97$
in panel (c).  The $t^{-3}$ law (black solid line)
indicates the trend of free-free flux for a freely expanding
nebula with no mass supply.  See text for the sources of the observed data.
\label{v382_vel_v1500_cyg_v_color_logscale_no2}}
\end{figure}

\begin{figure}
\epsscale{1.15}
\plotone{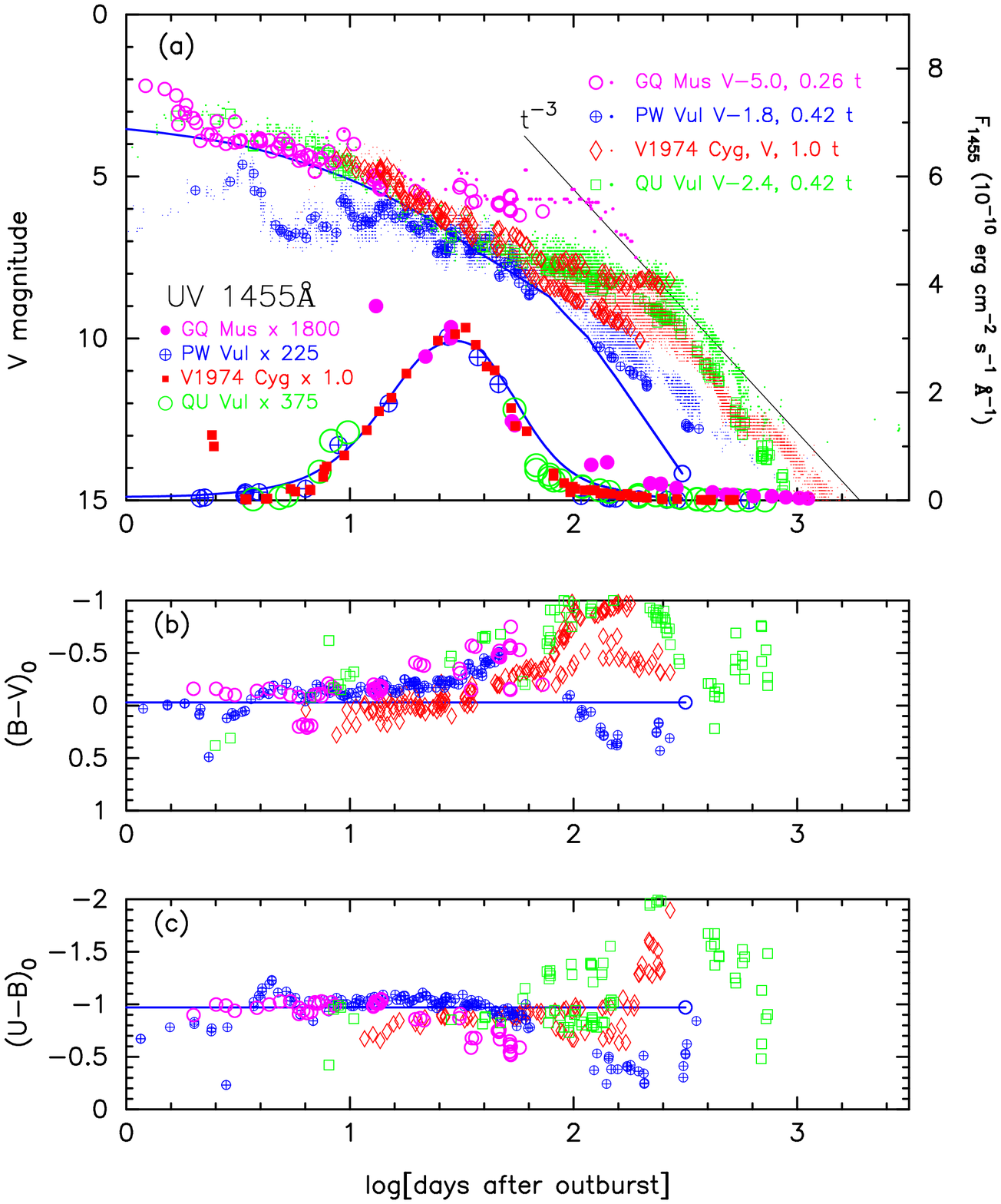}
\caption{
Four novae follow a universal decline law (blue solid line).
The UV~1455~\AA\  light curves are also overlapped each other
with the same stretching factor as that of the optical light curve. 
(a) $V$ band and UV 1455~\AA\  band light curves,
(b) $(B-V)_0$, and (c) $(U-B)_0$ color curves, for QU~Vul
(green open squares), PW~Vul (blue open circles with a plus sign inside), 
GQ~Mus (magenta open circles), and V1974~Cyg (red open diamonds).
To make them overlap in the early decline phase,
we shift horizontally their logarithmic time of QU~Vul, GQ~Mus, and PW~Vul
against V1974~Cyg, by $-0.38=\log 0.42$, $-0.585=\log 0.26$,
and $-0.38=\log 0.42$, and vertically their magnitudes
by $-2.4$, $-5.0$, and $-1.8$ mag, respectively, as indicated in the figure.
The UV~1455~\AA\  light curves are also rescaled against that of
V1974~Cyg as indicated in the figure.  We also added our free-free emission
and UV~1455\AA\  model light curves (blue solid lines) of $0.83~M_\sun$ WD
with CO Nova 4 (as the best-fit model of PW~Vul). 
See text for the sources of the observational data.
\label{qu_vul_pw_vul_gq_mus_v_bv_ub_color_logscale_no2}}
\end{figure}

The optical light curves of novae show overall similarity despite 
their wide variety of timescales and shapes \citep[e.g.,][]{pay57,
due81, str10, hac14k}.  
Optical and near infrared (NIR) spectra of some novae are consistent
with that of
free-free emission in the early decline phase \citep[e.g.,][]{gal76}.
Kato and her collaborators developed
optically thick wind theory as summarized in \citet{kat94h}, 
in which they calculated hydrogen-rich envelope models in the decay
phase of novae for various WD masses and chemical compositions.
\citet{hac06kb} calculated free-free emission model light curves
on the basis of Kato \& Hachisu's wind mass loss solutions for
various sets of WD masses and chemical compositions.  They found
a universal decline law from theoretical analysis and showed that
several well observed novae follow the law in the optical and 
infrared (IR) light curves.  We show such examples of light curve fitting
between the universal decline law (theoretical) and the observed ones
(Figures \ref{v382_vel_v1500_cyg_v_color_logscale_no2} and
\ref{qu_vul_pw_vul_gq_mus_v_bv_ub_color_logscale_no2}), which will be
described in more detail in later.
These figures suggest that there is a similarity in the light curves
both for optical and UV as well as color evolutions.
They found that the time-normalized light curves are independent of
the WD mass, chemical composition of ejecta, and wavelength.  
They also showed that the UV 1455 ~\AA\  light curves \citep{cas02},
interpreted as photospheric blackbody emission, can also be time-normalized
by the same factor as in the optical and IR
(Figures \ref{v382_vel_v1500_cyg_v_color_logscale_no2} and
\ref{qu_vul_pw_vul_gq_mus_v_bv_ub_color_logscale_no2}).
This strongly support the reliability of the universal decline law,
as it appears to be independent of the emission mechanism, 
be that free-free (optical/IR) or blackbody emission (UV).
Because the time-stretching factor is closely related to the WD mass,
the authors determined the WD masses and other parameters from
the light curve fittings for a number of relatively well-observed novae,
e.g., V1500~Cyg, V1668~Cyg, V1974~Cyg, V838~Her,
V598~Pup, V382~Vel, V4743~Sgr, V1281~Sco, V597~Pup, V1494~Aql,
V2467~Cyg, V5116~Sgr, V574~Pup, V458~Vul
\citep[see, e.g.,][]{hac07k, hac10k, hac14k, hac15k, hac08kc, kat07h, kat09hc}.

In fast novae, free-free emission dominates the spectrum in optical and
NIR bands.  The free-free emission is radiated from optically thin plasma
outside the photosphere and its flux is calculated from
$\dot M_{\rm wind}^2/v_{\rm ph}^2R_{\rm ph}$ in \citet{hac06kb},
where $\dot M_{\rm wind}$ is the wind mass-loss rate,
$v_{\rm ph}$ the photospheric velocity, and $R_{\rm ph}$ the
photospheric radius of the nova envelope.  Faster novae blow stronger
winds with larger mass-loss rates \citep{kat94h}.
So faster novae show brighter optical maxima.
In slower novae, their wind mass-loss rates
are smaller than those of fast novae, so the brightness of free-free
emission is much fainter than that of fast novae.
As a result, the brightness of free-free emission
becomes as faint as or fainter than
that of photospheric emission.  So we must take into account
photospheric emission, which is calculated from the blackbody emission
at the photosphere \citep[e.g.,][]{kat94h}.
\citet{hac15k} calculated three model light curves of free-free emission
(radiated from optically thin plasma outside the pseudophotosphere),
photospheric emission (blackbody radiation from the pseudophotosphere),
and their sum for various WD masses with various chemical compositions,
and fitted the total $V$ flux with the observed data for several slower
novae, e.g., PW~Vul, V705~Cas, GQ~Mus, V723~Cas, HR~Del, V5558~Sgr,
and RR~Pic.  Thus, in these novae, the photospheric emission cannot be
neglected compared with free-free emission to accurately determine
the WD mass and other properties of a nova.

In the present paper, we select five neon novae, QU~Vul, V351~Pup, 
V382~Vel, V693~CrA, and V1974~Cyg with a supplemental analysis of
V1668~Cyg, and analyze in detail,
including both the free-free emission and photospheric emission. 
These novae are selected from the following reasons:
(1) well observed in optical/IR; (2) chemical composition is known;
(3) {\it International Ultraviolet Explore (IUE)} UV data or
supersoft X-ray/on or X-ray/off epoch data are obtained, which
specifies the evolution timescale for different emission mechanisms
of individual novae.  We need the chemical composition of ejecta
to accurately determine the WD mass because the nova model light curves
depends not only on the WD mass but also slightly on the chemical
composition.  The chemical compositions of eleven neon novae were obtained
as listed in Table \ref{chemical_abundance_neon_novae}. 
The UV~1455\AA\  light curves are extremely useful for our light curve
analysis \citep[see, e.g.,][]{hac06kb, hac15k}.  
Five out of the eleven neon novae, i.e., QU~Vul, V351~Pup, V838~Her, 
V1974~Cyg, and V693~CrA were well observed with the {\it IUE} satellite
that provided plenty of UV~1455\AA\  light curve data \citep{cas02}.
Among them, V838~Her was already analyzed by \citet{kat09hc} and
its WD mass was estimated to be $\gtrsim1.35~M_\sun$, so we exclude V838~Her
in the present analysis.  
We include V382~Vel in our analysis because 
the end of hydrogen shell-burning of V382~Vel was detected with X-ray
\citep[e.g.,][]{nes05}, supporting our multiwavelength light curve analysis.
Although V1668~Cyg was not identified as a neon nova, 
we reanalyze V1668~Cyg because its WD mass is close to
the lower mass boundary of ONe WDs and the optical, especially 
medium-band $y$ magnitude, and {\it IUE} UV data are so rich. 

Our method for nova light curve analysis is introduced in
Section \ref{light_curve_v1668_cyg} for V1668~Cyg, which is a
carbon-oxygen (CO) nova but observed well in multiwavelength bands.
Then, we analyze the neon nova QU~Vul in Section \ref{qu_vul}.
Subsequently, we examine V351~Pup in Section \ref{v351_pup},
V382~Vel in Section \ref{v382_vel}, 
V693~CrA in Section \ref{v693_cra}, and
V1974~Cyg in Section \ref{light_curve_v1974_cyg}.
Discussion and conclusions follow in Sections \ref{discussion} and
\ref{conclusions}, respectively.
Nova evolution is discussed using the color-color (Section
\ref{color_color_diagrams}) and color-magnitude (Section
\ref{hr_diagrams}) diagrams. 
Our time-stretching method for obtaining the distance modulus of a nova
is also described in Appendix \ref{time_stretching_method_novae}.
The absolute magnitudes of the nova light curves are estimated from our
universal decline law in Appendix \ref{time_normalized_free_free}.

\begin{figure}
\epsscale{1.15}
\plotone{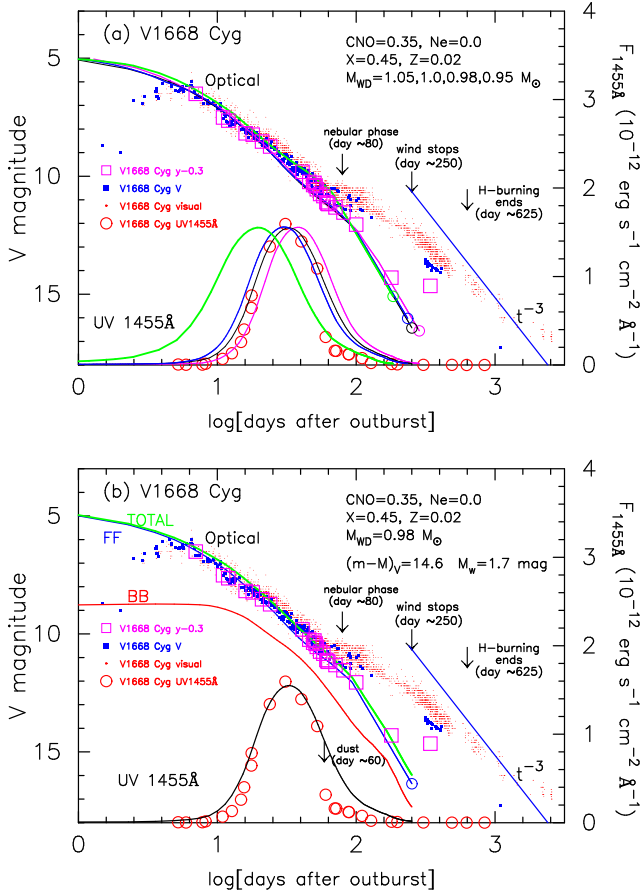}
\caption{
(a) Free-free emission and UV~1455~\AA\  model light curves
of $0.95~M_\sun$ (magenta solid lines), 
$0.98~M_\sun$ (black thin solid lines), $1.0~M_\sun$ (blue solid lines), 
and $1.05~M_\sun$ (green solid lines) WDs with CO Nova 3 as well as 
visual (small red dots), $V$ (blue filled sqaures), 
$y-0.3$ (magenta open squares), and UV~1455 ~\AA\  (red open circles)
light curves of V1668~Cyg.
(b) Assuming that $(m-M)_V=14.6$, we plot three model light curves
of the $0.98~M_\sun$ WD.  Here, we assume that $M_{\rm w}=1.7$ mag for
the free-free emission model light curve.  The green, blue, and red
solid lines show the total (labeled ``TOTAL''), free-free
(labeled ``FF''), and blackbody (labeled ``BB'') $V$ fluxes.
The black solid line denotes the UV~1455 ~\AA\  flux.
An optically thin dust shell formed $\sim60$ days after the outburst
\citep[e.g.,][]{geh80}.  
Optically thick winds and hydrogen shell-burning end approximately 
250 days and 625 days after the outburst, respectively,
for the $0.98~M_\sun$ WD model.  
The $t^{-3}$ law (blue solid line)
indicates the trend of free-free flux for a freely expanding
nebula with no mass supply.
See text for the sources of the observational data.
\label{all_mass_v1668_cyg_x45z02c15o20}}
\end{figure}


\begin{figure}
\epsscale{0.75}
\plotone{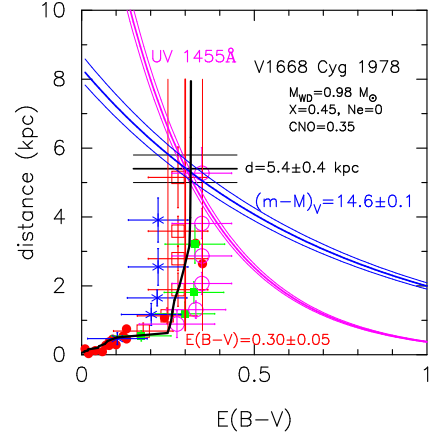}
\caption{
Distance-reddening relations toward V1668~Cyg.
The blue solid lines show the distance-reddening relations
calculated from Equation (\ref{qu_vul_distance_modulus_eq1}) with 
$(m-M)_V=14.6\pm0.1$.
The magenta solid lines show the distance-reddening relations
calculated from Equation (\ref{qu_vul_uv1455_fit_eq2}) with
the UV 1455~\AA\   flux fitting in Figure 
\ref{all_mass_v1668_cyg_x45z02c15o20}(b), i.e., for CO nova 3.
We also plot other three distance-reddening relations
toward V1668~Cyg.  One is taken from \citet{slo79} (red filled
circles) and the others are relations given by \citet{mar06}:
distance-reddening relations toward
four directions close to V1668~Cyg, $(l, b)=(90\fdg8373,-6\fdg7598)$;
that is, $(l, b)=(90\fdg75,-6\fdg75)$ (red open squares),
$(91\fdg00, -6\fdg75)$ (green filled squares),
$(90\fdg75,  -7\fdg00)$ (blue asterisks),
and $(91\fdg00,  -7\fdg00)$ (magenta open circles),
and given by \citet{gre15} (black solid line).
\label{v1668_cyg_distance_reddening_x45z02c15o20}}
\end{figure}


\begin{figure}
\epsscale{1.15}
\plotone{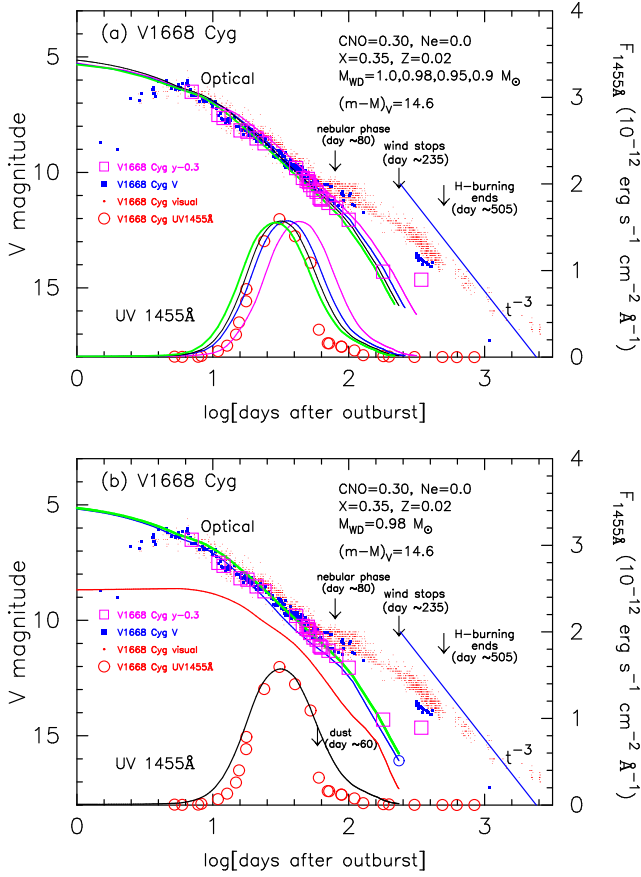}
\caption{
Similar to Figure \ref{all_mass_v1668_cyg_x45z02c15o20}, but for CO Nova 2.
(a) Assuming that $(m-M)_V=14.6$, we plot total $V$ and UV~1455~\AA\ 
model light curves of $0.9~M_\sun$ (magenta solid lines), 
$0.95~M_\sun$ (blue solid lines), $0.98~M_\sun$ (black thin solid lines), 
and $1.0~M_\sun$ (green solid lines) WDs. 
(b) Assuming also that $(m-M)_V=14.6$, we plot three (total, free-free,
blackbody) model light curves
of the $0.98~M_\sun$ WD.  Optically thick winds and hydrogen shell-burning
end approximately 235 days and 505 days after the outburst, respectively,
for the $0.98~M_\sun$ WD.
\label{all_mass_v1668_cyg_x35z02c10o20}}
\end{figure}


\begin{figure}
\epsscale{1.15}
\plotone{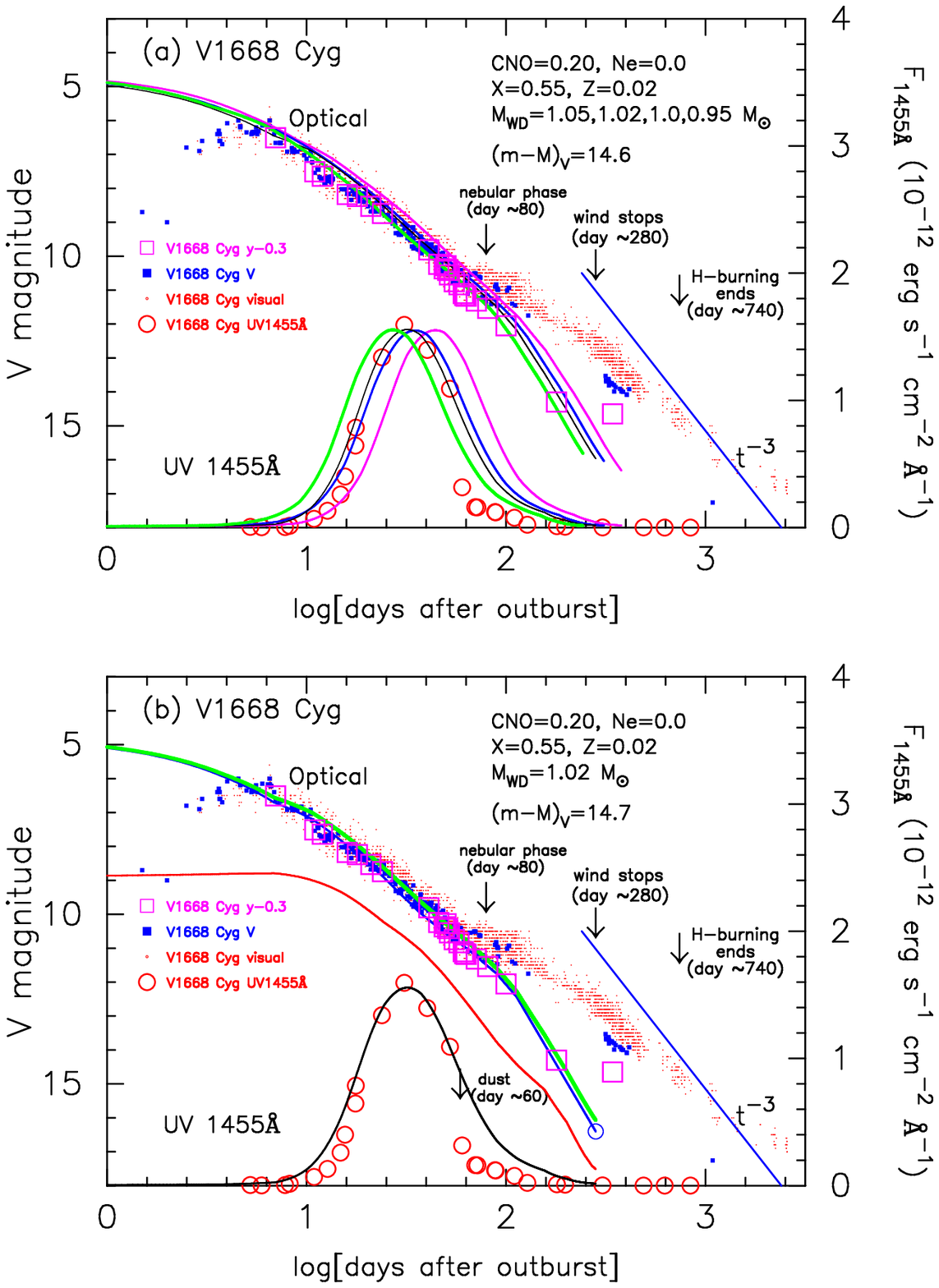}
\caption{
Same as Figure \ref{all_mass_v1668_cyg_x35z02c10o20}, but for
CO Nova 4. 
(a) Assuming that $(m-M)_V=14.6$, we plot model light curves of 
$0.95~M_\sun$ (magenta solid lines), $1.0~M_\sun$ (blue solid lines), 
$1.02~M_\sun$ (black thin solid lines),
and $1.05~M_\sun$ (green solid lines) WDs.
(b) Assuming that $(m-M)_V=14.7$, we plot three model light curves
of the $1.02~M_\sun$ WD.  Optically thick winds and hydrogen shell-burning
end approximately 280 days and 740 days after the outburst, respectively,
for the $1.02~M_\sun$ WD.
\label{all_mass_v1668_cyg_x55z02c10o10}}
\end{figure}


\begin{figure}
\epsscale{1.15}
\plotone{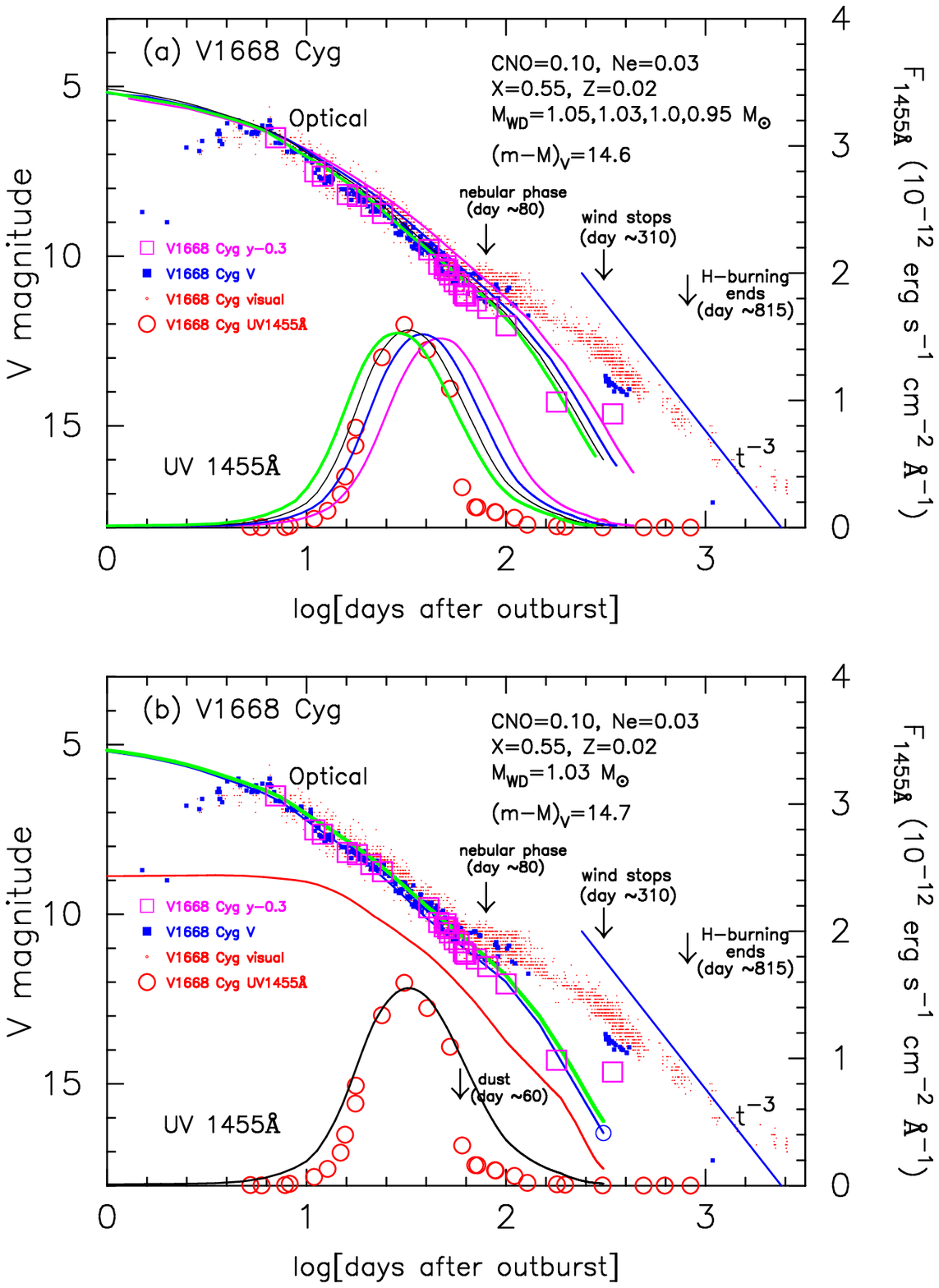}
\caption{
Same as Figure \ref{all_mass_v1668_cyg_x35z02c10o20}, but for
Ne Nova 2.  (a) Assuming that $(m-M)_V=14.6$, we plot model light curves
of $0.95~M_\sun$ (magenta solid lines), $1.0~M_\sun$ (blue solid lines), 
$1.03~M_\sun$ (black thin solid lines), and $1.05~M_\sun$ (green solid lines)
WDs.  (b) Assuming that $(m-M)_V=14.7$, we plot three model light curves
of the $1.03~M_\sun$ WD.  Optically thick winds and hydrogen shell-burning
end approximately 310 days and 815 days after the outburst, respectively,
for the $1.03~M_\sun$ WD.
\label{all_mass_v1668_cyg_x55z02o10ne03}}
\end{figure}


\begin{figure}
\epsscale{1.15}
\plotone{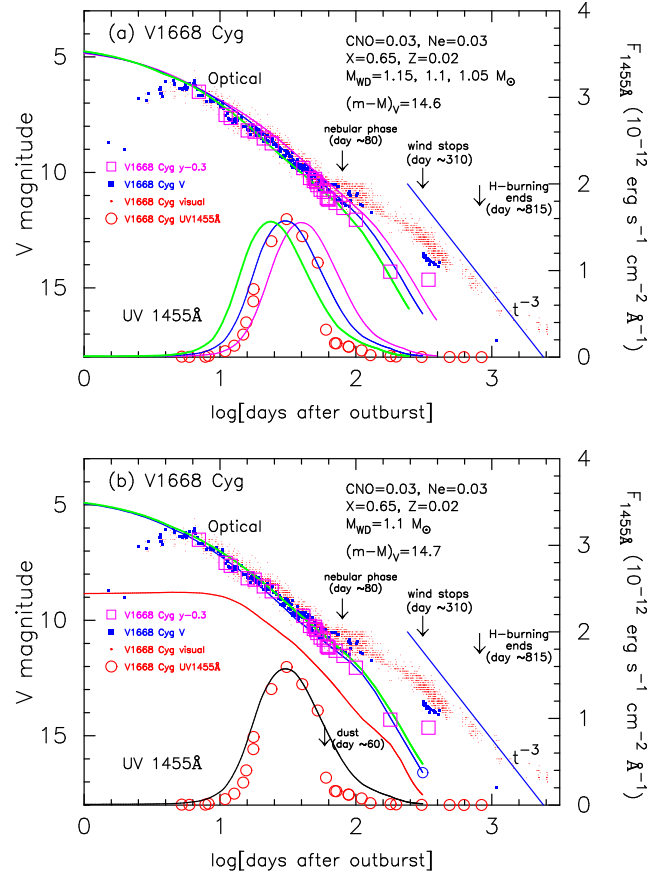}
\caption{
Same as Figure \ref{all_mass_v1668_cyg_x35z02c10o20}, but for
Ne Nova 3.  (a) Assuming that $(m-M)_V=14.6$, we plot model light curves of
$1.05~M_\sun$ (magenta solid lines), $1.1~M_\sun$ (blue solid lines),
and $1.15~M_\sun$ (green solid lines) WDs.
(b) Assuming that $(m-M)_V=14.7$, we plot three model light curves of
the $1.1~M_\sun$ WD.  Optically thick winds and hydrogen shell-burning
end approximately 310 days and 815 days after the outburst, respectively,
for the $1.1~M_\sun$ WD.
\label{all_mass_v1668_cyg_x65z02o03ne03}}
\end{figure}


\begin{figure}
\epsscale{1.15}
\plotone{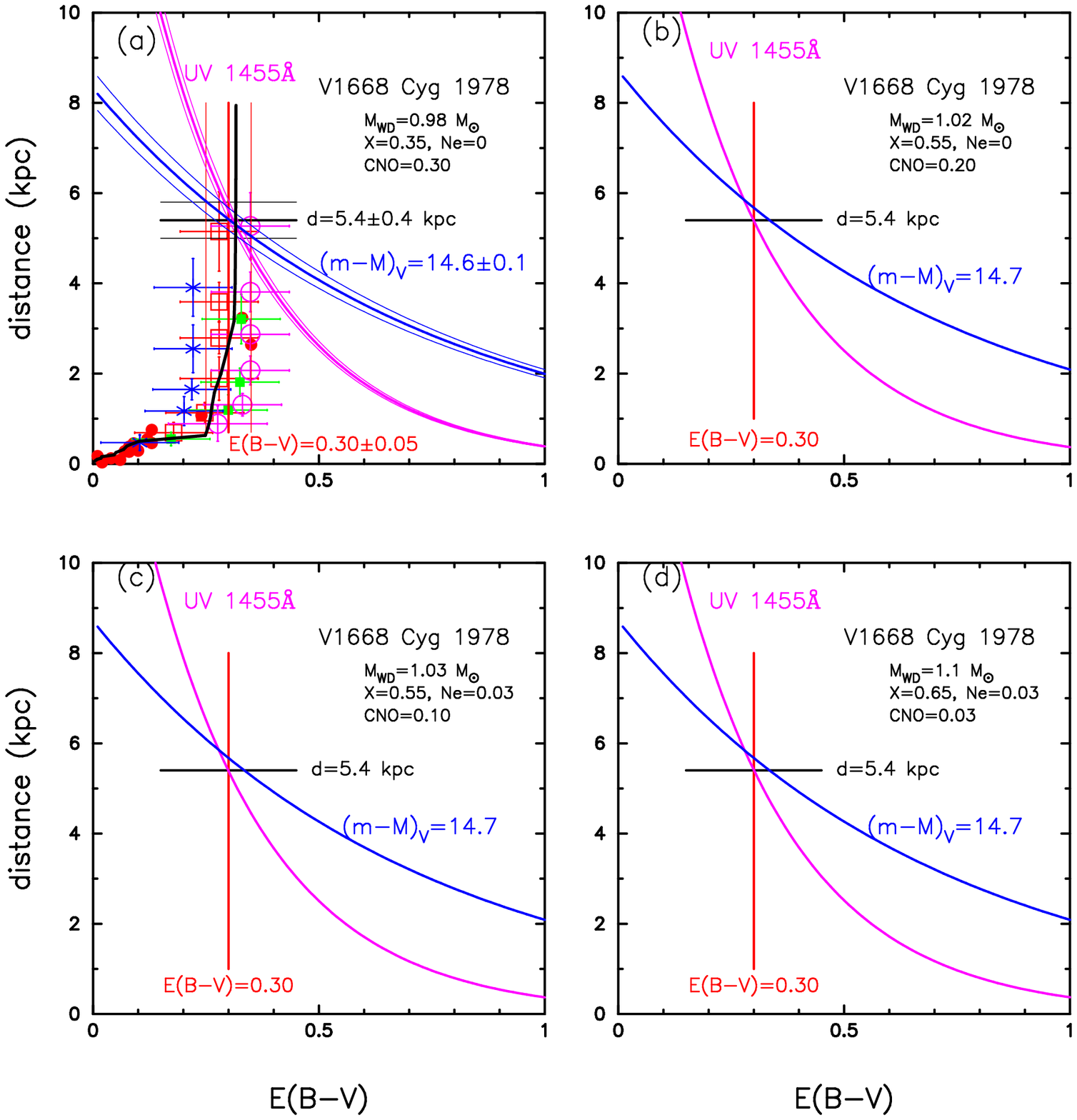}
\caption{
Same as Figure \ref{v1668_cyg_distance_reddening_x45z02c15o20}, but
for (a) CO nova 2 in Figure \ref{all_mass_v1668_cyg_x35z02c10o20},
(b) CO nova 4 in Figure \ref{all_mass_v1668_cyg_x55z02c10o10},
(c) Ne nova 2 in Figure \ref{all_mass_v1668_cyg_x55z02o10ne03},
and (d) Ne nova 3 in Figure \ref{all_mass_v1668_cyg_x65z02o03ne03}.
Symbols and lines are the same as those in Figure 
\ref{v1668_cyg_distance_reddening_x45z02c15o20}.
\label{v1668_cyg_distance_reddening_x35_x55_x55_x65_4figure}}
\end{figure}

\section{Numerical Method and Reanalysis of V1668~Cyg}
\label{light_curve_v1668_cyg}
V1668~Cyg was well observed with {\it IUE} and the chemical composition
of ejecta was obtained as listed in Table \ref{chemical_abundance_neon_novae}.
\citet{hac06kb} calculated free-free emission model light curves for
various WD masses and chemical compositions and fitted
their free-free emission model light curves with the $V$ and $y$
light curves of V1668~Cyg.  They further 
fitted their blackbody emission UV~1455\AA\  model light curves
with the {\it IUE} UV~1455\AA\   observation, and estimated the WD mass.
In this section, we reanalyze the light curves of V1668~Cyg
including contribution of photospheric emission, which is not
considered in the previous analysis of V1668~Cyg.
V1668~Cyg is a CO nova but its WD mass is close to
the lower mass bound of natal pure ONe WD.
We will discuss its implication on binary evolution
in Section \ref{wd_mass_neon_novae}.

It has been argued that nova envelopes reach steady-state 
after optical maximum.  Using this assumption, 
\citet{kat94h} calculated evolution of nova envelopes
for various WD masses and several sets of chemical compositions.
They numerically obtained optically thick winds of nova envelopes.
The wind mass loss rate, $\dot M_{\rm wind}$, is obtained 
as an eigen-value of the steady-state envelope solution, whose
hydrogen-rich envelope mass is $M_{\rm env}$.
The time evolution of a hydrogen-rich nova envelope is calculated
from $\dot M_{\rm env}= \dot M_{\rm acc} - \dot M_{\rm wind} 
- \dot M_{\rm nuc}$, where $\dot M_{\rm env}$ is the mass decreasing
rate of the hydrogen-rich envelope, $\dot M_{\rm acc}$
the mass accretion rate to the WD, $\dot M_{\rm wind}$ the wind mass loss
rate, and $\dot M_{\rm nuc}\equiv L_{\rm n}/X Q$ the mass decreasing rate
by nuclear burning, $L_{\rm n}$ the hydrogen nuclear burning luminosity,
$X$ the hydrogen content, and $Q=6.4\times10^{18}$~erg~s$^{-1}$ the
energy generation per unit mass by hydrogen burning. 
\citet{kat94h} obtained a sequence of envelope solutions, in which
various physical values such as $\dot M_{\rm wind}$ and $\dot M_{\rm nuc}$
are given as a function of the envelope mass $M_{\rm env}$.
If the initial envelope mass $M_{\rm env,0}$ is given, we follow
the evolution of the hydrogen-rich envelope mass by integrating
$\dot M_{\rm env}= \dot M_{\rm acc} - \dot M_{\rm wind} 
- \dot M_{\rm nuc}$,  and obtain the time
evolutions of various physical values such as $\dot M_{\rm wind}$, the
photospheric radius $R_{\rm ph}$, photospheric temperature $T_{\rm ph}$, 
and photospheric velocity $v_{\rm ph}$.  The mass accretion rate
is usually very small compared with the wind mass loss rate or
the nuclear burning rate, i.e., $\dot M_{\rm acc} \ll \dot M_{\rm wind} 
+ \dot M_{\rm nuc}$ in classical novae, so we set $\dot M_{\rm acc}=0$
in our model light curves. 
Because Kato \& Hachisu's solutions are applied only
to a steady-state envelope, that is, in the decay phase of a nova after 
optical maximum, we use our model light curves only for the decay
phase of novae.

The optical $V$ flux is calculated from the sum of
the free-free and photospheric blackbody emission.
The blackbody emission was described in \citet{kat94h}, and
the free-free emission was calculated from 
$F_\nu\propto\dot M_{\rm wind}^2/v_{\rm ph}^2R_{\rm ph}$ in \citet{hac06kb}
(see also Equation (\ref{free-free-wind}) in
Appendix \ref{time_normalized_free_free}).
The UV~1455~\AA\  band flux is useful for determining the WD mass.
This is a narrow-band (1445--1465~\AA) flux that represents well
the continuum fluxes of novae \citep{cas02}.
The flux in the UV~1455~\AA\  band was calculated from the blackbody
emission at the pseudophotosphere of the optically thick
wind solutions of \citet{kat94h} and is presented in
\citet{hac06kb} for various WD masses and chemical compositions.

\subsection{Light curve fitting of V1668~Cyg}
\label{light_curve_fit_v1668_cyg}
Figure \ref{all_mass_v1668_cyg_x45z02c15o20}
show the optical and UV light curves of V1668~Cyg and our model light curves.
The visual magnitudes are taken from the archive of the American
Association of Variable Star Observers (AAVSO) 
and those of $V$ magnitudes are from \citet{dip81}, \citet{pic84},
\citet{hop79}, \citet{kol80}, \citet{mal79}, and \citet{due80}.
The data of the $y$ magnitudes are from \citet{gal80}.
The UV~1455~\AA\  band, a 20\AA\  width at the center of
1455\AA, was defined by \citet{cas02} as one of the two
UV narrow bands that can represent the UV continuum flux.  The UV~1455~\AA\ 
fluxes are taken from \citet{cas02}.
An optically thin dust shell formed $\sim60$ days after the outburst
\citep[e.g.,][]{geh80}, so we observed a drop of the UV~1455\AA\  
flux after this epoch as indicated by the downward arrow
in Figure \ref{all_mass_v1668_cyg_x45z02c15o20}(b)
\citep[see also][]{hac06kb}.  

Table \ref{chemical_composition_model} shows our model composition
designed for various classical novae \citep[see, e.g.,][]{hac06kb}.
In Table \ref{chemical_composition_model},
$X$ is the mass fraction of hydrogen, $Y$ the mass fraction of
helium, $X_{\rm CNO}$ and $X_{\rm Ne}$ are the mass
fractions of carbon--nitrogen--oxygen and neon, respectively,
and $Z= 0.02$ is the mass fraction of heavy elements with the same
ratio as the solar abundance.
Model light curves of free-free emission and UV~1455~\AA\  were
already published in \citet{hac06kb} for these sets of chemical
composition.  The flux of free-free emission is calculated from
$F_\nu \propto \dot M_{\rm wind}^2/v_{\rm ph}^2R_{\rm ph}$
(or $F_\nu = C[ \dot M_{\rm wind}^2/v_{\rm ph}^2R_{\rm ph}]$)
in \citet{hac06kb},
where $\dot M_{\rm wind}$ is the wind mass-loss rate,
$v_{\rm ph}$ the photospheric velocity, and $R_{\rm ph}$ the
photospheric radius of the nova envelope, and are obtained from
the optically thick wind solutions of novae \citep{kat94h}.
We must know the proportionality constant $C$ to evaluate the
absolute magnitude of free-free emission light curve.
This was done by comparing the model light curve with a nova
whose distance modulus is known
(see also Appendix \ref{time_normalized_free_free}).
The absolute magnitudes of the free-free emission model light curves
were published for ``CO nova 2'' and ``Ne nova 2'' in \citet{hac10k} 
and for ``CO nova 4'' in \citet{hac15k}.   
For the other chemical compositions, the absolute magnitudes of
the free-free emission model light curves are calibrated in the present
paper (see Table \ref{chemical_composition_model}).

For each set of chemical composition,
we calculate WD mass models in coarser steps of
$0.05~M_\sun$ and then in finer steps of $0.01~M_\sun$,
and select one by eye as the best-fit WD mass model among them.
The light curve flux fit in the vertical direction is usually done
by the least square method with the already given best-fit WD mass model,
but sometimes by eye if the least square fit is not good.

First we adopt CO nova 3 as the chemical
composition of V1668~Cyg, because it is close to the observational
estimate of chemical compositions 
(see Table \ref{chemical_abundance_neon_novae}).
We plot our free-free and UV~1455~\AA\   model
light curves in Figure \ref{all_mass_v1668_cyg_x45z02c15o20}(a).
The free-free emission model light curves are calculated from
$F_\nu\propto\dot M_{\rm wind}^2/v_{\rm ph}^2R_{\rm ph}$ (see Equation
(\ref{free-free-wind}) in Appendix \ref{time_normalized_free_free}).
Because we have not yet calibrated the absolute magnitude of free-free
emission light curves for the chemical composition of CO nova 3,
we vertically move the free-free emission model light curves as it lies
just on the $V$ observation.  The shapes of free-free emission model
light curves follow the observed $V$ and especially $y$ magnitudes.
However, we cannot select the best-fit one only from fitting with 
the $V$ or $y$ observation because the shapes of model light curves
are very similar to each other among the 0.95, 0.98, 1.0, and 
1.05$~M_\sun$ WDs.  So, we use the UV~1455\AA\  flux.    
The UV~1455\AA\  flux is calculated from the blackbody emission
at the pseudophotosphere \citep[see][]{kat94h, hac06kb}.
Figure \ref{all_mass_v1668_cyg_x45z02c15o20}(a) shows that
the $0.98~M_\sun$ model reproduces well the UV~1455~\AA\  light curve.
The accuracy of our mass determination depends on how we select
the UV~1455\AA\  data points involved in our fitting.
In Figure \ref{all_mass_v1668_cyg_x45z02c15o20}(a), we omit the 
UV~1455\AA\  data later than 60 days after the outburst because the UV flux
could be cut by dust.  We estimated that the accuracy is as good as
$0.01~M_\sun$, that is, $M_{\rm WD}=0.98\pm0.01~M_\sun$.
This means that we accept $M_{\rm WD}=0.99~M_\sun$ but do not
$M_{\rm WD}=1.0~M_\sun$ from fitting in the figure.
Our flux fitting of UV~1455\AA\  in the vertical direction
typically has a $\sim10$\% accuracy
because the UV~1455\AA\  observed fluxes typically have a $\sim10$\%
error.

We must calibrate the absolute magnitude of free-free emission 
model light curve for CO nova 3.
We estimated the distance modulus in the $V$ band to V1668~Cyg as 
$\mu_V=(m-M)_V=14.6$ in Appendix \ref{time_stretching_method_novae}
based on the time-stretching method. 
Assuming a trial value for the proportionality constant $C$ in Equation
(\ref{free-free_calculation_original}) of Appendix
\ref{free-free_emission_light_curve}, we obtain the absolute
magnitude of the free-free model light curve (blue solid line
labeled ``FF'')  of the $0.98~M_\sun$ WD
in Figure \ref{all_mass_v1668_cyg_x45z02c15o20}(b).
We also calculate the blackbody light curve model of the $0.98~M_\sun$ WD
in the $V$ band (red solid line labeled ``BB'') in Figure 
\ref{all_mass_v1668_cyg_x45z02c15o20}(b).
The total absolute flux (green solid line labeled ``TOTAL'')
is the sum of these two fluxes.  However, this total flux
generally does not fit the observed data well.
We assume a different value of $C$ until it reproduces 
the observed $V$ flux.
Figure \ref{all_mass_v1668_cyg_x45z02c15o20}(b)
shows our final flux fit.

  We directly read $m_{\rm w}=16.3$
from Figure \ref{all_mass_v1668_cyg_x45z02c15o20}(b),
where $m_{\rm w}$ is the apparent magnitude at the end of the wind phase
(open circle at the end of the blue solid line labeled ``FF'').
Then, we obtain $M_{\rm w}= m_{\rm w} - (m-M)_V= 16.3 - 14.6=1.7$,
where $M_{\rm w}$ is the absolute magnitude of the free-free model
light curve at the end of the wind phase.
Thus, we can specify the proportionality constant by $M_{\rm w}=1.7$
for the $0.98~M_\sun$ WD model of V1668~Cyg.  In other words,
this value fixes the proportionality constant
$C$ in Equation (\ref{free-free_calculation_original}) of
Appendix \ref{free-free_emission_light_curve}.
Then, we obtain $M_{\rm w}$ for
the other WD mass models from Equation (\ref{real_timescale_flux}).
The absolute magnitudes are specified by the value of 
$M_{\rm w}$ and are listed in Table \ref{light_curves_of_novae_co3}
for $0.55$--$1.2~M_\sun$ WDs in steps of $0.05~M_\sun$.
Once the calibrated free-free model light curves are given
by the absolute magnitude of $M_{\rm w}$ at the end point of the
wind phase, we calculate the absolute magnitudes of the
total $V$ flux (free-free plus blackbody emission). 

In Figure \ref{all_mass_v1668_cyg_x45z02c15o20}(b),
optically thick winds and hydrogen shell burning end approximately
250 and 625 days after the outburst, respectively.
The total $V$ light curve model agrees well
the $V$ magnitude observation and the shape of UV~1455~\AA\  light curve
model fits the observation well.
After the wind stops, the ejecta mass $M_{\rm ej}$ is constant in time
because of no mass supply from the WD.  We estimate the free-free flux by
assuming that the ejecta expand homologously.  
Then the free-free emission light curve changes
its decline shape as expressed by Equation (\ref{free-free-wind_wind_stop}),
i.e., $\propto t^{-3}$.  We also plot this trend by the black solid line
labeled ``$t^{-3}$'' in Figure \ref{all_mass_v1668_cyg_x45z02c15o20}.

We should note that the strong emission lines such as [\ion{O}{3}] 
significantly contribute to the $V$ and visual magnitudes in the
nebular phase.  V1668~Cyg entered the nebular phase at $m_V\approx10.4$,
about 80 days after the outburst \citep[e.g.,][]{kla80}.
Our model light curves (free-free plus blackbody) 
begins to deviate from the observed $V$ and visual
magnitude as shown in Figure \ref{all_mass_v1668_cyg_x45z02c15o20}
\citep[see also Figure 11 of][for a schematic illustration]{hac06kb}. 
The intermediate $y$ band (magenta open squares),
which is designed to avoid such strong emission lines, 
reasonably follows our model light curve which represents
the continuum spectrum.

\subsection{Distance and reddening toward V1668~Cyg}
\label{distance_reddening_v1668_cyg}
We obtain two distance-reddening relations toward V1668~Cyg from
the two model light curve fittings ($V$ and UV~1455~\AA) of
CO nova 3,  which are shown in
Figure \ref{v1668_cyg_distance_reddening_x45z02c15o20}.
The distance modulus in the $V$ band, $\mu_V = (m-M)_V$,
is obtained using the time-stretching method \citep{hac10k}
for nova light curves, which yielded a value of 
$(m-M)_{V,{\rm V1668~Cyg}}=14.6\pm0.1$
(see Appendix \ref{time_stretching_method_novae}). 
Adopting $(m-M)_V=14.6\pm0.1$ for V1668~Cyg, 
we obtain a distance-reddening relation of
\begin{equation}
(m-M)_V = 5 \log \left({d \over{\rm 10~pc}}\right) + 3.1 E(B-V),
\label{qu_vul_distance_modulus_eq1}
\end{equation}
and plot this in Figure \ref{v1668_cyg_distance_reddening_x45z02c15o20}
by a thick blue line flanked by thin blue lines.

The UV~1455~\AA\  light curve fitting gives another
distance-reddening relation of
\begin{eqnarray}
& & 2.5 \log F_{1455}^{\rm mod} 
- 2.5 \log F_{1455}^{\rm obs} \cr
&=& 5 \log \left({d \over {10\mbox{~kpc}}} \right)  + 8.3 \times E(B-V).
\label{qu_vul_uv1455_fit_eq2}
\end{eqnarray}
The magenta solid lines show Equation (\ref{qu_vul_uv1455_fit_eq2})
with $F_{1455}^{\rm mod}= 11.8 \times 
10^{-12}$~erg~cm$^{-2}$~s$^{-1}$~~\AA$^{-1}$ as the calculated model
flux of the upper bound of Figure \ref{all_mass_v1668_cyg_x45z02c15o20}(b)
at the distance of 10~kpc and $F_{1455}^{\rm obs}= (4.0\pm0.3) \times 
10^{-12}$~erg~cm$^{-2}$~s$^{-1}$~~\AA$^{-1}$ as the observed flux 
corresponding to that of the upper bound in 
Figure \ref{all_mass_v1668_cyg_x45z02c15o20}(b).
Here we assume an absorption of
$A_\lambda= 8.3 \times E(B-V)$ at $\lambda= 1455$~\AA\   \citep{sea79}.

We further plot other distance-reddening relations toward V1668~Cyg
in Figure \ref{v1668_cyg_distance_reddening_x45z02c15o20}:
that given by \citet{slo79} (large red filled circles),
that given by \citet{mar06}, and that given by \citet{gre15}.
The galactic distance-reddening relation was presented by \citet{mar06}
with a grid of $\Delta l =0\fdg25$ and $\Delta b=0\fdg25$ for
$-100\fdg0 \le l \le 100\fdg0$ and $-10\fdg0 \le b \le 10\fdg0$,
respectively.
The galactic coordinates of V1668~Cyg are $(l, b)=(90\fdg8373,-6\fdg7598)$.
Here we plot four nearby directions using data from \citet{mar06}:
$(l, b)= (90\fdg75,-6\fdg75)$ (red open squares),
$(91\fdg00,-6\fdg75)$ (green filled squares), 
$(90\fdg75,-7\fdg00)$ (blue asterisks), and
$(91\fdg00,-7\fdg00)$ (magenta open circles).
\citet{gre15} published the data of the galactic extinction map
which covers a much wider range of the galactic
coordinates (over three quarters of the sky) with much more finer
grids of 3\farcm4 to 13\farcm7 and a maximum distance resolution of 25\%.
We added Green et al.'s distance-reddening line (best fitted one
of their examples) by the black thick solid line in
Figure \ref{v1668_cyg_distance_reddening_x45z02c15o20}.

All these trends/lines cross at $d\sim 5.4$~kpc and
$E(B-V)\sim 0.30$.  Therefore, we conclude that 
$d\sim 5.4\pm0.4$~kpc and $E(B-V)\sim 0.30\pm0.05$ for V1668~Cyg.
This is consistent with the galactic dust absorption map
of $E(B-V)=0.29 \pm 0.02$ in the direction toward V1668~Cyg
at the NASA/IPAC Infrared Science
Archive\footnote{http://irsa.ipac.caltech.edu/applications/DUST/},
which is calculated on the basis of data from \citet{schl11}.
\citet{hac14k} proposed a new method of determining the 
reddening of classical novae.  They identified a general course of
dereddened $UBV$ color-color evolution and determined the reddening
for a number of novae by comparing the observed track of a target nova
with the general course (see Section \ref{color_color_diagrams} below).
Their value is $E(B-V)\sim 0.35\pm0.05$
for V1668~Cyg, being also consistent with our new value of
$E(B-V)\sim 0.30\pm0.05$.

\subsection{Light curve fitting for other various chemical compositions}
For the other chemical compositions, we plot the $V$ and UV~1455~\AA\  
model light curves as shown in Figures
\ref{all_mass_v1668_cyg_x35z02c10o20} -- 
\ref{all_mass_v1668_cyg_x65z02o03ne03}.
Here we include two sets of chemical composition of neon novae 
to see the dependence of neon abundance.  Our theoretical
light curves are insensitive to the neon content. 
The wind acceleration occurs mainly near the iron-peak of
OPAL opacity ($\log T ({\rm K}) \sim 5.2$), which hardly changes
if we replace neon by oxygen.  On the other hand, nuclear burning
is dominated by the CNO cycle, which depends on the hydrogen mass
fraction $X$ and the CNO mass fraction $X_{\rm CNO}$,
but not on the neon mass fraction $X_{\rm Ne}$.   
Thus, the nova model light curves depends not only on the WD mass
but also slightly on the chemical composition of $X$ and $X_{\rm CNO}$.
We must know the chemical composition of hydrogen-rich envelope
to accurately determine the WD mass.

Using the absolute magnitude of model light curves in Figures 
\ref{all_mass_v1668_cyg_x35z02c10o20} -- 
\ref{all_mass_v1668_cyg_x65z02o03ne03}, 
we calculated the distance modulus in the $V$
band, $\mu_V=(m-M)_V$, for each best-fit model.
We first determine the WD mass from the UV~1455\AA\  light curve fitting
as shown in Figures \ref{all_mass_v1668_cyg_x35z02c10o20}(a) -- 
\ref{all_mass_v1668_cyg_x65z02o03ne03}(a)
and, for this WD mass model, 
we calculate the absolute $V$ magnitude of the total $V$ flux 
(free-free plus blackbody) from the absolute magnitude of $M_{\rm w}$
in tables of \citet{hac10k}, \citet{hac15k}, and the present paper
as shown in Figures
\ref{all_mass_v1668_cyg_x35z02c10o20}(b) -- 
\ref{all_mass_v1668_cyg_x65z02o03ne03}(b).
Then, we obtain the distance modulus in the $V$ band,
$(m-M)_V$, by fitting our total $V$ model light curve
with the observed $V$ magnitudes.
Thus, we chose the $M_{\rm WD}=0.98~M_\sun$ and
$(m-M)_V=14.6$ for CO nova 2
in Figure \ref{all_mass_v1668_cyg_x35z02c10o20}(b),
$M_{\rm WD}=1.02~M_\sun$ and $(m-M)_V=14.7$ for CO nova 4
in Figure \ref{all_mass_v1668_cyg_x55z02c10o10}(b),
$M_{\rm WD}=1.03~M_\sun$ and $(m-M)_V=14.7$ for Ne nova 2
in Figure \ref{all_mass_v1668_cyg_x55z02o10ne03}(b),
and $M_{\rm WD}=1.1~M_\sun$ and $(m-M)_V=14.7$ for Ne nova 3
in Figure \ref{all_mass_v1668_cyg_x65z02o03ne03}(b).
Optically thick winds and hydrogen shell burning end approximately
235 and 505 days after the outburst, respectively,
for the $0.98~M_\sun$ WD with CO nova 2,  
280 days and 740 days, respectively, for the $1.02~M_\sun$ WD with
CO nova 4,  310 days and 815 days, respectively,
for the $1.03~M_\sun$ WD with Ne nova 2, and 310 days 815 days, respectively,
for the $1.1~M_\sun$ WD with Ne nova 3.

Note that we obtain the best fit at $(m-M)_V=14.7$
for the $1.02~M_\sun$ WD with CO nova 4,  and for the $1.03~M_\sun$ WD with
Ne nova 2, and for the $1.1~M_\sun$ WD with Ne nova 3, which is
slightly brighter than at $(m-M)_V=14.6$ of CO nova 2.
All the total $V$ light curve models agree well with 
the $V$ observation and all the shapes of UV~1455~\AA\  light curve
models fit the observation well.
The photospheric emission contributes slightly to the total $V$
light curve and it is $\sim 0.2$ -- $0.4$ mag at most.

For these chemical compositions,
we also obtained two distance-reddening relations from
the model light curve fittings of $V$ and UV~1455~\AA, which are shown in
Figure \ref{v1668_cyg_distance_reddening_x35_x55_x55_x65_4figure}.
Figure \ref{v1668_cyg_distance_reddening_x35_x55_x55_x65_4figure}(a)
is for the two distance-reddening relations for CO nova 2.
The blue solid lines show the distance-reddening relation of Equation 
(\ref{qu_vul_distance_modulus_eq1}) with
$(m-M)_V=14.6\pm 0.1$.  The magenta solid lines show
the distance-reddening relation of Equation (\ref{qu_vul_uv1455_fit_eq2})
with $F_{1455}^{\rm mod}= 11.3 \times 
10^{-12}$~erg~cm$^{-2}$~s$^{-1}$~~\AA$^{-1}$ and
$F_{1455}^{\rm obs}= (4.0\pm0.3) \times 
10^{-12}$~erg~cm$^{-2}$~s$^{-1}$~~\AA$^{-1}$.
These trends/lines cross approximately at $d\sim 5.4$~kpc and
$E(B-V)\sim 0.30$, which are consistent with
the conclusion that $d\sim 5.4\pm0.4$~kpc and $E(B-V)\sim 0.30\pm0.05$
obtained for CO nova 3 in Section \ref{distance_reddening_v1668_cyg}.

The distance-reddening relations for CO nova 4, Ne nova 2, 
and Ne nova 3 are plotted in  
Figures \ref{v1668_cyg_distance_reddening_x35_x55_x55_x65_4figure}(b),
\ref{v1668_cyg_distance_reddening_x35_x55_x55_x65_4figure}(c), and
\ref{v1668_cyg_distance_reddening_x35_x55_x55_x65_4figure}(d).
All the cross points of the two lines deviate slightly from 
$d=5.4$~kpc and $E(B-V)=0.30$, but within $d=5.4\pm0.4$~kpc
and $E(B-V)=0.30\pm0.05$.

To summarize, our model light curve fittings give a reasonable set of 
solutions for
the distance and reddening toward V1668~Cyg even if the chemical
composition is not so accurately constrained.  Therefore, we safely
conclude that $(m-M)_V=14.6\pm0.1$, $d=5.4\pm0.4$~kpc,
and $E(B-V)=0.30\pm0.05$.  If we adopt CO nova 3, then we obtain
the WD mass of $M_{\rm WD}=0.98~M_\sun$ and the hydrogen-rich envelope
mass of $M_{\rm env}=1.38\times10^{-5}~M_\sun$
at optical maximum (see Table \ref{nova_parameters_results}).
We suppose that the envelope mass at the optical maximum approximately
represents the ignition mass of the nova.

\subsection{Comparison with previous results}
\label{comparison_previous_v1668_cyg}
The timescale of a nova evolution is governed by the wind mass loss rate
because the decreasing rate of the hydrogen-rich envelope mass is
determined mainly by the wind mass loss rate.  \citet{kat94h} 
calculated optically thick wind solutions as the structure of
hydrogen-rich envelope of novae.  Then, they applied their model
to light curve analysis of novae.
The WD mass of V1668~Cyg was estimated by \citet{kat94h}
to be $0.9$--$1.0~M_\sun$ from optical and UV light curve
fittings based on their optically thick wind solutions and
blackbody photospheric emission models.
Their result is not so different from our result ($0.98~M_\sun$)
partly because the timescale of nova evolution is the same.
However, they adopted the blackbody emission flux for optical
light curve, which is too faint to reproduce the optical 
light curves of V1668~Cyg (labeled ``BB'') as shown in Figure
\ref{all_mass_v1668_cyg_x45z02c15o20}(b).
As a result, their distance estimate is 2.87~kpc and it is too short
to be compared with our new estimate of $d=5.4$~kpc.  

\citet{hac06kb} proposed the free-free emission model for optical
light curves and obtained the WD mass of $0.95~M_\sun$ based mainly
on the UV~1455\AA\  light curve fitting.  They could not fix 
the proportionality constant $C$ in their models and did not
estimate the distance modulus.  Using the time-stretching method
of the universal decline law, \citet{hac10k} obtained the absolute
magnitude of their free-free emission model light curves and
applied their light curve model to V1668~Cyg and obtained 
the distance modulus in the $V$ band as $(m-M)_V=14.25$
and the WD mass of $0.95~M_\sun$.  Including the effect of
photospheric emission as well as free-free emission, 
we reanalyzed the optical light curve of V1668~Cyg and
now fix $M_{\rm WD}=0.98~M_\sun$ and $(m-M)_V=14.6$ for the
chemical composition of CO nova 3.


\begin{deluxetable*}{llllll}
\tabletypesize{\scriptsize}
\tablecaption{Chemical abundances of selected neon novae
\label{chemical_abundance_neon_novae}}
\tablewidth{0pt}
\tablehead{
\colhead{object} & 
\colhead{H} & 
\colhead{CNO} & 
\colhead{Ne} &
\colhead{Na-Fe} &
\colhead{reference} \\ 
\colhead{} & 
\colhead{} & 
\colhead{} & 
\colhead{} & 
\colhead{} & 
\colhead{}  
} 
\startdata
Sun (solar) &  0.71 & 0.014 & 0.0018 & 0.0034 & \citet{gre89} \\
V1370 Aql 1982 & 0.053 & 0.23 & 0.52 & 0.11 & \citet{sni87} \\
V1370 Aql 1982 & 0.044 & 0.28 & 0.56 & 0.017 & \citet{and94} \\
V1370 Aql 1982 & 0.065 & 0.13 & 0.69 & \nodata & \citet{ark97} \\
V723 Cas 1995\tablenotemark{a} & 0.52 & 0.064 & 0.052  & 0.042 & \citet{iij06} \\
V693 CrA 1981 & 0.29 & 0.25 & 0.17  & 0.016 & \citet{wil85} \\
V693 CrA 1981 & 0.16 & 0.36 & 0.26  & 0.030 & \citet{and94} \\
V693 CrA 1981 & 0.40 & 0.14 & 0.23  & \nodata & \citet{van97} \\
V693 CrA 1981 & 0.29 & 0.21 & 0.17  & \nodata & \citet{ark97} \\
CP Cru 1996 & 0.47 & 0.18 & 0.047 & 0.0026 & \citet{lyk03} \\
V1500 Cyg 1975 & 0.57 & 0.149 & 0.0099 & \nodata & \citet{lan88} \\
V1500 Cyg 1975 & 0.49 & 0.275 & 0.023 & \nodata & \citet{fer78} \\
V1668 Cyg 1978\tablenotemark{b} & 0.45 & 0.33 & \nodata  & \nodata & \citet{and94} \\
V1668 Cyg 1978\tablenotemark{b} & 0.45 & 0.32 & 0.0068  & \nodata & \citet{sti81} \\
V1974 Cyg 1992 & 0.19 & 0.375 & 0.11 & 0.0051 & \citet{aus96} \\
V1974 Cyg 1992 & 0.30 & 0.14 & 0.037 & 0.075 & \citet{hay96} \\
V1974 Cyg 1992 & 0.23 & 0.14 & 0.43 & 0.0018 & \citet{ark97} \\
V1974 Cyg 1992 & 0.55 & 0.12 & 0.06 & \nodata & \citet{van05} \\
V838 Her 1991 & 0.78 & 0.041 & 0.081 & 0.0003 & \citet{van96} \\
V838 Her 1991 & 0.59 & 0.030 & 0.067 & 0.0003 & \citet{van97} \\
V838 Her 1991 & 0.56 & 0.038 & 0.070 & 0.015 & \citet{sch07} \\
V351 Pup 1991 & 0.37 & 0.26 & 0.13 & \nodata & \citet{sai96} \\
V977 Sco 1989 & 0.51 & 0.072 & 0.26 & 0.0027 & \citet{and94} \\
V4160 Sgr 1991 & 0.465 & 0.125 & 0.065 & 0.011 & \citet{sch07} \\
V382 Vel 1999 & 0.47 & 0.0018 & 0.0099 & 0.0069 & \citet{aug03} \\
V382 Vel 1999 & 0.66 & 0.043 & 0.027 & 0.0030 & \citet{sho03} \\
QU Vul 1984 \#2 & 0.30 & 0.06 & 0.040 & 0.0049 & \citet{sai92} \\
QU Vul 1984 \#2 & 0.33 & 0.25 & 0.086 & 0.063 & \citet{and94} \\
QU Vul 1984 \#2 & 0.36 & 0.26 & 0.18 & 0.0014 & \citet{aus96} \\
QU Vul 1984 \#2 & 0.63 & 0.029 & 0.032 & 0.009 & \citet{sch02} \\
\enddata
\tablenotetext{a}{V723~Cas was not identified as a neon nova, but
the neon lines were very strong and the neon abundance obtained by
\citet{iij06} is as large as those of other neon novae. See 
Section \ref{wd_mass_neon_novae}.}
\tablenotetext{b}{V1668~Cyg was not identified as a neon nova.
See Section \ref{light_curve_v1668_cyg}.}
\end{deluxetable*}


\begin{deluxetable*}{lllllllll}
\tabletypesize{\scriptsize}
\tablecaption{Chemical compositions of nova models
\label{chemical_composition_model}}
\tablewidth{0pt}
\tablehead{
\colhead{novae case} & 
\colhead{$X$} & 
\colhead{$Y$} & 
\colhead{$X_{\rm CNO}$} & 
\colhead{$X_{\rm Ne}$} & 
\colhead{$Z$\tablenotemark{a}}  & 
\colhead{mixing\tablenotemark{b}}  & 
\colhead{calibration\tablenotemark{c}}  & 
\colhead{comments}
} 
\startdata
Ne nova 1 & 0.35 & 0.33 & 0.20 & 0.10 & 0.02 & 100\% & V351 Pup & this work; Table \ref{light_curves_of_novae_ne1} \\
Ne nova 2\tablenotemark{d}
 & 0.55 & 0.30 & 0.10 & 0.03 & 0.02 & 25\% & V1500 Cyg & \citet{hac10k} \\
Ne nova 3 & 0.65 & 0.27 & 0.03 & 0.03 & 0.02 & 8\% & QU Vul & this work; Table \ref{light_curves_of_novae_ne3} \\
CO nova 1
 & 0.35 & 0.13 & 0.50 & 0.0 & 0.02 & 100\% & \nodata & \nodata \\
CO nova 2\tablenotemark{e} & 0.35 & 0.33 & 0.30 & 0.0 & 0.02 & 100\% & GQ Mus & \citet{hac10k} \\
CO nova 3 & 0.45 & 0.18 & 0.35 & 0.0 & 0.02 & 55\% & V1668 Cyg & this work; Table \ref{light_curves_of_novae_co3} \\ 
CO nova 4\tablenotemark{f} & 0.55 & 0.23 & 0.20 & 0.0 & 0.02 & 25\% & PW Vul & \citet{hac15k} \\
Solar & 0.70 & 0.28 & 0.0 & 0.0 & 0.02 & 0\% & \nodata & \nodata  
\enddata
\tablenotetext{a}{Carbon, nitrogen, oxygen, and neon are also
included in $Z=0.02$ with the same ratio as the solar abundance
\citep{gre89}.}
\tablenotetext{b}{Mixing between the helium layer + core material
and the accreted matter with solar abundances, which is calculated from
$\eta_{\rm mix}=(0.7/X)-1$.}
\tablenotetext{c}{Absolute magnitudes of free-free emission model
light curve are calibrated with each nova whose distance modulus is
known.  Adopted chemical composition of each nova in 
the present paper (V1668~Cyg, V351~Pup, QU~Vul) or in our separate
papers (V1500~Cyg, GQ~Mus, PW~Vul).}
\tablenotetext{d}{Free-free light curves of Ne nova 2 are tabulated
in Table 3 of \citet{hac10k}.}
\tablenotetext{e}{Free-free emission light curves of CO nova 2 are
tabulated in Table 2 of \citet{hac10k}.}
\tablenotetext{f}{Free-free light curves of CO nova 4 are tabulated
in Table 4 of \citet{hac15k}.}
\end{deluxetable*}


\begin{deluxetable*}{lllllllllllllll}
\tabletypesize{\scriptsize}
\tablecaption{Free-free Light Curves of CO Novae 3\tablenotemark{a}
\label{light_curves_of_novae_co3}}
\tablewidth{0pt}
\tablehead{
\colhead{$m_{\rm ff}$} &
\colhead{0.55$M_\sun$} &
\colhead{0.6$M_\sun$} &
\colhead{0.65$M_\sun$} &
\colhead{0.7$M_\sun$} &
\colhead{0.75$M_\sun$} &
\colhead{0.8$M_\sun$} &
\colhead{0.85$M_\sun$} &
\colhead{0.9$M_\sun$} &
\colhead{0.95$M_\sun$} &
\colhead{1.0$M_\sun$} &
\colhead{1.05$M_\sun$} &
\colhead{1.1$M_\sun$} &
\colhead{1.15$M_\sun$} &
\colhead{1.2$M_\sun$} \\
\colhead{(mag)} &
\colhead{(day)} &
\colhead{(day)} &
\colhead{(day)} &
\colhead{(day)} &
\colhead{(day)} &
\colhead{(day)} &
\colhead{(day)} &
\colhead{(day)} &
\colhead{(day)} &
\colhead{(day)} &
\colhead{(day)} &
\colhead{(day)} &
\colhead{(day)} &
\colhead{(day)} 
}
\startdata
  4.000     & 0.0 & 0.0 & 0.0 & 0.0 & 0.0 & 0.0 & 0.0 & 0.0 & 0.0 & 0.0 & 0.0 & 0.0 & 0.0 & 0.0 \\
  4.250     &  8.030     &  4.350     &  3.320     &  32.72     &  2.110     &  1.530     &  1.200     &  1.020     & 0.8900     & 0.7900     & 0.7020     & 0.6480     & 0.6160     & 0.5600     \\
  4.500     &  17.32     &  10.02     &  6.680     &  35.35     &  4.270     &  3.250     &  2.470     &  2.070     &  1.810     &  1.610     &  1.413     &  1.309     &  1.198     &  1.110     \\
  4.750     &  27.26     &  17.54     &  10.17     &  39.40     &  6.470     &  4.960     &  3.810     &  3.180     &  2.770     &  2.440     &  2.163     &  1.990     &  1.771     &  1.670     \\
  5.000     &  38.18     &  25.06     &  16.25     &  43.67     &  8.800     &  6.950     &  5.180     &  4.390     &  3.800     &  3.280     &  2.973     &  2.700     &  2.424     &  2.230     \\
  5.250     &  49.80     &  33.44     &  22.70     &  48.15     &  12.41     &  9.430     &  6.920     &  5.670     &  4.870     &  4.300     &  3.870     &  3.427     &  3.164     &  2.810     \\
  5.500     &  64.69     &  43.53     &  29.55     &  53.15     &  16.41     &  12.03     &  9.270     &  7.070     &  6.290     &  5.570     &  5.168     &  4.562     &  4.036     &  3.460     \\
  5.750     &  81.31     &  54.09     &  36.91     &  58.81     &  20.68     &  14.97     &  11.76     &  8.970     &  8.040     &  6.980     &  6.446     &  5.806     &  5.092     &  4.340     \\
  6.000     &  99.21     &  65.91     &  45.96     &  64.84     &  25.23     &  18.10     &  14.12     &  11.13     &  9.770     &  8.290     &  7.506     &  6.675     &  5.980     &  5.320     \\
  6.250     &  118.4     &  79.36     &  55.85     &  71.45     &  30.18     &  21.40     &  16.55     &  13.16     &  11.26     &  9.640     &  8.596     &  7.543     &  6.710     &  6.030     \\
  6.500     &  138.8     &  94.36     &  66.44     &  79.26     &  35.47     &  25.23     &  19.17     &  15.18     &  12.86     &  11.03     &  9.716     &  8.453     &  7.460     &  6.660     \\
  6.750     &  159.2     &  108.9     &  77.80     &  87.86     &  41.38     &  29.32     &  22.19     &  17.35     &  14.64     &  12.51     &  10.91     &  9.413     &  8.240     &  7.300     \\
  7.000     &  181.6     &  124.1     &  89.70     &  97.18     &  47.95     &  33.79     &  25.45     &  19.77     &  16.62     &  14.08     &  12.21     &  10.44     &  9.070     &  7.950     \\
  7.250     &  206.5     &  140.1     &  102.5     &  106.9     &  55.08     &  38.74     &  29.00     &  22.44     &  18.76     &  15.84     &  13.64     &  11.56     &  9.950     &  8.610     \\
  7.500     &  233.9     &  157.3     &  115.3     &  116.8     &  62.70     &  44.16     &  32.90     &  25.31     &  21.08     &  17.77     &  15.20     &  12.79     &  10.88     &  9.310     \\
  7.750     &  265.0     &  176.6     &  129.2     &  127.4     &  70.47     &  50.04     &  37.16     &  28.39     &  23.71     &  19.88     &  16.89     &  14.11     &  11.89     &  10.03     \\
  8.000     &  300.6     &  199.7     &  144.5     &  139.0     &  78.80     &  56.12     &  41.76     &  31.75     &  26.42     &  22.09     &  18.55     &  15.42     &  12.92     &  10.82     \\
  8.250     &  344.3     &  225.9     &  163.4     &  151.5     &  88.12     &  62.39     &  46.70     &  35.37     &  29.22     &  24.32     &  20.33     &  16.81     &  13.98     &  11.65     \\
  8.500     &  394.4     &  257.0     &  184.8     &  166.0     &  98.52     &  69.25     &  51.88     &  39.28     &  32.15     &  26.73     &  22.25     &  18.32     &  15.13     &  12.49     \\
  8.750     &  445.3     &  292.4     &  209.1     &  183.8     &  110.2     &  77.58     &  57.46     &  43.42     &  35.48     &  29.24     &  24.42     &  19.98     &  16.40     &  13.40     \\
  9.000     &  503.2     &  331.7     &  237.3     &  203.9     &  124.5     &  86.85     &  64.37     &  47.93     &  39.24     &  32.29     &  27.08     &  22.12     &  17.97     &  14.39     \\
  9.250     &  544.6     &  371.8     &  270.4     &  226.0     &  140.7     &  98.15     &  72.08     &  53.82     &  43.82     &  35.79     &  30.06     &  24.50     &  19.83     &  15.52     \\
  9.500     &  589.1     &  411.5     &  299.3     &  251.5     &  158.9     &  111.6     &  81.43     &  60.45     &  49.23     &  40.28     &  33.79     &  27.48     &  22.10     &  16.94     \\
  9.750     &  639.5     &  447.2     &  324.1     &  280.9     &  179.8     &  126.4     &  92.64     &  68.51     &  55.48     &  45.31     &  38.00     &  30.82     &  24.76     &  18.60     \\
  10.00     &  692.0     &  483.0     &  352.4     &  300.0     &  204.0     &  142.0     &  104.8     &  78.04     &  63.14     &  51.51     &  43.40     &  35.12     &  28.00     &  20.73     \\
  10.25     &  735.4     &  518.2     &  384.6     &  321.2     &  219.4     &  160.1     &  117.9     &  88.29     &  71.90     &  58.72     &  49.47     &  39.93     &  31.79     &  23.19     \\
  10.50     &  780.8     &  551.1     &  421.2     &  345.1     &  235.6     &  175.6     &  132.8     &  99.09     &  81.57     &  66.84     &  55.41     &  44.75     &  35.72     &  26.15     \\
  10.75     &  829.0     &  585.4     &  450.9     &  372.0     &  253.7     &  187.9     &  146.8     &  111.3     &  91.87     &  75.19     &  61.93     &  49.91     &  39.78     &  29.41     \\
  11.00     &  879.9     &  622.0     &  479.5     &  401.6     &  273.8     &  201.8     &  159.9     &  121.2     &  100.8     &  83.10     &  69.01     &  55.57     &  44.16     &  32.78     \\
  11.25     &  933.9     &  660.7     &  509.7     &  425.1     &  296.4     &  217.1     &  170.8     &  129.6     &  107.8     &  90.50     &  74.09     &  60.00     &  48.46     &  36.18     \\
  11.50     &  990.9     &  701.8     &  541.8     &  450.3     &  315.3     &  234.1     &  182.0     &  138.9     &  115.4     &  98.00     &  79.36     &  64.32     &  52.40     &  39.53     \\
  11.75     &  1052.     &  745.3     &  575.7     &  477.0     &  334.7     &  250.7     &  193.9     &  149.1     &  123.4     &  105.2     &  85.05     &  68.93     &  56.15     &  42.72     \\
  12.00     &  1116.     &  791.4     &  611.6     &  505.3     &  355.3     &  266.7     &  206.4     &  160.1     &  131.8     &  112.1     &  91.15     &  73.87     &  59.83     &  45.85     \\
  12.25     &  1184.     &  840.1     &  649.7     &  535.3     &  377.0     &  283.8     &  219.7     &  171.4     &  140.8     &  119.4     &  97.63     &  79.12     &  63.75     &  48.99     \\
  12.50     &  1256.     &  891.9     &  690.1     &  567.0     &  400.1     &  301.7     &  233.8     &  182.8     &  150.3     &  127.1     &  104.0     &  84.35     &  67.84     &  52.34     \\
  12.75     &  1332.     &  946.6     &  732.8     &  600.6     &  424.5     &  320.8     &  248.7     &  194.9     &  160.3     &  135.3     &  110.4     &  89.53     &  72.07     &  56.17     \\
  13.00     &  1413.     &  1005.     &  778.0     &  636.2     &  450.4     &  341.0     &  264.5     &  207.7     &  171.0     &  144.0     &  117.2     &  95.02     &  76.59     &  60.23     \\
  13.25     &  1498.     &  1066.     &  825.9     &  673.9     &  477.9     &  362.4     &  281.2     &  221.3     &  182.3     &  153.2     &  124.4     &  100.8     &  81.40     &  64.54     \\
  13.50     &  1589.     &  1132.     &  876.7     &  713.9     &  506.9     &  385.1     &  298.9     &  235.7     &  194.2     &  162.9     &  132.0     &  107.0     &  86.50     &  69.10     \\
  13.75     &  1685.     &  1200.     &  930.5     &  756.2     &  537.7     &  409.0     &  317.7     &  250.9     &  206.9     &  173.2     &  140.1     &  113.5     &  91.91     &  73.93     \\
  14.00     &  1787.     &  1274.     &  987.1     &  801.0     &  570.2     &  434.5     &  337.5     &  267.0     &  220.3     &  184.2     &  148.6     &  120.4     &  97.62     &  79.05     \\
  14.25     &  1894.     &  1351.     &  1048.     &  848.6     &  604.8     &  461.4     &  358.6     &  284.1     &  234.5     &  195.7     &  157.7     &  127.7     &  103.7     &  84.47     \\
  14.50     &  2008.     &  1433.     &  1112.     &  898.9     &  641.3     &  489.9     &  380.9     &  302.2     &  249.5     &  208.0     &  167.3     &  135.5     &  110.1     &  90.18     \\
  14.75     &  2129.     &  1520.     &  1179.     &  952.1     &  680.1     &  520.1     &  404.5     &  321.3     &  265.4     &  221.0     &  177.4     &  143.7     &  116.9     &  96.28     \\
  15.00     &  2257.     &  1611.     &  1251.     &  1009.     &  721.1     &  552.1     &  429.5     &  341.6     &  282.3     &  234.7     &  188.2     &  152.4     &  124.1     &  102.8     \\

\hline
X-ray\tablenotemark{b}
 & 5075  & 3827 & 2942  & 2278 & 1643   & 1186   & 847   & 604   & 443   & 331   & 225   & 151  & 102 & 67.5 \\
\hline
$\log f_{\rm s}$\tablenotemark{c}
 & 0.84  & 0.73 & 0.63  & 0.57 & 0.44   & 0.33   & $0.23$   & $0.17$   & $0.055$   & $-0.03$   & $-0.13$   & $-0.23$  & $-0.30$ & $-0.42$ \\
\hline
$M_{\rm w}$\tablenotemark{d}
 & 5.2 & 4.7  & 4.3 & 3.7   & 3.2   & 2.7   & 2.4   & 2.1   & 1.9   & 1.6   & 1.45  & 1.3 & 1.0 & 0.7 
\enddata
\tablenotetext{a}{Chemical composition of
the envelope is assumed
to be that of CO nova 3 in Table \ref{chemical_composition_model}.}
\tablenotetext{b}{Duration of supersoft X-ray phase in units of days.}
\tablenotetext{c}{Stretching factor against V1668~Cyg
 UV 1455~\AA\  observation
in Figure \ref{all_mass_v1668_cyg_x45z02c15o20_calib_universal}.}
\tablenotetext{d}{Absolute magnitudes at the bottom point in Figure
\ref{all_mass_v1668_cyg_x45z02c15o20_real_scale_universal} by assuming
$(m-M)_V = 14.6$  (V1668~Cyg).}
\end{deluxetable*}


\begin{figure}
\epsscale{1.15}
\plotone{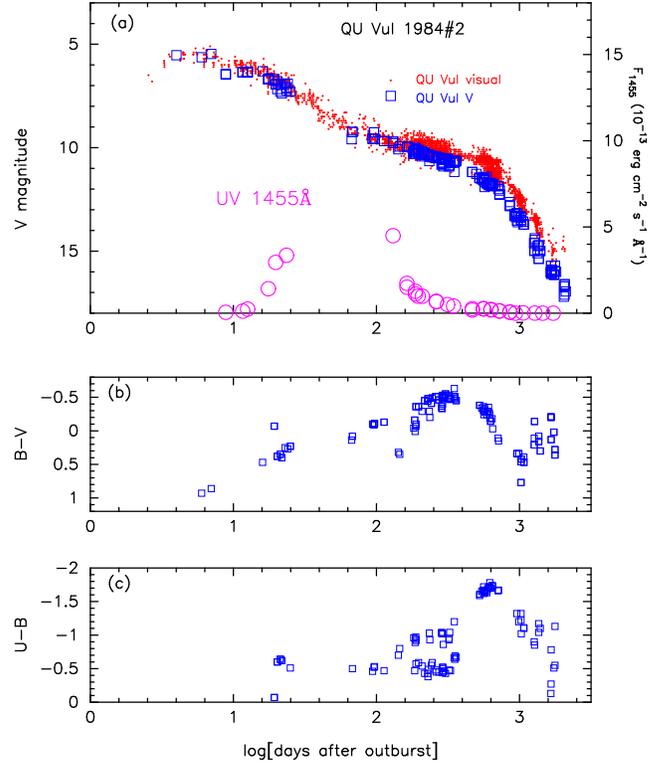}
\caption{
(a) $V$ (blue open squares), visual (red dots), and UV 1455~\AA\ 
(magenta open circles) band light curves, (b) $B-V$, and 
(c) $U-B$ color curves, for QU~Vul.
Here, we assume the start of the day ($t=0$) as JD 2446054.0 for QU~Vul.
See text for the sources of the data.
\label{qu_vul_ubv_color_logscale_no2}}
\end{figure}

\section{QU~Vul 1984\#2}
\label{qu_vul}
QU~Vul was discovered by P. Collins on UT 1984 December 22.13 
at about 6.8 mag \citep{col84}.  We plot the $V$ and visual light curves
and $B-V$ and $U-B$ color evolutions in Figure
\ref{qu_vul_ubv_color_logscale_no2}.
The optical data are taken from IAU Circular No. 4033, \citet{kol88},
\citet{ber88}, \citet{ros92}, and the AAVSO archive.
The UV~1455~\AA\  data are compiled from the {\it IUE} Newly Extracted
Spectra (INES) archive data
sever\footnote{http://sdc.cab.inta-csic.es/ines/index2.html}
\citep[see][for the UV~1455~\AA\  band]{cas02}.
It rose to $m_V\approx5.5$ at maximum on UT December 27  
\citep[e.g.,][]{ros92}.   
Then it gradually declined, with $t_2=22$~days and $t_3=49$~days
\citep[e.g.,][]{dow00}.  
The nova was identified as a fast neon nova by \citet{geh85}.
An orbital period of 2.68~hr was detected by \citet{shaf95}.


\begin{figure*}
\epsscale{0.95}
\plotone{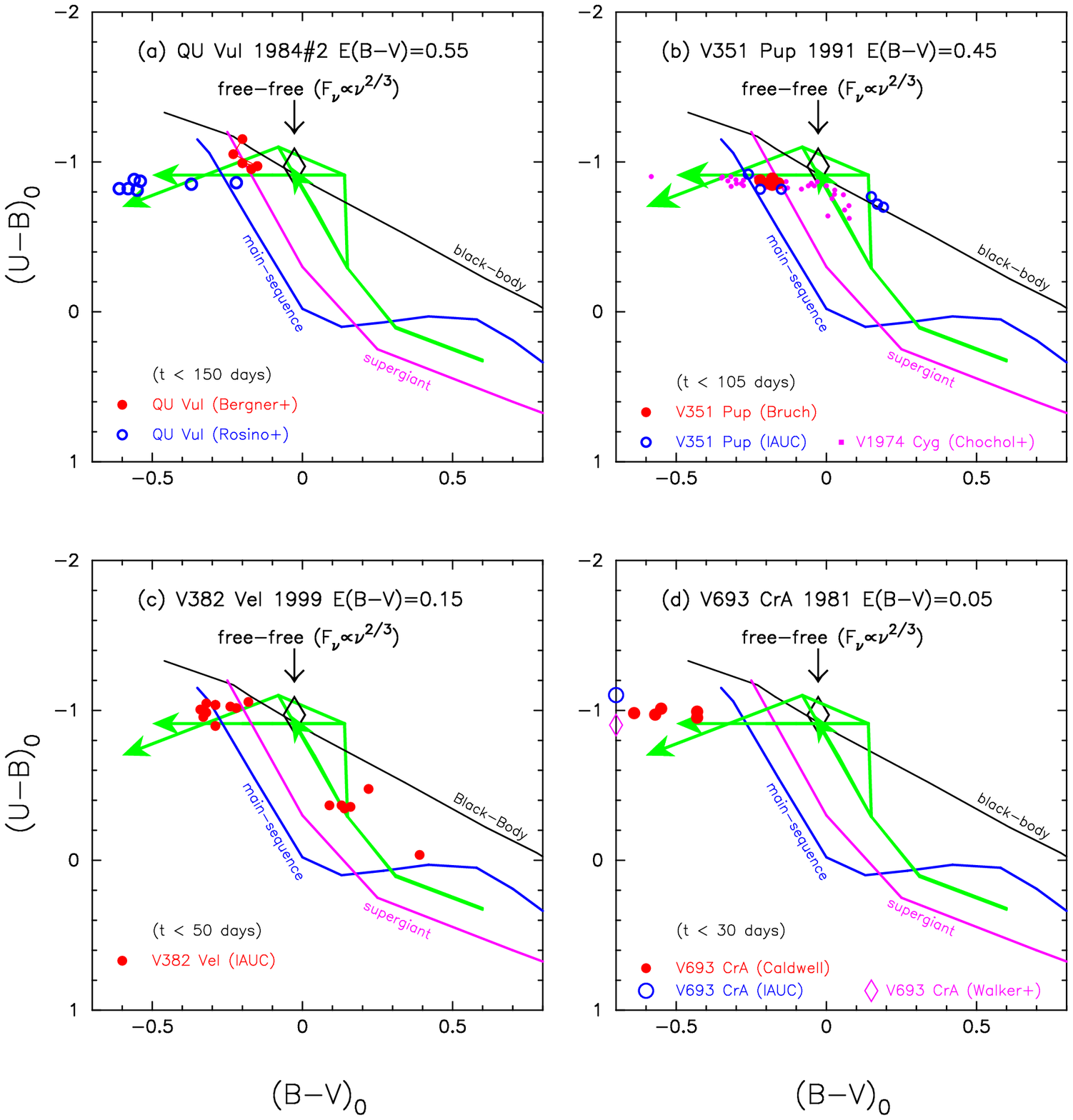}
\caption{
Color-color diagram for five neon novae in outburst as well as the general
course of novae (green solid lines), blackbody (black solid line),
main-sequence (blue solid line), and supergiant (magenta solid line)
sequences.  The data of these sequences are the same as those of
Figure 28 of \citet{hac14k}. 
(a) $(B-V)_0$ versus $(U-B)_0$ color-color diagram of QU~Vul
for a reddening of $E(B-V)=0.55$.
The data of QU~Vul are taken from \citet{ros92} and \citet{ber88}.
(b) V351~Pup for $E(B-V)=0.45$ as well as V1974~Cyg for $E(B-V)=0.30$.
The data of V351~Pup are taken from \citet{bru92} and IAU Circulars.  
The data of V1974~Cyg are taken from \citet{cho93}.
(c) V382~Vel for $E(B-V)=0.15$.
The data of V382~Vel are taken from IAU Circulars.
(d) V693~CrA for $E(B-V)=0.05$.  The data of V693~CrA are taken from 
\citet{cal81}, \citet{wal82}, and IAU Circulars.
\label{color_color_diagram_qu_vul_v351_pup_v382_vel_v693_cra}}
\end{figure*}

\subsection{Reddening and distance}
\label{reddening_distance_qu_vul}
The reddening toward QU~Vul was estimated as 
$E(B-V)=A_V/3.1\sim1.0/3.1=0.3$ by \citet{geh86a} from the galactic
absorption, as $E(B-V)=0.5$ by \citet{ros92} 
from the spectral type of the nova at optical maximum, 
and as $E(B-V)=0.61\pm0.1$ by \citet{sai92} from 
the \ion{He}{2} 1640/4686~\AA\  ratio and from the 2200~\AA\  UV
spectral feature.  \citet{del97} adopted $E(B-V)=A_V/3.1\sim1.7/3.1=0.55$
for dereddening the spectra.  Using Saizar et al.'s $E(B-V)=0.61$,
\citet{sch02} obtained UV fluxes that were consistent with the 
optical fluxes.

For the galactic extinction, the NASA/IPAC galactic dust absorption map
gives $E(B-V)=0.55 \pm 0.03$ in the direction of QU~Vul, whose
galactic coordinates are $(l, b)= (68\fdg5108,-6\fdg0263)$.
\citet{hac14k} proposed a new method of determining the 
reddening of classical novae.  They identified a general course of
dereddened $UBV$ color-color evolution and determined the reddening
for a number of novae by comparing the observed track of a target nova
with the general course.
Their obtained value is $E(B-V)=0.55\pm0.05$ for QU~Vul 
\citep[see Figure 31(c) of][]{hac14k} which is shown again in Figure 
\ref{color_color_diagram_qu_vul_v351_pup_v382_vel_v693_cra}(a).
These estimates strongly suggest that $E(B-V)=0.55 \pm 0.05$.
Therefore, we adopt $E(B-V)=0.55\pm0.05$ in this paper.

The distance toward QU~Vul was estimated as $d=3.6$~kpc 
by \citet{tay87} using the radio image expansion parallax method, and as
$d=2.6$~kpc by \citet{del97} using the optical shell expansion parallax
method.  On the basis of the shell structure observed by {\it Hubble Space
Telescope (HST)}, \citet{dow00} obtained $d=1.5$~kpc using the expansion
velocity of 1500~km~s$^{-1}$ and $d=2.0$~kpc using the velocity observed
by \citet{and91}.  \citet{dow00} summarized the value as 
$d=1.75\pm0.25$~kpc and
concluded that this range of the distance is in reasonable agreement
with estimates based on other methods ($d=1.9\pm0.5$~kpc).
The expansion parallax method of nova shells depends basically on
the assumptions that (1) nova shells expand isotropically and (2) 
the expansion velocity is constant in time.  Even if the shell size
is the same, this method gives different distances with different
expansion velocities and anisotropic velocity fields.  
The arithmetic mean of the above four values is 2.4~kpc.

\begin{figure}
\epsscale{0.90}
\plotone{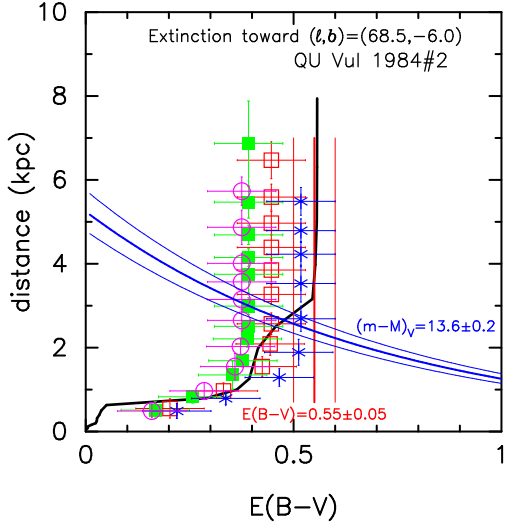}
\caption{
Various distance-reddening relations toward QU~Vul, whose
galactic coordinates are $(l, b)= (68\fdg5108,-6\fdg0263)$.
Blue solid line flanking two thin solid lines:
the distance-reddening relation calculated from Equation 
(\ref{qu_vul_distance_modulus_eq1}) with $(m-M)_V=13.6\pm0.2$.
Red vertical solid line flanking two thin solid lines:
the reddening estimate of $E(B-V)=0.55\pm0.05$.
We add the galactic distance-reddening relations
toward four directions close to QU~Vul:
$(l, b)= (68\fdg50,-6\fdg00)$ (red open squares),
$(68\fdg75,-6\fdg00)$ (green filled squares), 
$(68\fdg50,-6\fdg25)$ (blue asterisks), and
$(68\fdg75,-6\fdg25)$ (magenta open circles), the data for which
are taken from \citet{mar06}.
We further add Green et al.'s (2015) relation (black solid line).
\label{qu_vul_distance_reddening_stretching}}
\end{figure}

The distance modulus in the $V$ band, $\mu_V = (m-M)_V$,
is obtained using the time-stretching method \citep{hac10k}
for nova light curves, which yielded a value of 
$(m-M)_{V,{\rm QU~Vul}}=13.6\pm0.2$
(see Appendix \ref{time_stretching_method_novae}). 
We plot various distance-reddening relations toward QU~Vul in Figure
\ref{qu_vul_distance_reddening_stretching}.
Adopting $(m-M)_V=13.6\pm0.2$ for QU~Vul, 
we plot the relation of Equation (\ref{qu_vul_distance_modulus_eq1})
in Figure \ref{qu_vul_distance_reddening_stretching} by a thick blue line
flanked by thin blue lines.
We also plot distance-reddening relations at
four nearby directions using data from \citet{mar06}:
$(l, b)= (68\fdg50,-6\fdg00)$ (red open squares),
$(68\fdg75,-6\fdg00)$ (green filled squares), 
$(68\fdg50,-6\fdg25)$ (blue asterisks), and
$(68\fdg75,-6\fdg25)$ (magenta open circles).
We also add Green et al.'s (2015) relation (black solid line).

All the three trends, $E(B-V)=0.55\pm0.05$ (vertical red solid lines),
$(m-M)_V = 13.6\pm0.2$ (blue solid lines), 
and the distance-reddening relation of \citet{mar06},
approximately cross at $E(B-V)\sim0.55$ and $d\sim2.4$~kpc
in Figure \ref{qu_vul_distance_reddening_stretching},
although Green et al.'s relation deviates slightly from this cross point.
This is consistent with the NASA/IPAC galactic dust absorption map
of $E(B-V)=0.55 \pm 0.03$ in the direction toward QU~Vul.
Thus, we adopt $E(B-V)=0.55\pm0.05$ and $d=2.4\pm0.3$~kpc in this paper.


\subsection{Chemical abundance of ejecta}
\label{chemical_composition_qu_vul}
The chemical abundance of the ejecta was estimated by several groups.
The estimated values are summarized in Table
\ref{chemical_abundance_neon_novae}.
WD core material such as carbon, oxygen, neon, and magnesium
is generally mixed into hydrogen-rich envelopes before nova explosions and
ejected in nova winds.  If we assume that the abundance of
the accreted matter is solar ($X=0.70$, $Y=0.28$, and $Z=0.02$),
we roughly calculate the degree of mixing as $\eta_{\rm mix}=(0.7/X)-1$.
Here, $X$, $Y$, and $Z$ are the mass fractions of hydrogen, helium,
and heavy elements, respectively, in the hydrogen-rich envelopes of novae.

\citet{sai92}, \citet{and94}, and \citet{aus96} obtained chemical
compositions with a relatively high degree of mixing.
The average value of these three estimates is $X\sim0.33$, i.e., 
$\eta_{\rm mix}\sim1.1$ (110\%).
On the other hand, \citet{sch02} presented a low degree of mixing 
of $X\sim0.63$, i.e., $\eta_{\rm mix}\sim0.11$ (11\%).

First, we adopt the chemical abundance of Ne nova 2
and examine the light curves of QU~Vul because we have the absolute
magnitudes of optical light curves for Ne nova 2.
This case corresponds to a relatively low degree of mixing,
$X=0.55$, i.e., $\eta_{\rm mix}=0.25$ (25\%), which is lower than
the estimates made by the above three groups \citep{sai92, and94, aus96}
but higher than the estimate of \citet{sch02}.  Second, for comparison,
assuming different degrees of mixing for CO novae, i.e., 100\% of 
CO nova 2,  and 25\% of CO nova 4, 
we obtain model light curves and compare them with the light curves
of QU~Vul.  Because Schwarz's estimate is close to that of Ne nova 3,
i.e., a low degree of mixing, $X=0.65$ ($\eta_{\rm mix}=0.08$),
we finally examine this case of Ne nova 3, assuming that
$(m-M)_V=13.6$.

\subsection{Light curve analysis of QU~Vul}
\label{lc_analysis_qu_vul}
\begin{figure}
\epsscale{1.15}
\plotone{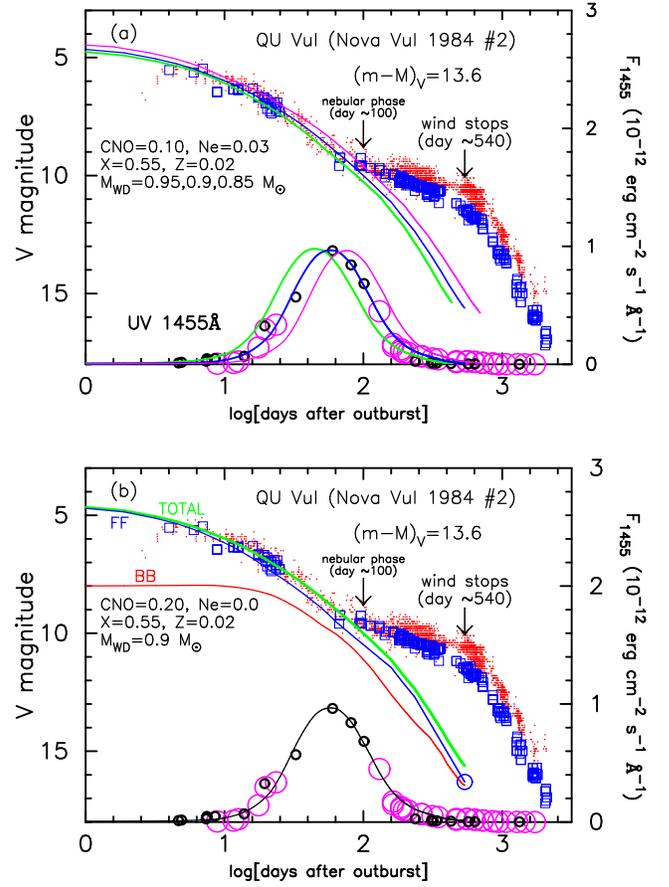}
\caption{
(a) Model light curves of $0.85~M_\sun$ (magenta solid lines), 
$0.9~M_\sun$ (blue solid lines), and $0.95~M_\sun$ (green solid lines)
WDs with the chemical composition of Ne Nova 2 for the distance
modulus of $(m-M)_V=13.6$ as well as
$V$ band (blue open squares), visual (red dots), and UV~1455 ~\AA\  
(large magenta open circles) light curves of QU~Vul.
The $V$, visual, and UV~1455 ~\AA\  light curves of QU~Vul
are the same as those in Figure \ref{qu_vul_ubv_color_logscale_no2},
but we added UV~1455~\AA\  fluxes of PW~Vul (small black open circles)
to estimate the peak flux of QU~Vul. 
(b) Model light curves of the $0.9~M_\sun$ WD for the distance modulus
of $(m-M)_V=13.6$.      Optically thick winds stop
approximately 540 days after the outburst for the $0.9~M_\sun$ WD.
The green solid line labeled ``TOTAL'' shows the total $V$ flux of
free-free plus blackbody, whereas the blue (labeled ``FF'')
and red (labeled ``BB'') solid lines show the $V$ fluxes
of free-free emission and blackbody emission, respectively.
The black solid line shows the UV~1455 ~\AA\  flux. 
\label{all_mass_qu_vul_x55z02o10ne03_absolute_mag}}
\end{figure}

We calculate three model light curves of free-free emission,
photospheric emission (blackbody approximation), and their sum
for various WD masses with various chemical compositions,
and fitted the total $V$ flux with the observed data.
The blackbody emission was described in \citet{kat94h} and
the free-free emission in \citet{hac06kb}.
The UV~1455~\AA\  band flux is also useful for determining the WD mass.
The flux in the UV~1455~\AA\  band was calculated from the blackbody
emission of a pseudophotosphere based on the optically thick
wind solutions of \citet{kat94h} and \citet{hac06kb, hac10k, hac15k}.
Unfortunately, no {\it IUE} observation is available around the 
UV~1455~\AA\   peak as shown in Figure \ref{qu_vul_ubv_color_logscale_no2}(a).
Thus, we added the UV~1455~\AA\  light curve  of PW~Vul in Figure 
\ref{all_mass_qu_vul_x55z02o10ne03_absolute_mag} (black open circles).
The shapes of the UV~1455~\AA\  and optical light curves of novae
show a remarkable similarity among different timescales, 
if we can squeeze/stretch the light curves in the direction of time
as demonstrated in Figures
\ref{v382_vel_v1500_cyg_v_color_logscale_no2} and 
\ref{qu_vul_pw_vul_gq_mus_v_bv_ub_color_logscale_no2}.
Using this time-scaled similarity, we fit our model light curves
with the UV~1455~\AA\  data for both QU~Vul and PW~Vul.


\begin{figure}
\epsscale{1.15}
\plotone{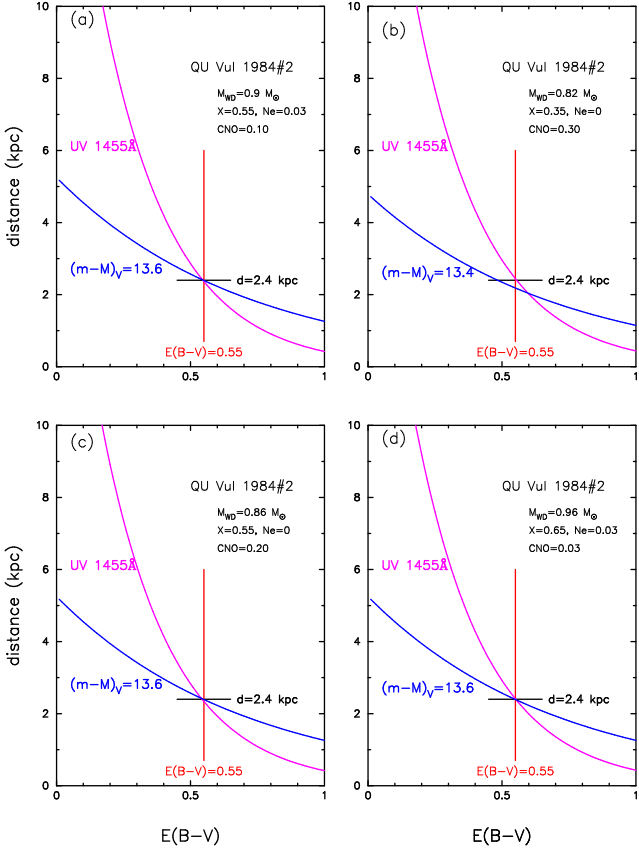}
\caption{
Distance-reddening relations toward QU~Vul for the $V$ and UV~1455~\AA\  
light curve fittings of (a) $0.9~M_\sun$ WD with the chemical
composition of Ne nova 2, (b) $0.82~M_\sun$ WD with CO nova 2,
(c) $0.86~M_\sun$ WD with CO nova 4, and (d) $0.96~M_\sun$ WD with Ne nova 3.
The blue solid lines show the results obtained from the $V$ model light
curve fittings.  The magenta solid lines show the results calculated from
the UV 1455~\AA\   flux fittings.  Two other constraints are also plotted;
the distance of $d=2.4$~kpc and the reddening of $E(B-V)=0.55$, which
are estimated in Section \ref{reddening_distance_qu_vul}.
\label{qu_vul_distance_reddening_x35_x55_x55_x65_4figure}}
\end{figure}

\subsubsection{Ne nova 2}
\label{qu_vul_ne_nova2}
     First, we consider a relatively low degree of mixing, 
25\%, i.e., Ne nova 2 in Table \ref{chemical_composition_model}.
The absolute magnitudes of the model light curves were published 
in \citet{hac10k}. 
     In Figure \ref{all_mass_qu_vul_x55z02o10ne03_absolute_mag}(a),
we plot our total $V$ and UV~1455~\AA\  model light curves
for WDs with 0.85 (magenta solid lines),
0.9 (blue solid lines), and $0.95~M_\sun$ (green solid lines),
assuming that the distance modulus in the $V$ band is $\mu_V=(m-M)_V=13.6$.
The model of $M_{\rm WD} = 0.9~M_\sun$ agrees well with
the $V$ and UV~1455~\AA\    observations.   In Figure
\ref{all_mass_qu_vul_x55z02o10ne03_absolute_mag}(b), 
we plot three model light curves of the blackbody emission (red solid
line), free-free emission (blue solid line), and total flux
(green solid line) for the $0.9~M_\sun$ WD.

Figure \ref{qu_vul_distance_reddening_x35_x55_x55_x65_4figure}(a) shows
the distance-reddening relation of Equation
(\ref{qu_vul_distance_modulus_eq1}) for
$(m-M)_V=13.6$ (blue solid line labeled ``$(m-M)_V=13.6$'').
The UV~1455~\AA\  light curve fitting gives another
distance-reddening relation of Equation (\ref{qu_vul_uv1455_fit_eq2}),
where we adopted $F_{1455}^{\rm mod}= 11.3 \times 
10^{-12}$~erg~cm$^{-2}$~s$^{-1}$~~\AA$^{-1}$ as the calculated model
flux of the upper bound of Figure
\ref{all_mass_qu_vul_x55z02o10ne03_absolute_mag}(a)
at a distance of 10~kpc and $F_{1455}^{\rm obs}= 3.0 \times 
10^{-12}$~erg~cm$^{-2}$~s$^{-1}$~~\AA$^{-1}$ as the observed flux 
corresponding to that of the upper bound in Figure 
\ref{all_mass_qu_vul_x55z02o10ne03_absolute_mag}(a).
This distance-reddening relation is also plotted as a magenta solid line
(labeled ``UV1455~\AA'') in Figure
\ref{qu_vul_distance_reddening_x35_x55_x55_x65_4figure}(a).
The two distance-reddening relations
cross each other at $E(B-V) \approx 0.55$ and 
$d \approx 2.4$~kpc.  These two values are consistent
with our estimates in Section \ref{reddening_distance_qu_vul}.

It should be noted that our model light curve fits the early $V$
and visual light curves well but deviates from
the $V$ and visual observation in the later phase, that is,
in the nebular phase.
QU~Vul entered the nebular phase in April 1985 at $m_V\approx9.6$
(about 100 days after the outburst) \citep{ros87,ros92}.
This deviation is due to strong emission lines such as [\ion{O}{3}],
which are not included in our model light curves \citep{hac06kb}.
The medium-band $y$ is designed to avoid such strong emission lines.
If $y$ magnitude data of QU~Vul were available, like
in Figures \ref{all_mass_v1668_cyg_x45z02c15o20}, and 
\ref{all_mass_v1668_cyg_x35z02c10o20}  --
\ref{all_mass_v1668_cyg_x65z02o03ne03} for V1668~Cyg,
our model light curves could follow the observed $y$ light curve
even in the nebular phase of QU~Vul.


\begin{figure}
\epsscale{1.15}
\plotone{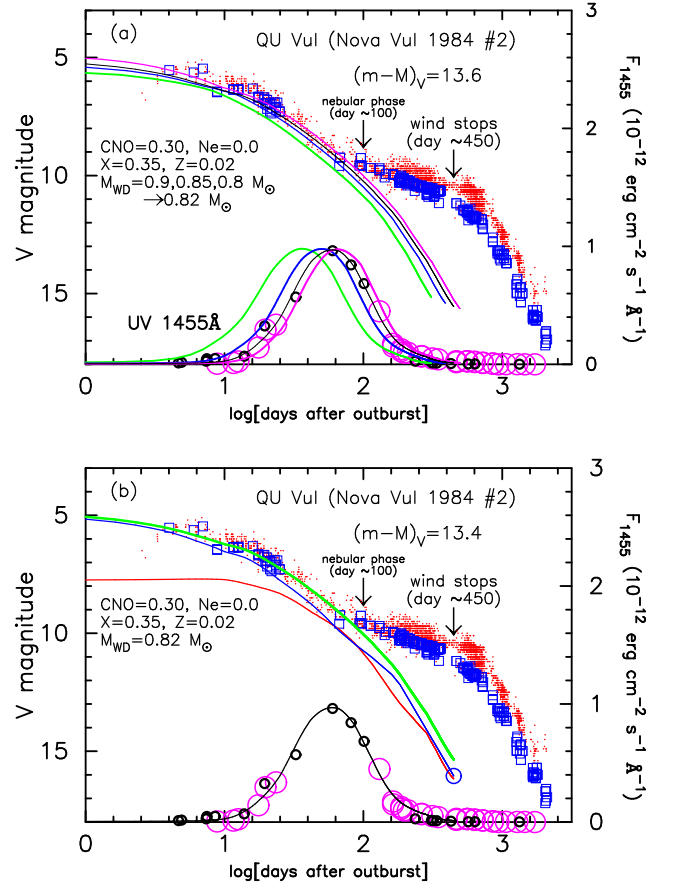}
\caption{
Same as Figure
\ref{all_mass_qu_vul_x55z02o10ne03_absolute_mag}, but for CO nova 2.
(a) Model light curves of $0.8~M_\sun$ (magenta solid lines), 
$0.82~M_\sun$ (black thin solid lines), $0.85~M_\sun$ (blue solid lines),
and $0.9~M_\sun$ (green solid lines) WDs with CO Nova 2 for $(m-M)_V=13.6$.
(b) Model light curves of the $0.82~M_\sun$ WD for $(m-M)_V=13.4$.
Optically thick winds stop approximately 450 days after the outburst
for the $0.82~M_\sun$ WD.  The green solid line shows the total $V$ flux,
whereas the blue and red solid lines show the $V$ fluxes of free-free
emission and blackbody emission, respectively.   The black solid line
shows the UV~1455 ~\AA\  flux. 
\label{all_mass_qu_vul_x35z02c10o20_absolute_mag}}
\end{figure}

\subsubsection{CO nova 2}
\label{qu_vul_co_nova2}
     Next, we consider a relatively high degree of mixing, 100\%, i.e., 
CO nova 2 in Table \ref{chemical_composition_model} for comparison.
The absolute magnitudes of the model light curves for this chemical 
composition were already calibrated in \citet{hac10k}, i.e., we know
the constant $C$ in Equation (\ref{free-free_calculation_original}).
Figure \ref{all_mass_qu_vul_x35z02c10o20_absolute_mag}(a)
shows our model light curves for WDs with 0.8, 0.82, 0.85, 
and $0.9~M_\sun$, assuming that $(m-M)_V=13.6$.
We changed the WD mass first in coarser steps of $0.05~M_\sun$ and then
in finer steps of $0.01~M_\sun$.  In Figure
\ref{all_mass_qu_vul_x35z02c10o20_absolute_mag}(a),
the $0.82~M_\sun$ WD model shows reasonable
agreement with the UV~1455~\AA\   observation.
However, the $V$ model light curve (black thin solid line)
is slightly darker than the $V$ and visual observation in Figure 
\ref{all_mass_qu_vul_x35z02c10o20_absolute_mag}(a).
Therefore, we must adopt a smaller value for
the distance modulus, i.e., $(m-M)_V=13.4$, to fit 
the model light curve with the $V$ and visual observation,
as shown in Figure \ref{all_mass_qu_vul_x35z02c10o20_absolute_mag}(b).

In Figure \ref{all_mass_qu_vul_x35z02c10o20_absolute_mag}(b),
we plot three model light curves of the blackbody emission, 
free-free emission, and total flux.
The distance-reddening relation of Equation
(\ref{qu_vul_distance_modulus_eq1}) is plotted by a blue solid
line (labeled ``$(m-M)_V=13.4$'') in Figure 
\ref{qu_vul_distance_reddening_x35_x55_x55_x65_4figure}(b).
We also plot the distance-reddening relation calculated from our 
UV 1455~\AA\  flux fitting (magenta solid line labeled ``UV 1455 ~\AA\  ''),
i.e., Equation (\ref{qu_vul_uv1455_fit_eq2}) 
with $F_{\lambda 1455}^{\rm mod}= 12.0 \times 
10^{-12}$~erg~cm$^{-2}$~s$^{-1}$~~\AA$^{-1}$.
The two distance-reddening relations cross each other at 
$E(B-V) \approx 0.58$ and $d \approx 2.1$~kpc, which deviate slightly
from but still consistent with the reddening of $E(B-V) = 0.55\pm0.05$
and distance of $d=2.4\pm0.3$~kpc 
in Section \ref{reddening_distance_qu_vul}.


\begin{figure}
\epsscale{1.15}
\plotone{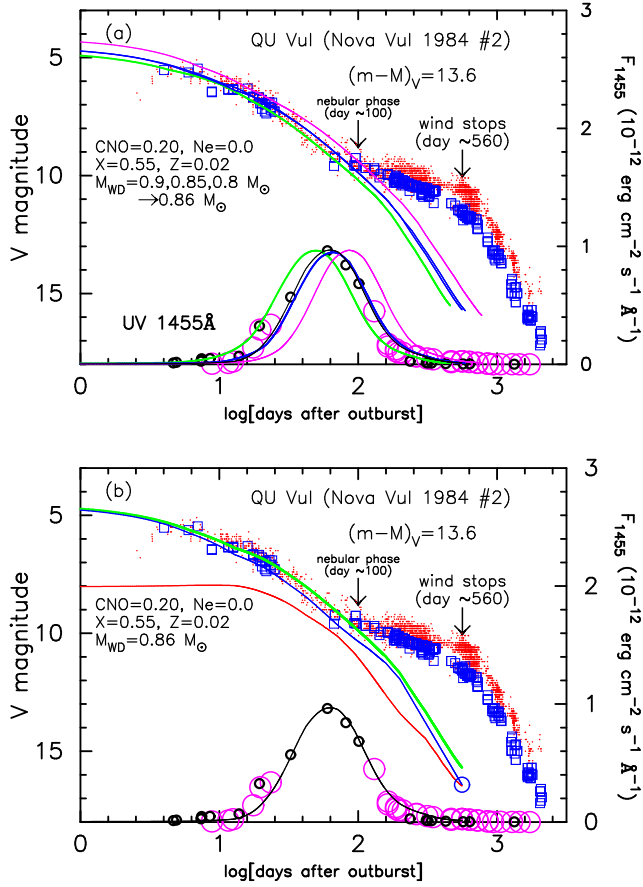}
\caption{
Same as Figure
\ref{all_mass_qu_vul_x55z02o10ne03_absolute_mag}, but for CO nova 4.
(a) We plot four model light curves of $0.8~M_\sun$ (magenta solid lines),
$0.85~M_\sun$ (blue solid lines), $0.86~M_\sun$ (black thin solid lines),
and $0.9~M_\sun$ (green solid lines) WDs.
(b) Model light curves of the $0.86~M_\sun$ WD for $(m-M)_V=13.6$.
Optically thick winds stop approximately 560 days
after the outburst for the $0.86~M_\sun$ WD.
\label{all_mass_qu_vul_x55z02c10o10_absolute_mag}}
\end{figure}

\subsubsection{CO nova 4}
\label{qu_vul_co_nova4}
     We further consider a relatively low degree of mixing, 
25\%, i.e., CO nova 4 in Table \ref{chemical_composition_model}.
The absolute magnitudes of the model light curves were already 
calibrated in \citet{hac15k}. 
Figure \ref{all_mass_qu_vul_x55z02c10o10_absolute_mag}(a)
shows our model light curves for 0.8, 0.85, 0.86, and $0.9~M_\sun$ WDs.
In this case, we chose a best-fit model of $M_{\rm WD} = 0.86~M_\sun$.
The UV 1455~\AA\  flux and total (free-free plus blackbody) emission
of the $0.86~M_\sun$ WD agree well with the observational data.
Assuming that $(m-M)_V = 13.6$, we plot three model light curves of the 
blackbody emission, free-free emission, and
total flux for the $0.86~M_\sun$ WD in Figure
\ref{all_mass_qu_vul_x55z02c10o10_absolute_mag}(b).

The distance-reddening relation of Equation
(\ref{qu_vul_distance_modulus_eq1}) is plotted by a blue solid
line (labeled ``$(m-M)_V=13.6$'') in Figure 
\ref{qu_vul_distance_reddening_x35_x55_x55_x65_4figure}(c).
We also plot the distance-reddening relation calculated from our 
UV 1455~\AA\  flux fitting (magenta solid line labeled ``UV 1455 ~\AA\  ''),
i.e., Equation (\ref{qu_vul_uv1455_fit_eq2}) 
with $F_{\lambda 1455}^{\rm mod}= 11.0 \times 
10^{-12}$~erg~cm$^{-2}$~s$^{-1}$~~\AA$^{-1}$.
Note that $F_{\lambda 1455}^{\rm obs}$ is the same as
$F_{1455}^{\rm obs}= 3.0 \times 
10^{-12}$~erg~cm$^{-2}$~s$^{-1}$~~\AA$^{-1}$ in Section 
\ref{qu_vul_co_nova2}.
The two distance-reddening relations 
cross each other at $E(B-V) \approx 0.55$ and 
$d \approx 2.4$~kpc.  These two values are consistent
with our estimates in Section \ref{reddening_distance_qu_vul}.


\begin{figure}
\epsscale{1.15}
\plotone{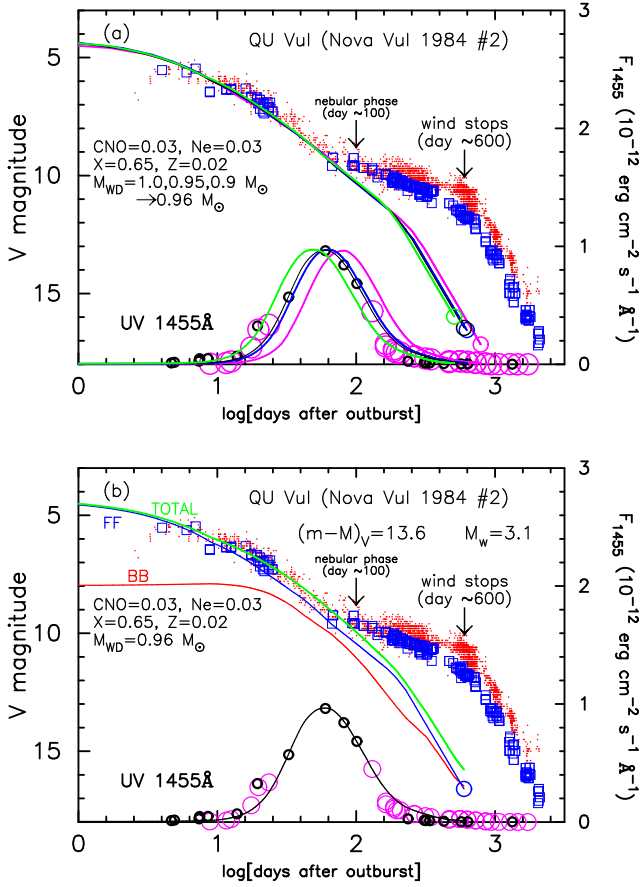}
\caption{
Similar to Figure \ref{all_mass_qu_vul_x55z02o10ne03_absolute_mag},
but for Ne nova 3.  (a) We plot four model light curves of 
$0.9~M_\sun$ (magenta solid), 0.95 (blue solid), 0.96 (black thin solid),
and 1.0 (green solid line) WDs.  Optical model light curves
are calculated in terms of free-free emission while UV~1455~\AA\  light
curves are calculated in terms of blackbody emission.  We chose
the $M_{\rm WD} = 0.96~M_\sun$ model as the best-fit model for a fine
mass-grid in steps of $0.01~M_\sun$ because our model light curve
for this WD fits the UV 1455~\AA\  observation well.
The free-free model light curves are arbitrarily shifted in the vertical
direction because the absolute magnitudes
of free-free model light curves are not yet calibrated for Ne nova 3. 
(b) We plot three model light curves of blackbody (red solid line labeled
``BB''), free-free (blue solid line labeled ``FF''), and total flux
(green solid line labeled ``TOTAL'') for the $0.96~M_\sun$ WD.
Assuming that $(m-M)_V=13.6$, we calibrated the absolute magnitudes
of the free-free emission light curves in such a way that the total $V$
flux fits the observed visual and $V$ light curves well.
Then, we obtained the absolute magnitude of $M_{\rm w}=3.1$ at the end of
the free-free model light curve (open circle at the right end of
the blue solid line).  Winds end approximately 600 days after the outburst
for the $0.96~M_\sun$ WD.
\label{all_mass_qu_vul_m0960_x65z02o03ne03_compsite}}
\end{figure}

\subsubsection{Ne nova 3}
\label{qu_vul_ne_nova3}
     Figure \ref{all_mass_qu_vul_m0960_x65z02o03ne03_compsite}(a)
shows light curve fittings of our model UV 1455~\AA\  flux with 
the {\it IUE} data and our free-free emission flux with the optical
data for 0.9, 0.95, 0.96, and $1.0~M_\sun$ WDs.  We chose the 
$M_{\rm WD} = 0.96~M_\sun$ model as the best-fit model for a fine
mass-grid in steps of $0.01~M_\sun$ because our model light curve
for this WD fits the UV 1455~\AA\  observation well.
The free-free model light curves are arbitrarily shifted in the vertical
direction, and their absolute magnitudes are determined below.

We plot the distance-reddening relation calculated from our UV 1455~\AA\
flux fitting (magenta solid line labeled ``UV 1455 ~\AA\  ''), i.e.,
Equation (\ref{qu_vul_uv1455_fit_eq2}) with
$F_{\lambda 1455}^{\rm mod}= 11.7 \times 
10^{-12}$~erg cm$^{-2}$ s$^{-1}$ ~\AA$^{-1}$,
as the calculated model flux for the upper bound of Figure
\ref{all_mass_qu_vul_m0960_x65z02o03ne03_compsite}(a).
This distance-reddening relation crosses the vertical line of
$E(B-V)=0.55$ at a distance of $d \approx 2.4$~kpc in Figure
\ref{qu_vul_distance_reddening_x35_x55_x55_x65_4figure}(d), which is
consistent with the distance estimate of $d=2.4$~kpc in
Section \ref{reddening_distance_qu_vul}.

Unlike the UV~1455~\AA\  blackbody flux, our free-free model light curves
of Ne Nova 3 composition are not yet calibrated.
The $V$ light curves in Figure
\ref{all_mass_qu_vul_m0960_x65z02o03ne03_compsite}(a)
are shifted up and down to fit them with 
the observed $V$ because the proportionality constant $C$ 
in Equation (\ref{free-free_calculation_original}) is not yet
determined for Ne nova 3.
We determine the absolute magnitude of each light curve from
the distance modulus of QU~Vul as follows.

First, we calculate the $0.96~M_\sun$ WD model
and obtain the blackbody light curve in the $V$ band
(red solid line labeled ``BB''), as shown in Figure 
\ref{all_mass_qu_vul_m0960_x65z02o03ne03_compsite}(b).
Here, we adopt a distance modulus of $(m-M)_V=13.6$.
With a trial value of $C$, we obtain the absolute
magnitude of the free-free model light curve (blue solid line
labeled ``FF'').  The total flux (green solid line labeled ``TOTAL'')
is the sum of these two fluxes.  This total $V$ magnitude
light curve generally does not fit the observed $V$.
Then, we change the proportionality constant $C$ and calculate
the light curve until the total $V$ flux approaches the observed data.
Figure \ref{all_mass_qu_vul_m0960_x65z02o03ne03_compsite}(b)
shows our final model.  We obtain $m_{\rm w}=16.7$
from Figure \ref{all_mass_qu_vul_m0960_x65z02o03ne03_compsite}(b),
where $m_{\rm w}$ is the apparent magnitude of
free-free emission light curve at the end of the wind phase
(open circle at the end of the blue solid line labeled ``FF'').
Then, we obtain $M_{\rm w}= m_{\rm w} - (m-M)_V= 16.7 - 13.6=3.1$,
where $M_{\rm w}$ is the absolute magnitude of the free-free model
light curve at the end of the wind phase.
Thus, the proportionality constant can be specified as $M_{\rm w}=3.1$
for the $0.96~M_\sun$ WD model of QU~Vul.

Using Equation (\ref{real_timescale_flux}),
we obtain the absolute
magnitudes of free-free light curves for the other WD masses with
the chemical composition of Ne nova 3 (Appendix 
\ref{time_normalized_free_free}).  We can specify the absolute
magnitude by the value of $M_{\rm w}$ and are listed
in Table \ref{light_curves_of_novae_ne3}
for $0.7$--$1.3~M_\sun$ WDs in steps of $0.05~M_\sun$.
Once the calibrated free-free model light curves are given
by the absolute magnitude of $M_{\rm w}$ at the end point of the
wind phase, we calculate the absolute magnitudes of the
total $V$ flux (free-free plus blackbody emission).


\begin{deluxetable*}{llllllllllllll}
\tabletypesize{\scriptsize}
\tablecaption{Free-free Light Curves of Neon Novae 3\tablenotemark{a}
\label{light_curves_of_novae_ne3}}
\tablewidth{0pt}
\tablehead{
\colhead{$m_{\rm ff}$} &
\colhead{0.7$M_\sun$} &
\colhead{0.75$M_\sun$} &
\colhead{0.8$M_\sun$} &
\colhead{0.85$M_\sun$} &
\colhead{0.9$M_\sun$} &
\colhead{0.95$M_\sun$} &
\colhead{1.0$M_\sun$} &
\colhead{1.05$M_\sun$} &
\colhead{1.1$M_\sun$} &
\colhead{1.15$M_\sun$} &
\colhead{1.2$M_\sun$} &
\colhead{1.25$M_\sun$} &
\colhead{1.3$M_\sun$} \\
\colhead{(mag)} &
\colhead{(day)} &
\colhead{(day)} &
\colhead{(day)} &
\colhead{(day)} &
\colhead{(day)} &
\colhead{(day)} &
\colhead{(day)} &
\colhead{(day)} &
\colhead{(day)} &
\colhead{(day)} &
\colhead{(day)} &
\colhead{(day)} &
\colhead{(day)} 
}
\startdata
  3.000     & 0.0 & 0.0 & 0.0 & 0.0 & 0.0 & 0.0 & 0.0 & 0.0 & 0.0 & 0.0 & 0.0 & 0.0 & 0.0 \\

  3.250     &  3.180     &  2.570     &  2.110     &  1.550     &  1.310     &  1.148     & 0.9900     & 0.8280     & 0.7320     & 0.6630     & 0.5970     & 0.6520     & 0.5580   \\
  3.500     &  6.390     &  5.500     &  4.220     &  3.250     &  2.660     &  2.320     &  2.010     &  1.681     &  1.468     &  1.311     &  1.191     &  1.362     &  1.119   \\
  3.750     &  9.650     &  9.050     &  6.610     &  5.370     &  4.200     &  3.577     &  3.060     &  2.555     &  2.228     &  1.970     &  1.770     &  2.162     &  1.694   \\
  4.000     &  14.30     &  13.37     &  9.080     &  7.490     &  5.840     &  4.929     &  4.140     &  3.454     &  2.992     &  2.645     &  2.368     &  2.952     &  2.300   \\
  4.250     &  21.59     &  18.65     &  12.65     &  9.910     &  7.530     &  6.324     &  5.310     &  4.384     &  3.778     &  3.321     &  2.987     &  3.632     &  2.945   \\
  4.500     &  30.51     &  24.36     &  16.99     &  12.39     &  9.270     &  7.762     &  6.570     &  5.444     &  4.613     &  4.020     &  3.620     &  4.352     &  3.593   \\
  4.750     &  42.18     &  30.92     &  21.80     &  15.60     &  11.79     &  9.817     &  7.870     &  6.604     &  5.605     &  4.796     &  4.352     &  5.172     &  4.294   \\
  5.000     &  54.67     &  38.87     &  26.98     &  19.26     &  14.74     &  12.11     &  9.420     &  7.814     &  6.697     &  5.830     &  5.211     &  6.052     &  5.024   \\
  5.250     &  68.18     &  48.77     &  33.15     &  23.41     &  17.89     &  14.56     &  11.48     &  9.304     &  7.867     &  6.870     &  6.087     &  6.962     &  5.724   \\
  5.500     &  84.88     &  59.67     &  40.32     &  28.13     &  21.25     &  17.28     &  13.66     &  10.97     &  9.115     &  7.900     &  6.959     &  7.932     &  6.414   \\
  5.750     &  103.1     &  72.33     &  48.06     &  34.01     &  24.84     &  20.16     &  15.97     &  12.76     &  10.46     &  8.950     &  7.869     &  8.972     &  7.134   \\
  6.000     &  123.1     &  86.68     &  57.12     &  40.51     &  29.61     &  23.44     &  18.44     &  14.69     &  11.93     &  10.08     &  8.839     &  10.06     &  7.904   \\
  6.250     &  145.1     &  102.9     &  68.08     &  47.65     &  34.80     &  27.42     &  21.19     &  16.79     &  13.56     &  11.34     &  9.879     &  11.22     &  8.714   \\
  6.500     &  168.5     &  118.8     &  79.98     &  55.82     &  40.47     &  31.77     &  24.37     &  19.20     &  15.36     &  12.75     &  10.97     &  12.45     &  9.574   \\
  6.750     &  191.1     &  135.9     &  92.60     &  64.87     &  46.71     &  36.59     &  27.85     &  21.89     &  17.33     &  14.32     &  12.13     &  13.75     &  10.48   \\
  7.000     &  215.8     &  153.7     &  104.6     &  74.78     &  53.65     &  41.87     &  31.74     &  24.83     &  19.53     &  16.06     &  13.41     &  15.13     &  11.44   \\
  7.250     &  244.3     &  172.6     &  117.7     &  84.16     &  61.27     &  47.74     &  36.08     &  27.89     &  21.89     &  17.88     &  14.85     &  16.58     &  12.47   \\
  7.500     &  278.3     &  193.4     &  132.9     &  94.16     &  68.74     &  54.06     &  40.84     &  31.26     &  24.43     &  19.72     &  16.41     &  18.13     &  13.55   \\
  7.750     &  316.5     &  219.9     &  149.8     &  106.0     &  76.73     &  60.26     &  45.88     &  34.93     &  27.22     &  21.72     &  18.00     &  19.76     &  14.70   \\
  8.000     &  357.2     &  250.4     &  168.8     &  119.5     &  86.38     &  67.04     &  51.14     &  38.94     &  30.24     &  23.93     &  19.68     &  21.50     &  15.92   \\
  8.250     &  403.0     &  282.8     &  190.6     &  134.6     &  97.58     &  75.70     &  56.87     &  43.30     &  33.51     &  26.51     &  21.60     &  23.33     &  17.20   \\
  8.500     &  455.4     &  316.2     &  215.3     &  152.5     &  110.3     &  85.49     &  64.26     &  48.49     &  37.35     &  29.42     &  23.97     &  25.28     &  18.57   \\
  8.750     &  509.7     &  354.5     &  240.9     &  172.6     &  124.8     &  96.65     &  72.72     &  54.93     &  42.13     &  33.06     &  26.68     &  27.34     &  20.01   \\
  9.000     &  566.7     &  395.7     &  269.6     &  193.9     &  141.2     &  109.3     &  82.25     &  62.15     &  47.45     &  37.32     &  29.95     &  29.52     &  21.54   \\
  9.250     &  622.9     &  430.3     &  302.3     &  217.4     &  158.7     &  123.4     &  92.97     &  70.31     &  53.63     &  42.09     &  33.61     &  31.83     &  23.16   \\
  9.500     &  679.8     &  470.0     &  336.8     &  244.1     &  178.4     &  139.3     &  105.0     &  79.44     &  60.58     &  47.43     &  37.94     &  34.28     &  24.88   \\
  9.750     &  744.4     &  515.2     &  375.2     &  267.3     &  200.7     &  157.3     &  118.3     &  89.79     &  68.43     &  53.39     &  42.75     &  36.87     &  26.70   \\
  10.00     &  792.7     &  566.5     &  405.6     &  289.7     &  221.9     &  175.7     &  133.3     &  101.5     &  77.44     &  60.53     &  48.17     &  39.62     &  28.63   \\
  10.25     &  841.1     &  612.5     &  436.4     &  315.1     &  244.8     &  189.9     &  149.3     &  114.6     &  87.50     &  68.51     &  54.30     &  42.53     &  30.67   \\
  10.50     &  892.3     &  649.5     &  470.8     &  343.6     &  263.0     &  205.6     &  164.1     &  126.2     &  97.41     &  76.51     &  61.02     &  45.61     &  32.83   \\
  10.75     &  946.6     &  688.7     &  499.8     &  375.4     &  281.7     &  223.1     &  179.3     &  138.8     &  107.1     &  82.96     &  67.43     &  48.88     &  35.12   \\
  11.00     &  1004.     &  730.3     &  530.5     &  399.3     &  302.4     &  242.5     &  191.9     &  149.1     &  116.4     &  89.97     &  73.00     &  52.33     &  37.55   \\
  11.25     &  1065.     &  774.3     &  563.1     &  423.9     &  321.3     &  261.6     &  205.7     &  159.8     &  124.8     &  97.52     &  78.94     &  56.00     &  40.11   \\
  11.50     &  1130.     &  820.9     &  597.6     &  449.9     &  341.3     &  277.2     &  220.0     &  171.4     &  133.9     &  105.7     &  85.33     &  59.88     &  42.84   \\
  11.75     &  1198.     &  870.2     &  634.2     &  477.6     &  362.5     &  293.7     &  233.8     &  182.0     &  142.9     &  114.0     &  91.96     &  63.99     &  45.72   \\
  12.00     &  1270.     &  922.5     &  672.9     &  506.8     &  384.9     &  311.3     &  248.4     &  193.3     &  151.9     &  121.0     &  97.97     &  68.35     &  48.77   \\
  12.25     &  1347.     &  977.9     &  714.0     &  537.8     &  408.7     &  329.8     &  263.9     &  205.2     &  161.7     &  128.5     &  104.0     &  72.96     &  52.01   \\
  12.50     &  1428.     &  1037.     &  757.4     &  570.6     &  433.9     &  349.5     &  280.3     &  217.7     &  172.3     &  136.4     &  110.4     &  77.85     &  55.43   \\
  12.75     &  1514.     &  1099.     &  803.5     &  605.4     &  460.6     &  370.3     &  297.7     &  231.1     &  183.5     &  144.8     &  117.1     &  83.02     &  59.07   \\
  13.00     &  1606.     &  1165.     &  852.2     &  642.2     &  488.8     &  392.4     &  316.1     &  245.2     &  194.4     &  153.7     &  124.3     &  88.50     &  62.91   \\
  13.25     &  1702.     &  1234.     &  903.8     &  681.2     &  518.7     &  415.7     &  335.6     &  260.2     &  206.0     &  163.1     &  131.8     &  94.30     &  66.98   \\
  13.50     &  1804.     &  1308.     &  958.6     &  722.5     &  550.4     &  440.5     &  356.3     &  276.0     &  218.3     &  173.1     &  139.9     &  100.5     &  71.29   \\
  13.75     &  1913.     &  1387.     &  1017.     &  766.3     &  584.0     &  466.7     &  378.1     &  292.8     &  231.2     &  183.6     &  148.3     &  107.0     &  75.87   \\
  14.00     &  2027.     &  1469.     &  1078.     &  812.7     &  619.6     &  494.4     &  401.3     &  310.6     &  245.0     &  194.8     &  157.3     &  113.9     &  80.71   \\
  14.25     &  2149.     &  1557.     &  1143.     &  861.8     &  657.2     &  523.8     &  425.8     &  329.5     &  259.6     &  206.7     &  166.9     &  121.2     &  85.83   \\
  14.50     &  2278.     &  1650.     &  1212.     &  913.8     &  697.2     &  555.0     &  451.8     &  349.4     &  275.0     &  219.2     &  176.9     &  128.9     &  91.26   \\
  14.75     &  2414.     &  1749.     &  1285.     &  968.9     &  739.4     &  588.0     &  479.4     &  370.6     &  291.4     &  232.5     &  187.6     &  137.1     &  96.98   \\
  15.00     &  2559.     &  1853.     &  1362.     &  1027.     &  784.2     &  623.0     &  508.5     &  393.0     &  308.7     &  246.6     &  199.0     &  145.8     &  103.1   \\

\hline
X-ray\tablenotemark{b}
 & 7229  & 5205 & 3791  & 2715 & 1958   & 1468   & 1090   & 749   & 511   & 346   & 233   & 136  & 71.5 \\
\hline
$\log f_{\rm s}$\tablenotemark{c}
 & 0.55  & 0.44 & 0.33  & 0.22 & 0.12   & 0.02   & $-0.06$   & $-0.18$   & $-0.28$   & $-0.40$   & $-0.49$   & $-0.62$  & $-0.78$ \\
\hline
$M_{\rm w}$\tablenotemark{d}
 & 5.5 & 4.9  & 4.4 & 3.9   & 3.5   & 3.1   & 2.7   & 2.3   & 1.9   & 1.6   & 1.3  & 1.0 & 0.7 
\enddata
\tablenotetext{a}{Chemical composition of the envelope is assumed
to be that of Ne nova 3 in Table \ref{chemical_composition_model}.}
\tablenotetext{b}{Duration of supersoft X-ray phase in units of days.}
\tablenotetext{c}{Stretching factor against QU~Vul
 UV 1455~\AA\  observation
in Figure \ref{all_mass_qu_vul_x65z02o03ne03_calib_universal}.}
\tablenotetext{d}{Absolute magnitudes at the bottom point in Figure
\ref{all_mass_qu_vul_x65z02o03ne03_absolute_mag} by assuming
$(m-M)_V = 13.6$  (QU~Vul).}
\end{deluxetable*}

\subsection{Dependence on chemical composition}
\label{dependence_chemical_composition}
Our light curve fittings yield different results for four different
chemical compositions.  For example, the chemical composition of
CO nova 2 ($X=0.35$) results in a WD mass of $0.82~M_\sun$,
whereas that of Ne nova 2 ($X=0.55$) yields $0.9~M_\sun$.  
A nova on a less massive WD generally evolves more slowly.
On the other hand, a nova evolves faster at a lower hydrogen content $X$,
even if the WD mass is the same \citep[see][]{kat94h, hac01kb}.
Thus, a combination of smaller $X$ and smaller WD mass ultimately
results in a similar timescale.  The distance modulus is affected
by the WD mass.  The wind mass-loss rate is smaller for a less
massive WD \citep{kat94h, hac01kb}.  Thus, the lower the wind
mass-loss rate the fainter the free-free emission.  As a result, 
the fainter the total $V$ light curve.  This is the reason that $(m-M)_V=13.4$
for CO nova 2 ($X=0.35$), i.e., slightly smaller than 
$(m-M)_V=13.6$ for the higher value of $X=0.55$.

For the chemical compositions of Ne nova 2 ($X_{\rm CNO}=0.10$)
and CO nova 4 ($X_{\rm CNO}=0.20$),
the hydrogen content of $X=0.55$ is the same, but the CNO abundance,
$X_{\rm CNO}$, is different.
The CNO abundance is relevant to the nuclear
burning rate, and a lower value of $X_{\rm CNO}=0.10$ makes
the evolution slower.  This requires a more massive
WD ($0.9~M_\sun$) than the $0.86~M_\sun$ WD of CO nova 4,
as shown in Figures
\ref{all_mass_qu_vul_x55z02o10ne03_absolute_mag}
and \ref{all_mass_qu_vul_x55z02c10o10_absolute_mag}.
A more massive WD generally blows stronger winds, resulting
in a brighter free-free emission and a brighter
total $V$ light curve.  However, the wind mass loss rate of 
the $0.9~M_\sun$ WD for Ne nova 2 is very similar to that of the 
$0.86~M_\sun$ WD for CO nova 4.   This is partly because
the larger carbon content contributes to the opacity and accelerates
the wind mass loss more strongly for CO nova 4.
This is the reason that the value of $(m-M)_V=13.6$ is the same for
CO nova 4 and Ne nova 2.  Note that small enrichment of
neon with unchanged hydrogen mass fraction $X$
and CNO mass abundance $X_{\rm CNO}$
affects the nova light curves very little in our model light curves
because neon is not relevant to either nuclear burning (the CNO cycle)
or the opacity \citep[e.g.,][]{kat94h,hac06kb,hac10k}.
In this sense, we cannot distinguish the neon abundance from our 
light curve analysis.

The consistency between the reddening and the distance could support
a lower degree of mixing, $\eta_{\rm mix}\sim0.25$, i.e., $X\sim0.55$,
rather than a higher degree of mixing, $\eta_{\rm mix}\sim1.0$,
i.e., $X\sim0.35$, as shown in
Figure \ref{qu_vul_distance_reddening_x35_x55_x55_x65_4figure}.
\citet{sch02} presented a very low degree of mixing, 
$X\sim0.63$, i.e., $\eta_{\rm mix}\sim0.11$ (11\%), as listed
in Table \ref{chemical_abundance_neon_novae}.  We prefer 
Ne nova 3 among six chemical compositions listed in Table
\ref{chemical_composition_model}.

\subsection{Summary of QU~Vul}
To summarize, our model light curves fit the optical $V$ and UV~1455~\AA\  
fluxes of QU~Vul well.  Our fitting results give reasonable
values for the distance, $d=2.4$~kpc, and the reddening, $E(B-V)=0.55$,
but give relatively small WD masses of $M_{\rm WD}=0.82$ -- $0.96~M_\sun$.
The distance estimate is consistent with the arithmetic mean of various
estimates based on the expansion parallax method (Section 
\ref{reddening_distance_qu_vul}).
As mentioned in Section \ref{introduction}, natal ONe WDs
with a CO mantle are more massive than $\sim1.0~M_\sun$.  
Such a mass range of $0.82$ -- $0.96~M_\sun$
suggests that the WD in QU~Vul would have lost 
a mass of $\sim 0.1~M_\sun$ or more since its birth.
Our best-fit model of Ne nova 3 gives $(m-M)_V=13.6$, $E(B-V)=0.55$, 
$d=2.4$~kpc, $M_{\rm WD}=0.96~M_\sun$, and
$M_{\rm env}=2.44\times10^{-5}~M_\sun$ at optical maximum,
where $M_{\rm env}$ is the mass of the hydrogen-rich
envelope, as summarized in Table \ref{nova_parameters_results}.
The envelope mass at the optical maximum approximately represents
the ignition mass of the nova.

     We calculated the contribution of photospheric emission to the $V$
magnitude of QU~Vul as well as that of free-free emission.  Figures
\ref{all_mass_qu_vul_x55z02o10ne03_absolute_mag}(b),
\ref{all_mass_qu_vul_x35z02c10o20_absolute_mag}(b) --
\ref{all_mass_qu_vul_m0960_x65z02o03ne03_compsite}(b)
show photospheric emission light curves, which are calculated
on the basis of the blackbody emission at the pseudophotosphere.
The blackbody emission flux is always smaller than
the free-free emission flux, but it contributes to the total
brightness by 0.4 -- 0.8 mag at most.  The overall features and shape of 
the total $V$ light curves are essentially the same as those of
the free-free light curve.

\subsection{Comparison with previous results}
\label{comparison_previous_qu_vul}
\citet{wan99} presented yields of nuclear burning on the ONe core
and estimated the WD mass from fitting their results with
the abundance pattern observed by \citet{sai92}.
They suggested the WD mass of $1.05$--$1.1~M_\sun$.  
\citet{dow13} also estimated the WD mass of QU~Vul by comparing their
theoretical yields of nuclear burning (abundance ratio)
with the observed ones.  They obtained $M_{\rm WD}<1.2~M_\sun$ for QU~Vel.
Downen et al.'s value is consistent with 
our new estimate of $0.96~M_\sun$ for the chemical composition of
Ne nova 3, but Wanajo et al.'s value is $\sim0.1$--$0.2~M_\sun$
more massive than ours.  It should be noted that, however,
Wanajo et al.'s model is based on a one-zone model
of a hydrogen-rich envelope and this is too simplified to correctly
predict the abundance ratio.


\begin{figure}
\epsscale{1.15}
\plotone{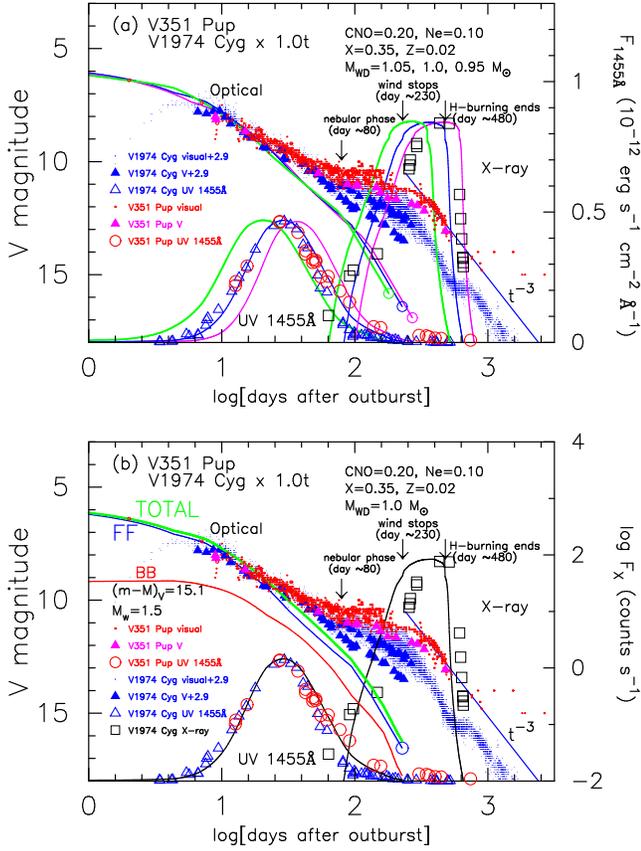}
\caption{
Same as Figure
\ref{all_mass_qu_vul_m0960_x65z02o03ne03_compsite}, but for V351~Pup.
(a) Model light curves of $0.95~M_\sun$ (magenta solid lines), 
$1.0~M_\sun$ (blue solid lines), and $1.05~M_\sun$ (green solid lines)
WDs with the chemical composition of Ne Nova 1
as well as visual (small red open circles),
$V$ (magenta filled triangles), and UV~1455 ~\AA\  
(large red open circles) light curves of V351~Pup.
We also added $V$ (small blue dots) and UV~1455 ~\AA\  
(large open blue triangles), and X-ray (large black open squares)
light curves of V1974~Cyg, because the timescale of V1974~Cyg is almost
the same as that of V351~Pup.  Optical model light curves are calculated
in terms of free-free emission while UV~1455~\AA\  and X-ray light curves
are calculated in terms of blackbody emission.  We chose the $1.0~M_\sun$ WD
as the best-fit model because the UV~1455~\AA\  model light curve fits
the observation very well.  The free-free model light curves are
arbitrarily shifted in the vertical direction because the absolute
magnitudes of free-free model light curves are not yet calibrated
for Ne nova 1. 
(b) Assuming that $(m-M)_V=15.1$ and $M_{\rm w}=1.5$ mag for
the $1.0~M_\odot$ free-free emission model light curve,
we plot three model light curves of the blackbody (red solid line),
free-free (blue solid), and total flux (green solid) of
free-free plus blackbody.  The UV~1455~\AA\  and X-ray model light curves
(black solid lines) are also added.
\label{all_mass_v351_pup_v1974_cyg_x35z02o20ne10}}
\end{figure}


\begin{figure}
\epsscale{0.85}
\plotone{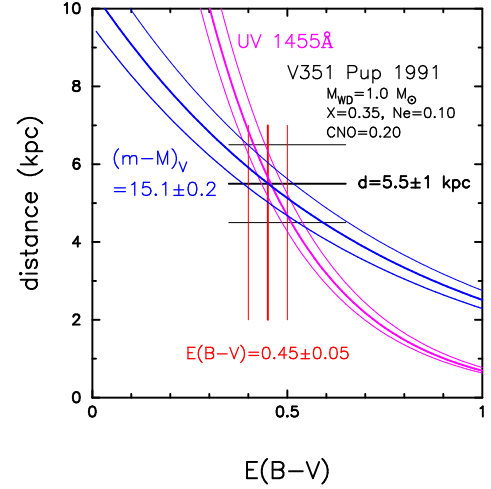}
\caption{
Various distance-reddening relations toward V351~Pup, whose
galactic coordinates are $(l, b)=(252\fdg7225, -0\fdg7329)$. 
The blue solid line flanking two thin solid lines shows 
the distance-reddening relation calculated from Equation 
(\ref{qu_vul_distance_modulus_eq1}) with $(m-M)_V=15.1\pm0.2$.
The red vertical solid line flanking two thin solid lines shows
the reddening estimate of $E(B-V)=0.45\pm0.05$.
The black horizontal solid line flanking two thin solid lines show 
the distance estimate of $d=5.5\pm1$~kpc.
We also add another distance-reddening relation of V351~Pup
for the UV 1455~\AA\   flux fitting of a $1.0~M_\sun$ WD with Ne nova 1,
that is, the magenta solid lines calculated from Equation
(\ref{qu_vul_uv1455_fit_eq2}).  See text for more detail.
\label{v351_pup_distance_reddening_x35z02o20ne10}}
\end{figure}


\begin{figure}
\epsscale{1.15}
\plotone{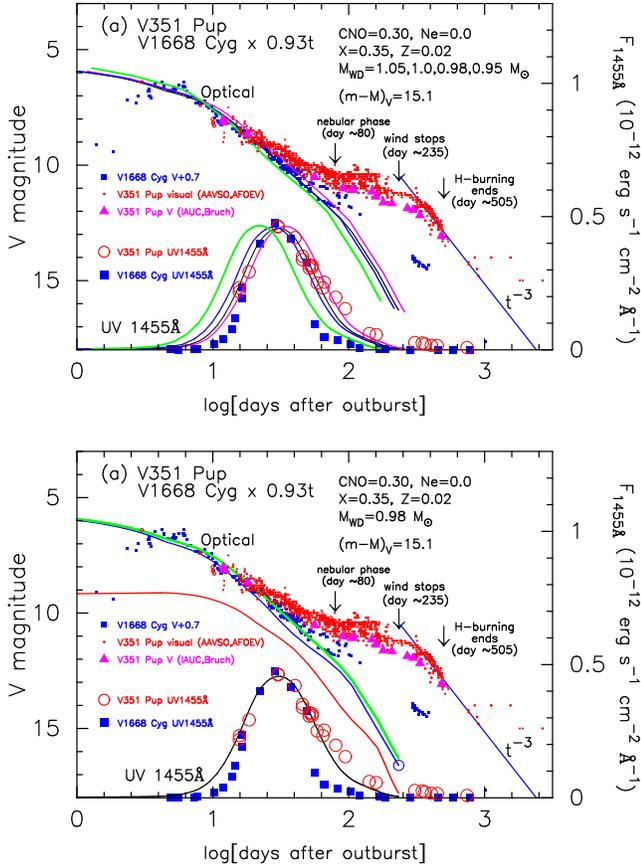}
\caption{
Same as Figure 
\ref{all_mass_qu_vul_x55z02o10ne03_absolute_mag},
but for V351~Pup and the chemical composition of CO nova 2.
(a) Model light curves of $0.95~M_\sun$ (magenta solid lines), 
$0.98~M_\sun$ (black thin solid lines), $1.0~M_\sun$ (blue solid lines), 
and $1.05~M_\sun$ (green solid lines) WDs for $(m-M)_V=15.1$.
We added $V$ (small blue filled squares) and UV~1455 ~\AA\  
(large blue filled squares) light curves of V1668~Cyg.  The timescale
of V1668~Cyg is squeezed by a factor of 0.93.
We plot the total $V$ flux in optical.
(b) Model light curves of the $0.98~M_\sun$ WD for the same distance modulus
of $(m-M)_V=15.1$.  Optically thick winds and hydrogen shell burning 
end approximately 235 days and 505 days after the outburst,
respectively, for the $0.98~M_\sun$ WD.
\label{all_mass_v351_pup_v1668_cyg_x35z02c10o20}}
\end{figure}


\begin{figure}
\epsscale{1.15}
\plotone{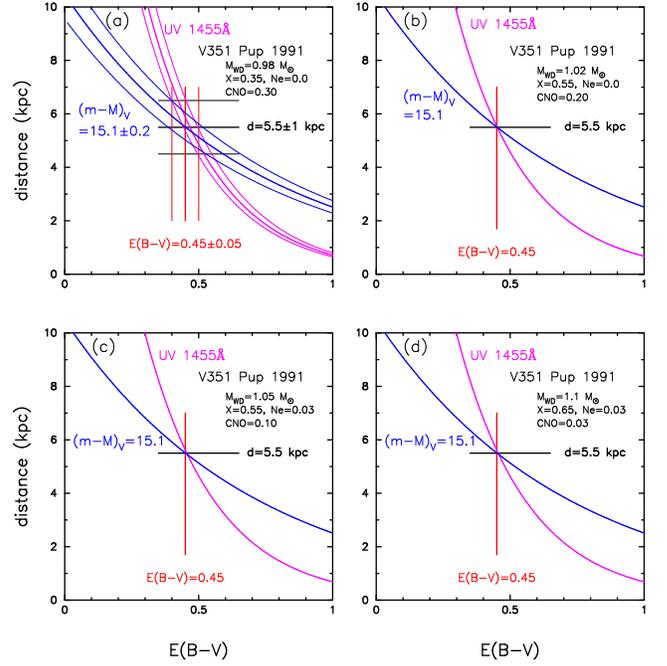}
\caption{
Same as Figure \ref{qu_vul_distance_reddening_x35_x55_x55_x65_4figure},
but for V351~Pup.  Two distance-reddening relations toward V351~Pup
are calculated from the $V$ (blue solid lines) and UV~1455~\AA\   
(magenta solid lines) light curve fittings of
(a) the $0.98~M_\sun$ WD model with CO nova 2,
(b) $1.02~M_\sun$ WD model with CO nova 4,
(c) $1.05~M_\sun$ WD model with Ne nova 2, and
(d) $1.1~M_\sun$ WD model with Ne nova 3.
Two other constraints are also plotted: one is the distance estimate of
$d=5.5\pm1$~kpc and the other is the reddening estimate 
of $E(B-V)=0.45\pm0.05$.
\label{v351_pup_distance_reddening_x35_x55_x55_x65_4comp}}
\end{figure}


\begin{figure}
\epsscale{1.15}
\plotone{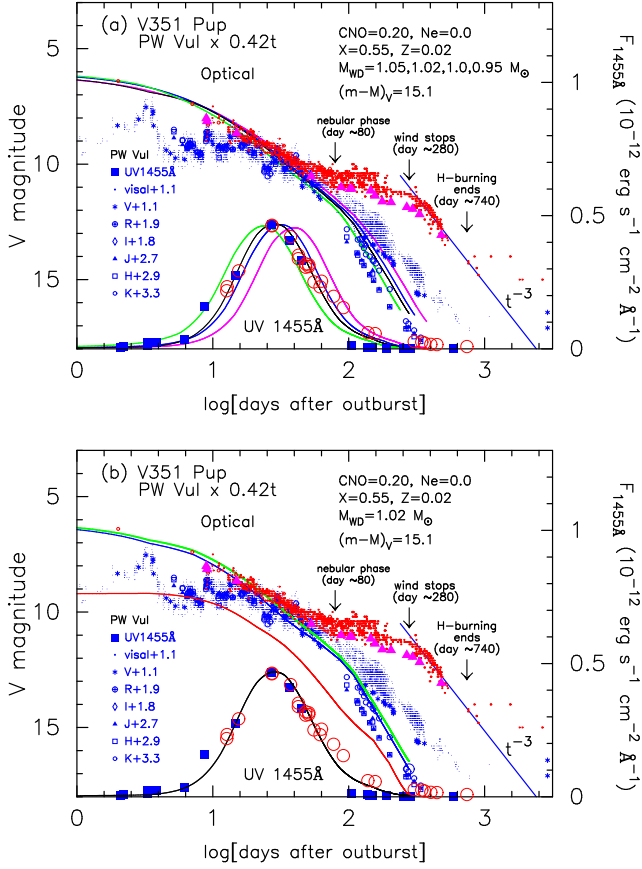}
\caption{
Same as Figure \ref{all_mass_v351_pup_v1668_cyg_x35z02c10o20},
but for CO nova 4.
(a) Assuming that $(m-M)_V=15.1$, we plot four model light curves of
$0.95~M_\sun$ (magenta solid line), $1.0~M_\sun$ (blue solid line),
$1.02~M_\sun$ (black thin solid line),
and $1.05~M_\sun$ (green solid line) WDs.
We added the $V$ (blue asterisks) and UV~1455 ~\AA\  
(large blue filled squares) light curves of PW~Vul.  The timescale of 
PW~Vul is squeezed by a factor of 0.42.
(b) Assuming again that $(m-M)_V=15.1$, we plot the blackbody emission
(red solid line), free-free emission (blue solid line), and
total emission (green thick solid line) light curves of $1.02~M_\sun$ WD.
Optically thick winds and hydrogen shell burning end approximately
280 days and 740 days after the outburst, respectively, 
for the $1.02~M_\sun$ WD. 
\label{all_mass_v351_pup_pw_vul_x55z02c10o10}}
\end{figure}


\begin{figure}
\epsscale{1.15}
\plotone{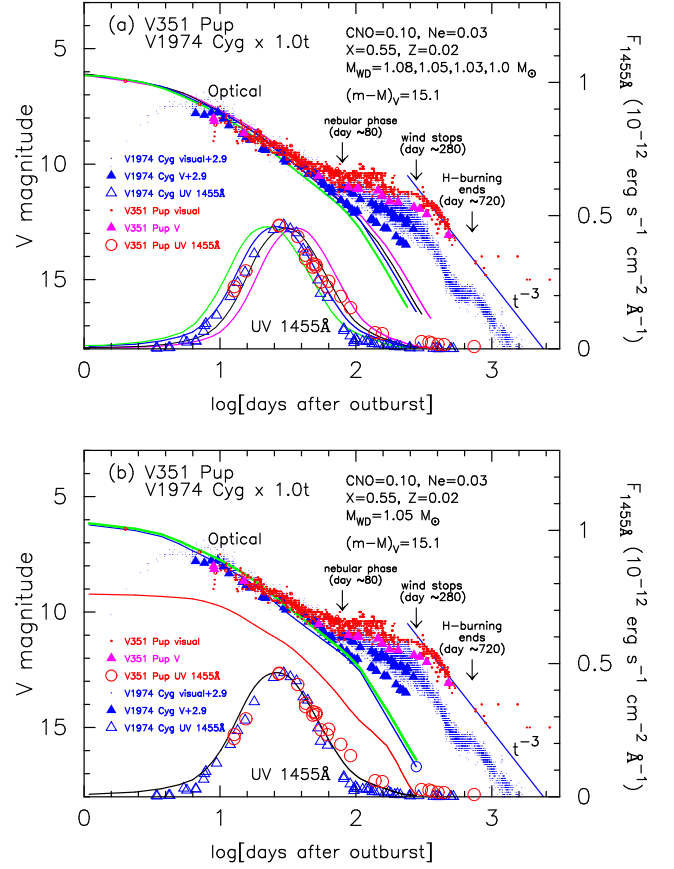}
\caption{
Same as Figure \ref{all_mass_v351_pup_v1668_cyg_x35z02c10o20},
but for Ne nova 2.
(a) Assuming that $(m-M)_V=15.1$, we plot four model light curves of
$1.0~M_\sun$ (magenta solid line), $1.03~M_\sun$ (black thin solid line),
$1.05~M_\sun$ (blue solid line), and $1.08~M_\sun$ (green solid line) WDs.
We added the $V$ (blue asterisks) and UV~1455 ~\AA\  
(large blue filled squares) light curves of V1974~Cyg.
The timescale of V1974~Cyg is the same as that of V351~Pup.
(b) Assuming that $(m-M)_V=15.1$, we plot the blackbody emission
(red solid line), free-free emission (blue solid line), and total emission 
(green thick solid line) light curves of $1.05~M_\sun$ WD.
Optically thick winds and hydrogen shell burning end approximately
280 days and 720 days after the outburst, respectively, 
for the $1.05~M_\sun$ WD. 
\label{all_mass_v351_pup_v1974_cyg_x55z02o10ne03}}
\end{figure}


\begin{figure}
\epsscale{1.15}
\plotone{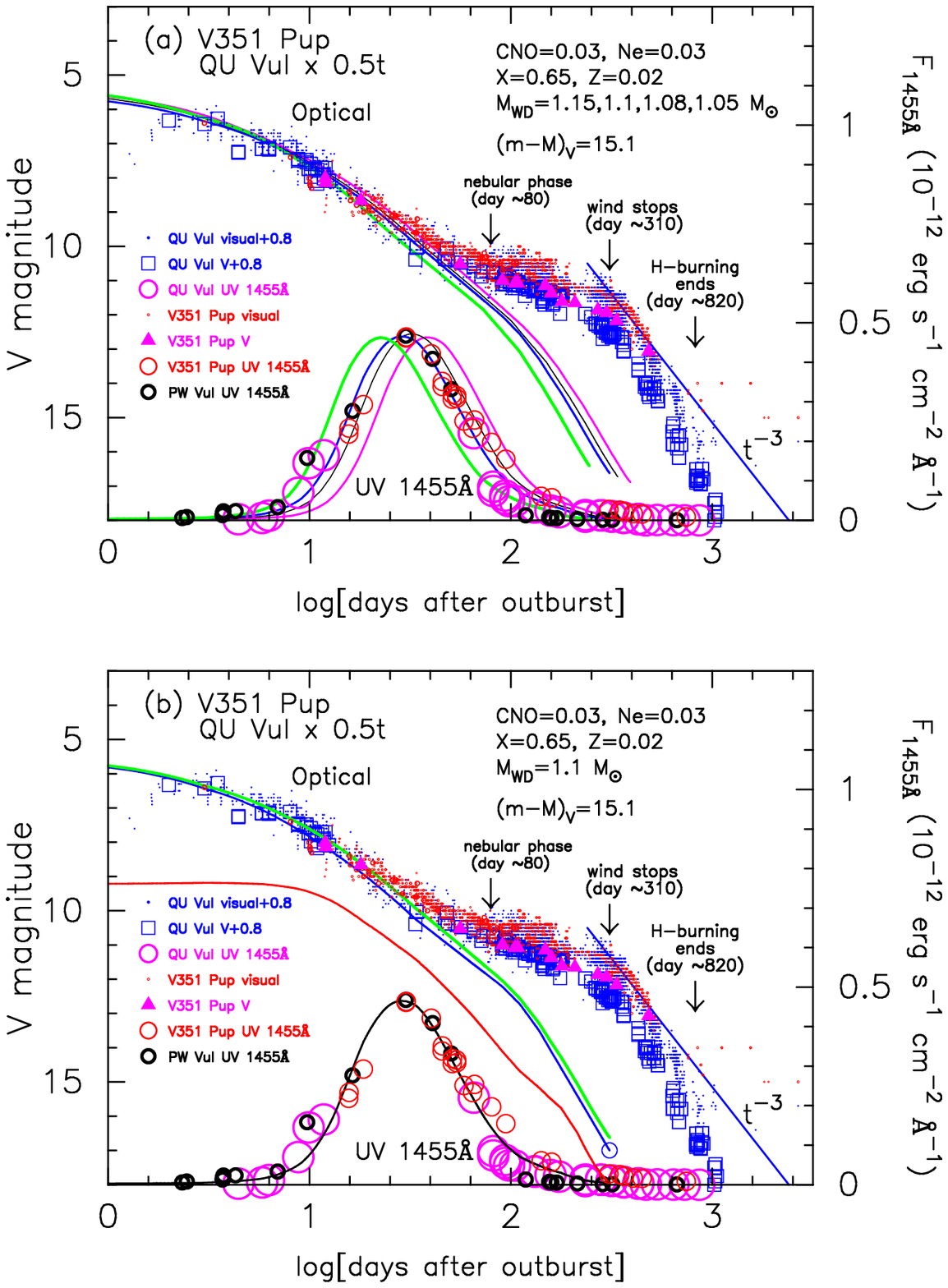}
\caption{
Same as Figure \ref{all_mass_v351_pup_v1668_cyg_x35z02c10o20},
but for Ne nova 3.   (a) Assuming that $(m-M)_V=15.1$,
we plot four model light curves of
$1.05~M_\sun$ (magenta solid line), $1.08~M_\sun$ (thin black solid line),
$1.1~M_\sun$ (blue solid line), and $1.15~M_\sun$ (green solid line) WDs.
We added the $V$ (blue open squares), visual (blue dots),
and UV~1455 ~\AA\  (large magenta open circles) light curves of
QU~Vul.  The timescale of QU~Vul is squeezed by a factor of 0.5.
We also added the UV~1455 ~\AA\  (black open circles) light curve
of PW~Vul, the same as that in Figure
\ref{all_mass_v351_pup_pw_vul_x55z02c10o10}.
(b) Assuming that $(m-M)_V=15.1$, we plot the blackbody emission
(red solid line), free-free emission (blue solid line), and total emission 
(green thick solid line) light curves of the $1.1~M_\sun$ WD.
Optically thick winds and hydrogen shell burning end approximately
310 days and 820 days after the outburst, respectively, 
for the $1.1~M_\sun$ WD. 
\label{all_mass_v351_pup_qu_vul_x65z02o03ne03}}
\end{figure}

\section{V351 Pup 1991}
\label{v351_pup}


V351 Pup was discovered on UT 1991 December 27 by P. Camilleri 
near maximum \citep{cam92}.  It reached $m_v=6.4$ at optical maximum 
\citep[e.g.,][]{dow00}. The light curves of V351~Pup are shown in Figure
\ref{all_mass_v351_pup_v1974_cyg_x35z02o20ne10}
together with those of V1974~Cyg.
Optical data are poor, and no $V$ magnitude data were
reported around the maximum.
Its decline rates were estimated to be $t_2=10$ and $t_3=26$ days
by \citet{dow00}, so V351~Pup belongs to the fast nova class.
\citet{sai96} reported that V351~Pup shows a pattern of enhanced abundances
(see Table \ref{chemical_abundance_neon_novae})
consistent with an ONe WD whose mass is close to the
Chandrasekhar limit.  An orbital period of $P_{\rm orb}=2.83$~hr
was obtained by \citet{wou01}.

\subsection{Reddening and distance}
\label{reddening_v351_pup}
The reddening toward V351~Pup was estimated as $E(B-V)=0.3\pm0.1$ 
by \citet{ori96} from the Balmer decrement observed in 1994 March
and as $E(B-V)=0.72\pm0.10$ by \citet{sai96} from the line ratio of 
\ion{He}{2}~$\lambda4686/$\ion{He}{2}~$\lambda1640$.
These two values are very different, and their
arithmetic mean is $E(B-V)= 0.51\pm0.1$.
In Figure \ref{color_color_diagram_qu_vul_v351_pup_v382_vel_v693_cra}(b)
(see also Section \ref{color_color_diagrams}),
we obtain $E(B-V)=0.45\pm0.05$ using
the $UBV$ color-color evolution method proposed by \citet{hac14k}.
This value is consistent with the above arithmetic mean.
Therefore, we adopt $E(B-V)=0.45\pm0.05$ in this paper.

The distance toward V351~Pup was estimated from the nebular expansion
parallax method by \citet{dow00} to be $d=2.7\pm0.10$~kpc 
with an expansion velocity of 1200~km~s$^{-1}$ and by \citet{ori96}
to be $d=4.7\pm0.6$~kpc with an expansion velocity of 2000~km~s$^{-1}$.
A distance modulus in the $V$ band of $(m-M)_{V,{\rm V351~Pup}}=15.1\pm0.2$
is obtained using the time-stretching method 
of nova light curves \citep{hac10k} described
in Appendix \ref{time_stretching_method_novae}.
Figure \ref{v351_pup_distance_reddening_x35z02o20ne10} shows
various distance-reddening relations toward V351~Pup.
Red vertical lines denote the reddening estimate of $E(B-V)=0.45\pm0.05$.
Blue lines show Equation (\ref{qu_vul_distance_modulus_eq1})
with $(m-M)_V=15.1\pm0.2$.
The two trends cross at $E(B-V)\sim0.45$ and $d\sim5.5$~kpc.
Our distance estimate is close to that given by \citet{ori96} but
twice larger than that by \citet{dow00}. 
In this paper, we adopt $E(B-V)=0.45\pm0.05$ and $d=5.5\pm1.0$~kpc.
In Section \ref{hr_diagrams}, we will discuss the nova outburst track
in the color-magnitude diagram and show that the track of V351~Pup 
almost overlaps that of a similar type of nova, V1974~Cyg, 
which supports the set of $E(B-V)=0.45$ and $(m-M)_V=15.1$ for V351~Pup.

\subsection{Light curve fitting for various chemical compositions}
\label{light_curve_fit_v351_pup}
\citet{sai96} estimated the chemical composition of the ejecta
from the {\it IUE} spectra, i.e., $X=0.37$, $Y=0.24$,
$X_{\rm CNO}=0.26$, and $X_{\rm Ne}=0.13$, as listed in Table
\ref{chemical_abundance_neon_novae}.  This estimate is close to that
of our model chemical composition of Ne nova 1 and is also close to that
of CO nova 2, except for the neon abundance of $X_{\rm Ne}=0.13$.
We found no other estimates for V351~Pup in the literature. 
Therefore, we examine first the light curves of Ne nova 1 and
then those of CO nova 2, CO nova 4, Ne nova 2, and Ne nova 3
in this order, in Figures
\ref{all_mass_v351_pup_v1974_cyg_x35z02o20ne10},
\ref{all_mass_v351_pup_v1668_cyg_x35z02c10o20} --
\ref{all_mass_v351_pup_qu_vul_x65z02o03ne03}
together with the light curves of V1974~Cyg, V1668~Cyg, PW~Vul, V1974~Cyg,
and QU~Vul, respectively,
which reflects the order of the degree of mixing,
because the nova chemical composition values estimated by various authors
usually have a large scatter, as shown in Table 
\ref{chemical_abundance_neon_novae}.
Note again that enrichment of neon with unchanged hydrogen and
CNO mass fractions affects the nova light curves very little in our
model light curves because neon is not relevant to nuclear burning
(the CNO cycle) or increasing the opacity 
\citep[e.g.,][]{kat94h, hac06kb, hac10k}.
Therefore, our model light curves could follow the observation well
even for the chemical composition of the same $X$ and $X_{\rm CNO}$ but
different $X_{\rm Ne}$ with $Y + X_{\rm Ne}$ being the same.

\subsubsection{Ne nova 1}
Figure \ref{all_mass_v351_pup_v1974_cyg_x35z02o20ne10}(a)
shows our model UV 1455~\AA\  flux and free-free emission flux, in
addition to the visual, $V$, and UV~1455~\AA\  light curves
of V351~Pup.  It also shows observational data for V1974~Cyg,
whose WD mass is estimated to be similar to that of V351~Pup.
For the V351~Pup data, the visual magnitudes are taken from the archives of
AAVSO and Association Fran\c{c}aise des Observateurs d'\'Etoiles Variables;
the $V$ magnitudes are from IAU Circ. Nos. 5422, 5427,
5430, 5455, 5483, 5493, 5527, 5538, 5552, 5566, 5628, 5655, 5666, and 5781,
and from \citet{bru92}.  The UV~1455~\AA\  fluxes are compiled from the INES
archive data \citep{cas02}.  We chose the $M_{\rm WD} = 1.0~M_\sun$ model
as the best-fit model because our model light curve for the $1.0~M_\sun$ WD
fits the UV 1455~\AA\  observation in Figure 
\ref{all_mass_v351_pup_v1974_cyg_x35z02o20ne10}(a) well.  Optically
thick winds and hydrogen shell burning end approximately 230 days and 
480 days after the outburst, respectively, for the $1.0~M_\sun$ WD.
We have no absolute magnitude calibration of the free-free
model light curves for the chemical composition of Ne nova 1,
so we tentatively shift the free-free model light curves up and down
to fit them with the optical observation in Figure 
\ref{all_mass_v351_pup_v1974_cyg_x35z02o20ne10}(a).

From our UV 1455~\AA\  flux fitting in Figure
\ref{all_mass_v351_pup_v1974_cyg_x35z02o20ne10}(a), we obtain
$F_{\lambda 1455}^{\rm mod}= 4.6 \times 10^{-12}$~erg cm$^{-2}$
s$^{-1}$ ~\AA$^{-1}$ and $F_{\lambda 1455}^{\rm obs}= (4.6 \pm 0.49)
\times 10^{-13}$~erg~cm$^{-2}$~s$^{-1}$~~\AA$^{-1}$ at the UV~1455~\AA\  
flux maximum.
Substituting these values into Equation (\ref{qu_vul_uv1455_fit_eq2}),
we plot the distance-reddening relation (magenta solid line) in Figure
\ref{v351_pup_distance_reddening_x35z02o20ne10}.  This line
crosses the blue solid line of
$(m-M)_V=15.1$ at a reddening of $E(B-V)=0.46$ and
a distance of $d \approx 5.4$~kpc, which is
consistent with our estimates of $E(B-V)=0.45\pm0.05$ and
$d=5.5\pm1$~kpc in Section \ref{reddening_v351_pup}.  This may support
our choice of these two values.

Our free-free emission model light curves are not yet calibrated
for the chemical composition of Ne nova 1.
In the same way as in Sections \ref{light_curve_v1668_cyg} and \ref{qu_vul},
we determine the absolute magnitude of the free-free emission model
light curve for the $1.0~M_\sun$ WD.
The total $V$ magnitudes are calculated as the sum of the
free-free and blackbody emission, which are shown in 
Figure \ref{all_mass_v351_pup_v1974_cyg_x35z02o20ne10}(b).
We directly read $m_{\rm w}=16.6$
from Figure \ref{all_mass_v351_pup_v1974_cyg_x35z02o20ne10}(b)
and obtain $M_{\rm w}= m_{\rm w} - (m-M)_V= 16.6 - 15.1=1.5$.
Thus, the proportionality constant is specified by $M_{\rm w}=1.5$
for the $1.0~M_\sun$ WD with the Ne nova 1 composition.
Using Equation (\ref{real_timescale_flux}),
we obtain the absolute magnitudes
of the free-free model light curves for the other WD masses with
the chemical composition of Ne nova 1, which are
listed in Table \ref{light_curves_of_novae_ne1}.

Our model light curve fits the early $V$ and visual light curves
but deviates from the $V$ and visual observation in the later phase,
that is, in the nebular phase \citep{wil94,sai96},  as shown in Figure
\ref{all_mass_v351_pup_v1974_cyg_x35z02o20ne10}.
In the figure, we suppose that V351~Pup entered the nebular phase
at $m_V\approx11.0$, about 80 days after the outburst.
This is because strong emission lines such as [\ion{O}{3}]
contribute to the $V$ magnitude but
our model light curves do not include such emission lines
as discussed in Section \ref{light_curve_fit_v1668_cyg}.

\subsubsection{CO nova 2}
Figure \ref{all_mass_v351_pup_v1668_cyg_x35z02c10o20} shows our model
light curve fittings for the chemical composition of CO nova 2.
The absolute magnitudes of the model light curves for this chemical 
composition were already calibrated in \citet{hac10k}.
Assuming that $(m-M)_V=15.1$, we obtain the
$0.98~M_\sun$ WD as the best-fit model, as shown in
Figure \ref{all_mass_v351_pup_v1668_cyg_x35z02c10o20}(a), 
from the UV~1455~\AA\  light curve fitting.
We plot the total, free-free emission,
and blackbody emission fluxes for the $0.98~M_\sun$ WD
model in Figure \ref{all_mass_v351_pup_v1668_cyg_x35z02c10o20}(b).

Optically thick winds and hydrogen shell burning end approximately 235 days
and 505 days after the outburst, respectively, for the $0.98~M_\sun$ WD.
For $(m-M)_V=15.1$, the total $V$ flux light curve follows the observed
$V$ magnitudes of V351~Pup reasonably well in the early decline phase.
It deviates from the observed $V$ in the middle and later phases,
mainly because strong emission lines dominate the spectrum
in the nebular phase that started on day $\sim 80$.
These strong emission lines are not included in our light curve model.

We obtain two distance-reddening relations from the $V$ and UV~1455~\AA\ 
light curve fittings of the $0.98~M_\sun$ WD, 
i.e., Equation (\ref{qu_vul_distance_modulus_eq1})
together with $(m-M)_V=15.1$ and Equation (\ref{qu_vul_uv1455_fit_eq2}) 
together with $F_{\lambda 1455}^{\rm mod}= 5.0 \times 
10^{-12}$~erg~cm$^{-2}$~s$^{-1}$~~\AA$^{-1}$ and
$F_{\lambda 1455}^{\rm obs}= (4.6\pm0.49) \times 
10^{-13}$~erg~cm$^{-2}$~s$^{-1}$~~\AA$^{-1}$
at the UV~1455~\AA\  flux maximum in Figure
\ref{all_mass_v351_pup_v1668_cyg_x35z02c10o20}(b).
The two distance-reddening relations are plotted in Figure
\ref{v351_pup_distance_reddening_x35_x55_x55_x65_4comp}(a); they
cross each other at $E(B-V) \approx 0.48$ and 
$d \approx 5.3$~kpc, which are consistent with our estimates of 
$E(B-V)=0.45\pm0.05$ and $d=5.5\pm1$~kpc in Section \ref{reddening_v351_pup}.

\subsubsection{CO nova 4}
Figure \ref{all_mass_v351_pup_pw_vul_x55z02c10o10}(a) shows that
the $1.02~M_\sun$ WD model is the best-fit model because of
its agreement with the UV~1455~\AA\   observed fluxes.
The absolute magnitudes of the free-free model light curves were already 
calibrated in \citet{hac15k}. 
We plot the total, free-free emission, and blackbody emission
fluxes of the $1.02~M_\sun$ WD model in Figure 
\ref{all_mass_v351_pup_pw_vul_x55z02c10o10}(b).
In this case, optically thick winds and hydrogen shell burning
end approximately 280 days and 740 days after the outburst, respectively.
For $(m-M)_V=15.1$, the total $V$ flux light curve follows the observed 
$V$ magnitudes of V351~Pup reasonably well in the early decline phase.

We obtain two distance-reddening relations from the $V$ and UV~1455~\AA\ 
light curve fittings of the $1.02~M_\sun$ WD, 
i.e., Equation (\ref{qu_vul_distance_modulus_eq1})
with $(m-M)_V=15.1$ and Equation (\ref{qu_vul_uv1455_fit_eq2}) 
with $F_{\lambda 1455}^{\rm mod}= 4.4 \times 
10^{-12}$~erg~cm$^{-2}$~s$^{-1}$~~\AA$^{-1}$ and
$F_{\lambda 1455}^{\rm obs}= 4.6 \times 
10^{-13}$~erg~cm$^{-2}$~s$^{-1}$~~\AA$^{-1}$
at the UV~1455~\AA\  flux maximum in Figure
\ref{all_mass_v351_pup_pw_vul_x55z02c10o10}(b).
The two distance-reddening relations are plotted in Figure
\ref{v351_pup_distance_reddening_x35_x55_x55_x65_4comp}(b); they
cross each other at $E(B-V) \approx 0.45$ and 
$d \approx 5.5$~kpc.  These values are consistent
with our estimates of $E(B-V)=0.45\pm0.05$ and $d=5.5\pm1$~kpc
in Section \ref{reddening_v351_pup}.

\subsubsection{Ne nova 2}
Figure \ref{all_mass_v351_pup_v1974_cyg_x55z02o10ne03}(a) shows that 
the $1.05~M_\sun$ WD model is the best-fit model because of
its agreement with the UV~1455~\AA\   observed fluxes.
The absolute magnitudes of the free-free model light curves were published 
in \citet{hac10k}. 
We plot the total, free-free emission, and blackbody emission
fluxes of the $1.05~M_\sun$ WD model in Figure 
\ref{all_mass_v351_pup_v1974_cyg_x55z02o10ne03}(b).
Optically thick winds and hydrogen shell burning end approximately 280 days
and 720 days after the outburst, respectively, for the $1.05~M_\sun$ WD.
For $(m-M)_V=15.1$, the total $V$ flux light curve follows the observed
$V$ magnitudes of V351~Pup reasonably well in the early decline phase.

We obtain two distance-reddening relations from the $V$ and UV~1455~\AA\ 
light curve fittings of the $1.05~M_\sun$ WD, 
i.e., Equation (\ref{qu_vul_distance_modulus_eq1})
with $(m-M)_V=15.1$ and Equation (\ref{qu_vul_uv1455_fit_eq2}) 
with $F_{\lambda 1455}^{\rm mod}= 4.6 \times 
10^{-12}$~erg~cm$^{-2}$~s$^{-1}$~~\AA$^{-1}$ and
$F_{\lambda 1455}^{\rm obs}= 4.6 \times 
10^{-13}$~erg~cm$^{-2}$~s$^{-1}$~~\AA$^{-1}$
at the UV~1455~\AA\  flux maximum in Figure
\ref{all_mass_v351_pup_v1974_cyg_x55z02o10ne03}(b).
The two distance-reddening relations are plotted in Figure
\ref{v351_pup_distance_reddening_x35_x55_x55_x65_4comp}(c); they
cross each other at $E(B-V) \approx 0.46$ and 
$d \approx 5.4$~kpc.  These values are consistent
with our estimates of $E(B-V)=0.45\pm0.05$ and $d=5.5\pm1$~kpc
in Section \ref{reddening_v351_pup}.

\subsubsection{Ne nova 3}
Figure \ref{all_mass_v351_pup_qu_vul_x65z02o03ne03}(a) shows that
the $1.1~M_\sun$ WD model is the best-fit model because of
its agreement with the UV~1455~\AA\  observed fluxes.
The absolute magnitudes of the free-free model light curves are
determined in Section \ref{qu_vul}.
We plot the total, free-free emission, and blackbody
emission fluxes for the $1.1~M_\sun$ WD model in Figure 
\ref{all_mass_v351_pup_qu_vul_x65z02o03ne03}(b).  
Optically thick winds and hydrogen shell burning end approximately
310 days and 820 days after the outburst, respectively.
For $(m-M)_V=15.1$, the total $V$ flux light curve follows the observed
$V$ magnitudes of V351~Pup reasonably well in the early decline phase.

We obtain two distance-reddening relations from the $V$ and UV~1455~\AA\ 
light curve fittings of the $1.1~M_\sun$ WD, 
i.e., Equation (\ref{qu_vul_distance_modulus_eq1})
with $(m-M)_V=15.1$ and Equation (\ref{qu_vul_uv1455_fit_eq2}) 
with $F_{\lambda 1455}^{\rm mod}= 4.55 \times 
10^{-12}$~erg~cm$^{-2}$~s$^{-1}$~~\AA$^{-1}$ and
$F_{\lambda 1455}^{\rm obs}= 4.6 \times 
10^{-13}$~erg~cm$^{-2}$~s$^{-1}$~~\AA$^{-1}$
at the UV~1455~\AA\  flux maximum in Figure
\ref{all_mass_v351_pup_qu_vul_x65z02o03ne03}(b).
The two distance-reddening relations are plotted in Figure
\ref{v351_pup_distance_reddening_x35_x55_x55_x65_4comp}(d); they
cross each other at $E(B-V) \approx 0.46$ and $d \approx 5.4$~kpc.
These values are consistent with our estimates of 
$E(B-V)=0.45\pm0.05$ and $d=5.5\pm1$~kpc in Section \ref{reddening_v351_pup}.

\subsection{Summary of V351~Pup}
Our model light curves fit the optical $V$ and UV~1455~\AA\   fluxes
of V351~Pup well and give a reasonable set of 
the distance, $d=5.5$~kpc, and the reddening, $E(B-V)=0.45$.
Our best-fit models give relatively small WD masses of
0.98 -- $1.1~M_\sun$ as an ONe WD.  As mentioned in Section \ref{introduction},
the WD should be a naked ONe WD because the ejecta are enriched with neon.
A natal ONe WD had a CO mantle.  Therefore, this WD may have lost
a mass of $\sim 0.1~M_\sun$ CO mantle before the naked ONe core was
revealed.  If we adopt Ne nova 1 for the chemical composition
of V351~Pup, we can summarize the results of our light curve
fittings (see also Table \ref{nova_parameters_results}) as follows:
$(m-M)_V=15.1$, $E(B-V)=0.45$, $d=5.5$~kpc, $M_{\rm WD}=1.0~M_\sun$, 
and $M_{\rm env}=1.98\times10^{-5}~M_\sun$ at optical maximum.

\subsection{Comparison with previous results}
\label{comparison_previous_v351_pup}
\citet{wan99} estimated the WD mass from fitting their abundance
pattern model with that observed by \citet{sai96} and 
suggested $M_{\rm WD} \gtrsim 1.25~M_\sun$.
\citet{kat07h} estimated the WD mass and distance of V351~Pup to be
$1.0~M_\sun$ and 2.1~kpc, respectively,
based on their optically thick wind model
\citep{kat94h}.  They fitted their UV~1455\AA\  model
light curves with the observation.  Because the timescale of the nova
evolution is the same between their models and ours, the obtained
WD mass is the same.  However, they determined the distance only
from the UV~1455\AA\  fitting.  They adopted the reddening of
$E(B-V)=0.72$ \citep{sai96}
and obtained the distance of 2.1~kpc, which is much
shorter than our new estimate of 5.5~kpc.


\begin{deluxetable*}{llllllllllllll}
\tabletypesize{\scriptsize}
\tablecaption{Free-free Light Curves of Neon Novae 1\tablenotemark{a}
\label{light_curves_of_novae_ne1}}
\tablewidth{0pt}
\tablehead{
\colhead{$m_{\rm ff}$} &
\colhead{0.7$M_\sun$} &
\colhead{0.75$M_\sun$} &
\colhead{0.8$M_\sun$} &
\colhead{0.85$M_\sun$} &
\colhead{0.9$M_\sun$} &
\colhead{0.95$M_\sun$} &
\colhead{1.0$M_\sun$} &
\colhead{1.05$M_\sun$} &
\colhead{1.1$M_\sun$} &
\colhead{1.15$M_\sun$} &
\colhead{1.2$M_\sun$} &
\colhead{1.25$M_\sun$} &
\colhead{1.3$M_\sun$} \\
\colhead{(mag)} &
\colhead{(day)} &
\colhead{(day)} &
\colhead{(day)} &
\colhead{(day)} &
\colhead{(day)} &
\colhead{(day)} &
\colhead{(day)} &
\colhead{(day)} &
\colhead{(day)} &
\colhead{(day)} &
\colhead{(day)} &
\colhead{(day)} &
\colhead{(day)} 
}
\startdata
  3.500     & 0.0 & 0.0 & 0.0 & 0.0 & 0.0 & 0.0 & 0.0 & 0.0 & 0.0 & 0.0 & 0.0 & 0.0 & 0.0 \\
  3.750     &  1.704     &  1.540     &  1.320     &  1.050     & 0.940     & 0.838     & 0.776     & 0.699     & 0.620     & 0.583     & 0.557     & 0.510     & 0.511     \\
  4.000     &  3.446     &  3.330     &  2.670     &  2.130     &  1.880     &  1.683     &  1.549     &  1.395     &  1.241     &  1.161     &  1.089     &  1.030     & 0.993     \\
  4.250     &  5.849     &  5.160     &  4.040     &  3.260     &  2.870     &  2.546     &  2.327     &  2.077     &  1.862     &  1.727     &  1.627     &  1.540     &  1.473     \\
  4.500     &  8.850     &  7.050     &  5.590     &  4.460     &  3.890     &  3.416     &  3.122     &  2.767     &  2.482     &  2.297     &  2.162     &  2.050     &  1.948     \\
  4.750     &  11.93     &  8.970     &  7.320     &  5.680     &  4.920     &  4.296     &  3.922     &  3.489     &  3.117     &  2.890     &  2.710     &  2.560     &  2.428     \\
  5.000     &  15.12     &  12.21     &  9.120     &  7.050     &  6.030     &  5.266     &  4.772     &  4.231     &  3.777     &  3.507     &  3.290     &  3.080     &  2.898     \\
  5.250     &  19.01     &  15.80     &  10.98     &  8.800     &  7.220     &  6.306     &  5.742     &  5.129     &  4.447     &  4.134     &  3.874     &  3.600     &  3.398     \\
  5.500     &  25.07     &  19.62     &  14.65     &  10.61     &  8.600     &  7.376     &  6.752     &  6.086     &  5.317     &  5.043     &  4.634     &  4.160     &  3.978     \\
  5.750     &  31.59     &  24.06     &  18.70     &  12.90     &  10.63     &  8.996     &  8.202     &  7.336     &  6.277     &  6.004     &  5.604     &  4.830     &  4.738     \\
  6.000     &  38.10     &  28.73     &  22.03     &  16.21     &  13.03     &  11.00     &  10.22     &  9.106     &  7.477     &  7.084     &  6.514     &  5.750     &  5.478     \\
  6.250     &  44.82     &  33.82     &  25.47     &  19.28     &  15.46     &  12.81     &  11.66     &  10.29     &  8.957     &  8.164     &  7.424     &  6.730     &  6.248     \\
  6.500     &  52.06     &  39.37     &  29.18     &  21.96     &  17.78     &  14.64     &  13.04     &  11.40     &  9.797     &  8.864     &  8.094     &  7.360     &  6.828     \\
  6.750     &  60.05     &  45.31     &  33.65     &  24.83     &  20.02     &  16.51     &  14.54     &  12.61     &  10.67     &  9.614     &  8.704     &  7.850     &  7.208     \\
  7.000     &  69.13     &  51.76     &  38.51     &  28.26     &  22.43     &  18.44     &  16.20     &  13.90     &  11.63     &  10.42     &  9.354     &  8.350     &  7.598     \\
  7.250     &  78.94     &  58.74     &  43.75     &  32.14     &  25.35     &  20.51     &  18.00     &  15.31     &  12.71     &  11.29     &  10.05     &  8.870     &  8.008     \\
  7.500     &  89.61     &  66.33     &  49.10     &  36.37     &  28.57     &  22.92     &  20.05     &  16.88     &  13.89     &  12.25     &  10.80     &  9.410     &  8.418     \\
  7.750     &  99.71     &  74.54     &  54.89     &  40.49     &  32.00     &  25.57     &  22.29     &  18.58     &  15.20     &  13.24     &  11.61     &  9.980     &  8.858     \\
  8.000     &  110.1     &  82.59     &  61.14     &  44.87     &  35.26     &  28.44     &  24.72     &  20.46     &  16.57     &  14.33     &  12.45     &  10.59     &  9.298     \\
  8.250     &  121.3     &  90.77     &  67.67     &  49.59     &  38.77     &  31.23     &  26.94     &  22.23     &  17.93     &  15.49     &  13.29     &  11.20     &  9.718     \\
  8.500     &  133.8     &  99.47     &  74.19     &  54.68     &  42.54     &  34.22     &  29.31     &  24.14     &  19.37     &  16.63     &  14.15     &  11.81     &  10.15     \\
  8.750     &  147.5     &  110.5     &  81.23     &  59.98     &  46.61     &  37.33     &  31.87     &  26.17     &  20.83     &  17.86     &  15.05     &  12.42     &  10.58     \\
  9.000     &  162.7     &  122.4     &  89.70     &  65.64     &  51.25     &  40.66     &  34.63     &  28.35     &  22.41     &  19.18     &  16.09     &  13.06     &  11.01     \\
  9.250     &  180.1     &  135.2     &  99.20     &  72.25     &  56.27     &  44.54     &  38.09     &  31.11     &  24.22     &  20.62     &  17.25     &  13.80     &  11.52     \\
  9.500     &  199.5     &  149.3     &  109.7     &  80.18     &  61.87     &  48.76     &  41.89     &  34.19     &  26.34     &  22.53     &  18.62     &  14.67     &  12.12     \\
  9.750     &  220.1     &  164.8     &  121.3     &  88.87     &  68.70     &  53.87     &  46.28     &  37.75     &  28.90     &  24.62     &  20.21     &  15.68     &  12.83     \\
  10.00     &  243.2     &  182.4     &  134.2     &  98.17     &  76.14     &  59.70     &  51.21     &  41.76     &  31.87     &  27.01     &  22.01     &  16.85     &  13.64     \\
  10.25     &  269.4     &  202.6     &  148.9     &  108.6     &  84.29     &  66.12     &  56.76     &  46.29     &  35.18     &  29.63     &  24.10     &  18.18     &  14.55     \\
  10.50     &  293.1     &  225.5     &  166.0     &  120.3     &  93.24     &  73.26     &  63.09     &  51.64     &  38.96     &  32.84     &  26.42     &  19.72     &  15.61     \\
  10.75     &  314.4     &  246.9     &  185.5     &  134.5     &  103.8     &  81.11     &  70.04     &  57.42     &  43.37     &  36.65     &  29.49     &  21.42     &  16.77     \\
  11.00     &  338.5     &  269.5     &  200.2     &  150.5     &  116.3     &  90.57     &  78.34     &  64.19     &  48.22     &  40.84     &  32.84     &  23.76     &  18.41     \\
  11.25     &  365.8     &  286.5     &  213.8     &  165.7     &  130.3     &  101.4     &  87.68     &  71.78     &  54.61     &  45.61     &  36.93     &  26.33     &  20.24     \\
  11.50     &  396.5     &  303.9     &  228.9     &  180.0     &  139.6     &  113.5     &  97.74     &  79.95     &  60.57     &  50.83     &  40.92     &  29.32     &  22.31     \\
  11.75     &  421.1     &  323.0     &  245.7     &  192.4     &  148.9     &  122.8     &  104.1     &  86.11     &  66.49     &  55.74     &  44.83     &  32.31     &  24.43     \\
  12.00     &  446.2     &  338.6     &  264.3     &  203.0     &  158.4     &  132.8     &  111.0     &  92.76     &  71.85     &  59.67     &  48.10     &  35.21     &  26.51     \\
  12.25     &  472.8     &  355.0     &  277.9     &  214.6     &  168.4     &  141.2     &  118.4     &  98.29     &  77.06     &  63.49     &  51.45     &  37.75     &  28.33     \\
  12.50     &  501.0     &  373.1     &  290.6     &  225.8     &  178.9     &  150.0     &  126.0     &  103.9     &  82.04     &  67.52     &  54.78     &  40.26     &  30.10     \\
  12.75     &  530.8     &  393.7     &  304.5     &  236.4     &  190.1     &  159.3     &  133.9     &  109.9     &  87.31     &  71.80     &  58.21     &  42.83     &  31.96     \\
  13.00     &  562.4     &  417.7     &  323.0     &  251.4     &  202.0     &  169.1     &  142.3     &  115.9     &  92.89     &  76.33     &  62.05     &  45.37     &  34.33     \\
  13.25     &  595.8     &  443.2     &  342.8     &  267.2     &  214.6     &  179.5     &  151.1     &  122.1     &  98.77     &  81.12     &  66.12     &  47.98     &  36.84     \\
  13.50     &  631.3     &  470.1     &  363.8     &  284.0     &  227.9     &  190.6     &  160.6     &  129.6     &  105.1     &  86.20     &  70.43     &  51.69     &  39.50     \\
  13.75     &  668.9     &  498.7     &  386.1     &  301.7     &  242.0     &  202.3     &  170.5     &  137.5     &  111.7     &  91.59     &  74.99     &  55.65     &  42.32     \\
  14.00     &  708.7     &  528.9     &  409.6     &  320.5     &  257.0     &  214.7     &  181.1     &  145.9     &  118.8     &  97.27     &  79.82     &  59.86     &  45.31     \\
  14.25     &  750.8     &  561.0     &  434.6     &  340.5     &  272.8     &  227.8     &  192.2     &  154.9     &  126.2     &  103.4     &  84.95     &  64.30     &  48.47     \\
  14.50     &  795.4     &  594.9     &  461.0     &  361.6     &  289.5     &  241.8     &  204.1     &  164.3     &  134.1     &  109.8     &  90.37     &  69.01     &  51.82     \\
  14.75     &  842.7     &  630.9     &  489.0     &  384.0     &  307.3     &  256.5     &  216.6     &  174.3     &  142.5     &  116.5     &  96.10     &  74.00     &  55.37     \\
  15.00     &  892.8     &  669.0     &  518.6     &  407.6     &  326.1     &  272.1     &  229.9     &  184.9     &  151.3     &  123.7     &  102.2     &  79.29     &  59.13     \\
\hline
X-ray\tablenotemark{b}
 & 1734  & 1250 & 900  & 645 & 457   & 338   & 248   & 170   & 114   & 78   & 52   & 30.5  & 16.3 \\
\hline
$\log f_{\rm s}$\tablenotemark{c}
 & 0.56  & 0.46 & 0.35  & 0.25 & 0.16   & 0.08   & $0.0$   & $-0.10$   & $-0.19$   & $-0.29$   & $-0.385$   & $-0.50$  & $-0.65$ \\
\hline
$M_{\rm w}$\tablenotemark{d}
 & 3.8 & 3.3  & 2.8 & 2.4   & 2.0   & 1.8   & 1.5   & 1.3   & 1.0   & 0.7   & 0.5  & 0.2 & 0.0 
\enddata
\tablenotetext{a}{Chemical composition of
the envelope is assumed
to be that of Ne nova 1 in Table \ref{chemical_composition_model}.}
\tablenotetext{b}{Duration of supersoft X-ray phase in units of days.}
\tablenotetext{c}{Stretching factor against V351~Pup
 UV 1455~\AA\  observation
in Figure \ref{all_mass_v351_pup_univeral_scale_x35z02o20ne10}.}
\tablenotetext{d}{Absolute magnitudes at the bottom point in Figure
\ref{all_mass_v351_pup_univeral_real_abs_x35z02o20ne10} by assuming
$(m-M)_V = 15.1$  (V351~Pup).}
\end{deluxetable*}


\begin{figure}
\epsscale{1.15}
\plotone{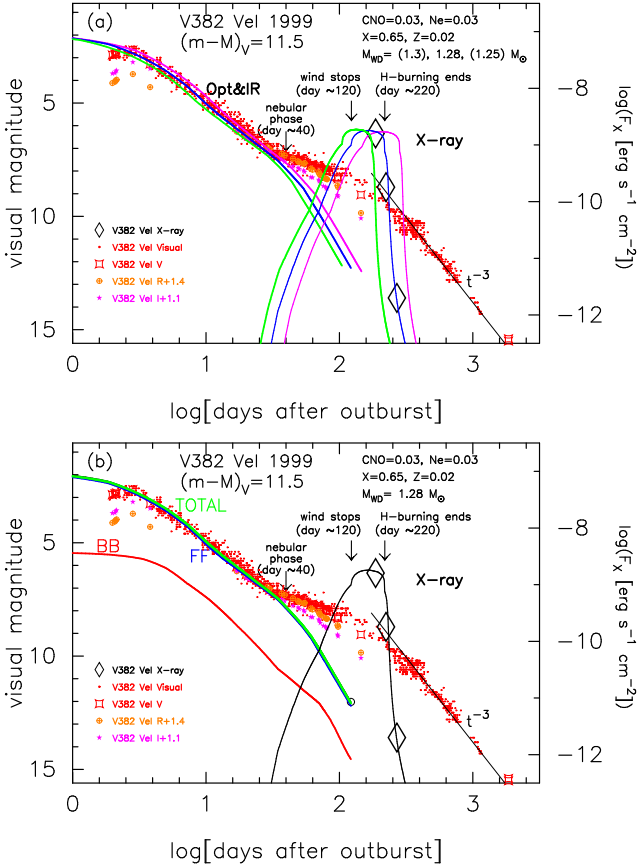}
\caption{
(a) Model light curves of $1.25~M_\sun$
(magenta solid lines), $1.28~M_\sun$ (blue solid),
and $1.3~M_\sun$ (green solid) WDs for Ne nova 3
as well as visual (small red dots), $V$ (red boxes with sharp corners),
$R$ (orange encircled pluses), near infrared $I$ (magenta filled 
star marks), and supersoft X-ray (black large open diamonds)
light curves of V382~Vel.
Assuming that $(m-M)_V=11.5$, we plot the $V$ model light curves.
(b) Assuming also that $(m-M)_V=11.5$, we plot
three $V$ model light curves for the $1.28~M_\sun$ WD:
the green, blue, and red solid lines show the total, free-free emission,
and blackbody emission $V$ fluxes, respectively.  
The black solid line denotes the supersoft X-ray fluxes. 
Optically thick winds and hydrogen shell burning end approximately
120 days and 220 days after the outburst, respectively, 
for the $1.28M_\sun$ WD.  The nebular phase starts
$\sim 40$ days after the outburst \citep{del02}.
See text for the sources of the observed data.
\label{all_mass_v382_vel_x65z02o03ne03_absolute_mag}}
\end{figure}


\begin{figure}
\epsscale{0.95}
\plotone{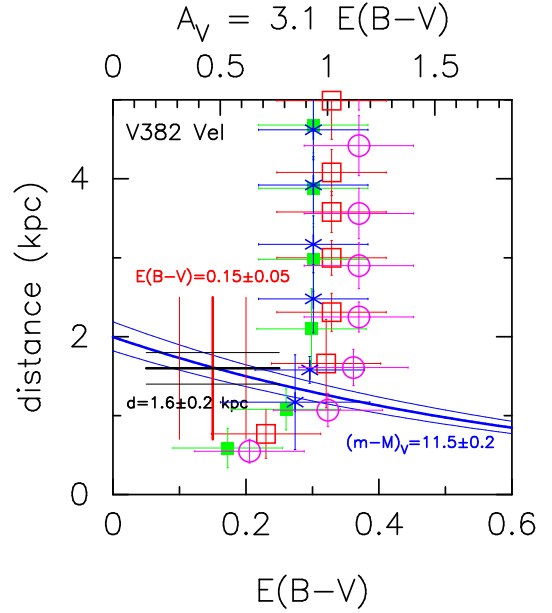}
\caption{
Various distance-reddening relations toward V382~Vel.
We plot Equation (\ref{qu_vul_distance_modulus_eq1}) with 
$(m-M)_V=11.5\pm0.2$ (blue solid lines) from our time-stretching method
in Appendix \ref{time_stretching_method_novae},
$E(B-V)=0.15\pm0.05$ (red thick solid lines) from our color-color diagram
method in Section \ref{color_color_diagrams},
$d=1.6\pm0.2$~kpc (black solid lines), and
four Marshall et al.'s (2006) distance-reddening relations toward
$(l, b)= (284\fdg00,+5\fdg75)$ (red open squares),
$(284\fdg25,+5\fdg75)$ (green filled squares),
$(284\fdg00,+6\fdg00)$ (blue asterisks), and
$(284\fdg25,+6\fdg00)$ (magenta open circles),
which are four directions close to V382~Vel, 
$(l,b)=(284\fdg1674, +5\fdg7715)$.
\label{v382_vel_distance_reddening_x55z02o10ne03_x65z02o03ne03_onefigure}}
\end{figure}


\begin{figure}
\epsscale{1.15}
\plotone{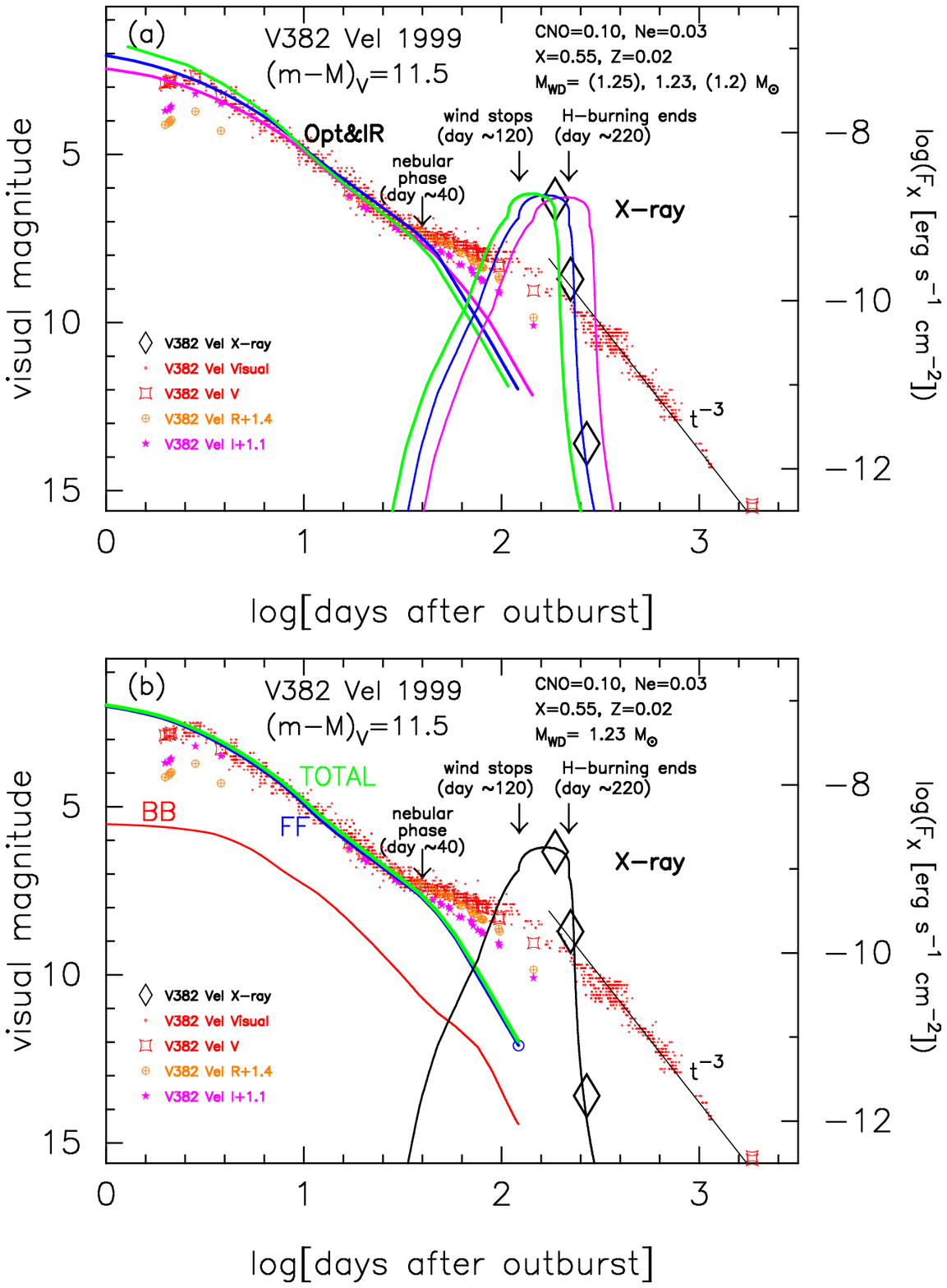}
\caption{
Same as Figure
\ref{all_mass_v382_vel_x65z02o03ne03_absolute_mag}, but for Ne nova 2.
(a) Model light curves of $1.2~M_\sun$ (magenta solid lines), 
$1.23~M_\sun$ (blue solid lines), and $1.25~M_\sun$ (green solid lines) WDs.
Assuming that $(m-M)_V=11.5$, we plot the $V$ model light curves.
(b) Assuming also that $(m-M)_V=11.5$, we plot three $V$ model light curves
of the $1.23~M_\sun$ WD.
Optically thick winds and hydrogen shell burning end approximately
120 days and 220 days after the outburst, respectively, 
for the $1.23M_\sun$ WD.
\label{all_mass_v382_vel_x55z02o10ne03_absolute_mag}}
\end{figure}


\begin{figure}
\epsscale{1.15}
\plotone{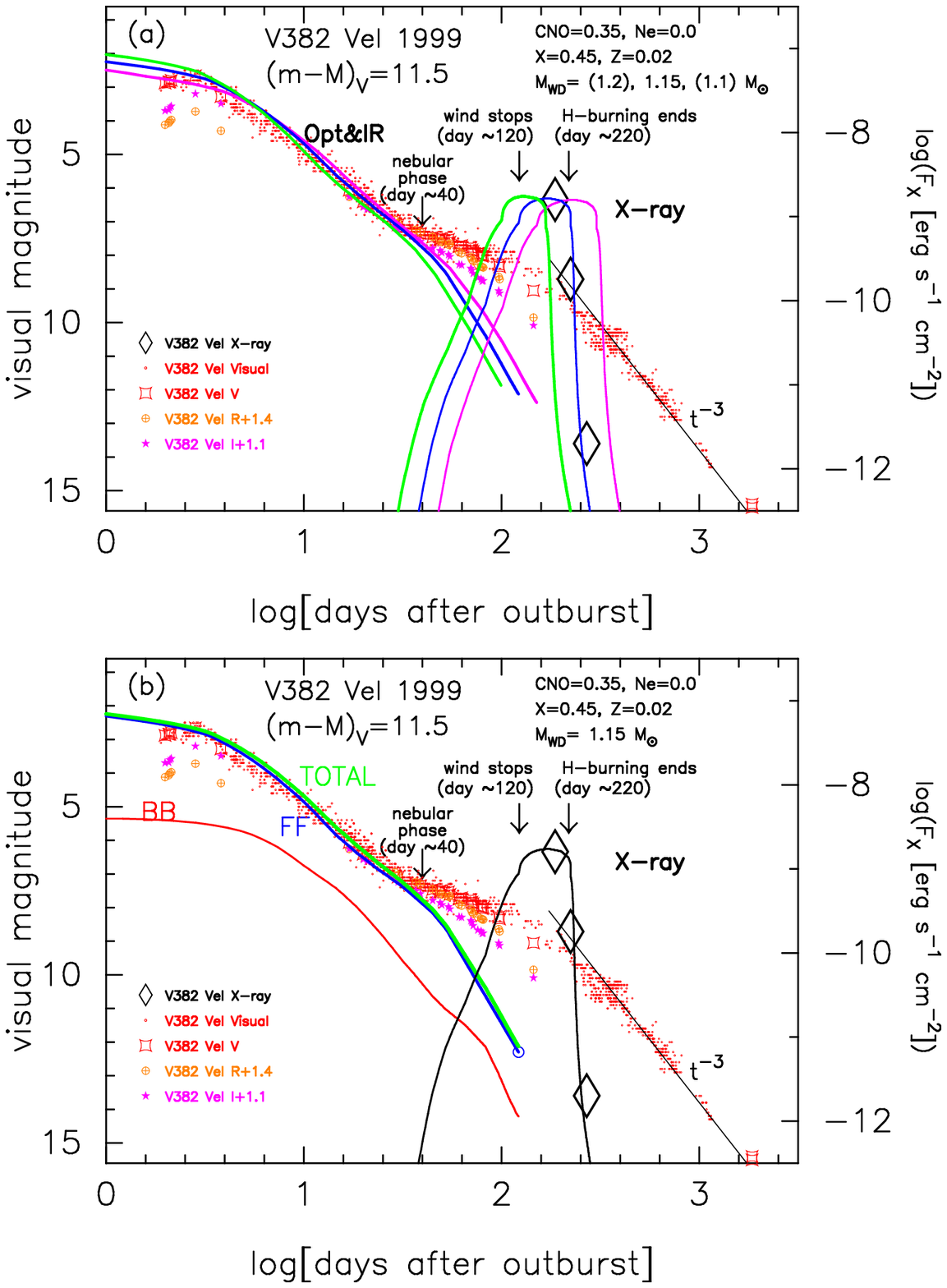}
\caption{
Same as Figure
\ref{all_mass_v382_vel_x65z02o03ne03_absolute_mag}, but for CO nova 3.
(a) Model light curves of $1.1~M_\sun$ (magenta solid lines), 
$1.15~M_\sun$ (blue solid lines), and $1.2~M_\sun$ (green solid lines) WDs.
Assuming that $(m-M)_V=11.5$, we plot the $V$ model light curves.
(b) Assuming also that $(m-M)_V=11.5$, we plot three $V$ model light curves
of the $1.15~M_\sun$ WD.  Optically thick winds and hydrogen shell burning end
approximately 120 days and 220 days after the outburst, respectively, 
for the $1.15M_\sun$ WD.
\label{all_mass_v382_vel_x45z02c15o20_absolute_mag}}
\end{figure}


\begin{figure}
\epsscale{1.15}
\plotone{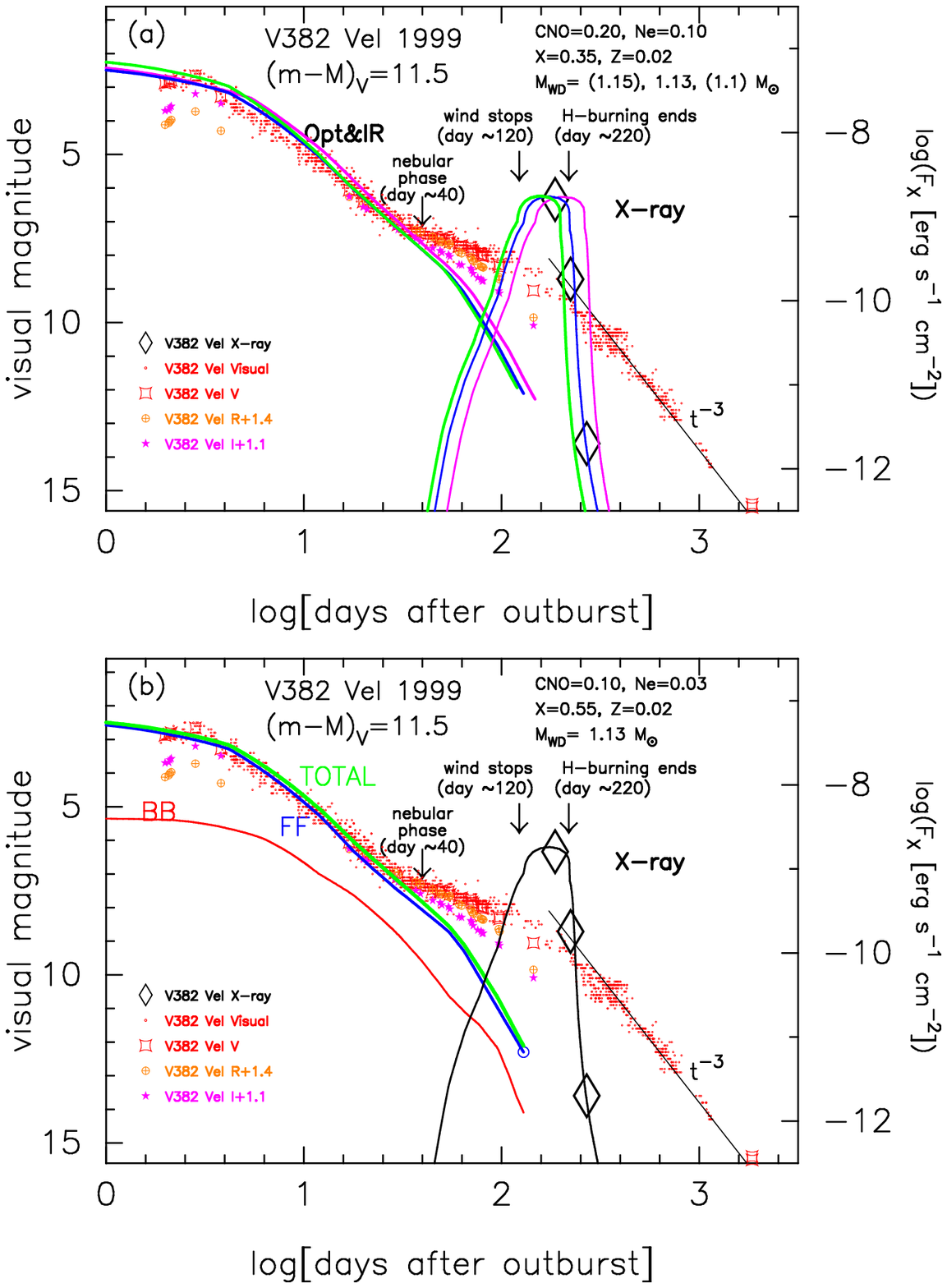}
\caption{
Same as Figure \ref{all_mass_v382_vel_x65z02o03ne03_absolute_mag}, but
for Ne Nova 1. 
(a) Model light curves of $1.1~M_\sun$ (magenta solid lines), 
$1.13~M_\sun$ (blue solid lines), and $1.15~M_\sun$ (green solid lines)
WDs with the chemical composition of ``Ne Nova 1.'' 
Assuming that $(m-M)_V=11.5$, we plot the $V$ model light curves.
(b) Assuming also that $(m-M)_V=11.5$, we plot three $V$ model light curves
of the $1.13~M_\sun$ WD.  Optically thick winds and hydrogen shell burning
end approximately 120 days and 220 days after the outburst, respectively, 
for the $1.13M_\sun$ WD. 
\label{all_mass_v382_vel_x35z02o20ne10_absolute_mag}}
\end{figure}

\section{V382 Vel 1999}
\label{v382_vel}
V382~Vel is a very fast nova identified as a neon nova \citep{woo99}.
The optical and NIR light curves and supersoft X-ray
count rates are plotted in Figure 
\ref{all_mass_v382_vel_x65z02o03ne03_absolute_mag}.
The nova reached $m_V = 2.7$ at maximum on UT 1999 May 23.
Supersoft X-rays were detected by {\it BeppoSAX} \citep{ori02}
about six months after the outburst, followed by a rapid decline,
as shown in Figure \ref{all_mass_v382_vel_x65z02o03ne03_absolute_mag}.
{\it Chandra} observations by \citet{bur02} and \citet{nes05}
suggest that hydrogen shell burning ended approximately 220 days after the
outburst.  We have already analyzed the light curves of V382~Vel
in \citet{hac10k} on the basis of our optical free-free emission and X-ray 
model light curves, but not included the effect of photospheric
emission.  Here, we calculate the total $V$ flux
model light curves and compare them with the optical observations.

\subsection{Reddening and distance}
\label{reddening_v382_vel}
The reddening toward V382~Vel was estimated as
$E(B-V)= 0.20$ by \citet{sho03}, and as $E(B-V)=0.05$--$0.099$
by \citet{del02} from various line ratios
and the \ion{Na}{1}~D interstellar absorption features.
\citet{hac14k} obtained $E(B-V)=0.15\pm0.05$ by the $UBV$ color-color
diagram fitting.  We again plot the color-color diagram of V382~Vel in
Figure \ref{color_color_diagram_qu_vul_v351_pup_v382_vel_v693_cra}(c).
\citet{muk01} obtained a hydrogen column density toward V382~Vel
of $N_{\rm H}= (1.01 \pm 0.05)\times 10^{21}$ from early hard
X-ray spectrum fittings.  This value can be converted to
$E(B-V)= N_{\rm H}/ 5.8 \times 10^{21} \approx 0.2$ \citep{boh78},
$E(B-V)= N_{\rm H}/ 6.8 \times 10^{21} \approx 0.15$ \citep{guv09},
or $E(B-V)= N_{\rm H}/ 8.3 \times 10^{21} \approx 0.12$ \citep{lis14},
which are consistent with Hachisu \& Kato's value of $E(B-V)=0.15\pm0.05$.
Therefore, we adopt $E(B-V)=0.15\pm0.05$ in this paper.

The distance of V382~Vel was also estimated to be $d=1.66 \pm 0.11$~kpc
by \citet{del02} from the maximum magnitude versus rate of decline
(MMRD) relation whereas \citet{sho03} obtained
a larger distance of $d = 2.5$ kpc, assuming that the UV fluxes of 
V382~Vel and V1974 Cyg are the same.  \citet{sho03} assumed a distance
of $d=3.1$~kpc to V1974 Cyg, but this value is much larger than the
reasonable one of $d=1.8$~kpc, as shown in Section \ref{light_curve_v1974_cyg}
\citep[see also, e.g.,][]{cho97a, hac10k, hac14k}.
If we take $d=1.8$~kpc instead of 3.1 kpc to V1974 Cyg,
Shore et al.'s method gives a much shorter distance of 1.5 kpc
to V382 Vel, which is consistent with della Valle et al.'s value.
\citet{hac10k} estimated the distance modulus of V382~Vel to be
$(m-M)_V = 11.5 \pm 0.1$ using four different methods, including
the free-free emission model light curve fitting method based on
the chemical composition Ne nova 2.
If we adopt $E(B-V)=0.15\pm0.05$, the distance of V382~Vel is calculated
to be $d=1.6\pm0.2$~kpc from Equation (\ref{qu_vul_distance_modulus_eq1}).
These values are all consistent with each other.  
Therefore, we adopt $d=1.6\pm0.2$~kpc in this paper.  In Figure 
\ref{v382_vel_distance_reddening_x55z02o10ne03_x65z02o03ne03_onefigure},
we plot these three constraints of the distance, reddening, and
distance modulus in the $V$ band as well as
Marshall et al.'s (2006) distance-reddening relations.

Marshall et al.'s relation suggests an extinction of as large as
$E(B-V)=0.3$--$0.4$ at the distance of 1.6~kpc.  
The NASA/IPAC galactic dust absorption map also
gives $E(B-V)=0.385 \pm 0.01$ in the direction of V382~Vel, whose
galactic coordinates are $(l,b)=(284\fdg1674, +05\fdg7715)$.
These two extinctions are consistent with each other, but 
disagree with our obtained value of $E(B-V)=0.15\pm0.05$.
Because various estimates on the absorption toward V382~Vel are
consistent with $E(B-V)=0.15\pm0.05$ as mentioned above,
this disagreement between the pin-point extinction
toward V382~Vel ($E(B-V)\sim0.15$) and Marshall et al.'s relation 
($E(B-V)\sim0.3$--$0.4$ at 1.6~kpc) suggests a low extinction hole
in front of V382~Vel.

\subsection{Light curve fitting of V382~Vel}
\label{light_curve_fit_v382_vel}
The chemical composition of V382~Vel was obtained as $X=0.47$, 
$X_{\rm CNO}=0.018$, and $X_{\rm Ne}=0.0099$ by \citet{aug03}, and 
as $X=0.66$, $X_{\rm CNO}=0.043$, and $X_{\rm Ne}=0.027$ by \citet{sho03}
(see Table \ref{chemical_abundance_neon_novae}).  Shore et al.'s estimate
is close to that for Ne nova 3.  We first examine the chemical composition
of Ne nova 3, then those of Ne nova 2, CO nova 3, and finally Ne nova 1
in order of the degree of mixing.
The absolute magnitudes of the free-free emission model light curves
are already calibrated for Ne nova 1 in Section \ref{v351_pup},
for CO nova 3 in Section \ref{light_curve_v1668_cyg},
for Ne nova 2 in \citet{hac10k}, and for Ne nova 3
in Section \ref{qu_vul}.
Our fitting results are shown in Figures 
\ref{all_mass_v382_vel_x65z02o03ne03_absolute_mag},
\ref{all_mass_v382_vel_x55z02o10ne03_absolute_mag} --
\ref{all_mass_v382_vel_x35z02o20ne10_absolute_mag}.  In these figures,
the visual magnitudes are taken from the AAVSO archive.
The $V$, $R$, and $I$ data are taken from IAU Circular Nos. 7176, 
7179, 7196, 7209, 7216, 7226, 7232, 7238, and 7277.
The faintest observational $V$ magnitudes at day $\sim 1850$
are taken from \citet{wou05}.  The X-ray data are taken from 
\citet{ori02} and \citet{bur02}.

\subsubsection{Ne nova 3}
Figure \ref{all_mass_v382_vel_x65z02o03ne03_absolute_mag}(a) shows that
the $1.28~M_\sun$ WD model is the best-fit model, especially because of
its agreement with the X-ray light curves.
We plot the total, free-free emission, and blackbody emission
fluxes of the $1.28~M_\sun$ WD model in Figure 
\ref{all_mass_v382_vel_x65z02o03ne03_absolute_mag}(b).
Optically thick winds and hydrogen shell burning end approximately
120 days and 220 days after the outburst, respectively,
for the $1.28~M_\sun$ WD.
For $(m-M)_V=11.5$, the total $V$ flux light curve follows the observed
$V$ magnitudes of V382~Vel reasonably well in the early decline phase.
We have obtained a hydrogen-rich envelope mass of $M_{\rm env}=
0.46\times 10^{-5} M_\sun$ at optical maximum, which corresponds
approximately to the ignition mass. 

Our model light curve fits the early $V$ and visual light curves
but deviates from the $V$ and visual observation in the later phase,
that is, in the nebular phase.  V382~Vel entered the nebular phase
at least by the end of June 1999, that is, 
at $m_V\approx 7.4$ and $\sim 40$ days after
the the optical maximum \citep{del02}, as shown in Figure
\ref{all_mass_v382_vel_x65z02o03ne03_absolute_mag}.
This is because strong emission lines such as [\ion{O}{3}]
contribute to the $V$ magnitude but
our model light curves do not include such emission lines
as discussed in Section \ref{light_curve_fit_v1668_cyg}.

\subsubsection{Ne nova 2}
Figure \ref{all_mass_v382_vel_x55z02o10ne03_absolute_mag}(a) shows that
the $1.23~M_\sun$ WD model is the best-fit model from the X-ray light
curve fitting.  We plot the total, free-free emission, and blackbody
emission fluxes of the $1.23~M_\sun$ WD model in Figure 
\ref{all_mass_v382_vel_x55z02o10ne03_absolute_mag}(b).
Optically thick winds and hydrogen shell burning end approximately
120 days and 220 days after the outburst, respectively.
For $(m-M)_V=11.5$, the total $V$ flux light curve follows the observed
$V$ magnitudes of V382~Vel reasonably well in the early decline phase.
We have obtained a hydrogen-rich envelope mass of $M_{\rm env}=
0.48\times 10^{-5} M_\sun$ at optical maximum.

\subsubsection{CO nova 3}
Figure \ref{all_mass_v382_vel_x45z02c15o20_absolute_mag}(a) shows that
the $1.15~M_\sun$ WD model is the best-fit model from the X-ray
light curve fitting.  We plot the total, free-free emission,
and blackbody emission fluxes of the $1.15~M_\sun$ WD model in Figure 
\ref{all_mass_v382_vel_x45z02c15o20_absolute_mag}(b).
Optically thick winds and hydrogen shell burning end approximately
120 days and 220 days after the outburst, respectively.
For $(m-M)_V=11.5$, the total $V$ flux light curve follows the observed
$V$ magnitudes of V382~Vel reasonably well in the early decline phase.
We have obtained a hydrogen-rich envelope mass of $M_{\rm env}=
0.70\times 10^{-5} M_\sun$ at optical maximum.

\subsubsection{Ne nova 1}
Figure \ref{all_mass_v382_vel_x35z02o20ne10_absolute_mag}(a) shows that
the $1.13~M_\sun$ WD model is the best-fit model from the X-ray light curve
fitting.  We plot the total, free-free emission,
and blackbody emission fluxes of the $1.13~M_\sun$ WD model in Figure 
\ref{all_mass_v382_vel_x35z02o20ne10_absolute_mag}(b).
Optically thick winds and hydrogen shell burning end approximately
120 days and 220 days after the outburst, respectively.
For $(m-M)_V=11.5$, the total $V$ flux light curve follows the observed
$V$ magnitudes of V382~Vel reasonably well in the early decline phase.
We have obtained a hydrogen-rich envelope mass of $M_{\rm env}=
0.85\times 10^{-5} M_\sun$ at optical maximum.

\subsection{Summary of V382~Vel}
Our model light curve fittings of Ne nova 3, Ne nova 2, CO nova 3,
and Ne nova 1 give WD masses of 1.28, 1.23, 1.15, and $1.13~M_\sun$,
respectively, with a distance modulus of $(m-M)_V=11.5$.
Here, we adopt Ne nova 2 for the chemical composition
of V382~Vel because it is close to the arithmetic mean of $X=0.47$ 
\citep{aug03} and $X=0.66$ \citep{sho03}, i.e., $X=0.56$.
We summarize the results of our light curve fittings
(see also Table \ref{nova_parameters_results}) as follows:
$(m-M)_V=11.5$, $E(B-V)=0.15$, $d=1.6$~kpc, 
$M_{\rm WD}=1.23~M_\sun$, and $M_{\rm env}=0.48\times10^{-5}~M_\sun$
at optical maximum.
The effect of photospheric emission is rather small and
it affects the light curve fitting very little. 
Therefore, we can reproduce the optical light curves of V382~Vel
using only the free-free emission light curves.
The reddening and distance are consistent with the previous estimates
mentioned in Section \ref{reddening_v382_vel}, but deviate from
the distance-reddening relation given by \citet{mar06}.

\subsection{Comparison with previous results}
\label{comparison_prvious_v382_vel}
Using the time-stretching method of the universal decline law, 
\citet{hac10k} obtained the absolute
magnitude of their free-free emission model light curves and
applied their light curve model to V382~Vel.  They obtained 
the distance modulus in the $V$ band of $(m-M)_V=11.5\pm0.1$
and the WD mass of $1.23\pm0.05~M_\sun$.
\citet{dow13} estimated the WD mass of V382~Vel by comparing their
model elemental abundance with the observed ratios.
They obtained $1.18$--$1.21~M_\sun$.
Our new estimate of $1.23~M_\sun$ is close to their upper value.


\begin{figure}
\epsscale{1.15}
\plotone{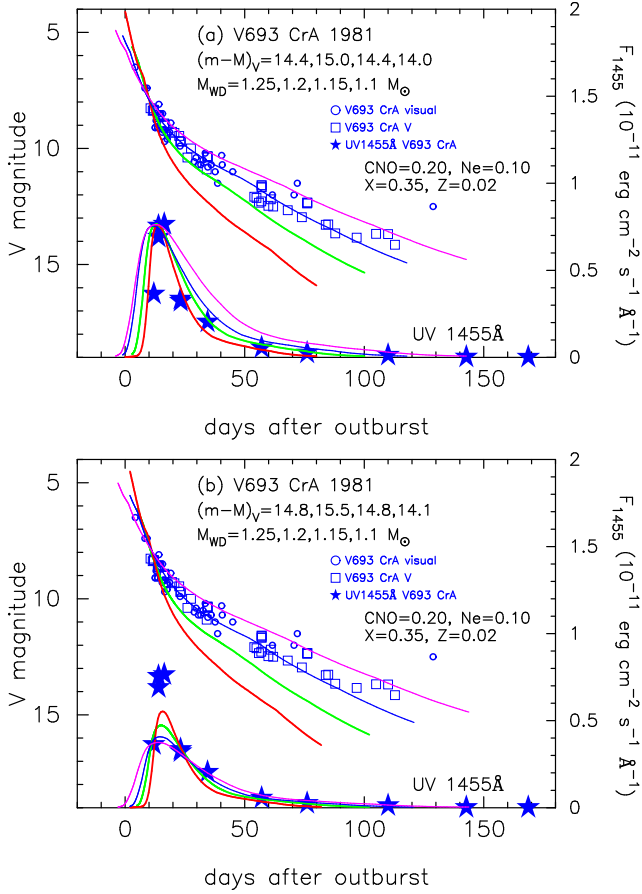}
\caption{
(a) Model light curves of $1.1~M_\sun$ (magenta solid lines), 
$1.15~M_\sun$ (blue solid lines), $1.2~M_\sun$ (green solid lines), 
and $1.25~M_\sun$ (red solid lines) WDs with the chemical composition
of Ne Nova 1 as well as $V$ band (blue open squares), visual (blue open
circles), and UV~1455 ~\AA\ (large blue filled star-marks)
light curves of V693~CrA.  We place the UV~1455\AA\  model light curves
to run through the observed peak.
(b) We place the UV~1455\AA\  model light curves to skip the observed peak
(three star-marks near the peak) but follow the other observations.
See text for details.
\label{all_mass_v693_cra_x35z02o20ne10_absolute_mag_linear}}
\end{figure}


\begin{figure}
\epsscale{1.15}
\plotone{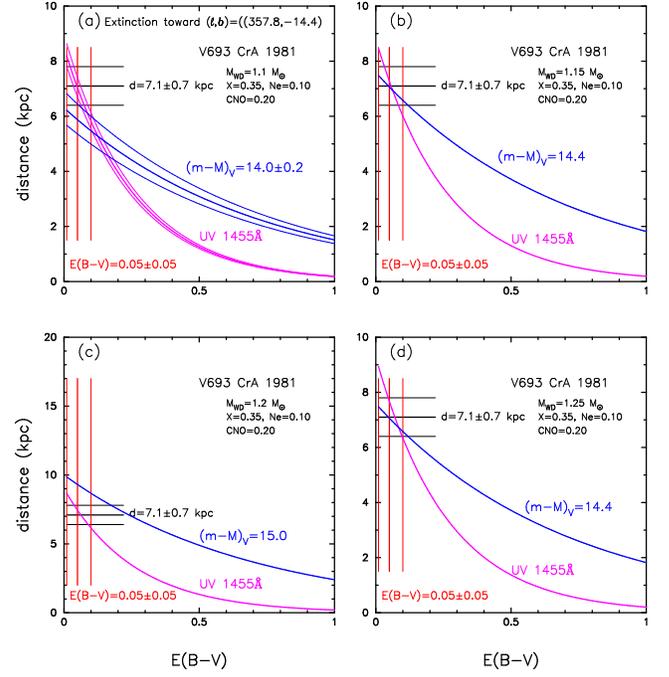}
\caption{
Various distance-reddening relations toward V693~CrA 
for various WD masses with the chemical composition of Ne nova 1.
In panel (a), we plot the results of $1.1~M_\sun$ WD obtained from
Equation (\ref{qu_vul_distance_modulus_eq1}) with
$(m-M)_V=14.0\pm0.2$ from our model $V$ light curve fitting
and Equation (\ref{qu_vul_uv1455_fit_eq2}) with our UV~1455\AA\ 
fitting in Figure
\ref{all_mass_v693_cra_x35z02o20ne10_absolute_mag_linear}(a).
Two other constraints are
also plotted, $E(B-V)=0.05\pm0.05$ and $d=7.1\pm0.7$~kpc.
In panel (b), the results of $1.15~M_\sun$ WD.
In panel (c), the results of $1.2~M_\sun$ WD.
In panel (d), the results of $1.25~M_\sun$ WD.
See text for details.
\label{v693_cra_distance_reddening_x35_m110_m115_m120_m125}}
\end{figure}


\begin{figure}
\epsscale{1.15}
\plotone{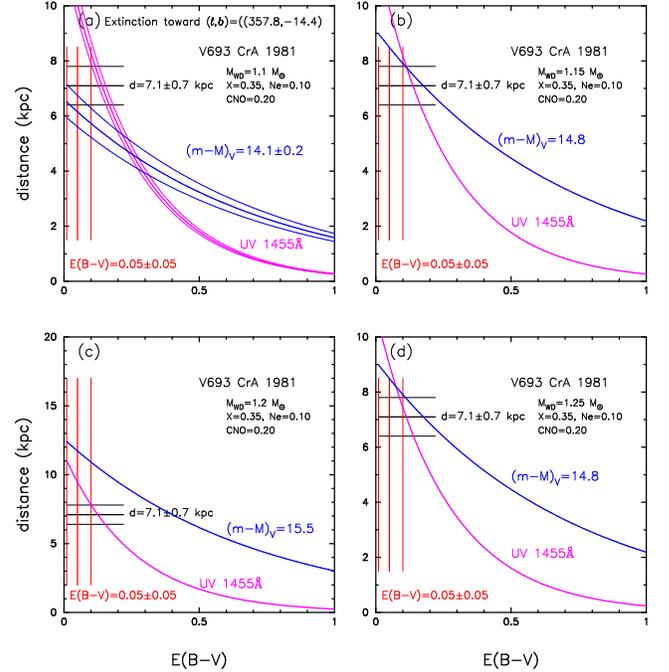}
\caption{
Same as Figure
\ref{v693_cra_distance_reddening_x35_m110_m115_m120_m125},
but for the fittings in Figure
\ref{all_mass_v693_cra_x35z02o20ne10_absolute_mag_linear}(b).
See text for details.
\label{v693_cra_distance_reddening_x35_m110_m115_m120_m125_no2}}
\end{figure}


\begin{figure}
\epsscale{1.15}
\plotone{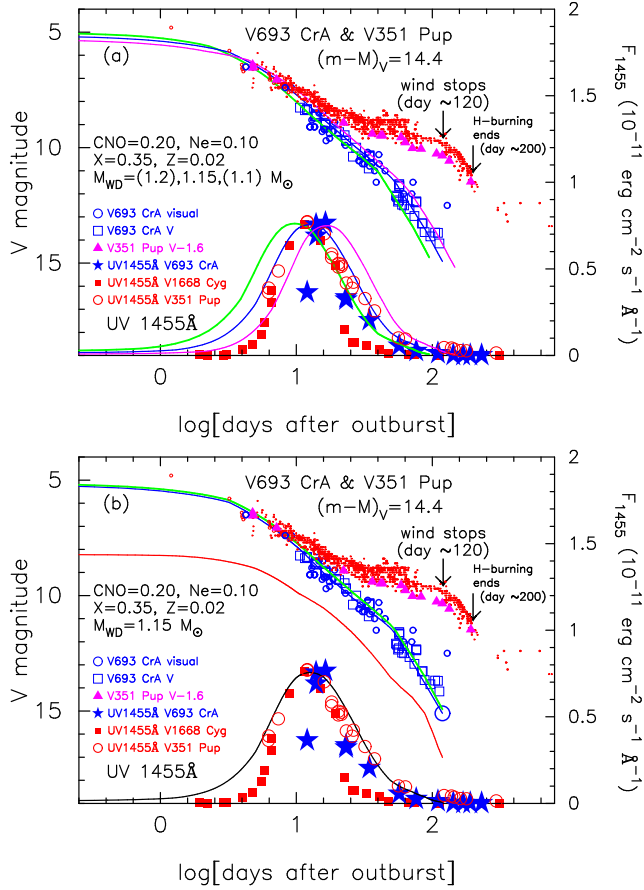}
\caption{
Same as Figure \ref{all_mass_v351_pup_v1668_cyg_x35z02c10o20}, but
for V693~CrA.
(a) Model light curves of $1.1~M_\sun$ (magenta solid lines), 
$1.15~M_\sun$ (blue solid lines), and $1.2~M_\sun$ (green solid lines)
WDs with Ne Nova 1 as well as visual (small blue open circles),
$V$ (blue open sqaures), and UV~1455 ~\AA\  (blue star marks) light curves
of V693~CrA.  We added the visual (red small dots), $V$ (magenta filled
triangles), UV~1455 ~\AA\  (red open circles) light curves of V351~Pup.
The timescale of V351~Pup is squeezed by a factor of $f_s=0.40$.
We also added the UV~1455 ~\AA\  (large red filled squares) light curve
of V1668~Cyg.  The timescale of V1668~Cyg is squeezed by a factor of
$f_s=0.37$.  Assuming that $(m-M)_V=14.4$, we plot the $V$ model light curves.
(b) Assuming also that $(m-M)_V=14.4$, we plot three $V$ model light curves
of the $1.15~M_\sun$ WD.
Optically thick winds end $\sim120$ days and hydrogen 
shell-burning stops $\sim200$ days after the outburst
for the $1.15~M_\sun$ WD.
\label{all_mass_v693_cra_x35z02o20ne10_absolute_mag}}
\end{figure}


\begin{figure}
\epsscale{1.15}
\plotone{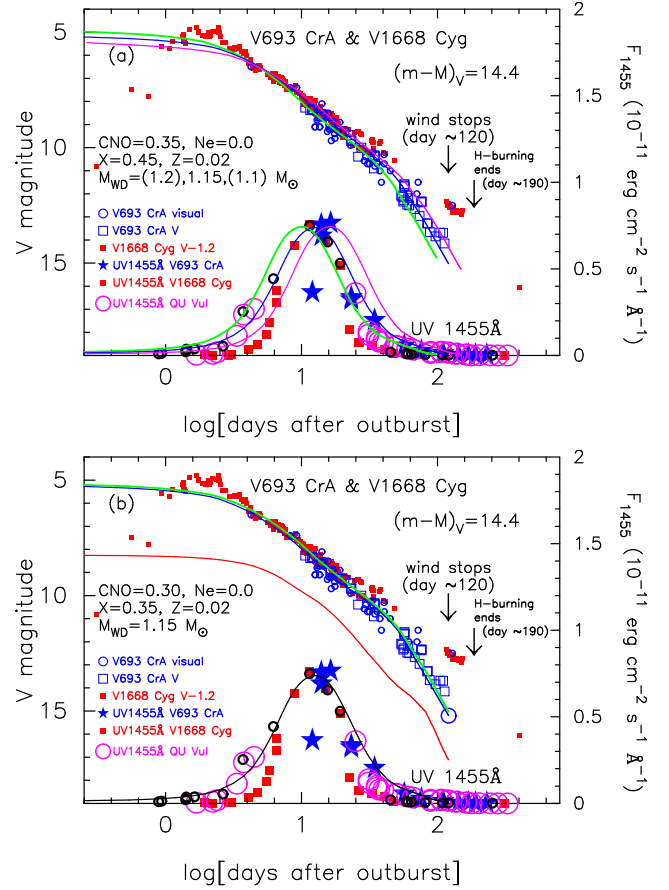}
\caption{
Same as Figure \ref{all_mass_v693_cra_x35z02o20ne10_absolute_mag},
but for CO nova 3.  We added the $V$ light curve (red filled squares)
of V1668~Cyg.  The timescale of V1668~Cyg is squeezed by a factor of
$f_s=0.37$.  We also added the UV~1455~\AA\  light curve
(magenta large open circles) of QU~Vul.  The timescale of QU~Vul
is squeezed by a factor of $f_s=0.19$.
(a) We plot three model light curves of $1.1~M_\sun$ (magenta solid lines), 
$1.15~M_\sun$ (blue solid lines), and $1.2~M_\sun$ (green solid lines) WDs.
Assuming that $(m-M)_V=14.4$, we plot the $V$ model light curves.
(b) Assuming also that $(m-M)_V=14.4$, we plot three $V$ model light curves
of the $1.15~M_\sun$ WD.
Optically thick winds end $\sim120$ days and hydrogen 
shell-burning stops $\sim190$ days after the outburst
for the $1.15~M_\sun$ WD.
\label{all_mass_v693_cra_x45z02c15o20_absolute_mag}}
\end{figure}


\begin{figure}
\epsscale{1.15}
\plotone{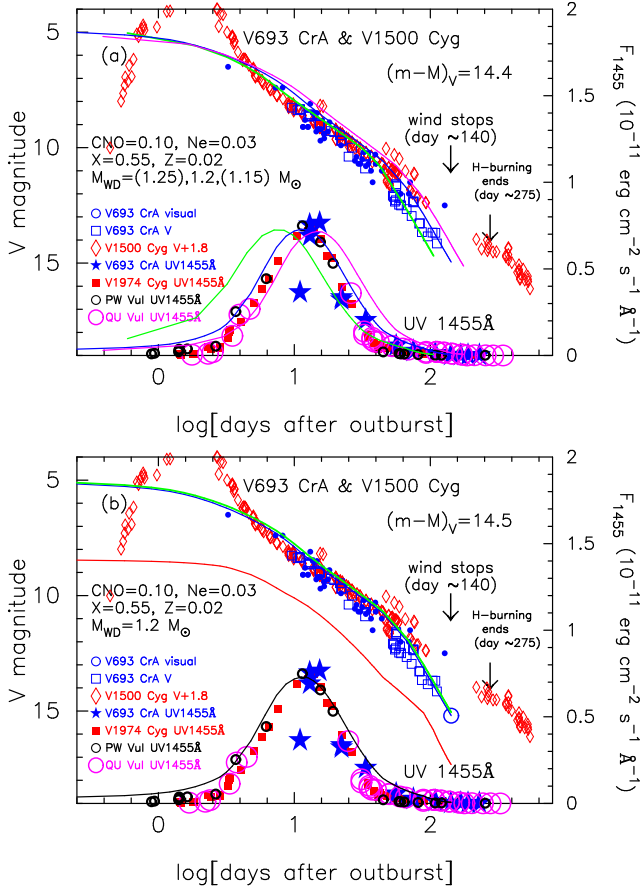}
\caption{
Same as Figure \ref{all_mass_v693_cra_x45z02c15o20_absolute_mag},
but for Ne nova 2.
We added the $V$ (red open diamonds) light curve of V1500~Cyg.
The timescale of V1500~Cyg is squeezed by a factor of $f_s=0.79$.
(a) We plot model light curves of $1.15~M_\sun$ (magenta solid lines), 
$1.2~M_\sun$ (blue solid lines), and $1.25~M_\sun$ (green solid lines)
WDs.  Assuming that $(m-M)_V=14.4$, we plot the $V$ model light curves. 
(b) Assuming that $(m-M)_V=14.5$, slightly larger than that 
in Figure \ref{all_mass_v693_cra_x45z02c15o20_absolute_mag},
we plot three $V$ model light curves
of the $1.2~M_\sun$ WD.
Optically thick winds end $\sim140$ days and hydrogen 
shell-burning stops $\sim275$ days after the outburst
for the $1.2~M_\sun$ WD.
\label{all_mass_v693_cra_x55z02o10ne03_absolute_mag}}
\end{figure}


\begin{figure}
\epsscale{1.15}
\plotone{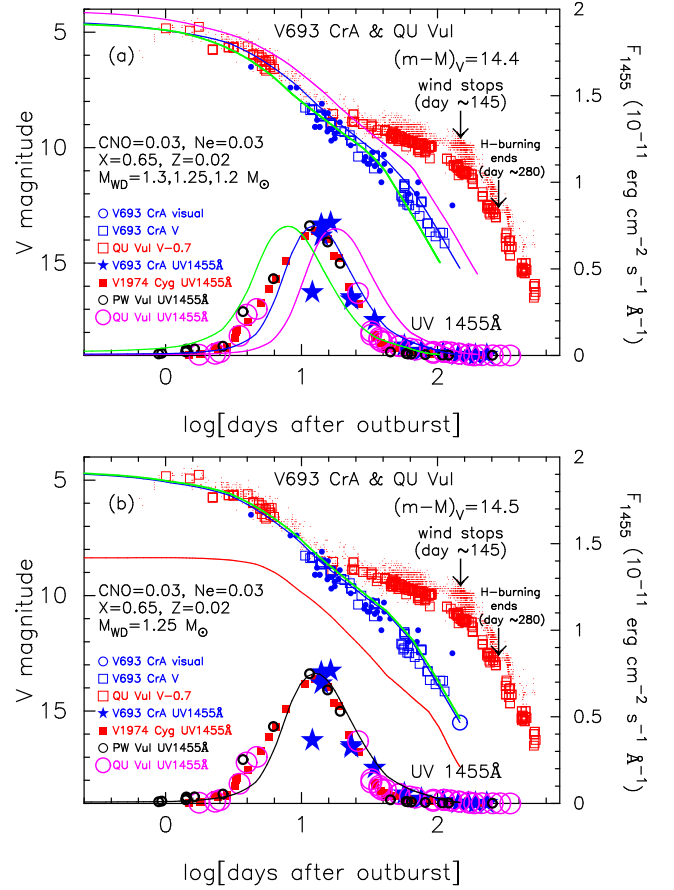}
\caption{
Same as Figure \ref{all_mass_v693_cra_x55z02o10ne03_absolute_mag},
but for Ne nova 3.
We added the $V$ (red open squares), visual (red dots), and UV~1455~\AA\  
(large magenta open circles) light curves of QU~Vul.
The timescale of QU~Vul is squeezed by a factor of $f_s=0.25$. 
(a) We plot three model light curves of $1.2~M_\sun$ (magenta solid lines), 
$1.25~M_\sun$ (blue solid lines), and $1.3~M_\sun$ (green solid lines)
WDs.  Assuming that $(m-M)_V=14.4$, we plot the $V$ model light curves. 
(b) Assuming that $(m-M)_V=14.5$, we plot three model light curves of 
the $1.25~M_\sun$ WD.
Optically thick winds end $\sim145$ days and hydrogen 
shell-burning stops $\sim280$ days after the outburst
for the $1.25~M_\sun$ WD.
\label{all_mass_v693_cra_x65z02o03ne03_absolute_mag}}
\end{figure}


\begin{figure}
\epsscale{1.15}
\plotone{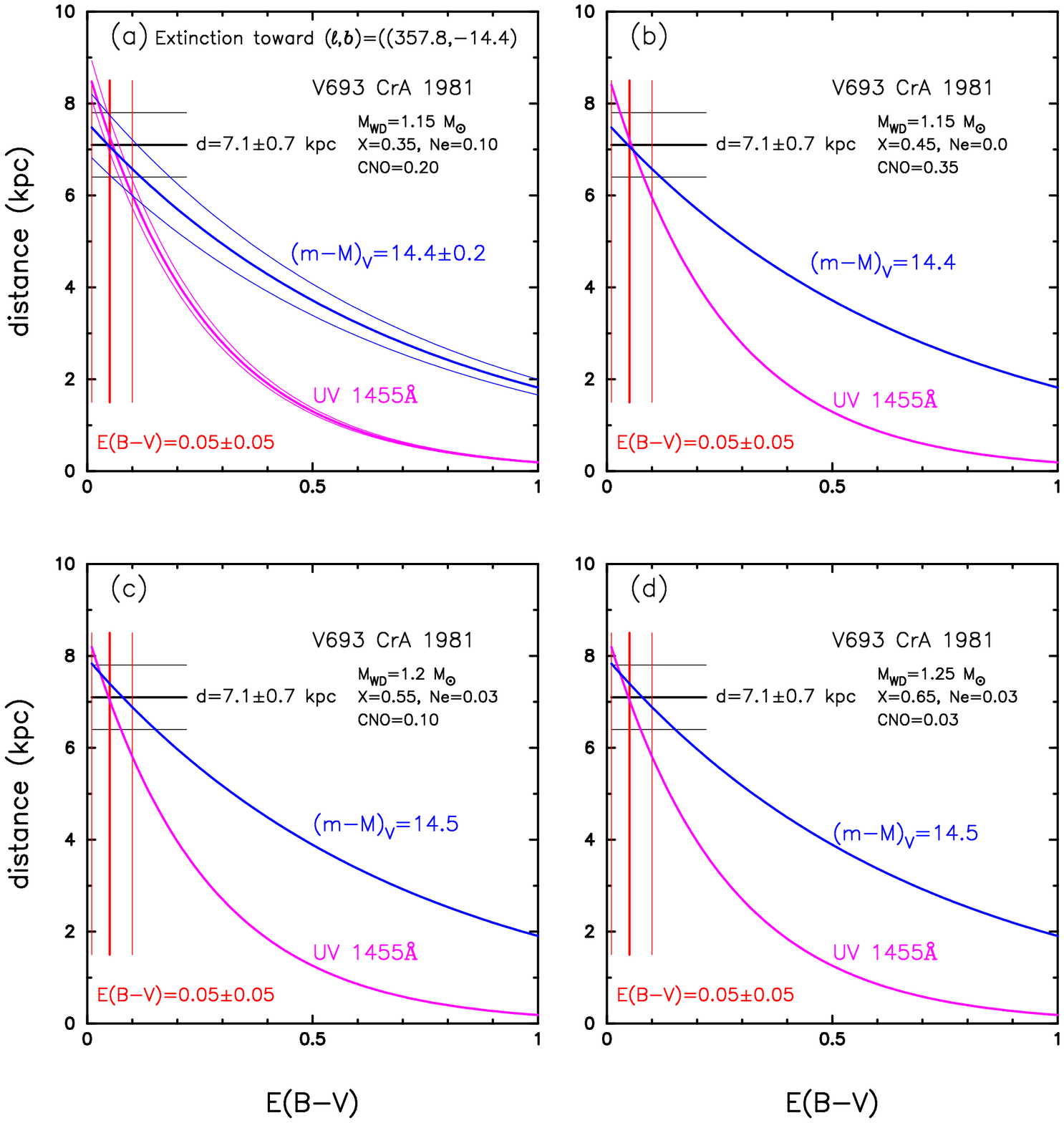}
\caption{
Various distance-reddening relations toward V693~CrA.
We plot distance-reddening relations for the chemical composition of
(a) Ne nova 1, (b) CO nova 3, (c) Ne nova 2, and (d) Ne nova 3.
In panel (a), we plot the results obtained from
Equation (\ref{qu_vul_distance_modulus_eq1}) with
$(m-M)_V=14.4\pm0.2$ from our model $V$ light curve fitting
and Equation (\ref{qu_vul_uv1455_fit_eq2}) with 
$F_{1455}^{\rm mod}= 1.55 \times 
10^{-11}$~erg~cm$^{-2}$~s$^{-1}$~~\AA$^{-1}$ as the calculated model
flux of the upper bound of Figure
\ref{all_mass_v693_cra_x35z02o20ne10_absolute_mag}(b)
at the distance of 10~kpc and $F_{1455}^{\rm obs}= (2.0\pm0.2) \times 
10^{-11}$~erg~cm$^{-2}$~s$^{-1}$~~\AA$^{-1}$ as the observed flux 
corresponding to that of the upper bound in Figure 
\ref{all_mass_v693_cra_x35z02o20ne10_absolute_mag}(b)
from our UV~1455~\AA\  flux fitting. Two other constraints are
also plotted, $E(B-V)=0.05\pm0.05$ and $d=7.1\pm0.7$~kpc.
See Figures \ref{all_mass_v693_cra_x45z02c15o20_absolute_mag}(b),
\ref{all_mass_v693_cra_x55z02o10ne03_absolute_mag}(b), and
\ref{all_mass_v693_cra_x65z02o03ne03_absolute_mag}(b)
for model fittings in panels (b), (c), and (d), respectively.
\label{v693_cra_distance_reddening_x35_x45_x55_x65_four}}
\end{figure}

\section{V693~CrA 1981}
\label{v693_cra}
V693~CrA is a very fast nova identified as a neon nova \citep{wil85}.
The optical and {\it IUE} UV light curves are plotted in Figure 
\ref{all_mass_v693_cra_x35z02o20ne10_absolute_mag_linear}.
The nova had already reached 7.0 mag at discovery by M. Honda \citep{koz81}
on UT 1981 April 2.75.    The discovery magnitude
was later revised by M. Honda to 6.5 mag \citep[see][]{cal82}.  
V693~CrA declined with $t_2=5.8$ and $t_3=12$ days
\citep[see, e.g., Table 5.2 of][]{war95}.

\subsection{Reddening and distance}
\label{reddening_distance_v693_cra}
The reddening toward V693~CrA was estimated
as $E(B-V)=A_V/3.1=1.7/3.1=0.55$ by \citet{bro81} from the Balmer decrement.
\citet{mir93} gave $E(B-V)=A_V/3.1=(1.65\pm0.05)/3.1=0.53\pm0.02$
toward V693~CrA. On the other hand, \citet{van97} obtained
$E(B-V)= 0.2\pm0.1$ using several independent methods, each of which
has a large uncertainty, as noted in their paper.
The Galactic dust absorption map of NASA/IPAC
gives $E(B-V)=0.099 \pm 0.001$ in the direction toward V693~CrA,
$(l,b)=(357\fdg8299, -14\fdg3912)$.  We prefer a very small value
of $E(B-V)$ because the undreddened $UBV$ color-color
evolution of V693~CrA is close to the general track of
the $UBV$ evolution, as shown in Figure
\ref{color_color_diagram_qu_vul_v351_pup_v382_vel_v693_cra}(d).
Therefore, we adopted $E(B-V)=0.05\pm0.05$ in this paper.

The distance of V693~CrA was estimated using the MMRD relations:
\citet{cal81} derived $d=7.9$--8.5~kpc, whereas
\citet{bro82} revised it to $d=12$~kpc using $m_{v,\rm max}=6.5$.
It is well known that the MMRD relations do not give an accurate estimate
of the distance for a single nova because of the large scatter of each
MMRD value \citep[e.g.,][]{dow00}.  Instead, we adopted the distance modulus
of $(m-M)_V=14.4\pm0.2$ obtained by the time-stretching
method \citep{hac10k} described in Appendix 
\ref{time_stretching_method_novae}.
Using Equation (\ref{qu_vul_distance_modulus_eq1}) with
$E(B-V)=0.05\pm0.05$, we obtain a distance of $d=7.1\pm0.7$~kpc.
We plot these constraints in Figures
\ref{v693_cra_distance_reddening_x35_m110_m115_m120_m125} and
\ref{v693_cra_distance_reddening_x35_m110_m115_m120_m125_no2}.

\subsection{Light curve fitting of V693~CrA}
\label{light_curve_fit_v693_cra}
The chemical composition of the ejecta was estimated by four groups 
\citep{wil85, and94, ark97,  van97} and their values
are summarized in Table \ref{chemical_abundance_neon_novae}.
There is a large scatter for each value, i.e., $X=0.16$ -- 0.40,
$X_{\rm CNO}=0.14$ -- 0.36, and $X_{\rm Ne}=0.17$ -- 0.26.
We obtain $X=0.29$, $X_{\rm CNO}=0.24$, and $X_{\rm Ne}=0.21$
as an arithmetic average.  We first examine the chemical composition
of Ne nova 1, then those of CO nova 3, Ne nova 2, 
and finally Ne nova 3 in the order of the degree of mixing.
Our light curve fittings are shown in Figures 
\ref{all_mass_v693_cra_x35z02o20ne10_absolute_mag_linear},
\ref{all_mass_v693_cra_x35z02o20ne10_absolute_mag} --
\ref{all_mass_v693_cra_x65z02o03ne03_absolute_mag}.
In these figures, the data of visual magnitudes are taken from 
IAU Circ. Nos. 3590, 3594 and the archive of AAVSO.  The
$V$ magnitudes are taken from \citet{cal81},
\citet{wal82}, IAU Circ. No. 3604, and the {\it IUE} VFES archive.
The UV~1455~\AA\  fluxes are taken from
the {\it IUE} archive \citep[see, e.g.,][]{cas02}.

\subsubsection{Ne nova 1}
Figure \ref{all_mass_v693_cra_x35z02o20ne10_absolute_mag_linear}
shows model light curves of the chemical composition of Ne nova 1
for four different WD masses, i.e., 1.1 (magenta), 1.15 (blue), 
1.2 (green), and $1.25~M_\sun$ (red), as well as observational data.
The absolute magnitudes of the free-free emission model light
curves are already calibrated in Section \ref{v351_pup}.
Note that we plot the light curve in a linear timescale.
In Figure \ref{all_mass_v693_cra_x35z02o20ne10_absolute_mag_linear}(a),
we place the UV~1455\AA\ 
model light curves to run through the observational peak (three 
blue filled star-marks).  The total $V$ model light curve of
$1.15~M_\sun$ WD reasonably fits to the $V$ and visual data, 
but the UV~1455\AA\  model light curve does deviate largely
from the first observational point. 

Although we successfully obtained light curve models 
for V1668~Cyg, QU~Vul, V351~Pup, and V1974~Cyg, 
it seems difficult for V693~CrA to obtain a model 
that reasonably reproduce both optical and UV~1455\AA\  light curves.
In our experience of GQ~Mus, we saw that
the UV~1455\AA\  light curve sometimes shows sharp pulses 
above a smooth light curve \citep[see][]{hac08kc}.  
If this UV peak in Figure
\ref{all_mass_v693_cra_x35z02o20ne10_absolute_mag_linear}(a)
is in such a pulse phase, we may ignore the points and fit
our model light curve with other points.   
Figure \ref{all_mass_v693_cra_x35z02o20ne10_absolute_mag_linear}(b)
shows such fittings with the same models as
in Figure \ref{all_mass_v693_cra_x35z02o20ne10_absolute_mag_linear}(a).
In this figure, we skip the three peak points (three filled star-marks)
but adopt the other data.

Now we examine which model is the best from
the distance-reddening relations in Figures
\ref{v693_cra_distance_reddening_x35_m110_m115_m120_m125}
and \ref{v693_cra_distance_reddening_x35_m110_m115_m120_m125_no2}.
The absolute magnitudes of model light curves are already calibrated
in Section \ref{v351_pup} for Ne Nova 1.  Therefore, we directly
obtain the distance modulus in $V$ band from fitting between our model
light curve and the observation, which are shown in Figure 
\ref{all_mass_v693_cra_x35z02o20ne10_absolute_mag_linear}, 
in the order of the WD mass.  Equation (\ref{qu_vul_distance_modulus_eq1})
is plotted in Figures
\ref{v693_cra_distance_reddening_x35_m110_m115_m120_m125}
and \ref{v693_cra_distance_reddening_x35_m110_m115_m120_m125_no2}
by blue solid lines.  On the other hand, the fitting with 
the UV~1455\AA\  data gives a relation of
Equation (\ref{qu_vul_uv1455_fit_eq2}), which are
also plotted in Figures
\ref{v693_cra_distance_reddening_x35_m110_m115_m120_m125}
and \ref{v693_cra_distance_reddening_x35_m110_m115_m120_m125_no2}
by magenta solid lines.  The intersection of these two distance-reddening
relations gives a set of (distance, reddening) for V693~CrA.
As mentioned in Section \ref{reddening_distance_v693_cra}, we
obtain a constraint of $E(B-V)=0.05\pm0.05$ and $d=7.1\pm0.7$~kpc.
The $1.15~M_\sun$ WD model can satisfy this constraint
both in Figures \ref{v693_cra_distance_reddening_x35_m110_m115_m120_m125}
and \ref{v693_cra_distance_reddening_x35_m110_m115_m120_m125_no2}.
Although the intersects for the $1.25~M_\sun$ WD reasonably satisfy
our constraint, its $V$ light curve decays too fast to be comparable
with the observation.  So, we exclude the $1.25~M_\sun$ WD.
We also exclude the two models of 1.1 and $1.2~M_\sun$ WDs, 
because they do not result in a good agreement in Figures
\ref{v693_cra_distance_reddening_x35_m110_m115_m120_m125} and
\ref{v693_cra_distance_reddening_x35_m110_m115_m120_m125_no2}.
Among the model fittings of the $1.15~M_\sun$ WD, Figure
\ref{all_mass_v693_cra_x35z02o20ne10_absolute_mag_linear}(b)
results in a distance modulus of $(m-M)_V=14.8$, which is larger than
$(m-M)_V=14.4\pm0.2$ obtained from the time-stretching method.
Therefore, we adopt the $1.15~M_\sun$ model in Figure
\ref{all_mass_v693_cra_x35z02o20ne10_absolute_mag_linear}(a).
Our UV~1455\AA\  model light curve runs through
the observational peak but skip the first observational point.
In what follows, we adopt a similar way, that is,
we do not include the first observational point in our UV~1455\AA\ 
fitting process of V693~CrA but regard the subsequent three points
as the peak of UV~1455\AA\  flux.

Figure \ref{all_mass_v693_cra_x35z02o20ne10_absolute_mag}(a) shows 
V693~CrA as well as other novae which we will discuss later. 
We plot the light curves in a logarithmic timescale.  We see that
the $1.15~M_\sun$ model is the best-fit model among the 1.1, 1.15, and
$1.2~M_\sun$ WD models.  In Figure 
\ref{all_mass_v693_cra_x35z02o20ne10_absolute_mag}(b),
we plot the total, free-free emission, and blackbody emission
fluxes of the $1.15~M_\sun$ WD model. 
Optically thick winds end $\sim120$ days and hydrogen shell-burning stops
$\sim200$ days after the outburst for the $1.15~M_\sun$ WD.
For $(m-M)_V=14.4$, the total $V$ flux light curve follows the observed
$V$ magnitudes of V693~CrA reasonably well.
We have obtained a hydrogen-rich envelope mass of $M_{\rm env}=
0.71\times 10^{-5} M_\sun$ at optical maximum, which corresponds
approximately to the ignition mass.

\subsubsection{CO nova 3}
Figure \ref{all_mass_v693_cra_x45z02c15o20_absolute_mag}(a) shows that
the $1.15~M_\sun$ model is the best-fit model among the 1.1, 1.15, and
$1.2~M_\sun$ WD models.
We plot the total, free-free emission, and blackbody emission
fluxes of the $1.15~M_\sun$ WD model in Figure 
\ref{all_mass_v693_cra_x45z02c15o20_absolute_mag}(b).
Optically thick winds end approximately 120 days and hydrogen
shell-burning stops $\sim190$ days 
after the outburst for the $1.15~M_\sun$ WD.
For $(m-M)_V=14.4$, the total $V$ flux light curve follows the observed
$V$ magnitudes of V693~CrA reasonably well.
We have obtained a hydrogen-rich envelope mass of $M_{\rm env}=
0.59\times 10^{-5} M_\sun$ at optical maximum.

\subsubsection{Ne nova 2}
Figure \ref{all_mass_v693_cra_x55z02o10ne03_absolute_mag}(a) shows that
the $1.2~M_\sun$ model is the best-fit model among the 1.15, 1.2,
and $1.25~M_\sun$ WD models. 
We plot the total, free-free emission, and blackbody emission
fluxes of the $1.2~M_\sun$ WD model in Figure 
\ref{all_mass_v693_cra_x55z02o10ne03_absolute_mag}(b).
Optically thick winds end approximately 140 days and hydrogen
shell-burning stops $\sim275$ days  
after the outburst for the $1.2~M_\sun$ WD.
For $(m-M)_V=14.5$, the total $V$ flux light curve follows the observed
$V$ magnitudes of V693~CrA reasonably well.
We have obtained a hydrogen-rich envelope mass of $M_{\rm env}=
0.53\times 10^{-5} M_\sun$ at optical maximum.

\subsubsection{Ne nova 3}
Figure \ref{all_mass_v693_cra_x65z02o03ne03_absolute_mag}(a) shows that
the $1.25~M_\sun$ model is the best-fit model among the 1.2, 1.25, and
$1.3~M_\sun$ WD models.
We plot the total, free-free emission, and blackbody emission
fluxes of the $1.25~M_\sun$ WD model in Figure 
\ref{all_mass_v693_cra_x65z02o03ne03_absolute_mag}(b).
Optically thick winds end approximately 145 days and hydrogen
shell-burning stops $\sim280$ days  
after the outburst for the $1.25~M_\sun$ WD.
For $(m-M)_V=14.5$, the total $V$ flux light curve follows the observed
$V$ magnitudes of V693~CrA reasonably well.
We have obtained a hydrogen-rich envelope mass of $M_{\rm env}=
0.58\times 10^{-5} M_\sun$ at optical maximum.

\subsection{Summary of V693~CrA}
Our model light curve fittings give a reasonable set of
distance and reddening, $(m-M)_V=11.4$, $E(B-V)=0.05$, and $d=7.1$~kpc,
for Ne nova 1 and CO nova 3, but the other two compositions, Ne nova 2
and Ne nova 3, result in a slightly larger distance modulus of $(m-M)_V=14.5$,
as shown in Figure \ref{v693_cra_distance_reddening_x35_x45_x55_x65_four},
but still consistent with $(m-M)_V=11.4\pm0.2$ and $E(B-V)=0.05\pm0.05$.
Because the chemical composition of Ne nova 1 is close to the arithmetic
average of the four groups' values mentioned above,
we adopt Ne nova 1 for the chemical composition
of V693~CrA.  Thus, we summarize the results of our light curve
fittings as follows (see also Table \ref{nova_parameters_results}):
$(m-M)_V=14.4$, $E(B-V)=0.05$, $d=7.1$~kpc, $M_{\rm WD}=1.15~M_\sun$, 
and $M_{\rm env}=0.71\times10^{-5}~M_\sun$ at optical maximum.
The contribution of photospheric emission is rather small, and
it affects the light curve fitting very little. 
Therefore, we can reproduce the optical light curves of V693~CrA
using only the free-free emission light curves.

\subsection{Comparison with previous results}
\label{comparison_previous_v693_cra}
\citet{wan99} estimated the WD mass from fitting their nuclear 
synthesis results with
the abundance pattern determined by \citet{van97}.
They suggested the WD mass of $1.05~M_\sun$.
\citet{dow13} estimated the WD mass of 
V693~CrA to be $M_{\rm WD}<1.3~M_\sun$ from the abundance
analysis similar to \citet{wan99}.

\citet{kat07h} estimated the WD mass and distance of V693~CrA to be
$1.3~M_\sun$ and 4.4~kpc based on the optically thick wind model.
They fitted their UV~1455\AA\  model light curves with the observation.
Because they adopted a super-Eddington luminosity model of
artificially reduced effective opacity (based on an assumption
of porous structure of 
hydrogen-rich envelope), the early timescale of the nova
evolution is not the same between their models and ours.
This is the reason that they adopted the $1.3~M_\sun$ WD, which
is much larger than our new estimate of $1.15~M_\sun$.
They adopted the reddening of $E(B-V)=0.20$ and obtained the distance
of 4.4~kpc, which is shorter than our new estimate of 7.1~kpc.


\begin{figure}
\epsscale{1.15}
\plotone{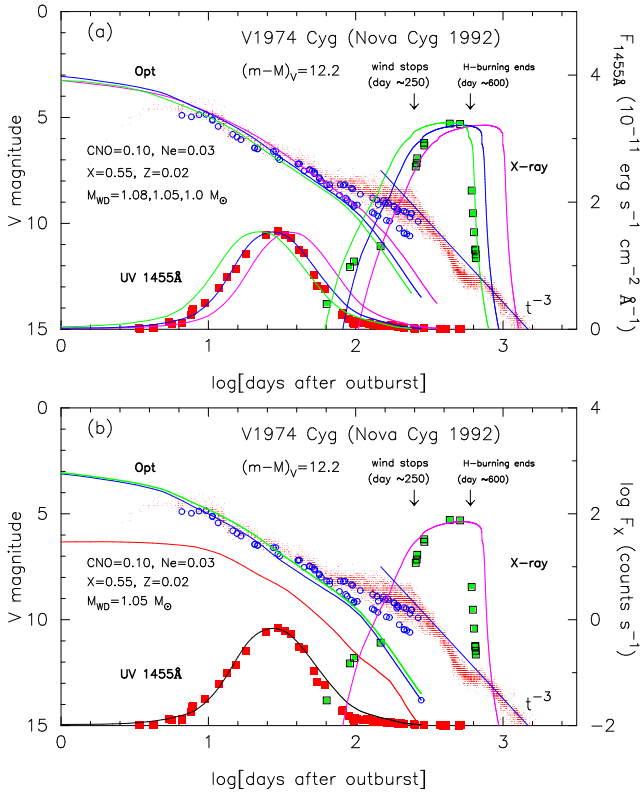}
\caption{
Same as Figure \ref{all_mass_qu_vul_x55z02o10ne03_absolute_mag},
but for V1974~Cyg.
(a) Model light curves of $1.0~M_\sun$ (magenta solid lines), 
$1.05~M_\sun$ (blue solid lines), 
and $1.08~M_\sun$ (green solid lines) WDs with Ne nova 2.
We indicate two epochs, which are observationally
suggested, by large downward arrows: the end of optically thick
winds and the end of hydrogen shell-burning.
Assuming that $(m-M)_V=12.2$, we plot the $V$ model light curves. 
(b) Assuming also that $(m-M)_V=12.2$, we plot three model light curves
of the $1.05~M_\sun$ WD.  The end of hydrogen shell-burning of the
$1.05~M_\sun$ WD model is not consistent with the observation 
($\sim600$~days).
\label{all_mass_v_uv_x_v1974_cyg_x55z02o10ne03}}
\end{figure}


\begin{figure}
\epsscale{1.15}
\plotone{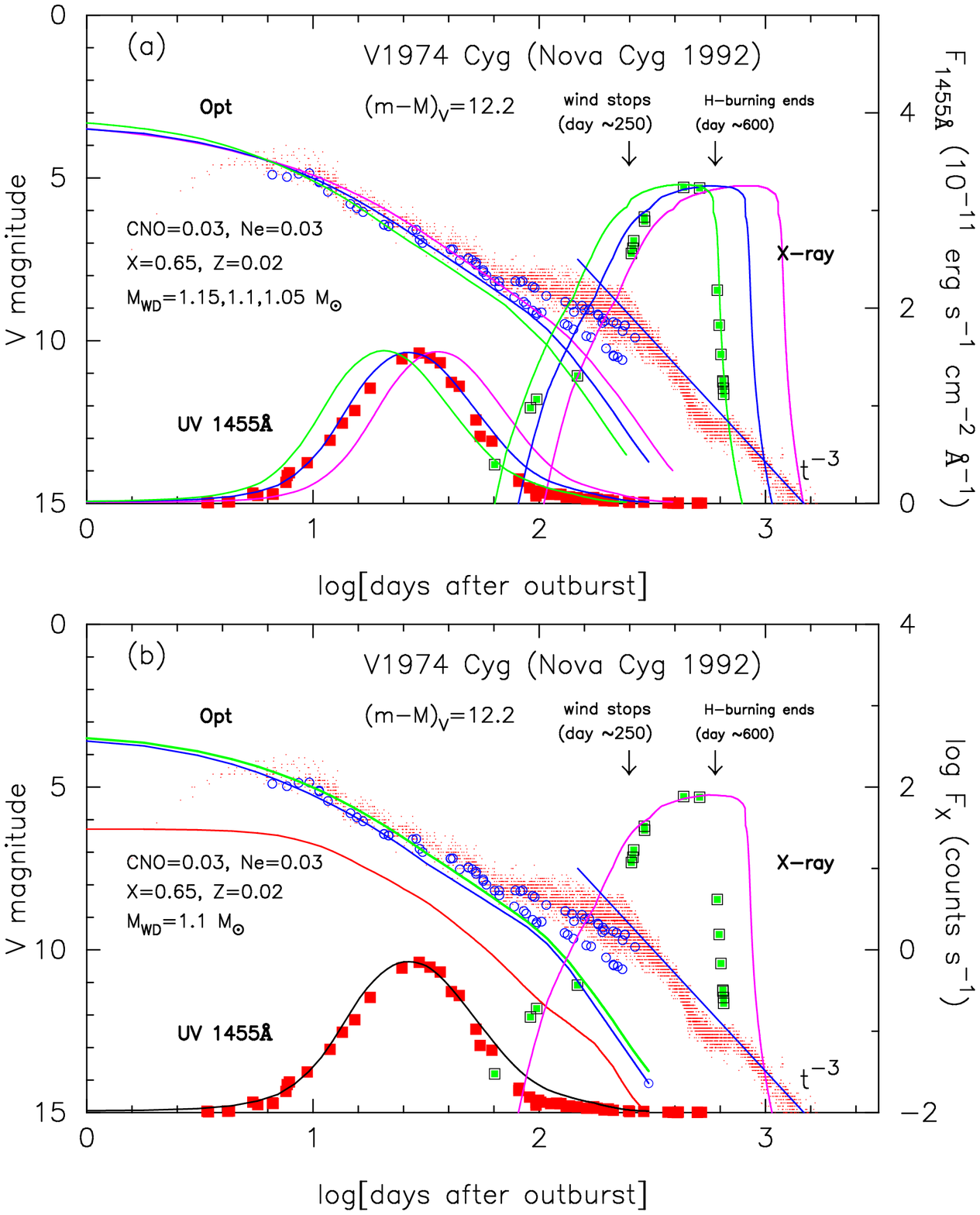}
\caption{
Same as Figure \ref{all_mass_v_uv_x_v1974_cyg_x55z02o10ne03}, but for
Ne nova 3.
(a) Model light curves of $1.05~M_\sun$ (magenta solid lines), 
$1.1~M_\sun$ (blue solid lines), 
and $1.15~M_\sun$ (green solid lines) WDs.
Assuming that $(m-M)_V=12.2$, we plot the $V$ model light curves. 
(b) Assuming also that $(m-M)_V=12.2$, we plot three model light curves
of the $1.1~M_\sun$ WD.  The end of hydrogen shell-burning
is not consistent with the observation.
\label{all_mass_v_uv_x_v1974_cyg_x65z02o03ne03}}
\end{figure}


\begin{figure}
\epsscale{1.15}
\plotone{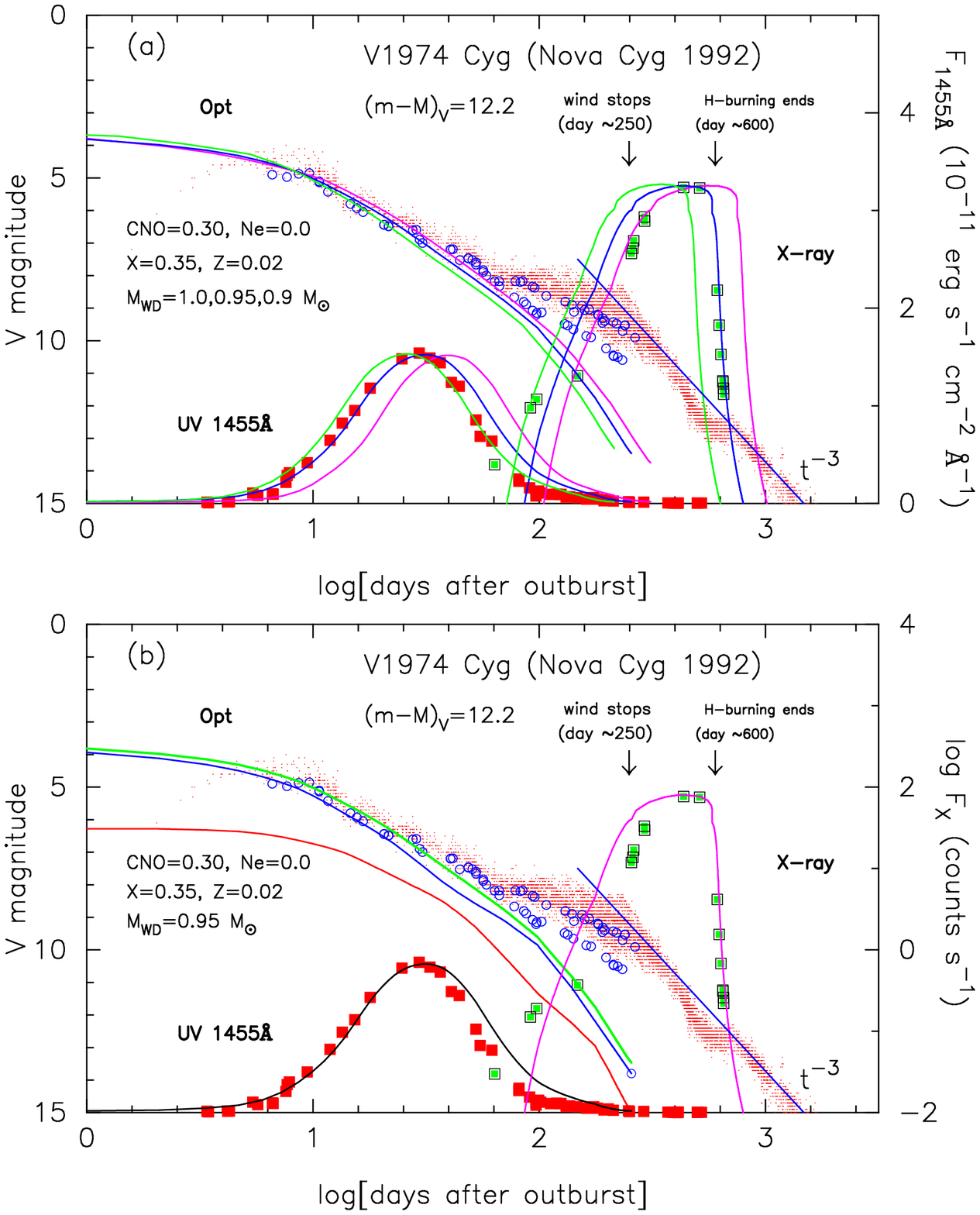}
\caption{
Same as Figure \ref{all_mass_v_uv_x_v1974_cyg_x55z02o10ne03}, but for
CO nova 2.
(a) Model light curves of $0.9~M_\sun$ (magenta solid lines), 
$0.95~M_\sun$ (blue solid lines), 
and $1.0~M_\sun$ (green solid lines) WDs.
Assuming that $(m-M)_V=12.2$, we plot the $V$ model light curves. 
(b) Assuming also that $(m-M)_V=12.2$, we plot three model light curves
of the $0.95~M_\sun$ WD.
Optically thick winds and hydrogen shell-burning end approximately
250 days and 600 days after the outburst, respectively,
for the $0.95~M_\sun$ WD, which are both consistent with the observation.
\label{all_mass_v_uv_x_v1974_cyg_x35z02c10o20}}
\end{figure}


\begin{figure}
\epsscale{1.15}
\plotone{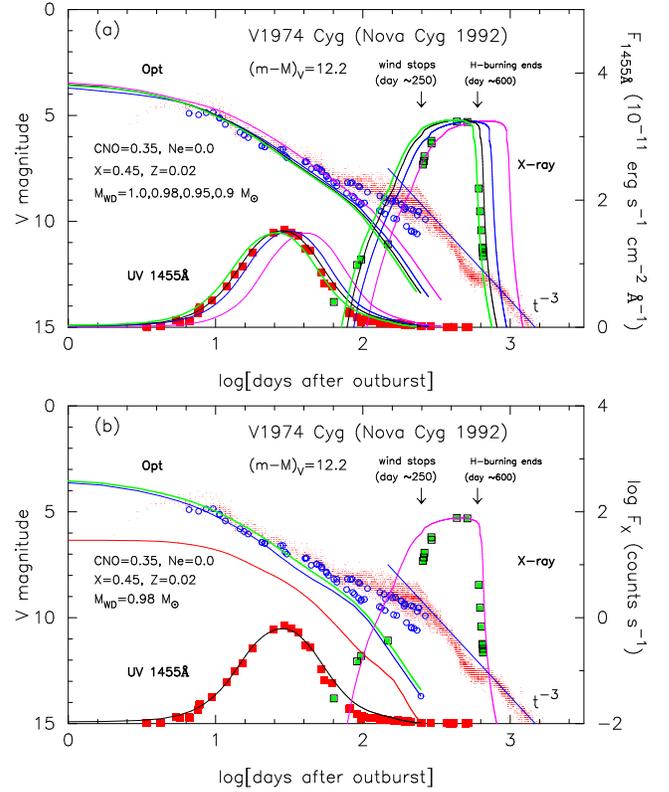}
\caption{
Same as Figure \ref{all_mass_v_uv_x_v1974_cyg_x55z02o10ne03},
but for CO Nova 3.
(a) Model light curves of $0.9~M_\sun$ (magenta solid lines), 
$0.95~M_\sun$ (blue solid lines), $0.98~M_\sun$ (black thin solid lines), 
and $1.0~M_\sun$ (green solid lines) WDs.
Assuming that $(m-M)_V=12.2$, we plot the $V$ model light curves. 
(b) Assuming also that $(m-M)_V=12.2$, we plot three model light curves
of the $0.98~M_\sun$ WD.  In this case, the two epochs are
consistent with the observation.
\label{all_mass_v_uv_x_v1974_cyg_x45z02c15o20}}
\end{figure}


\begin{figure}
\epsscale{1.15}
\plotone{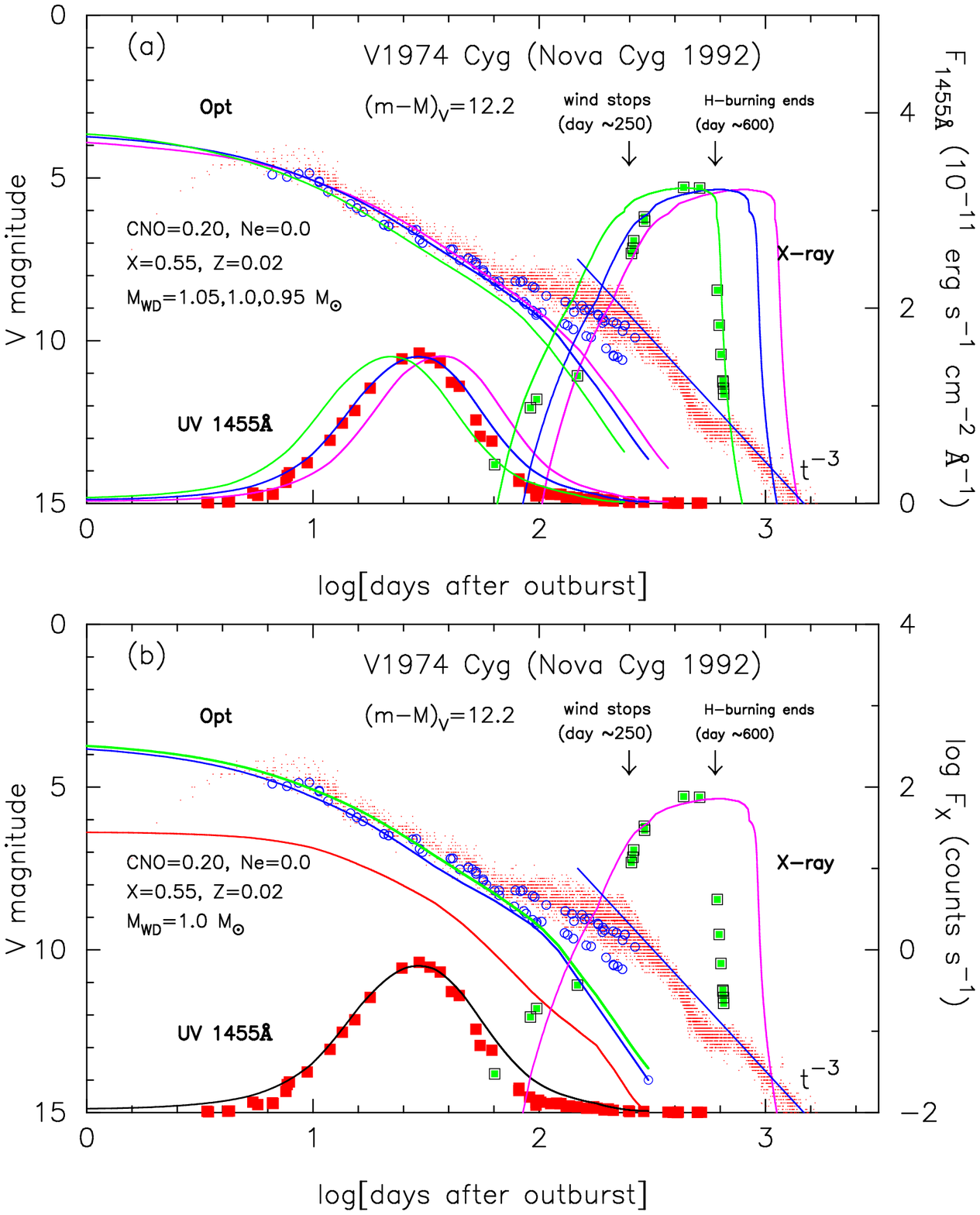}
\caption{
Same as Figure \ref{all_mass_v_uv_x_v1974_cyg_x55z02o10ne03}, but for
CO nova 4.
(a) Model light curves of $0.95~M_\sun$ (magenta solid lines), 
$1.0~M_\sun$ (blue solid lines), 
and $1.05~M_\sun$ (green solid lines) WDs.
Assuming that $(m-M)_V=12.2$, we plot the $V$ model light curves. 
(b) Assuming also that $(m-M)_V=12.2$, we plot three model light curves
of the $1.0~M_\sun$ WD.
In this case, however, the end of hydrogen shell-burning
is not consistent with the observation.
\label{all_mass_v_uv_x_v1974_cyg_x55z02c10o10}}
\end{figure}


\begin{figure}
\epsscale{0.85}
\plotone{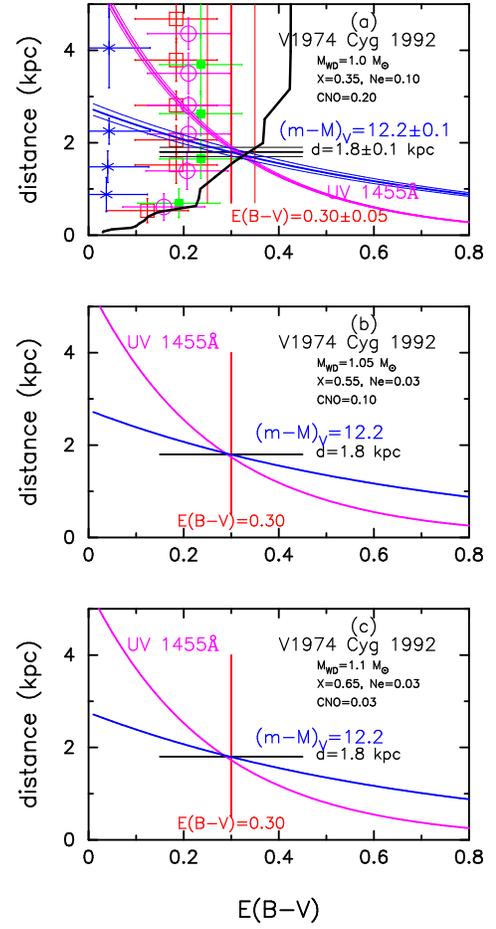}
\caption{
Distance-reddening relations toward V1974~Cyg calculated from the
$V$ (blue solid lines) and UV~1455\AA\   (magenta solid lines)
light curve fittings:
(a) $1.0~M_\sun$ WD with the chemical composition of Ne nova 1,
(b) $1.05~M_\sun$ WD with Ne nova 2,
and (c) $1.1~M_\sun$ WD with Ne nova 3.
The red vertical solid lines show the reddening estimate of
$E(B-V)=0.3\pm0.05$, whereas the black horizontal solid lines
correspond to the distance estimate of $d=1.8\pm0.1$~kpc.
Panel (a) also shows the distance-reddening relations
in four directions close to V1974~Cyg, $(l, b)=(89\fdg1338, 7\fdg8193)$:
$(l, b)=(89\fdg00,7\fdg75)$ (red open squares),
$(89\fdg25, 7\fdg75)$ (green filled squares),
$(89\fdg00,  8\fdg00)$ (blue asterisks),
and $(89\fdg25,  8\fdg00)$ (magenta open circles), the data for which
are taken from \citet{mar06}.  We also add Green et al.'s (2015)
relation (black solid line).  See text for more detail.
\label{v1974_cyg_distance_reddening_x35_x55_x65_3fig_ne}}
\end{figure}


\begin{figure}
\epsscale{0.85}
\plotone{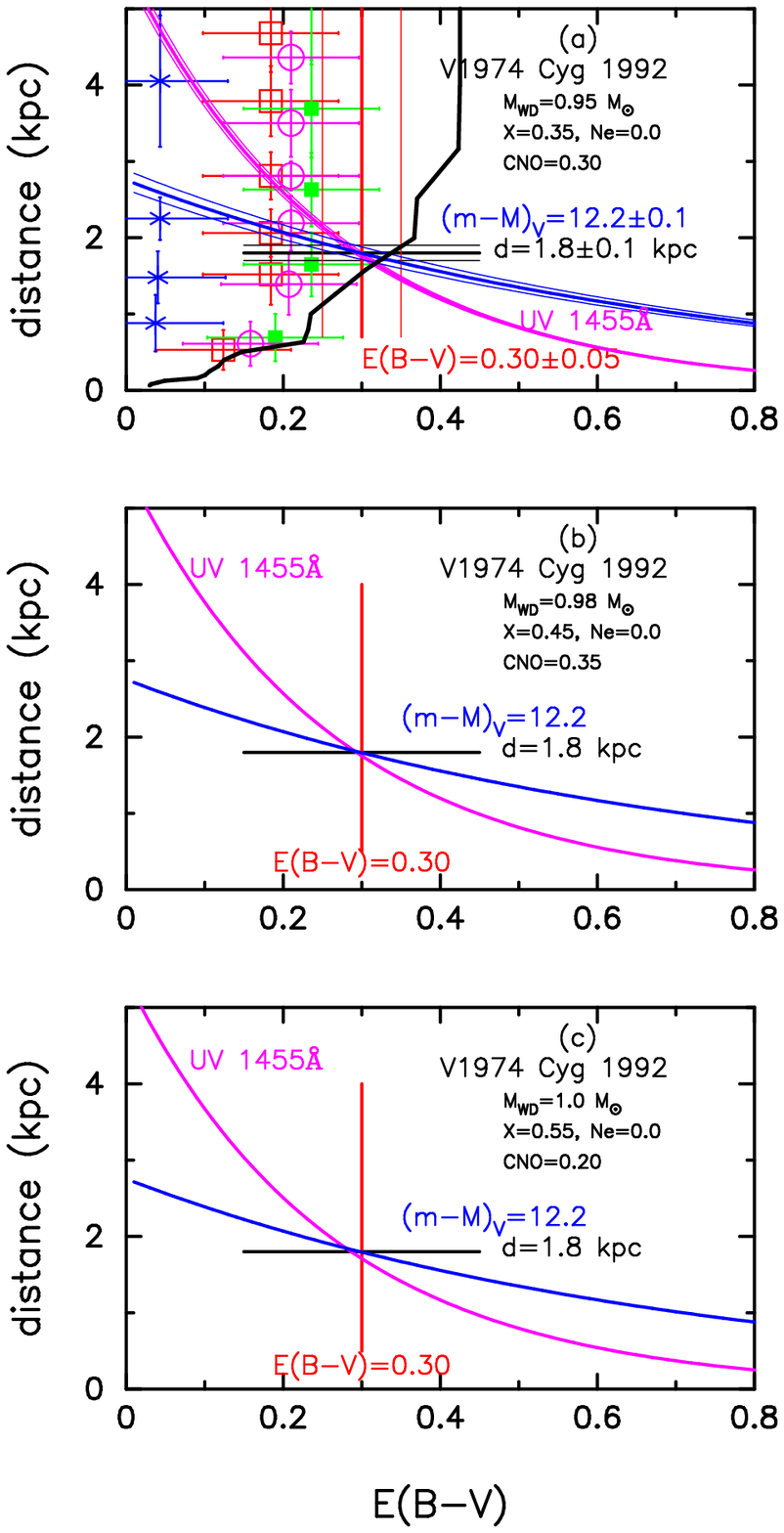}
\caption{
Same as Figure \ref{v1974_cyg_distance_reddening_x35_x55_x65_3fig_ne},
but for (a) $0.95~M_\sun$ WD with CO Nova 2,
(b) $0.98~M_\sun$ WD with CO Nova 3,
and (c) $1.0~M_\sun$ WD with CO Nova 4.
\label{v1974_cyg_distance_reddening_x35_x45_x55_3fig_co}}
\end{figure}

\section{V1974~Cyg 1992}
\label{light_curve_v1974_cyg}
We reanalyzed the light curves of V1974~Cyg
including the effect of photospheric emission, which is not considered in 
\citet{hac06kb, hac10k, hac14k}.
We obtained a distance modulus of $(m-M)_V=12.2\pm 0.1$ for V1974~Cyg
using our time-stretching method (see Appendix 
\ref{time_stretching_method_novae}) and other estimates cited in
\citet{hac14k}.  In the following analyses, we assume that $(m-M)_V=12.2$.
The chemical composition of V1974~Cyg was estimated by
\citet{aus96}, \citet{hay96}, \citet{ark97}, and \citet{van05}
to be $X=0.19$ -- 0.55 and $X_{\rm CNO}=0.12$ -- 0.375 
(see Table \ref{chemical_abundance_neon_novae}).  Because these values
are so scattered, we use all six of the chemical compositions in Table
\ref{chemical_composition_model}, i.e.,  Ne nova 1, Ne nova 2,
Ne nova 3, CO nova 2, CO nova 3, and CO nova 4.

\subsection{Light curve fittings of V1974~Cyg}
Figures 
\ref{all_mass_v351_pup_v1974_cyg_x35z02o20ne10},
\ref{all_mass_v351_pup_v1974_cyg_x55z02o10ne03},
\ref{all_mass_v_uv_x_v1974_cyg_x55z02o10ne03},
\ref{all_mass_v_uv_x_v1974_cyg_x65z02o03ne03},
\ref{all_mass_v_uv_x_v1974_cyg_x35z02c10o20},
\ref{all_mass_v_uv_x_v1974_cyg_x45z02c15o20}, and
\ref{all_mass_v_uv_x_v1974_cyg_x55z02c10o10}
show our model light curve fittings for six chemical compositions.
In Figures
\ref{all_mass_v_uv_x_v1974_cyg_x55z02o10ne03} --
\ref{all_mass_v_uv_x_v1974_cyg_x55z02c10o10},
the visual (red small dots) data are taken from
the AAVSO archive; the $V$ magnitudes (blue open circles) are from
\citet{cho93}; the supersoft X-ray data (green filled squares) are
from \citet{kra96}; the UV 1455 ~\AA\   data (red filled squares)
are from \citet{cas02}. 
The supersoft X-ray phase started $\sim250$~days after the outburst
and that is identified to be the epoch when optically
thick winds stop \citep{hac06kb, hac10k}.  
The hydrogen shell-burning ended $\sim600$~days
after the outburst, corresponding to the sharp decay of supersoft
X-ray flux.

Figure \ref{all_mass_v351_pup_v1974_cyg_x35z02o20ne10}(a) shows that
the timescale of V1974~Cyg is almost the same as that of V351~Pup and
we chose the $1.0~M_\sun$ WD model as the best-fit model 
from the UV~1455~\AA\   fitting (Ne nova 1).
However, Figure \ref{all_mass_v351_pup_v1974_cyg_x35z02o20ne10}(b)
shows that the hydrogen shell-burning on the $1.0~M_\sun$ WD model 
ended too earlier than the observation.
Therefore, we examine other cases of the chemical composition.

Figure \ref{all_mass_v_uv_x_v1974_cyg_x55z02o10ne03}(a) shows that
the $1.05~M_\sun$ WD model is the best-fit model from
the UV~1455~\AA\   model light curve fitting (Ne nova 2).
The total $V$ light curve model agrees well with 
the $V$ magnitude observation, and the shape of UV~1455~\AA\  light curve
model fits the observation well. In this case, however,
the end of hydrogen shell burning is not consistent with the observation
as shown in Figure \ref{all_mass_v_uv_x_v1974_cyg_x55z02o10ne03}(b).

Figure \ref{all_mass_v_uv_x_v1974_cyg_x65z02o03ne03}(a)
suggests that the $1.1~M_\sun$ WD model is the best-fit model 
from the UV~1455~\AA\   model light curve fitting (Ne nova 3).
In this case, too, the end of hydrogen shell burning is not consistent
with the observation as shown in Figure
\ref{all_mass_v_uv_x_v1974_cyg_x65z02o03ne03}(b).

Figure \ref{all_mass_v_uv_x_v1974_cyg_x35z02c10o20}(a) shows that
the $0.95~M_\sun$ WD model is the best-fit model from the UV~1455 ~\AA\  
light curve fitting (CO nova 2).
Figure \ref{all_mass_v_uv_x_v1974_cyg_x35z02c10o20}(b) shows that
optically thick winds and hydrogen shell burning end approximately 250 days 
and 600 days after the outburst, respectively, which is consistent with
the X-ray decay time.
The total $V$ light curve model agrees well with the $V$ magnitude
observation, and the shape of the UV~1455~\AA\  light curve
model fits the observation reasonably well.  

Figure \ref{all_mass_v_uv_x_v1974_cyg_x45z02c15o20}(a) shows that
the $0.98~M_\sun$ WD model is the best-fit model from
the UV~1455~\AA\   model light curve fitting (CO nova 3).
Figure \ref{all_mass_v_uv_x_v1974_cyg_x45z02c15o20}(b) shows that
the $V$ magnitude observation and the shape of the UV~1455~\AA\  light curve
model fit the observation well.  The two epochs are consistent
with the X-ray observation; that is, optically thick winds and hydrogen
shell burning end approximately 250 days and 600 days after the outburst,
respectively.

Figure \ref{all_mass_v_uv_x_v1974_cyg_x55z02c10o10}(a) shows that
the $1.0~M_\sun$ WD model is the best-fit model from
the UV~1455~\AA\   model light curve fitting (CO nova 4).
Figure \ref{all_mass_v_uv_x_v1974_cyg_x55z02c10o10}(b) shows that
the total $V$ light curve model agrees well with 
the $V$ magnitude observation, and the shape of the UV~1455~\AA\  light curve
model fits the observation well.  In this case, however,
the end of hydrogen shell burning is not consistent with the observation
as shown in Figure \ref{all_mass_v_uv_x_v1974_cyg_x55z02c10o10}(b).

It should be noted that our UV~1455\AA\  light curve represents 
the early phase development of the pseudophotosphere while the X-ray
light curve constrains the later phase development.
Therefore, if the X-ray data as well as UV~1455\AA\  data
are available, we could select
a better model to constrain the abundance.

\subsection{Distance and reddening toward V1974~Cyg}
\label{v1974_cyg_distance_reddening}
We obtain two distance-reddening relations to V1974~Cyg from
the two model light curve fittings ($V$ and UV~1455~\AA), which are plotted
in Figures \ref{v1974_cyg_distance_reddening_x35_x55_x65_3fig_ne}
and \ref{v1974_cyg_distance_reddening_x35_x45_x55_3fig_co}.
The thick blue solid line shows the result obtained using Equation 
(\ref{qu_vul_distance_modulus_eq1}) with
$(m-M)_V=12.2\pm 0.1$.  The thick magenta solid line shows the result
obtained using Equation (\ref{qu_vul_uv1455_fit_eq2})
with $F_{1455}^{\rm mod}= 1.53 \times 
10^{-11}$~erg~cm$^{-2}$~s$^{-1}$~~\AA$^{-1}$ as the 
upper bound of Figure 
\ref{all_mass_v351_pup_v1974_cyg_x35z02o20ne10}(b)
at a distance of 10~kpc and $F_{1455}^{\rm obs}= (5.0\pm0.3) \times 
10^{-11}$~erg~cm$^{-2}$~s$^{-1}$~~\AA$^{-1}$ as the 
corresponding observed flux.
These two relations are indicated in Figure
\ref{v1974_cyg_distance_reddening_x35_x55_x65_3fig_ne}(a)
by flanking thin solid lines.

Figure \ref{v1974_cyg_distance_reddening_x35_x55_x65_3fig_ne}(a)
shows another distance-reddening relation
toward V1974~Cyg: that given by \citet{mar06},
where the galactic coordinates of V1974~Cyg are
$(l, b)=(89\fdg1338, 7\fdg8193)$.  
The four sets of data points with error bars correspond to
the distance-reddening relations in four directions close to V1974~Cyg:
$(l, b)=(89\fdg00,7\fdg75)$ (red open squares),
$(89\fdg25, 7\fdg75)$ (green filled squares),
$(89\fdg00,  8\fdg00)$ (blue asterisks),
and $(89\fdg25,  8\fdg00)$ (magenta open circles),
the data for which are taken from \citet{mar06}.
We also add Green et al.'s (2015) relation (black thick solid line).
These trends are consistent with the cross at $d\approx 1.8$~kpc and
$E(B-V)\approx0.30$.

\citet{cho97a} estimated the distance to V1974~Cyg as
$d=1.77\pm 0.11$~kpc from an expansion parallax method.
Here, we adopted $d=1.8\pm0.1$~kpc (black horizontal solid line
flanked by thin solid lines) after Chochol et al.
The reddening was also estimated by many researchers.
\citet{aus96} obtained the reddening toward V1974~Cyg mainly on the basis
of the UV and optical line ratios for days 200 through 500, i.e.,
$E(B-V)=0.3 \pm 0.1$.  The NASA/IPAC galactic dust absorption map gives
$E(B-V)=0.35 \pm 0.01$ in the direction toward V1974~Cyg.
These are all consistent with the above observational
estimates, that is, $d=1.8\pm 0.1$~kpc and $E(B-V)=0.3 \pm 0.05$.

For Ne nova 2, Ne nova 3, CO nova 2, CO nova 3, and CO nova 4,
we obtain a reasonable set of $d\sim 1.8$~kpc and $E(B-V)\sim 0.30$,
as shown in Figures
\ref{v1974_cyg_distance_reddening_x35_x55_x65_3fig_ne}(b),
\ref{v1974_cyg_distance_reddening_x35_x55_x65_3fig_ne}(c),
\ref{v1974_cyg_distance_reddening_x35_x45_x55_3fig_co}(a),
\ref{v1974_cyg_distance_reddening_x35_x45_x55_3fig_co}(b),
and \ref{v1974_cyg_distance_reddening_x35_x45_x55_3fig_co}(c),
respectively.

\subsection{Summary of V1974~Cyg}
Our light curve fittings of the $V$
and UV~1455~\AA\  fluxes give a reasonable set of
(distance, reddening)=$(d,~E(B-V))=(1.8\pm0.1{\rm ~kpc}, 0.30\pm0.05)$
for all six chemical compositions.  The reddening and distance
are consistent with the results presented in the previous subsection
(Section \ref{v1974_cyg_distance_reddening}).  We obtain 
a chemical composition close to that of CO nova 2
or CO nova 3 from X-ray flux fittings because  
the two epochs of the model light curve are consistent
with the X-ray observation; that is,
optically thick winds and hydrogen shell burning end approximately
250 days and 600 days after the outburst, respectively.  
It should be noted again that enrichment of neon with unchanged hydrogen
and CNO mass fractions affects the nova model light curves very little.
Therefore, we cannot select the best-fit $X_{\rm Ne}$ only from
our model light curve analysis.  We suppose that our model
light curves do follow the observation well for other chemical
compositions, for example,
of $X=0.35$, $X_{\rm CNO}=0.30$, $Y=0.33 - X_{\rm Ne}$,
$Z=0.02$ (CO nova 2 type), and any $X_{\rm Ne}\approx0.0$--$0.1$,  
or of $X=0.45$, $X_{\rm CNO}=0.35$, $Y=0.18 - X_{\rm Ne}$, $Z=0.02$
(CO nova 3 type), and any $X_{\rm Ne}\approx0.0$--$0.1$.  
If we adopt the composition of CO nova 3 type, we obtain
$M_{\rm WD}=0.98~M_\sun$ and $M_{\rm env}=1.28\times10^{-5}~M_\sun$
at optical maximum (see Table \ref{nova_parameters_results}).

\subsection{Comparison with previous results}
\label{comparison_previous_v1974_cyg}
Several groups tried to estimate the WD mass of V1974~Cyg.
\citet{par95} presented a result of $0.75 - 1.1 ~M_\sun$
from various empirical relations on novae.
\citet{ret97} also obtained a mass range of 
$M_{\rm WD} = 0.75 - 1.07 ~M_\sun$
based on the precessing disk model of superhump phenomenon.
\citet{wan99} presented nuclear burning yields on the ONe core
and compared them with the observed pattern by \citet{aus96}.
They suggested the WD mass of $M_{\rm WD} \lesssim1.15~M_\sun$.

\citet{sal05} calculated static sequences of hydrogen shell-burning
and compared them with the evolutional speed of post-wind phase for V1974~Cyg.
They suggested that the WD mass is $0.9 ~M_\sun$ for 50\% mixing of
a solar composition envelope with an ONe core (i.e., $X=0.35$, 
$X_{\rm CNO}=0.25$, and $X_{\rm Ne}=0.16$), or $1.0 ~M_\sun$ for 
25\% mixing (i.e., $X=0.53$, $X_{\rm CNO}=0.13$, and $X_{\rm Ne}=0.08$).
Their two cases resemble our Ne nova 1 and Ne nova 2, respectively.
These WD masses are consistent with our new estimates.

\citet{hac06kb} proposed the free-free emission model for optical
light curves and applied to the V1974~Cyg light curve.
They obtained the WD mass of $0.95~M_\sun$ based
on the UV~1455\AA\  and supersoft X-ray light curve fittings.
They could not fix the proportionality constant $C$ in their models
and did not estimate the distance modulus.  Instead, they adopted
the distance of 1.8~kpc and the reddening of $E(B-V)=0.32$, 
taken from \citet{cho97a}.  

\citet{kat07h} estimated the WD mass and distance of V1974~Cyg to be
$1.05~M_\sun$ and 1.8~kpc, respectively,
based on the optically thick wind model.
They fitted their UV~1455\AA\  model light curves with the observation.
Because they adopted a super-Eddington luminosity model of
reduced effective opacity, the early timescale of the nova
evolution is not the same between their models and ours
as explained in Section \ref{comparison_previous_v693_cra}.
This is the reason that they adopted the $1.05~M_\sun$ WD, which
is slightly larger than our new estimate of $0.98~M_\sun$.
They adopted the reddening of $E(B-V)=0.32$ and obtained the distance
of 1.8~kpc.

Using the time-stretching method
of the universal decline law, \citet{hac10k} obtained the absolute
magnitude of their free-free emission model light curves and
applied their light curve model to V1974~Cyg and obtained 
the distance modulus in the $V$ band as $(m-M)_V=12.2$
and the WD mass of $0.95~M_\sun$.  Introducing the effect of
photospheric emission as well as free-free emission in the
present paper, we reanalyzed the optical light curve of V1974~Cyg and
fixed $M_{\rm WD}=0.98~M_\sun$ and $(m-M)_V=12.2$ for the
chemical composition of CO nova 3.


\begin{deluxetable*}{lllllll}
\tabletypesize{\scriptsize}
\tablecaption{Nova parameters
\label{nova_parameters_results}}
\tablewidth{0pt}
\tablehead{
\colhead{object name} & 
\colhead{$E(B-V)$} & 
\colhead{$(m-M)_V$} & 
\colhead{distance} & 
\colhead{WD mass} & 
\colhead{envelope mass\tablenotemark{a}} & 
\colhead{best-fit}\\
\colhead{} &
\colhead{} &
\colhead{} &
\colhead{(kpc)} &
\colhead{($M_\sun$)} &
\colhead{($10^{-5}M_\sun$)} &
\colhead{abundance} 
} 
\startdata
V1668~Cyg & 0.30 & 14.6 & 5.4 & 0.98 & 1.38 & CO nova 3 \\
V1974~Cyg & 0.30 & 12.3 & 1.8 & 0.98 & 1.28 & CO nova 3\tablenotemark{b} \\
QU~Vul & 0.55 & 13.6 & 2.4 & 0.96 & 2.44 & Ne nova 3 \\
V351~Pup & 0.45 & 15.1 & 5.5 & 1.0 & 1.98 & Ne nova 1 \\
V382~Vel & 0.15 & 11.5 & 1.6 & 1.23 & 0.48 & Ne nova 2 \\ 
V693~CrA & 0.05 & 14.4 & 7.1 & 1.15 & 0.71 & Ne nova 1 
\enddata
\tablenotetext{a}{Mass of the hydrogen-rich envelope at optical maximum
calculated from each nova model.}
\tablenotetext{b}{V1974~Cyg was identified as a neon nova. Note, however,
that enrichment of neon with unchanged hydrogen and CNO mass fractions
affects the nova light curves very little in our model light curves
because neon is not relevant to either nuclear burning (the CNO cycle)
or the opacity.}
\end{deluxetable*}


\begin{deluxetable}{llllll}
\tabletypesize{\scriptsize}
\tablecaption{Extinctions and Distances of Selected Novae
\label{extinction_distance_from_stretching}}
\tablewidth{0pt}
\tablehead{
\colhead{Object} & \colhead{Year} & \colhead{$E(B-V)$}
& \colhead{$(m-M)_V$} & \colhead{distance} & reference\tablenotemark{a} \\
  &  &  &  &  (kpc) 
}
\startdata
OS~And & 1986 & 0.15 & 14.7 & 7.0 & 1 \\
V603~Aql & 1918 & 0.07 & 7.2 & 0.25 & 2,3 \\
V1370~Aql & 1982 & 0.35 & 15.2 & 6.7 & 1 \\
V1419~Aql & 1993 & 0.50 & 14.6 & 4.1 &  1 \\
V705~Cas & 1993 & 0.45 & 13.4 & 2.5 & 4 \\
V723~Cas & 1995 & 0.35 & 14.0 & 3.85 & 4,5 \\
IV~Cep & 1971 & 0.70 & 14.7 & 3.2 & 1 \\
V693~CrA & 1981 & 0.05 & 14.4 & 7.1 & 6 \\
V1500~Cyg & 1975 & 0.45 & 12.3 & 1.5 & 1 \\
V1668~Cyg & 1978 & 0.30 & 14.6 & 5.4 & 6 \\
V1974~Cyg & 1992 & 0.30 & 12.2 & 1.8 & 1,6,7 \\
V2274~Cyg & 2001\#1 & 1.35 & 18.7 & 8.0 & 1 \\
HR~Del & 1967 & 0.15 & 10.4 & 0.97 & 1,4,8 \\
DQ~Her & 1934 & 0.10 & 8.2 & 0.39 & 3,9 \\
V446~Her & 1960 & 0.40 & 11.7 & 1.2 & 1 \\
V533~Her & 1963 & 0.05 & 10.8 & 1.3 & 1 \\
GQ~Mus & 1983 & 0.45 & 15.7 & 7.3 & 4 \\
RS~Oph  & 1958 & 0.65 & 12.8 & 1.4 & 1 \\
GK~Per  & 1901 & 0.30 & 9.3 & 0.48 & 3,10 \\
RR~Pic  & 1925 & 0.04 & 8.7 & 0.52 & 3 \\
V351~Pup  & 1991 & 0.45 & 15.1 & 5.5 & 6 \\
T~Pyx  & 1966 & 0.25 & 14.2 & 4.8 & 1,12 \\
V443~Sct & 1989 & 0.40 & 15.5 & 7.1 & 1 \\
V475~Sct & 2003 & 0.55 & 15.6 & 6.0 & 1 \\
FH~Ser & 1970 & 0.60 & 11.7 & 0.93 & 1 \\
V5114~Sgr & 2004 & 0.45 & 16.5 & 10.5 & 1 \\
V5558~Sgr & 2007 & 0.70 & 13.9 & 2.2 & 1 \\
V382~Vel & 1999 & 0.15 & 11.5 & 1.6 & 6 \\
LV~Vul & 1968\#1 & 0.60 & 11.9 & 1.0 & 1 \\
NQ~Vul & 1976 & 1.00 & 13.6 & 1.3 & 1 \\
PU~Vul & 1979 & 0.30 & 14.3 & 4.7 & 1,11 \\
PW~Vul & 1984\#1 & 0.55 & 13.0 & 1.8 & 4 \\
QU~Vul & 1984\#2 & 0.55 & 13.6 & 2.4 & 6 \\
QV~Vul & 1987 & 0.60 & 14.0 & 2.7 & 1 
\enddata
\tablenotetext{a}{1--\citet{hac14k}, 2--\citet{gal74}, 3--\citet{har13},
4--\citet{hac15k}, 5--\citet{lyk09}, 6--Present paper, 7--\citet{cho97a},
8--\citet{har03}, 9--\citet{ver87}, 10--\citet{wu89}, 11--\citet{kat12mh},
12--\citet{sok13}.}
\end{deluxetable}

\section{Discussion}
\label{discussion}
\subsection{White dwarf masses of neon novae}
\label{wd_mass_neon_novae}
We obtained the WD masses of neon novae, i.e., 
$M_{\rm WD}=0.82$ -- $0.96~M_\sun$ for QU~Vul in Section \ref{qu_vul},
$M_{\rm WD}=0.98$ -- $1.1~M_\sun$ for V351~Pup in Section \ref{v351_pup},
$M_{\rm WD}=1.13$ -- $1.28~M_\sun$ for V382~Vel in Section \ref{v382_vel}, 
$M_{\rm WD}=1.15$ -- $1.25~M_\sun$ for V693~CrA in Section \ref{v693_cra},
and $M_{\rm WD}=0.95$ -- $1.1~M_\sun$ for V1974~Cyg in Section 
\ref{light_curve_v1974_cyg}.
On the other hand, a lower mass bound of $\sim1.07~M_\sun$ for natal ONe WDs
with a CO-rich mantle (as well as a thin helium-rich layer on the CO mantle)
was obtained by \citet{ume99} from their evolution calculation.
A different estimate, $\sim1.0~M_\sun$, was also obtained by \citet{wei00}
from discussion on the initial/final mass relation of
remnants in open clusters.  \citet{gil03} presented their evolution
calculations showing that a CO-rich mantle is more massive for lower-mass
ONe cores, i.e., as massive as $\sim0.1~M_\sun$ for a $1.1~M_\sun$ WD.
Therefore, we suppose that the lower mass bound for natal
pure ONe cores without a CO-rich mantle is $\sim 1.0~M_\sun$.
So we conclude that the WD mass of QU~Vul, 0.82 -- $0.96~M_\sun$, 
is smaller than the minimum mass of ONe core.  This suggests that the WD
in QU~Vul has lost at least $\sim 0.1~M_\sun$ of mass since its birth.

The WD masses of V351~Pup and V1974~Cyg are close to this lower bound
of natal pure ONe cores, whereas those of V382~Vel and V693~CrA are
much above this lower bound.  The WD mass of V1668~Cyg (not a neon nova)
was also estimated in Section \ref{light_curve_v1668_cyg}
to be $M_{\rm WD}=0.98$ -- $1.1~M_\sun$, which is on the lower mass 
bound of natal pure ONe cores.  We point out two possibilities.
One is that at its birth, the WD mass was close to the upper
mass bound of natal CO cores, $< 1.07~M_\sun$ \citep[e.g.,][]{ume99},
and it decreased to the present value.  The other is that at its birth,
the WD mass was above $1.07~M_\sun$, and it decreased
to the present value, but the CO-rich mantle is still surrounding
the ONe core.

Note that strong neon emission lines were also observed in the very
slow nova V723~Cas, although it was not identified as a neon nova.
\citet{iij06} estimated its neon mass fraction to be $X_{\rm Ne}=0.052$,
which is comparable to the values of 0.06 for V1974~Cyg
\citep{van05}, 0.03--0.038 for V838~Her \citep{van97,sch07},
0.043 for V382~Vel \citep{sho03}, and 0.032--0.040 for QU~Vul
\citep{sai92, sch02} as listed
in Table \ref{chemical_abundance_neon_novae}.
Unlike that of the above novae, the WD mass of V723~Cas was 
estimated to be as small as $M_{\rm WD}=0.5$--$0.55~M_\sun$ from optical,
UV~1455~\AA, and X-ray light curve fittings \citep{hac15k}.
It could be a CO WD, if the strong neon emission lines 
are not direct evidence of an ONe core.

\citet{liv94} proposed an explanation of modest neon enrichments 
for such low mass WDs.
They grouped the 18 classical novae into three classes on the basis
of the abundance characteristics: class 1 that show a modest enrichment
in heavy elements but a large enrichment in helium; class 2 that
show a high enrichment in CNO nuclei and sometimes a modest enrichment
in neon; class 3 which show quite extreme enrichments in both 
neon and heavier elements (than neon).  Class 3 includes V693~CrA,
V1370~Aql, and QU~Vul.  They identified these class 3 novae unambiguously
as ``the neon novae,'' which require dredge-up from an underlying
ONe WD.  Class 2 includes DQ~Her, V1500~Cyg, V1668~Cyg, GQ~Mus,
PW~Vul, V842~Cen, V827~Her, and V2214~Oph.  These systems show
considerable CNO enrichment and sometimes a modest enrichment in neon
and a range in helium concentration from approximately solar to twice solar.
Class 1 includes T~Aur, RR~Pic, HR~Del, V977~Sco, V443~Sct,
and LMC~1990\#1.  These show a modest enrichment in CNO nuclei but
a large enrichment in helium, and a modest enrichment in neon appears
in some of these novae.  For class 1 and class 2 novae, \citet{liv94}
discussed that their modest neon enrichment can be explained
by a concentration of $^{22}$Ne in the underlying CO core.  
In the pre-nova evolution, intermediate mass stars convert
most of the initial carbon, nitrogen, and oxygen isotopes into $^{14}$N
in the CNO cycle during their main-sequence period.  During the ensuing 
helium burning phase, some of this $^{14}$N is transformed into $^{22}$Ne
by $^{14}$N$(\alpha,\gamma)^{18}$F$(e^+\nu)^{18}$O$(\alpha,\gamma)^{22}$Ne.
Dredge-up of the CO core material could accompany enrichment of $^{22}$Ne.
We suppose that V723~Cas belongs to the class 1 defined by Livio \& Truran,
because the optical light curve of V723~Cas is very similar to
those of RR~Pic and HR~Del \citep[e.g.,][]{hac15k}.

\subsection{Color-color diagram of nova outbursts}
\label{color_color_diagrams}
\citet{hac14k} proposed a new method of determining the
reddening of classical novae.  They identified a general course of the 
color evolution track in the $(B-V)_0$-$(U-B)_0$ diagram 
and determined the reddening of novae by comparing the track of a target
nova with the general course.  Our target neon novae also show a
similarity in the $B-V$ and $U-B$ color evolutions.
We made the color-color diagrams of our five neon novae in Figure
\ref{color_color_diagram_qu_vul_v351_pup_v382_vel_v693_cra}, that is,
\ref{color_color_diagram_qu_vul_v351_pup_v382_vel_v693_cra}(a) for QU~Vul,
\ref{color_color_diagram_qu_vul_v351_pup_v382_vel_v693_cra}(b)
for V351~Pup and V1974~Cyg,
\ref{color_color_diagram_qu_vul_v351_pup_v382_vel_v693_cra}(c) for V382~Vel,
and
\ref{color_color_diagram_qu_vul_v351_pup_v382_vel_v693_cra}(d) for V693~CrA.

\citet{hac14k} showed that 
novae evolve along the nova-giant sequence in the pre-maximum and 
near-maximum phases.  
This sequence is parallel to but $\Delta (U-B)\approx -0.2$ mag
bluer than the supergiant sequence, as indicated by green solid lines
in Figure \ref{color_color_diagram_qu_vul_v351_pup_v382_vel_v693_cra}.
After optical maximum, a nova quickly comes back blueward along
the nova-giant sequence and reaches the point of
free-free emission ($(B-V)_0=-0.03$, $(U-B)_0=-0.97$; this point is
denoted by a black open diamond), which coincides with
the intersection of the blackbody sequence and the nova-giant sequence,
and remains there for a while.  Then the color evolves leftward
(blueward in $B-V$ but almost constant in $U-B$), owing mainly to
the development of strong emission lines.

Figure \ref{color_color_diagram_qu_vul_v351_pup_v382_vel_v693_cra}(a)
shows the color-color evolution of QU~Vul.  The data are taken
from \citet{ber88} (red filled circles) and \citet{ros92}
(blue open circles).  We obtained $E(B-V)=0.55\pm0.05$.

Figure \ref{color_color_diagram_qu_vul_v351_pup_v382_vel_v693_cra}(b)
shows that of V351~Pup and V1974~Cyg.  We plot the data for V351~Pup taken
from \citet{bru92} (large red filled circles) and IAU Circulars
(blue open circles), as well as the data for V1974~Cyg
(small magenta dots), which are taken from \citet{cho93}.
The track of V351~Pup is similar to that of V1974~Cyg.
We obtained $E(B-V)=0.45\pm0.05$ for V351~Pup as well as
$E(B-V)=0.30\pm0.05$ for V1974~Cyg \citep{hac14k}.

Figure \ref{color_color_diagram_qu_vul_v351_pup_v382_vel_v693_cra}(c)
shows that of V382~Vel.  The color data for V382~Vel are taken
from IAU Circulars (red filled circles).  We obtained $E(B-V)=0.15\pm0.05$.

Figure \ref{color_color_diagram_qu_vul_v351_pup_v382_vel_v693_cra}(d)
shows that of V693~CrA.  The data for V693~CrA are taken
from \citet{cal81} (red filled circles), \citet{wal82} 
(magenta open diamond), and IAU Circulars (blue open circle).
We obtained $E(B-V)=0.05\pm0.05$.

These reddening values of $E(B-V)$ are very consistent with
those obtained in the previous sections.


\begin{figure}
\epsscale{1.15}
\plotone{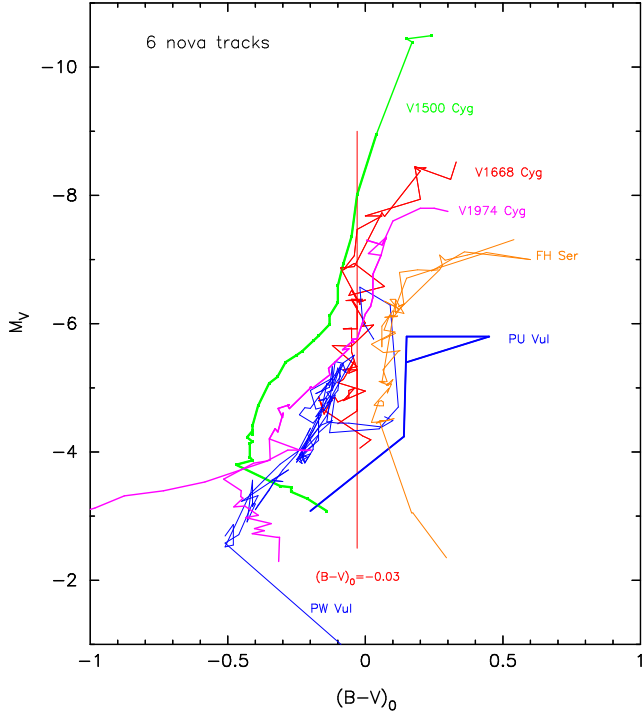}
\caption{
Color-magnitude diagrams of six well-observed novae in outburst, i.e.,
V1500~Cyg, V1668~Cyg, V1974~Cyg, PW~Vul, FH~Ser, and PU~Vul, in the order
of the nova speed class.
Here, $(B-V)_0$ is the dereddened color of $B-V$, and $M_V$ is
the absolute $V$ magnitude.  The red vertical line of $(B-V)_0=-0.03$
indicates the color of optically thick free-free emission.
The figure is a similar one to Figure 39 of \citet{hac14k},
but the tracks of V1974~Cyg and V1668~Cyg are revised. 
See text for more detail.
The sources of these six nova data were the same
as those cited in \citet{hac14k}.
We adopt $E(B-V)=0.45$, 0.30, 0.30, 0.55, 0.60, 0.30, and
$(m-M)_V=12.3$, 14.6, 12.2, 13.0, 11.7, 14.3, for V1500~Cyg, V1668~Cyg,
V1974~Cyg, PW~Vul, FH~Ser, and PU~Vul, respectively. 
\label{hr_diagram_6types_novae_one_no2}}
\end{figure}


\begin{figure}
\epsscale{1.15}
\plotone{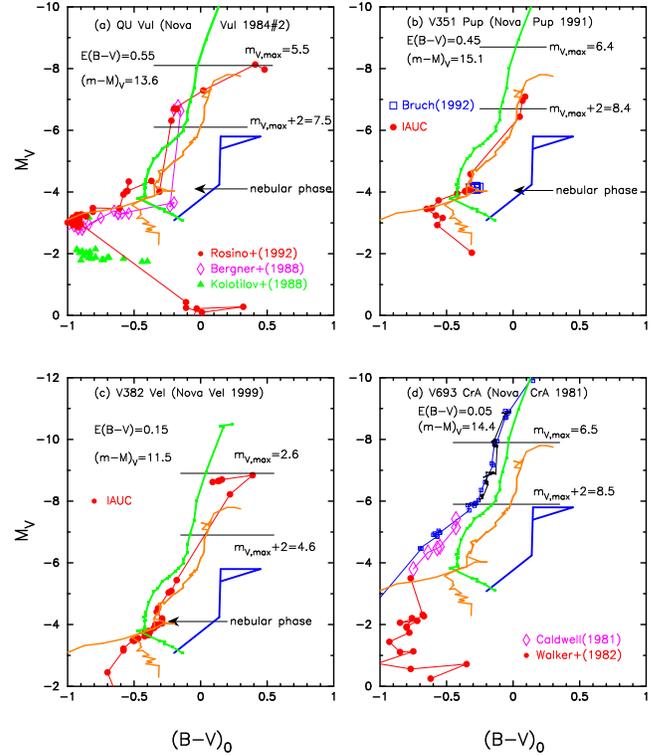}
\caption{
Color-magnitude diagrams of four neon novae in outburst as well as
the tracks of V1500~Cyg (green thick solid line), 
V1974~Cyg (orange thick solid line), and PU~Vul (blue thick solid line).
The data of V1500~Cyg, V1974~Cyg,
and PU~Vul are the same as those in Figure 
\ref{hr_diagram_6types_novae_one_no2}.
(a) QU~Vul for $(m-M)_V=13.6$ and $E(B-V)=0.55$.
The data of QU~Vul are taken from \citet{ber88}, \citet{kol88}, 
and \citet{ros92}. 
(b) V351~Pup for $(m-M)_V=15.1$ and $E(B-V)=0.45$.  The data of
V351~Pup are taken from \citet{bru92} and IAU Circulars.  
(c) V382~Vel for $(m-M)_V=11.5$ and $E(B-V)=0.15$.
The data of V382~Vel are taken from IAU Circulars.
(d) V693~CrA for $(m-M)_V=14.4$ and $E(B-V)=0.05$.
The data of V693~CrA are taken from \citet{cal81} and \citet{wal82}.
Other color-magnitude data of V1500~Cyg are added to show different
observed colors, i.e., taken from \citet[][small blue squares connected
by blue thin solid line]{pfa76}
and \citet[][small filled black circles connected by black thin solid 
line]{lin79}.  
\label{hr_diagram_qu_vul_v351_pup_v382_vel_v693_cra_outburst}}
\end{figure}

\subsection{Color-magnitude diagram of nova outbursts}
\label{hr_diagrams}
Figure \ref{hr_diagram_6types_novae_one_no2} summarizes
the color-magnitude diagrams for six well-observed novae.
The figure is similar to Figure 39 of \citet{hac14k}, but
we added the track of V1974~Cyg assuming that $(m-M)_V=12.2$ and 
$E(B-V)=0.30$.  We revised the distance modulus of
$(m-M)_V=14.6$ and the color excess of $E(B-V)=0.30$ for V1668~Cyg,
as obtained in Section \ref{light_curve_v1668_cyg}.
The other four novae are the same as those in Figure 39 of
\citet{hac14k}, i.e., $(m-M)_V=12.3$ and $E(B-V)=0.45$ for V1500~Cyg,
$(m-M)_V=13.0$ and $E(B-V)=0.55$ for PW~Vul,
$(m-M)_V=11.7$ and $E(B-V)=0.60$ for FH~Ser,
$(m-M)_V=14.3$ and $E(B-V)=0.30$ for PU~Vul.
The sources of these color-magnitude data are cited in \citet{hac14k}.
A vertical straight line of $(B-V)_0= -0.03$ indicates the 
intrinsic $B-V$ color of optically thick free-free emission \citep{hac14k}.
If optically thick free-free emission dominates the spectrum of a nova,
its track follows this straight line of $(B-V)_0= -0.03$.

\citet{van87} derived two general trends of color evolution in nova
light curves, i.e., $(B-V)_{0,\rm max} = 0.23\pm 0.06$ at maximum and 
$(B-V)_{0,t2} = -0.02\pm 0.04$ at day $t_2$.
However, the color at maximum, $(B-V)_{0,\rm max}$,
is not the same for all the novae, as clearly shown in 
Figure \ref{hr_diagram_6types_novae_one_no2}.
A nova rises to the peak magnitude (top of each line)
and then descends along each line.  V1500~Cyg, V1668~Cyg, and V1974~Cyg
show colors consistent with $(B-V)_{0, \rm max} = 0.23\pm 0.06$.
In this way, many fast and very fast novae are consistent with
van den Bergh \& Younger's $(B-V)_{0, \rm max} = 0.23\pm 0.06$.
The other types of novae, e.g., FH~Ser and PU~Vul,
are not consistent with this, because their long journeys
along the nova-giant sequence in the color-color diagram reach
$(B-V)_0\sim0.6$  \citep[see Figures 4 and 16 of][]{hac14k},
far beyond $(B-V)_0=0.23$.  Thus, van den Bergh and Younger's law of
$(B-V)_{0,\rm max} = 0.23\pm 0.06$ is usually not applicable to
slow/very slow novae.

The second law of van den Bergh \& Younger's, $(B-V)_{0,t2} = -0.02\pm0.04$
at day $t_2$, shows good agreement with the fast novae,
V1500~Cyg, V1668~Cyg, and V1974~Cyg, but is not consistent with
the slow novae, FH~Ser and PU~Vul.
Their value of $(B-V)_{0,t2} = -0.02\pm0.04$ is very close to the value of
$(B-V)_0=-0.03$ for optically thick free-free emission.
This is because fast novae usually remain at the point of the open diamond
in Figure \ref{color_color_diagram_qu_vul_v351_pup_v382_vel_v693_cra}
for a while when free-free emission dominates the spectrum
\citep{hac14k}.
However, some very slow novae, e.g., PU~Vul, V5558~Sgr, V723~Cas,
and HR~Del, remain at a different point of $(B-V)_0=+0.13$ and
$(U-B)_0=-0.82$ in the color-color 
diagram for a while, as shown by \citet{hac14k}; the color 
is not consistent with $(B-V)_{0,t2} = -0.02\pm0.04$.  Thus,
the PU~Vul and FH~Ser type tracks show much redder colors at day $t_2$.
This is because photospheric emission rather than free-free emission 
dominates the spectra of slow/very slow novae \citep{hac15k}.

Figure \ref{hr_diagram_qu_vul_v351_pup_v382_vel_v693_cra_outburst}
shows the color-magnitude diagrams of four neon novae in outburst.
Figure \ref{hr_diagram_qu_vul_v351_pup_v382_vel_v693_cra_outburst}(a)
is that of QU~Vul 1984\#2, assuming
$(m-M)_V=13.6$ and $E(B-V)=0.55$, as well as the tracks of
V1500~Cyg (thick green solid line), V1974~Cyg (orange solid line), 
and PU~Vul (blue solid line).  The thick solid lines for V1500~Cyg,
V1974~Cyg, and PU~Vul are the same as those in Figure 
\ref{hr_diagram_6types_novae_one_no2}.
The data for QU~Vul are taken from \citet{ber88}, \citet{kol88},
and \citet{ros92}.
The $B-V$ colors of \citet{ros92} are systematically $\sim0.2$ mag bluer
than those of \citet{ber88}, so that we shifted them rightward (redward)
by 0.2 mag.  
Although the color data of QU~Vul are scattered between the two groups,
i.e., \citet{ber88} and \citet{ros92}, the track of QU~Vul is 
close to those of V1500~Cyg and V1974~Cyg.

Figure \ref{hr_diagram_qu_vul_v351_pup_v382_vel_v693_cra_outburst}(b)
is that of V351~Pup 1991 for $(m-M)_V=15.1$ and $E(B-V)=0.45$.  The data
for V351~Pup are taken from \citet{bru92} and IAU Circulars.  
The track of V351~Pup is very close to that of V1974~Cyg.  
This may support our estimates of $(m-M)_V=15.1$ and $E(B-V)=0.45$.

Figure \ref{hr_diagram_qu_vul_v351_pup_v382_vel_v693_cra_outburst}(c)
is that of V382~Vel 1999 for $(m-M)_V=11.5$ and $E(B-V)=0.15$.
The data for V382~Vel are taken from IAU Circulars.
The track of V382~Vel is also close to that of V1974~Cyg.
This may also support our estimates of $(m-M)_V=11.5$ and $E(B-V)=0.15$.

Figure \ref{hr_diagram_qu_vul_v351_pup_v382_vel_v693_cra_outburst}(d) 
is that of V693~CrA 1981 for $(m-M)_V=14.4$ and $E(B-V)=0.05$.
The data for V693~CrA are taken from \citet{cal81} and \citet{wal82}.
It should be noted that the response functions of $B$ and $V$ filters
are sometimes slightly different among different observers.
If strong emission lines contribute to the edge of the filter,
this small difference makes a large difference in the $B-V$ colors.
For V1500~Cyg, we added such different data observed
by \citet[][small blue squares
connected by blue thin solid line]{pfa76} and \citet[][small filled
black circles connected by black thin solid line]{lin79}.
These new lines deviate somewhat from the green thick
solid line (same as that in Figure \ref{hr_diagram_6types_novae_one_no2}).
The track of V693~CrA is close to that of V1500~Cyg, that is,
closer to the color-magnitude tracks of V1500~Cyg taken from 
\citet{pfa76} and \citet{lin79} rather than that taken from \citet{kis77}
(green thick solid line).


\begin{figure}
\epsscale{1.15}
\plotone{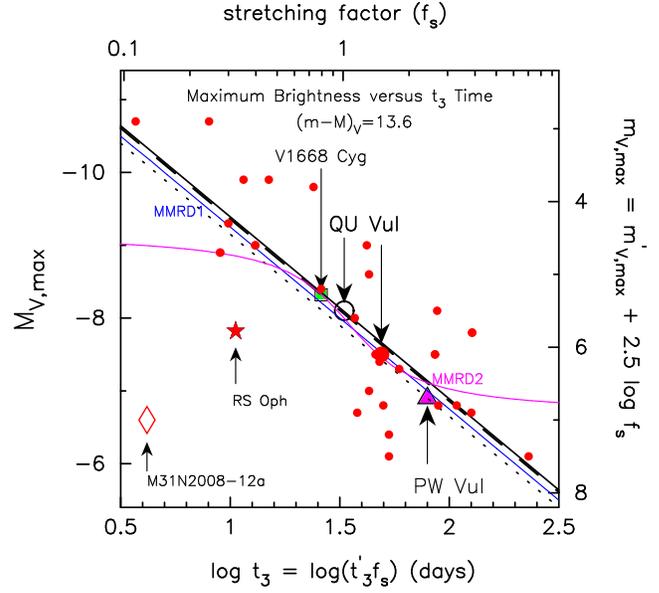}
\caption{
Various MMRD relations on the basis of our free-free emission model 
light curves.  The black solid line shows the theoretical MMRD relation
calibrated with QU~Vul in Section \ref{qu_vul}.
We also add observational MMRD points ($M_{V, \rm{max}}$ vs. $t_3$)
for individual novae (red filled circles), the data for which are
taken from Table 5 of \cite{dow00}.  
The black dashed line shows the theoretical MMRD relation 
calibrated with V1668~Cyg in Section \ref{light_curve_v1668_cyg}.
The black dotted line denotes the theoretical MMRD relation calibrated
with PW~Vul, which is taken from \citet{hac15k}.
The blue solid line labeled ``MMRD1'' represents the Kaler-Schmidt law
\citep{sch57}.  The magenta solid line labeled ``MMRD2'' corresponds to
della Valle \& Livio's law (1995).
The red filled star is the MMRD point of the recurrent nova RS Oph,
as an example for a very high mass accretion rate
and very short $t_3$ time (very small $f_{\rm s}$).
The red open diamond is the MMRD point of the 1-yr recurrence period
M31 nova, M31N2008-12a, taken from \citet{dar15}, i.e., 
$M_{V, \rm max}=-6.6$ and $t_{3,V}\approx3.8$ days.
The large red filed circle is the MMRD point of QU~Vul, the data for which
are taken from Table 5 of \citet{dow00}.
The large open black circle is the MMRD point of QU~Vul, which is
estimated from our free-free model light curve of the $0.96~M_\sun$ WD.
See text for more details.
\label{max_t3_point_B_scale_qu_vul}}
\end{figure}


\begin{figure}
\epsscale{1.15}
\plotone{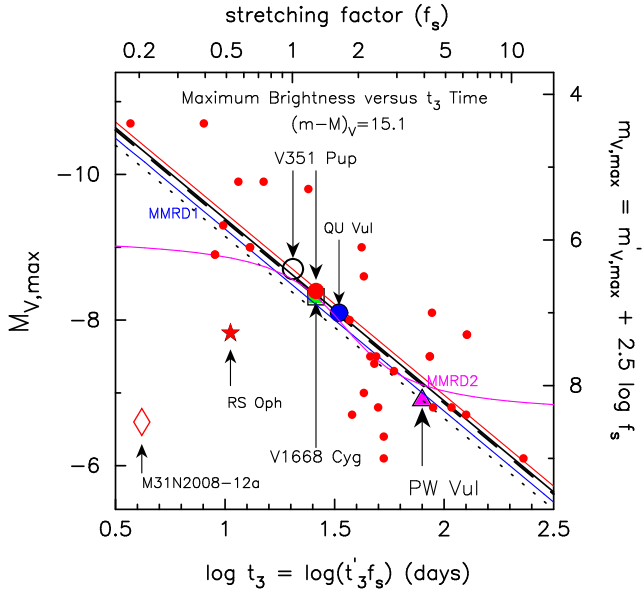}
\caption{
Same as Figure \ref{max_t3_point_B_scale_qu_vul}, but for
the theoretical MMRD relation (red solid line) on the basis of
our free-free model light curves calibrated with V351~Pup in
Section \ref{v351_pup}.
The large red filled circle denotes the MMRD point of V351~Pup,
which is taken from Table 5 of \citet{dow00}.
The large black open circle is the MMRD point of V351~Pup, which is
estimated from our free-free emission model light curve
of the $1.0~M_\sun$ WD with Ne nova 1.
The large blue filled circle is the MMRD point for QU~Vul,
which is the same point as the large black open circle in Figure
\ref{max_t3_point_B_scale_qu_vul}.
\label{max_t3_point_B_scale_v351_pup}}
\end{figure}


\begin{figure*}
\epsscale{0.85}
\plotone{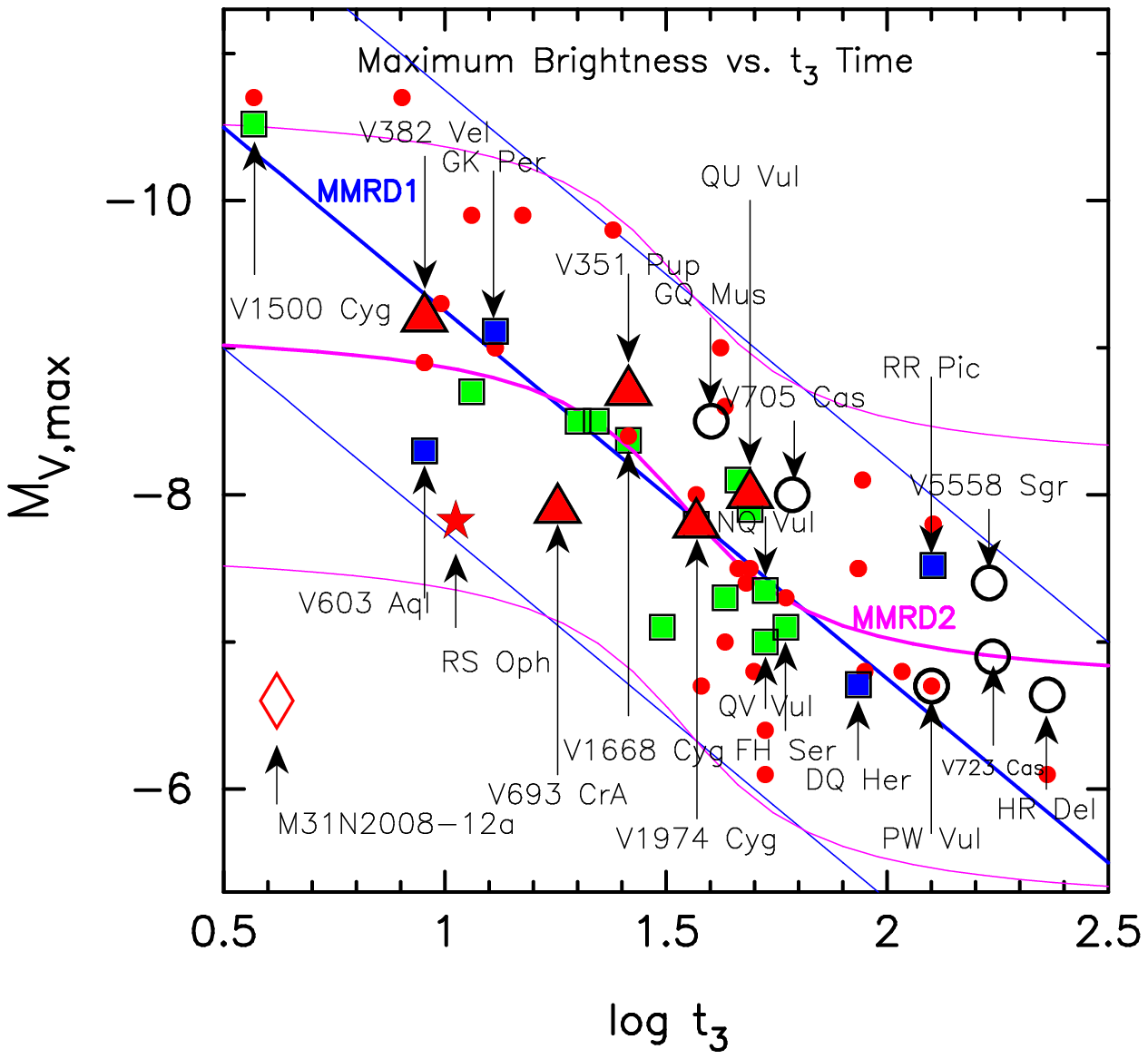}
\caption{
Various MMRD relations, where $t_3$ is measured on the observational data.
The red filled triangles show the five novae studied
in the present work, QU~Vul, V351~Pup, V382~Vel, V693~CrA, and V1974~Cyg.
The black open circles denote the six novae studied in \citet{hac15k},
PW~Vul, V705~Cas, GQ~Mus, V5558~Sgr, HR~Del, and V723~Cas.
The blue filled squares correspond to the four novae calibrated
with {\it HST} annual parallaxes, the data for which are 
taken from \citet{har13}.
The green filled squares indicate the novae calibrated
with the time-stretching method, the data for which are
taken from \citet{hac14k}.
The blue solid line labeled ``MMRD1'' is the Kaler-Schmidt law \citep{sch57}
and its $\pm1.5$~mag lines (see Equation (\ref{kaler-schmidt-law})).
The magenta solid line labeled ``MMRD2'' is 
della Valle \& Livio's law (1995) and its $\pm1.5$~mag lines
(see Equation (\ref{della-valle-livio-law})).
The other symbols are the same as those in Figures
\ref{max_t3_point_B_scale_qu_vul} and \ref{max_t3_point_B_scale_v351_pup}.
\label{max_t3_point_B_scale_qu_vul_no1}}
\end{figure*}

\subsection{MMRD relation of neon novae}
\label{mmrd_neon_novae}
The MMRD relations have frequently been used to estimate the distance
to a nova.  We call the relation between $t_3$ (or $t_2$)
and $M_{V, {\rm max}}$ the MMRD relation.
Here, $t_3$ ($t_2$) time is defined as the 3 mag (2 mag) decay time 
from maximum in units of days.   On the basis of our nova light curve
analyses, we have already calculated theoretical MMRD relations
for CO nova 2 and Ne nova 2 in \citet{hac10k},
and for CO nova 4 in \citet{hac15k}. 
In Appendix \ref{mmrd_relation_model_fit}, we obtained
theoretical MMRD relations for CO nova 3 based on the V1668~Cyg
observations, for Ne nova 3 based on QU~Vul, 
and for Ne nova 1 based on V351~Pup.

\subsubsection{MMRD relations based on V1668~Cyg, QU~Vul, and V351~Pup}
\label{mmrd_qu_vul_v351_pup_v1668_cyg}
Using theoretical light curves of free-free emission, we can derive
a relation between the maximum magnitude and $t_3$ time, i.e., the
MMRD relation (see Appendix \ref{mmrd_relation_model_fit}).
Figure \ref{max_t3_point_B_scale_qu_vul} shows the theoretical
MMRD relation calibrated with the data for QU~Vul as well as
those calibrated with V1668~Cyg and with PW~Vul \citep{hac15k}.
Each MMRD relation is a straight line that goes through the calibrated
object.  The location of three novae are highlighted by large symbols.
On the left ordinate, we convert the apparent magnitude to the absolute
magnitude using the distance modulus of QU~Vul, $(m-M)_V=13.6$.
We also show the stretching factor $f_{\rm s}$ on the upper axis 
of the same figure.

For comparison, two empirical MMRD relations
are plotted in Figure \ref{max_t3_point_B_scale_qu_vul},
i.e., the Kaler-Schmidt law 
\citep[blue solid line labeled ``MMRD1'',][]{sch57}:
\begin{equation}
M_{V, {\rm max}} = -11.75 + 2.5 \log t_3,
\label{kaler-schmidt-law}
\end{equation}
and della Valle \& Livio's law 
\citep[magenta solid line labeled ``MMRD2'',][]{del95}:
\begin{equation}
M_{V, {\rm max}} = -7.92 -0.81 \arctan \left(
{{1.32-\log t_2} \over {0.23}} \right),
\label{della-valle-livio-law}
\end{equation}
where we use the relation $t_2  \approx 0.6 \times t_3$ 
for the optical light curves that follow the free-free model
light curves \citep{hac06kb}.

\citet{dow00} also obtained an MMRD relation of
\begin{equation}
M_{V, {\rm max}} = (-11.99\pm0.56) + (2.54\pm0.35) \log t_3,
\label{downes-duerbeck-t3-law}
\end{equation}
on the basis of their data, although this MMRD relation is not shown
in the figure in order to avoid complexity.

Our MMRD relation based on the light curve of QU~Vul, i.e.,
Equation (\ref{theoretical_MMRD_relation_qu_vul}), is consistent with
the Kaler-Schmidt law in Equation (\ref{kaler-schmidt-law}) and
Downes \& Duerbeck's law of Equation (\ref{downes-duerbeck-t3-law}),
and lies between them.
\citet{dow00} estimated a $t_3$ time of 49~days
and $M_{V, \rm max}=-7.5$ for QU~Vul (large red filled circle 
in Figure \ref{max_t3_point_B_scale_qu_vul}).
Note that we remeasured the $t_3$ time
along our model light curve and obtained $t_3= 32.7$~days
and $M_{V, \rm max}= -8.1$ for QU~Vul (this data point corresponds to
a large black open circle above and to the left of Downes \& Duerbeck's one).

Figure \ref{max_t3_point_B_scale_v351_pup} shows the MMRD relation 
(red thin solid line with a large black open circle) based on
the light curve of V351~Pup, i.e.,
Equation (\ref{theoretical_MMRD_relation_v351_pup}) 
in Appendix \ref{mmrd_relation_model_fit}.  This relation is also 
consistent with the Kaler-Schmidt law of Equation (\ref{kaler-schmidt-law})
and Downes \& Duerbeck's law of Equation (\ref{downes-duerbeck-t3-law}),
and lies between them.

We plot the MMRD relation based on the data of V1668~Cyg in Figures 
\ref{max_t3_point_B_scale_qu_vul} and \ref{max_t3_point_B_scale_v351_pup}
(black thick dashed line with green filled square).  Our MMRD relation, i.e.,
Equation (\ref{theoretical_MMRD_relation_v1668_cyg}), is consistent with
the Kaler-Schmidt law of Equation (\ref{kaler-schmidt-law}) and
Downes \& Duerbeck's law of Equation (\ref{downes-duerbeck-t3-law}).

\citet{hac15k} also obtained an MMRD relation based on PW~Vul
(CO nova 4).  This result is also plotted in Figures
\ref{max_t3_point_B_scale_qu_vul} and \ref{max_t3_point_B_scale_v351_pup}.
The four MMRD relations based on QU~Vul (Ne nova 3), 
V351~Pup (Ne nova 1), V1668~Cyg (CO nova 3), and PW~Vul (CO nova 4)
are close enough that we may conclude that the theoretical MMRD relations
depend very little on the chemical composition.

\subsubsection{Physical properties of MMRD relations}
\label{mmrd_relation_physical_property}
Figure \ref{max_t3_point_B_scale_qu_vul_no1} shows our updated
MMRD points ($t_3$, $M_{V, \rm{max}}$) as well as Downes \& Duerbeck's
(2000) and Harrison et al.'s (2013) points.
Red large filled triangles represent five novae studied in this work,
QU~Vul, V351~Pup, V382~Vel, V693~CrA, and V1974~Cyg.
The locations of these neon novae in the MMRD diagram are
similar to those of other fast novae and have no noteworthy distribution.  
Black large open circles represent the six novae studied in \citet{hac15k},
PW~Vul, V705~Cas, GQ~Mus, V5558~Sgr, HR~Del, and V723~Cas.
Red filled circles are individual novae taken from Table 5 of \citet{dow00}.
Blue filled squares represent four novae calibrated with the annual
parallax method, which are taken from \citet{har13}.
Green filled squares depict novae calibrated with the time-stretching method,
taken from Table \ref{extinction_distance_from_stretching} 
\citep[see][]{hac14k}.
Blue solid line labeled ``MMRD1''  is the Kaler-Schmidt law \citep{sch57}
and the blue thin solid lines flanking it are its $\pm1.5$~mag lines
(see Equation (\ref{kaler-schmidt-law})).
Magenta solid line labeled ``MMRD2'' is 
della Valle \& Livio's (1995) law and is flanked by its $\pm1.5$~mag lines
(see Equation (\ref{della-valle-livio-law})).
Red filled star is the recurrent nova RS Oph as
an example of a very high mass accretion rate
and very short timescale $t_3$ \citep{hac14k}.
Red open diamond is the 1~yr recurrence period M31 nova, M31N2008-12a,
taken from \citet{dar15}, i.e., $M_{V, \rm max}\approx-6.6\pm0.2$
and $t_{3,V}\approx3.8\pm0.2$ days.

The data points are highly scattered evenly both above and below
the empirical formulae.
This simply means that there is a second (or even a third) parameter that
specifies the MMRD relation for individual novae.
\citet{hac10k} proposed that
the main parameter is the WD mass represented by the
stretching factor $f_{\rm s}$, and the second parameter
is the initial envelope mass (or the mass accretion rate to the WD).
This second parameter can reasonably explain the large scatter
of individual novae around the proposed MMRD relation of MMRD1
\citep[see, e.g.,][]{hac10k, hac15k}.
For a larger envelope mass at ignition, the wind mass loss starts
with a large mass-loss rate.  Then the nova is bright, as seen in Equation 
(\ref{free-free-wind}) in Appendix \ref{time_normalized_free_free}.
For a smaller envelope mass at ignition, the wind mass loss starts
with a smaller rate and, as a result, the nova is fainter even if
the WD mass is the same.

\citet{hac15k} further pointed out that the effect of
photospheric emission makes $t_3$ (or $t_2$) much longer than when
free-free emission dominates.  This happens in very
slow novae such as RR~Pic, V723~Cas, HR~Del, and V5558~Sgr.
These four novae are located far to the right of the line of MMRD1,
i.e., Equation (\ref{kaler-schmidt-law}).   This photospheric emission
effect could be a third parameter causing deviation from the MMRD relation.

As shown in the previous sections, the photospheric emission contributes
slightly to the total $V$ light curve in QU~Vul, V351~Pup, and V1974~Cyg,
but barely in V382~Vel and V693~CrA.  Thus, these five novae distribute
evenly above and below the MMRD relation of
Equation (\ref{kaler-schmidt-law}).

It is interesting to place the recurrent nova RS Oph and the 1-yr
recurrence period M31 nova, M31N2008-12a, in this diagram.
These two recurrent novae are located at the smallest $t_3$ time
and on the underluminous side of the $M_{V, {\rm max}}$--$\log t_3$ relation.
This indicates a relatively small envelope mass at the optical maximum,
suggesting that the mass accretion rate to the WD was very high
\citep{kat15sh}.
Such a situation is very consistent with a preoutburst picture of
recurrent novae \cite[see, e.g.,][for more details]{hac01kb,
hac10k, hac15k, kat14shn, kat15sh}.



\section{Conclusions}
\label{conclusions}
We analyzed five neon novae and reanalyzed a CO nova
on the basis of our model light curve 
fittings and obtained the following results.

\noindent
{\bf 1.} On the basis of an optically thick wind model \citep{kat94h},
we made free-free emission model light
curves of classical novae for various chemical compositions,
including neon enrichment.  During the wind phase, our model
light curves were calculated from free-free emission of
optically thin ejecta outside the photosphere of a nova envelope.
The absolute magnitudes of the free-free emission model light curves
were calibrated with the known distance and reddening of a target nova
in the same way as in \citet{hac10k, hac15k}.
We then calculated the total $V$ flux of free-free emission plus
photospheric emission and compared them with the light curves of the novae. 

\noindent
{\bf 2.} With the optical $V$ and UV~1455~\AA\  light curve fittings
to QU~Vul, we confirmed our basic picture of nova evolution; that is,
nova outburst evolution is governed by optically thick winds,
and the $V$ light curves can be reproduced well by free-free emission.
The photospheric emission contributes to
the optical light curve by at most 0.4 -- 0.8 mag.  This is because 
QU~Vul is relatively slow compared with other fast neon novae, and the
wind mass-loss rate is relatively small.  We have obtained a consistent
set of the distance, $d\sim2.4$~kpc, and reddening, $E(B-V)\sim0.55$.
The estimated WD mass is in the range of 0.82 -- $0.96~M_\sun$ for
various chemical compositions.  This small value suggests that the WD has
lost a mass of at least $\sim 0.1~M_\sun$ since its birth
if it was born as an ONe WD.

\noindent
{\bf 3.} In the same way as for QU~Vul, we obtained a consistent
distance and reddening of $d\sim5.5$~kpc and $E(B-V)\sim0.45$ for V351~Pup.
The estimated WD mass is in the range of 0.98 -- $1.1~M_\sun$ for
various chemical compositions.  In this case, the WD could have
lost a mass as large as $\sim 0.1~M_\sun$ or more since its birth if
it was born as an ONe WD.  The photospheric emission contributes
very little to the optical light curve, 0.2 -- 0.4 mag at most.
This is because V351~Pup is slightly faster than QU~Vul,
and the wind mass-loss rate is slightly larger than that of QU~Vul.
These stronger winds produce brighter free-free emission.

\noindent
{\bf 4.} For V382~Vel, we obtained a consistent set of 
distance and reddening, $d\sim1.6$~kpc and $E(B-V)\sim0.15$.
The estimated WD mass is in the range of 1.13 -- $1.28~M_\sun$ for
various chemical compositions.  In this case, the WD mass
is consistent with those of natal ONe WDs.
The photospheric emission does not contribute to the optical light curve.
This is because V382~Vel is a very fast nova, and the wind mass-loss rate
is much larger than that of QU~Vul.

\noindent
{\bf 5.}  For V693~CrA, we obtained a consistent set of
distance and reddening, $d\sim7.1$~kpc and $E(B-V)\sim0.05$.
The estimated WD mass is in the range of 1.15 -- $1.25~M_\sun$ for
various chemical compositions.
In this case, the WD mass is consistent with those of natal ONe WDs.
The photospheric emission does not contribute to the optical light curve
in the decay phase.  This is because V693~CrA is also a very fast nova,
and the wind mass-loss rate is much larger than that of QU~Vul.

\noindent
{\bf 6.}  For V1974~Cyg, we obtained a consistent set of
distance and reddening, $d\sim1.8$~kpc and $E(B-V)\sim0.30$.
The estimated WD mass is in the range of 0.95 -- $1.1~M_\sun$ for
various chemical compositions.
The WD could have lost as much mass as $\sim0.1~M_\sun$ or more
since its birth if it was born as an ONe WD.
The photospheric emission contributes very little to the optical light
curve, 0.2 -- 0.4 mag at most.  

\noindent
{\bf 7.} For V1668~Cyg, not a neon nova but a carbon-oxygen nova,
we reanalyzed light curves including the effect of photospheric emission.
We obtained a consistent set of
distance and reddening, $d\sim5.4$~kpc and $E(B-V)\sim0.30$.
The photospheric emission contributes slightly to the optical light
curve, 0.2 -- 0.4 mag at most.   
The estimated WD mass is in the range of 0.98 -- $1.1~M_\sun$ for
various chemical compositions, which is on the lower mass 
bound of natal pure ONe cores.  We point out two possibilities.
One is that at its birth, the WD mass was close to the upper
mass bound of natal CO cores, $< 1.07~M_\sun$ \citep[e.g.,][]{ume99},
and it decreased to the present value.  The other is that at its birth,
the WD mass was above $1.07~M_\sun$, and it decreased
to the present value, but the CO-rich mantle is still surrounding
the ONe core.

\noindent
{\bf 8.} On the basis of the universal decline law, we derived
three MMRD relations calibrated with the data of V1668~Cyg,
QU~Vul, and V351~Pup.
These are consistent with other known empirical MMRD relations.  The five
neon novae studied in the present paper locate evenly above and below
the empirical MMRD relations (as well as our theoretically obtained 
MMRD relations).

\noindent
{\bf 9.} Additionally, we calibrated the absolute magnitudes of the 
free-free emission model light curves for the
chemical compositions of CO nova 3, Ne nova 3, and Ne nova 1,
on the basis of the distances and reddenings of V1668~Cyg,
QU~Vul, and V351~Pup, respectively.
The obtained absolute magnitudes give reasonable fits to the other 
novae.

\acknowledgments
     We thank the late Angelo Cassatella for providing us with his
machine readable UV 1455 ~\AA ~data of QU~Vul, V351~Pup, V382~Vel,
V693~CrA, V1974~Cyg, and V1668~Cyg, and also 
the American Association of Variable Star Observers
(AAVSO) and Variable Star Observers League of Japan (VSOLJ)
for archival data on novae.  We are also grateful to the anonymous
referee for useful comments that improved the manuscript.
This research was supported in part by Grants-in-Aid for
Scientific Research (24540227, 15K05026) 
from the Japan Society for the Promotion of Science.



\appendix

\section{Distance Modulus Determined by the Time-stretching Method}
\label{time_stretching_method_novae}
\citet{hac10k, hac14k, hac15k} developed a method to estimate
the distance modulus in the $V$ band on the basis of the time-stretching
method \citep{hac10k}.  Nova light curves follow the universal
decline law when free-free emission dominates the spectrum in optical
and NIR regions \citep{hac10k, hac14k, hac15k}.
Using this property, \citet{hac10k} found that,
if two nova light curves overlap each other after one of
the two is squeezed/stretched by a factor
of $f_s$ ($t'=t/f_s$) in the time direction, the brightnesses of the
two novae obey the relation of
\begin{equation}
m'_V = m_V - 2.5 \log f_s.
\label{time-stretching_method}
\end{equation}
(This equation is essentially the same as Equation 
(\ref{simple_final_scaling_flux}) in Appendix \ref{absolute_magnitude}.)
This means that the light curves of two novae are connected by the
simple relation (\ref{time-stretching_method}).  Thus, if one is
a well-studied nova on the distance and reddening, we are able to get
information on the distance modulus of the other nova.
Applying this property to a target nova, we can estimate
the absolute magnitude of the target nova from the calibrated nova
with a known distance modulus.
 
Figure \ref{qu_vul_pw_vul_gq_mus_v_bv_ub_color_logscale_no2}
shows time-stretched light curves of QU~Vul, PW~Vul, and GQ~Mus, against
V1974~Cyg.  This figure is similar to Figure 50 of \citet{hac14k},
but with reanalyzed data.
Here, we adopt V1974~Cyg as a well-observed nova and the others are
the target novae.
The $UBV$ data for GQ~Mus are taken from \citet{bud83} and \citet{whi84},
whereas the $V$ data are from the Fine Error Sensor monitor on board {\it IUE}
and the visual data are those collected by the Royal
Astronomical Society of New Zealand and by the AAVSO
\citep[see][for more details of GQ~Mus light curve data]{hac08kc}.
The $UBV$ data for PW~Vul are taken from \citet{rob95} and the
$V$ data are the same as those in Figure 8 of \citet{hac14k}.
The data for QU~Vul and V1974~Cyg are the same as those in 
Sections \ref{qu_vul}   and  \ref{light_curve_v1974_cyg}. 

Figure \ref{qu_vul_pw_vul_gq_mus_v_bv_ub_color_logscale_no2} shows that 
all four novae well overlap in
the UV~1455~\AA\  and $V$ light curves, $B-V$, and $U-B$ color curves.
To overlap them, we shift the $V$ light curves up or down by $\Delta V$ 
against that of V1974~Cyg.  At the same time,
the stretching factors of each nova are obtained to be
$f_s=0.42$ for PW~Vul, $f_s=0.26$ for GQ~Mus, 
and $f_s=0.42$ for QU~Vul, against that of V1974~Cyg.
Then, we have relations between the light curves of 
QU~Vul, V1974~Cyg, PW~Vul, and GQ~Mus,
\begin{eqnarray}
(m-M)_{V, \rm V1974~Cyg} &=& 12.2\cr
&=&(m-M)_{V,\rm PW~Vul} + \Delta V - 2.5 \log 0.42 \cr
&\approx& 13.0 - 1.8 + 0.95 = 12.15\cr
&=& (m-M)_{V,\rm GQ~Mus} + \Delta V - 2.5 \log 0.26 \cr
&\approx& 15.7 - 5.0 + 1.46 = 12.16\cr
&=& (m-M)_{V,\rm QU~Vul} + \Delta V - 2.5 \log 0.42 \cr
&\approx& 13.6 - 2.4 + 0.95 = 12.15,
\label{time-stretching_brightness_qu_vul}
\end{eqnarray}
where we use the apparent distance moduli of
V1974~Cyg, PW~Vul, and GQ~Mus to be
$(m-M)_{V,\rm V1974~Cyg} = 12.2$ in Section \ref{light_curve_v1974_cyg},
and both $(m-M)_{V,\rm PW~Vul} = 13.0$ and
$(m-M)_{V,\rm GQ~Mus} = 15.7$ in \citet{hac15k}.
All these three are consistent with each other.
Considering ambiguity ($\pm0.2$ mag) of fitting accuracy,
we obtained $(m-M)_V=13.6\pm0.2$ for QU~Vul.

Similarly, for V351~Pup, we obtain
\begin{eqnarray}
(m-M)_{V, \rm V351~Pup} &=& 15.1\cr
&=&(m-M)_{V,\rm V1668~Cyg} + \Delta V - 2.5 \log 0.93 \cr
&\approx& 14.6 + 0.4 + 0.08 = 15.08\cr
&=&(m-M)_{V,\rm PW~Vul} + \Delta V - 2.5 \log 0.42 \cr
&\approx& 13.0 + 1.1 + 0.95 = 15.05\cr
&=& (m-M)_{V,\rm V1974~Cyg} + \Delta V - 2.5 \log 1.0 \cr
&\approx& 12.2 + 2.9 + 0.0 = 15.1\cr
&=& (m-M)_{V,\rm QU~Vul} + \Delta V - 2.5 \log 0.5 \cr
&\approx& 13.6 + 0.8 + 0.75 = 15.15,
\label{time-stretching_brightness_v351_pup}
\end{eqnarray}
from Figure \ref{all_mass_v351_pup_v1668_cyg_x35z02c10o20} (V1668~Cyg),
Figure \ref{all_mass_v351_pup_pw_vul_x55z02c10o10} (PW~Vul),
Figure \ref{all_mass_v351_pup_v1974_cyg_x55z02o10ne03} (V1974~Cyg),
and Figure \ref{all_mass_v351_pup_qu_vul_x65z02o03ne03} (QU~Vul).
Here, $\Delta V$ is the difference of brightness obtained in Figures
\ref{all_mass_v351_pup_v1668_cyg_x35z02c10o20},
\ref{all_mass_v351_pup_pw_vul_x55z02c10o10},
\ref{all_mass_v351_pup_v1974_cyg_x55z02o10ne03},
and \ref{all_mass_v351_pup_qu_vul_x65z02o03ne03}, respectively,
against that of V351~Pup.
The stretching factors are also obtained in each figure
against V351~Pup.  The apparent distance moduli of
V1668~Cyg, PW~Vul, V1974~Cyg, and QU~Vul were calibrated as
$(m-M)_{V,\rm V1668~Cyg} = 14.6$ in Section \ref{light_curve_v1668_cyg},
$(m-M)_{V,\rm PW~Vul} = 13.0$ in \citet{hac15k},
and $(m-M)_{V,\rm V1974~Cyg} = 12.2$ in Section \ref{light_curve_v1974_cyg},
and $(m-M)_{V,\rm QU~Vul} = 13.6$ in Section \ref{qu_vul}.
These are all consistent with each other.
Thus, we obtained $(m-M)_V=15.1\pm0.2$ for V351~Pup.

For V382~Vel, we obtain
\begin{eqnarray}
(m-M)_{V, \rm V1500~Cyg} &=& 12.3\cr
&=&(m-M)_{V,\rm V382~Vel} + \Delta V - 2.5 \log 0.89 \cr
&\approx& 11.5 + 0.7 + 0.13 = 12.33\cr
&=& (m-M)_{V,\rm GK~Per} + \Delta V - 2.5 \log 0.63 \cr
&\approx& 9.3 + 2.5 + 0.50 = 12.30\cr
&=& (m-M)_{V,\rm V1974~Cyg} + \Delta V - 2.5 \log 0.47 \cr
&\approx& 12.2 - 0.7 + 0.82 = 12.32\cr
&=& (m-M)_{V,\rm QU~Vul} + \Delta V - 2.5 \log 0.166 \cr
&\approx& 13.6 - 3.0 + 1.7 = 12.3,
\label{time-stretching_brightness_v382_vel}
\end{eqnarray}
from Figure \ref{v382_vel_v1500_cyg_v_color_logscale_no2}.
Here, $\Delta V$ is the difference of brightness obtained in Figure
\ref{v382_vel_v1500_cyg_v_color_logscale_no2}
against that of V1500~Cyg.
In Figure \ref{v382_vel_v1500_cyg_v_color_logscale_no2},
the $V$ data for V1500~Cyg are taken from \citet{loc76}
and \citet{tem79} and the $UBV$ data are from \citet{pfa76},
\citet{ark76}, \citet{due77}, and \citet{con80}.
The optical $V$ data for GK~Per are the same as
those in Figure 2 of \citet{hac07k}.
The stretching factors are also obtained against V1500~Cyg.
The apparent distance moduli of
V1500~Cyg, GK~Per, V1974~Cyg, and QU~Vul were calibrated as
$(m-M)_{V,\rm V1500~Cyg} = 12.3$ in \citet{hac14k},
$(m-M)_{V,\rm GK~Per} = 9.3$ in \citet{har13} from {\it HST}
annual parallaxes, 
$(m-M)_{V,\rm V1974~Cyg} = 12.2$ in Section \ref{light_curve_v1974_cyg},
and $(m-M)_{V,\rm QU~Vul} = 13.6$ in Section \ref{qu_vul}.
These are all consistent with each other.
Thus, we obtained $(m-M)_V=11.5\pm0.2$ for V382~Vel.

For V693~CrA, we obtain
\begin{eqnarray}
(m-M)_{V, \rm V693~CrA} &=& 14.4\cr
&=&(m-M)_{V,\rm V351~Pup} + \Delta V - 2.5 \log 0.40 \cr
&\approx& 15.1 - 1.6 + 1.0 = 14.5 \cr
&=&(m-M)_{V,\rm V1668~Cyg} + \Delta V - 2.5 \log 0.37 \cr
&\approx& 14.6 - 1.2 + 1.07 = 14.47 \cr
&=&(m-M)_{V,\rm V1500~Cyg} + \Delta V - 2.5 \log 0.79 \cr
&\approx& 12.3 + 1.8 + 0.25 = 14.35 \cr
&=&(m-M)_{V,\rm QU~Vul} + \Delta V - 2.5 \log 0.25 \cr
&\approx& 13.6 - 0.7 + 1.5 = 14.4,
\label{time-stretching_brightness_v693_cra}
\end{eqnarray}
where $\Delta V$ is the difference of brightness obtained in Figures
\ref{all_mass_v693_cra_x35z02o20ne10_absolute_mag},
\ref{all_mass_v693_cra_x45z02c15o20_absolute_mag},
\ref{all_mass_v693_cra_x55z02o10ne03_absolute_mag},
and \ref{all_mass_v693_cra_x65z02o03ne03_absolute_mag}
against that of V693~CrA.  The timescales are squeezed in these
figures as $f_s=0.40$ for V351~Pup (Figure
\ref{all_mass_v693_cra_x35z02o20ne10_absolute_mag}),
$f_s=0.37$ for V1668~Cyg (Figure
\ref{all_mass_v693_cra_x45z02c15o20_absolute_mag}),
$f_s=0.79$ for V1500~Cyg (Figure
\ref{all_mass_v693_cra_x55z02o10ne03_absolute_mag}), 
and $f_s=0.25$ for QU~Vul (Figure
\ref{all_mass_v693_cra_x65z02o03ne03_absolute_mag})
against V693~CrA.  The apparent distance moduli of V351~Pup,
V1668~Cyg, V1500~Cyg, and QU~Vul were calibrated as
$(m-M)_{V,\rm V351~Pup} = 15.1$ in Section \ref{v351_pup},
$(m-M)_{V,\rm V1668~Cyg} = 14.6$ in Section \ref{light_curve_v1668_cyg},
$(m-M)_{V,\rm V1500~Cyg} = 12.3$ in \citet{hac10k, hac14k},
and $(m-M)_{V,\rm QU~Vul} = 13.6$ in Section \ref{qu_vul}.
These three are all consistent with each other.
Thus, we obtained $(m-M)_V=14.4\pm0.2$ for V693~CrA.


\begin{figure}
\epsscale{1.15}
\plotone{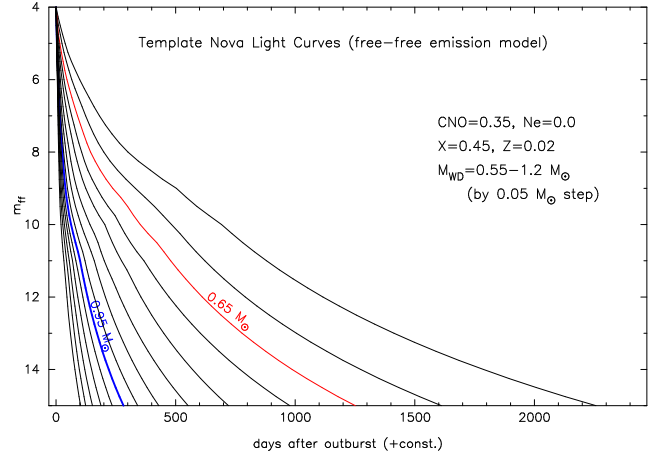}
\caption{
Magnitudes of our free-free emission model light curves 
for CO nova 3 and WD masses of $0.55 - 1.2~M_\sun$ in $0.05~M_\sun$ steps,
numerical data of which are tabulated
in Table \ref{light_curves_of_novae_co3}.
Two light curves are specified by the red thick solid line ($0.65 ~M_\sun$)
and blue thick solid line ($0.95 ~M_\sun$).
\label{light_curve_combine_x45z02c15o20}}
\end{figure}


\begin{figure}
\epsscale{1.15}
\plotone{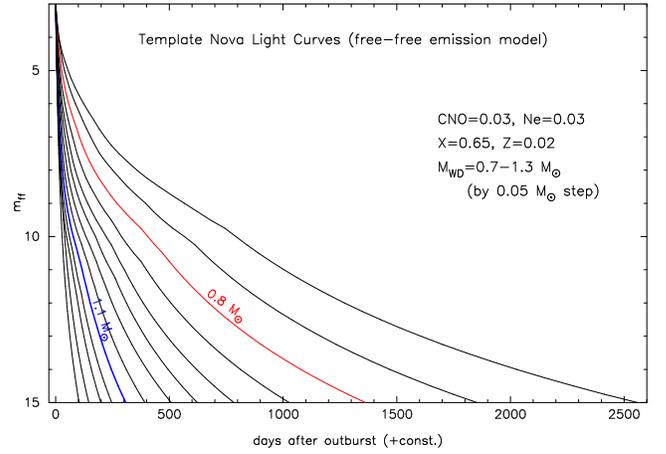}
\caption{
Same as Figure \ref{light_curve_combine_x45z02c15o20}, but
for Ne nova 3 and WD masses of $0.7 - 1.3~M_\sun$.
The numerical data are tabulated
in Table \ref{light_curves_of_novae_ne3}.
Two light curves are specified by the red thick solid line ($0.8 ~M_\sun$)
and blue thick solid line ($1.1 ~M_\sun$).
\label{all_mass_light_curve_model_x65z02o03ne03}}
\end{figure}


\begin{figure}
\epsscale{1.15}
\plotone{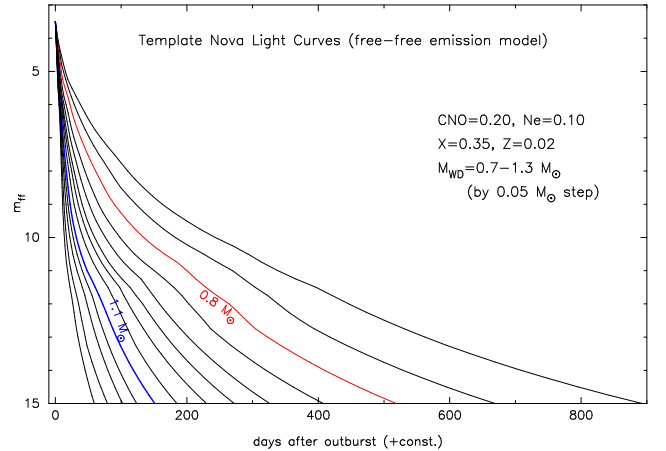}
\caption{
Same as Figure \ref{light_curve_combine_x45z02c15o20}, but
for Ne nova 1 and WD masses of $0.7 - 1.3~M_\sun$.
The numerical data are tabulated
in Table \ref{light_curves_of_novae_ne1}.
\label{all_mass_light_curve_model_x35z02o20ne10}}
\end{figure}


\begin{figure*}
\epsscale{0.95}
\plotone{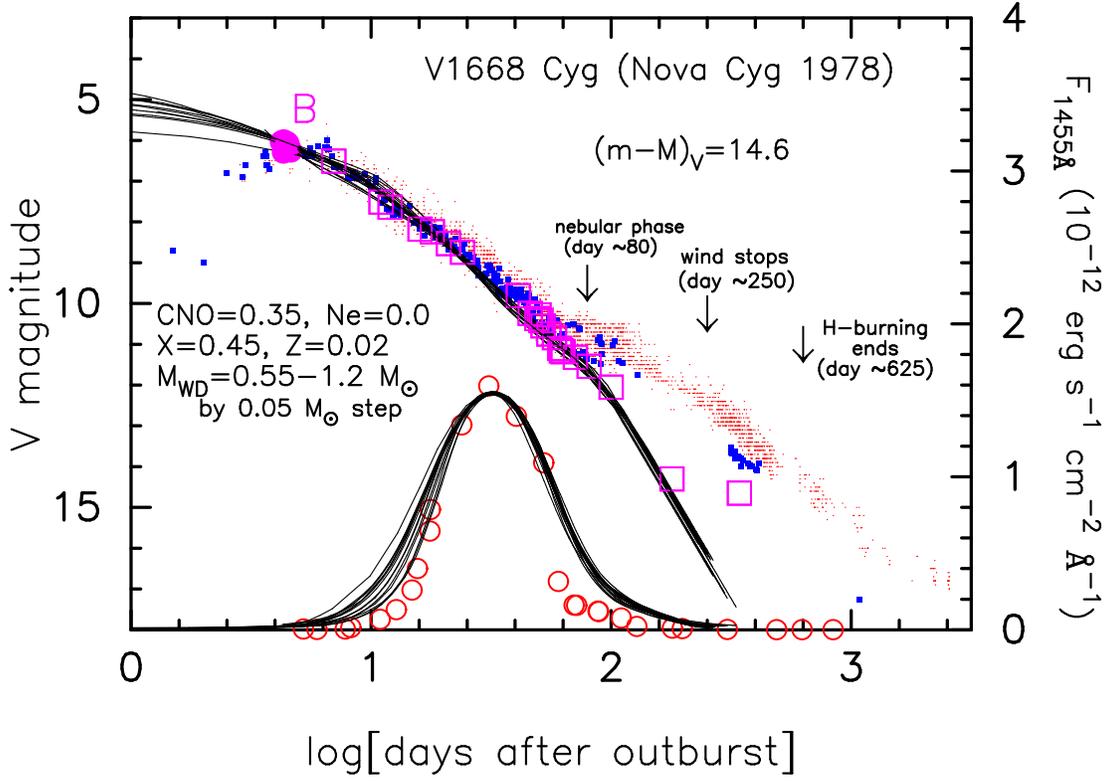}
\caption{
Same as the free-free emission model light curves in
Figure \ref{light_curve_combine_x45z02c15o20} (CO nova 3), but
all the light curves of free-free emission and UV~1455~\AA\  blackbody
emission are rescaled to overlap on the UV~1455~\AA\  observation 
(red open circles) of V1668~Cyg.  Each timescaling factor, $f_{\rm s}$,
is tabulated in Table \ref{light_curves_of_novae_co3}.
The right edge of each free-free emission model light curve
corresponds to the epoch when the optically thick winds stop, i.e.,
$\sim250$ days after the outburst for the $0.98~M_\sun$ WD (denoted
by an arrow).  Point B (magenta filled circle) corresponds to the peak
of the $V$ magnitude of V1668~Cyg.
\label{all_mass_v1668_cyg_x45z02c15o20_calib_universal}}
\end{figure*}



\begin{figure*}
\epsscale{0.95}
\plotone{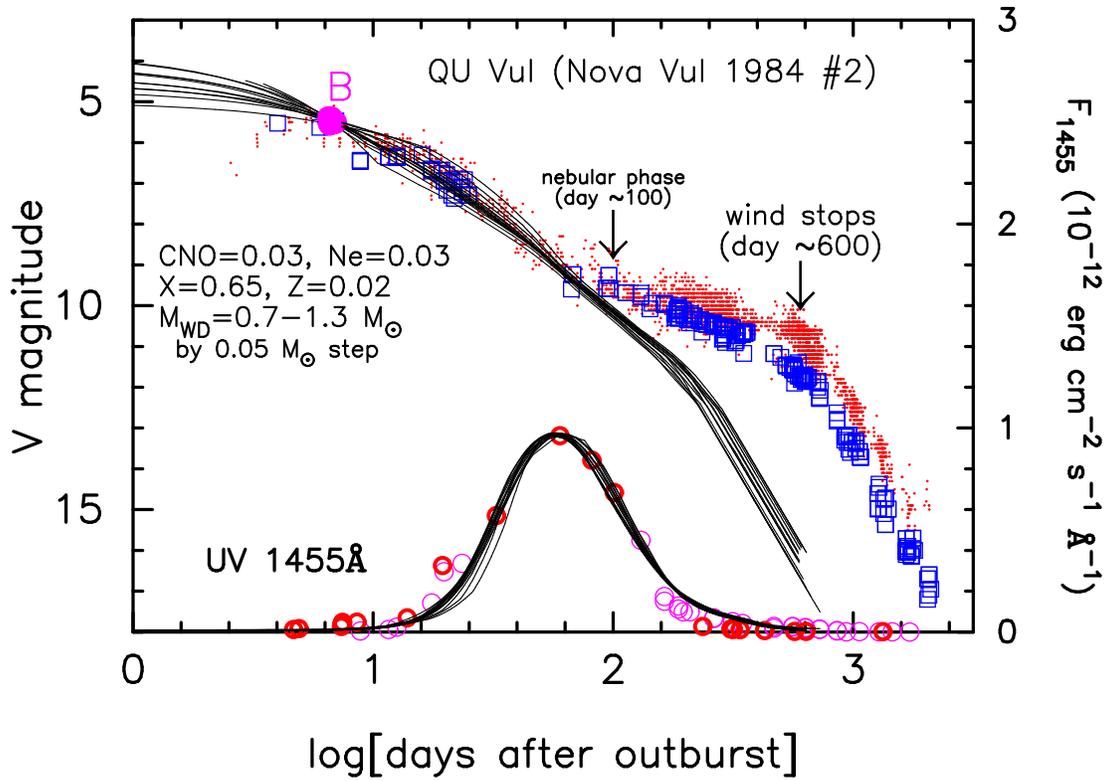}
\caption{
Same model light curves of free-free emission as those in
Figure \ref{all_mass_light_curve_model_x65z02o03ne03}
(Ne nova 3), but
all the light curves of free-free emission and UV~1455~\AA\  blackbody
emission are rescaled to overlap on the UV~1455~\AA\  observation of QU~Vul
(magenta open circles) and PW~Vul (red open circles).
Each timescaling factor, $f_{\rm s}$, is tabulated in Table
\ref{light_curves_of_novae_ne3}.
The right edge of each free-free emission model light curve
corresponds to the epoch when the optically thick winds stop, i.e.,
$\sim600$ days after the outburst for the $0.96~M_\sun$ WD model
(denoted by an arrow).  Point B (magenta filled circle)
corresponds to the peak of the $V$ magnitude.
\label{all_mass_qu_vul_x65z02o03ne03_calib_universal}}
\end{figure*}


\begin{figure*}
\epsscale{0.95}
\plotone{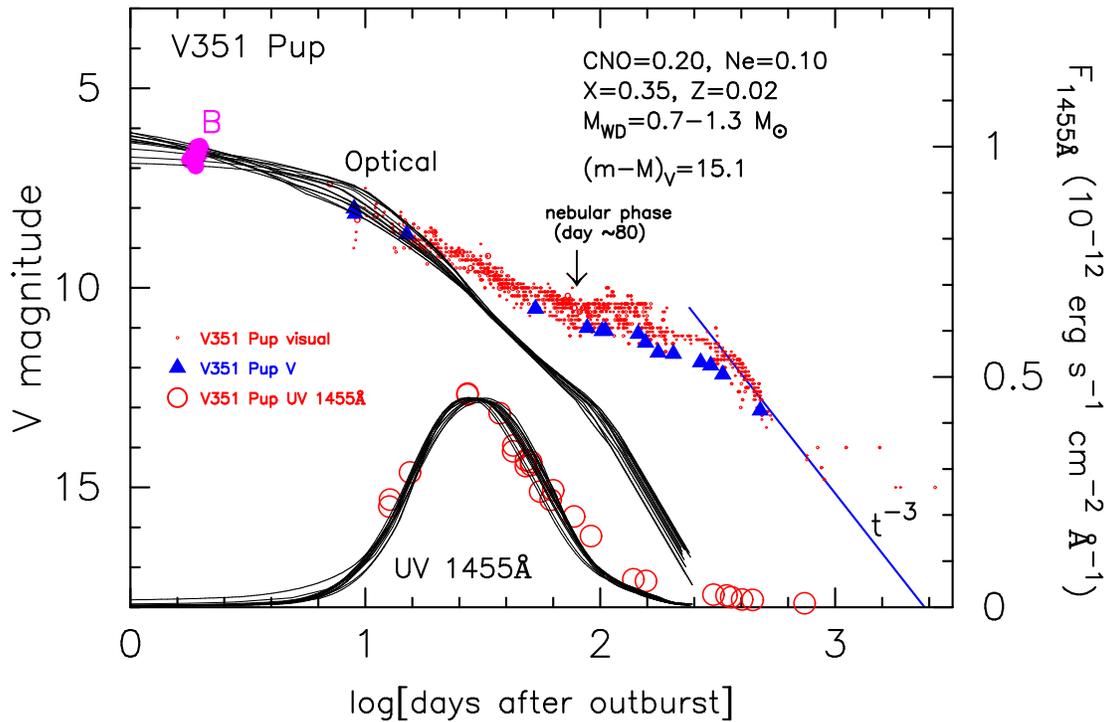}
\caption{
Same free-free emission model light curves as those in
Figure \ref{all_mass_light_curve_model_x35z02o20ne10}
(Ne nova 1), but
all the light curves of free-free emission and UV~1455~\AA\  blackbody
are rescaled to overlap on the UV~1455~\AA\  observation 
(large red open circles) of V351~Pup.
Each timescaling factor, $f_{\rm s}$, is tabulated in Table
\ref{light_curves_of_novae_ne1}.
Points B (magenta filled circles) correspond to the peak of
the $V$ magnitude.
\label{all_mass_v351_pup_univeral_scale_x35z02o20ne10}}
\end{figure*}


\begin{figure*}
\epsscale{0.90}
\plotone{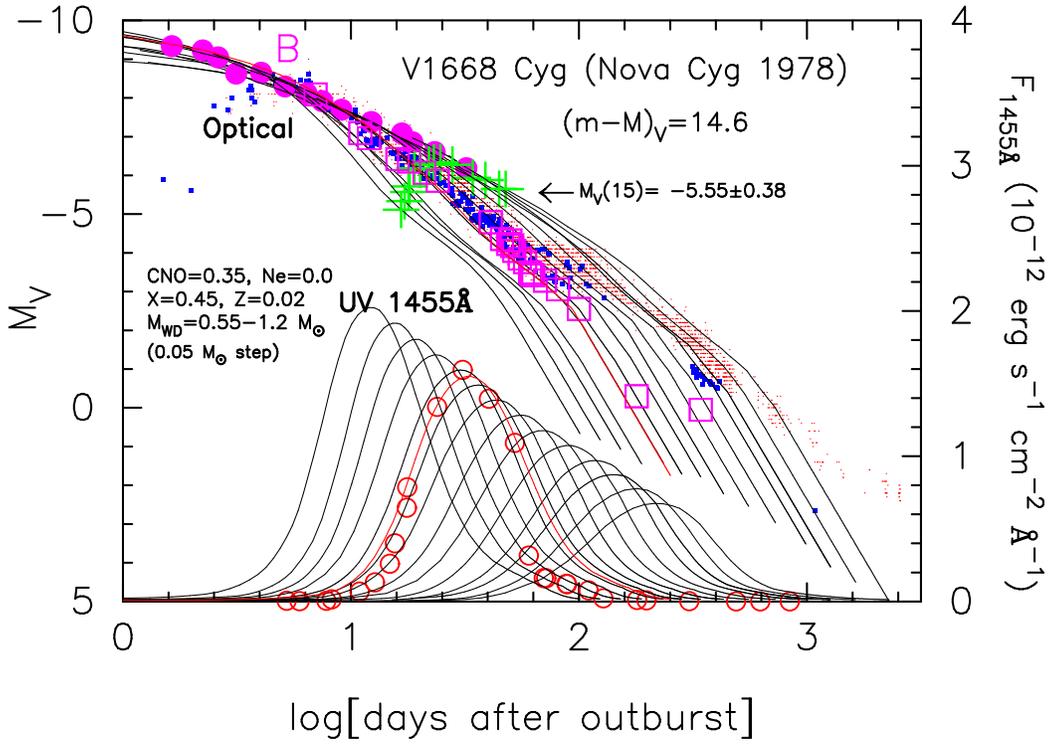}
\caption{
Same models as Figure \ref{all_mass_v1668_cyg_x45z02c15o20_calib_universal}
(CO nova 3), but for the absolute magnitudes and real timescales.
We calibrated the free-free model light curves by the distance modulus
of $(m-M)_V = 14.6$ and calculated the absolute magnitude of
each free-free emission light curve (labeled ``Optical'') from
Equation (\ref{real_timescale_flux}). 
The position at point B in Figure
\ref{all_mass_v1668_cyg_x45z02c15o20_calib_universal}
is indicated by a magenta filled circle for each light curve.
We also show the magnitude, $M_V(15)$, 15 days after the optical maximum,
by green crosses.  Their average value of $M_V(15) = -5.55 \pm 0.38$
is obtained for $0.55$--$1.2~M_\sun$ WDs.
The UV~1455~\AA\  model light curves are also rescaled to recover
the real timescale and flux.  The red solid lines denote those for
the $0.98~M_\sun$ WD model.
\label{all_mass_v1668_cyg_x45z02c15o20_real_scale_universal}}
\end{figure*}


\begin{figure*}
\epsscale{0.90}
\plotone{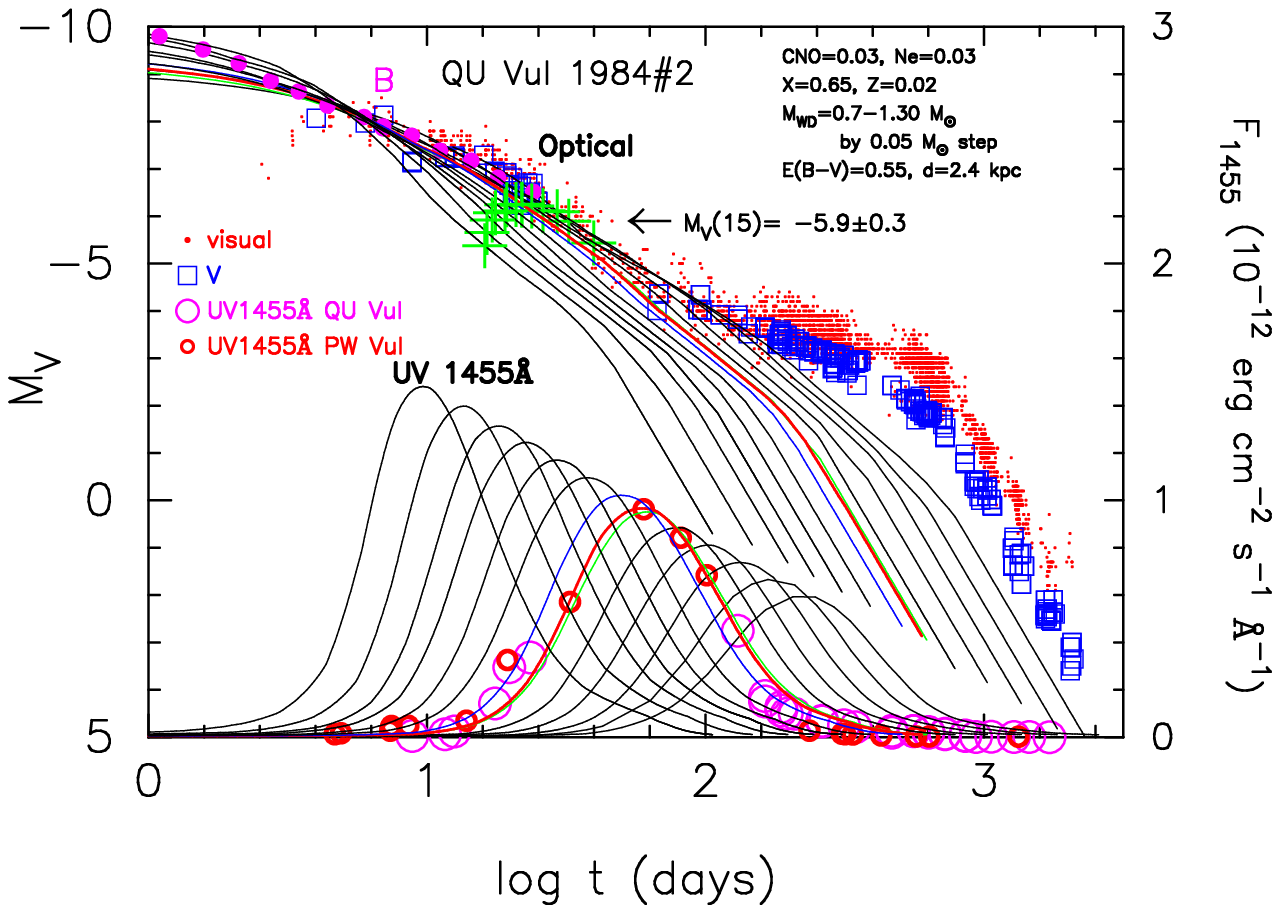}
\caption{
Same models as Figure \ref{all_mass_qu_vul_x65z02o03ne03_calib_universal}
(Ne nova 3),
but for the absolute magnitudes and real timescales.  We calibrated
the free-free model light curves with the distance modulus of
$(m-M)_V = 13.6$ in Figure
\ref{all_mass_qu_vul_m0960_x65z02o03ne03_compsite}(b)
and calculated the absolute magnitude of
each free-free emission light curve (labeled ``Optical'') 
from Equation (\ref{real_timescale_flux}). 
The position at point B in Figure
\ref{all_mass_qu_vul_x65z02o03ne03_calib_universal}
is indicated by a magenta filled circle for each light curve.
We also show the magnitude,
$M_V(15)$, 15 days after the optical maximum, by green crosses.
Their average value of $M_V(15) = -5.9 \pm 0.3$ is obtained
for $0.7$--$1.3~M_\sun$ WDs.  Our UV~1455~\AA\  model light curves
are also rescaled to recover the real timescales and fluxes.
The red thick, green thin, and blue thin solid lines denote
those for the $0.96~M_\sun$, $0.95~M_\sun$, and
$1.0~M_\sun$ WDs, respectively.
\label{all_mass_qu_vul_x65z02o03ne03_absolute_mag}}
\end{figure*}


\begin{figure*}
\epsscale{0.95}
\plotone{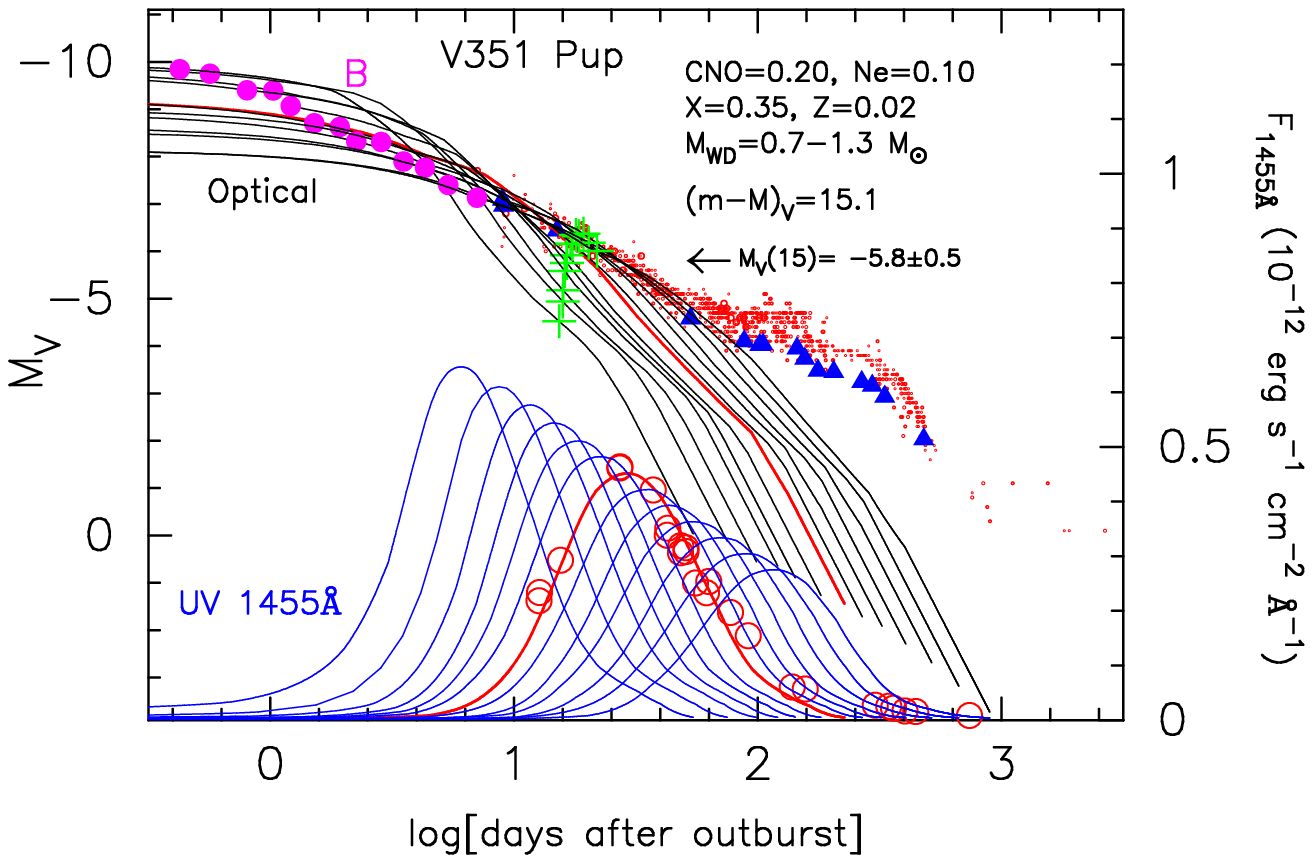}
\caption{
Same models as Figure \ref{all_mass_v351_pup_univeral_scale_x35z02o20ne10}
(Ne nova 1), but for the absolute magnitudes and real timescales.
We calibrated the free-free model light curves using the distance modulus
of $(m-M)_V = 15.1$ and restored the absolute magnitude of
each free-free emission model light curve (labeled ``Optical'') 
from Equation (\ref{real_timescale_flux}). 
Each position at point B in Figure
\ref{all_mass_v351_pup_univeral_scale_x35z02o20ne10}
is indicated by a magenta filled circle for each light curve.
We also show the magnitude, $M_V(15)$, 15 days after the optical maximum,
by green crosses.  Their average value of $M_V(15) = -5.8 \pm 0.5$
is obtained for $0.7$--$1.3~M_\sun$ WDs.
The UV~1455~\AA\  model light curves are also rescaled to recover
the real timescales and fluxes.
The red thick solid lines denote those for the $1.0~M_\sun$ WD model.
\label{all_mass_v351_pup_univeral_real_abs_x35z02o20ne10}}
\end{figure*}


\begin{figure}
\epsscale{0.95}
\plotone{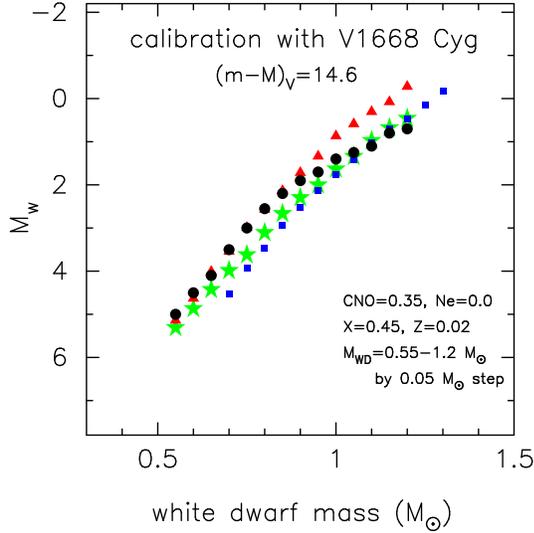}
\caption{
The absolute magnitudes at the end of the wind phase, $M_{\rm w}$,
against the WD mass for CO nova 3 (black filled circles) in the present work
as well as CO nova 2 (red filled triangles) and Ne nova 2 
(blue filled squares), the both of which are taken from \citet{hac10k},
and CO nova 4 (green filled star-marks), taken from \citet{hac15k}. 
\label{bottom_ff_line_x45z02c15o20_absolute_mag}}
\end{figure}


\begin{figure}
\epsscale{0.95}
\plotone{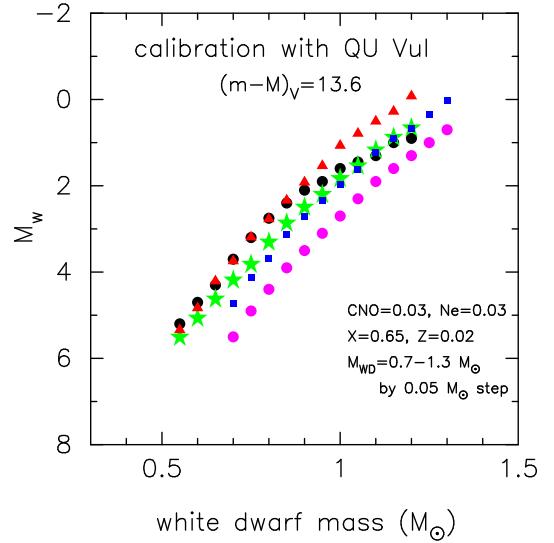}
\caption{
Same as Figure \ref{bottom_ff_line_x45z02c15o20_absolute_mag}, but we
added the end points of the wind phase for Ne nova 3 (magenta filled circles).
\label{bottom_ff_line_x65z02o03ne03_absolute_mag}}
\end{figure}


\begin{figure}
\epsscale{0.95}
\plotone{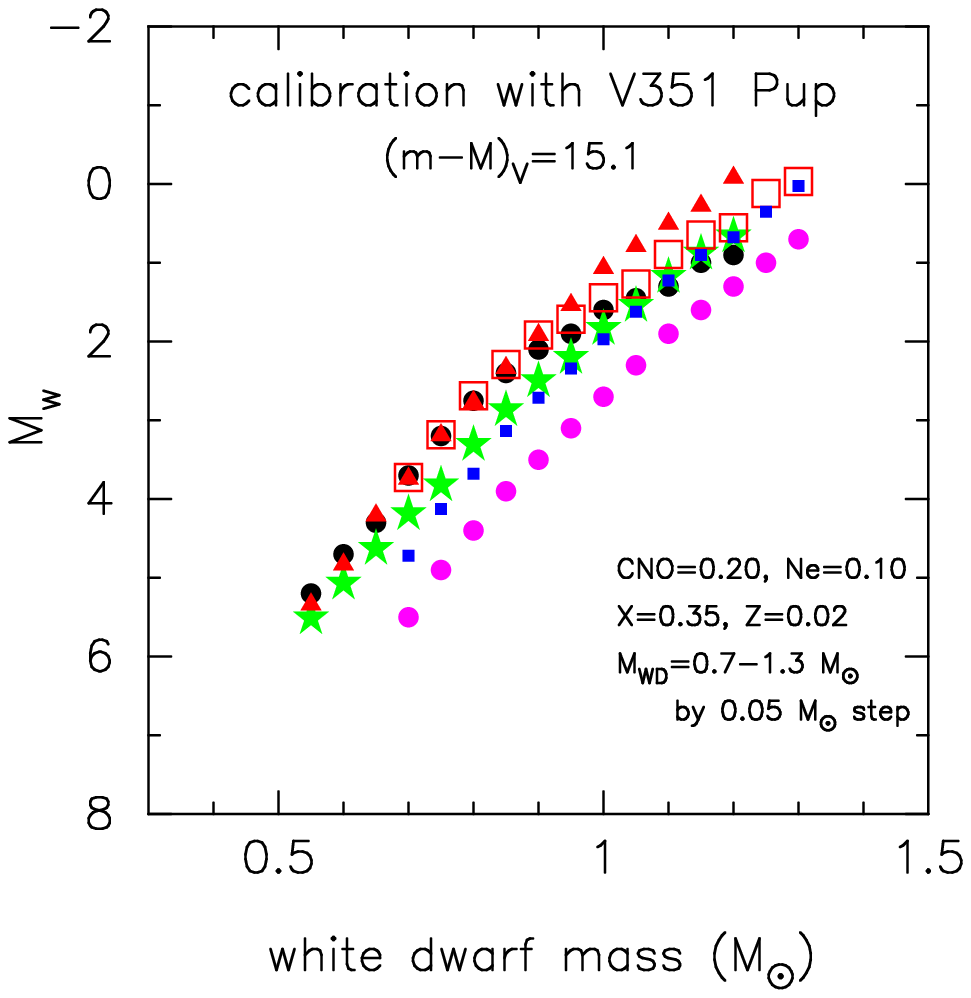}
\caption{
Same as Figure \ref{bottom_ff_line_x65z02o03ne03_absolute_mag}, but
we added the end points of the wind phase for Ne nova 1 (red open squares).
\label{bottom_ff_line_x35z02o20ne10_absolute_mag}}
\end{figure}

\section{Time-normalized Light Curves of Free-free Emission}
\label{time_normalized_free_free}

\subsection{Free-free emission model light curves}
\label{free-free_emission_light_curve}
\citet{hac06kb} made theoretical light curves of free-free
emission from optically thin ejecta outside the photosphere 
on the basis of the optically thick wind theory \citep{kat94h}.
They showed that these light curves reproduced the optical light curves
of novae reasonably well.  The flux of optically thin
free-free emission is given by
\begin{eqnarray}
j_\nu d \Omega d V d t d \nu  &=&  
{{16} \over 3} \left( {{\pi} \over 6} \right)^{1/2}
{{e^6 Z^2} \over {c^3 m_e^2}} \left( {{m_e} \over {k T_e}} \right)^{1/2} \cr
&\times& ~g ~\exp \left( -{{h \nu} \over {k T_e}} \right) N_{\rm e} N_{\rm i} 
d \Omega d V d t d \nu, 
\label{free-free-original}
\end{eqnarray}
where $j_\nu$ is the emissivity at the frequency $\nu$,
$\Omega$ the solid angle,
$V$ the volume, 
$t$ the time, 
$e$ the electron charge,
$Z$ the ion charge in units of $e$,
$c$ the speed of light, 
$m_e$ the electron mass,
$k$ the Boltzmann constant, 
$T_e$ the electron temperature, 
$g$ the Gaunt factor, 
$h$ the Planck constant, 
and $N_{\rm e}$ and $N_{\rm i}$ are the number densities of electrons
and ions \citep[][p.103]{all73}.

     Electron temperatures of nova ejecta were suggested to be
around $T_e \sim 10^4$~K, almost constant in time, during the nova outbursts
\citep[see, e.g.,][for \object{V1500 Cyg}]{enn77}.  If we further assume that
the ionization degree of ejecta is constant during the outburst,
we have the total flux of free-free emission, i.e.,
\begin{equation}
F_\nu \propto \int N_{\rm e} N_{\rm i} d V 
\propto \int_{R_{\rm ph}}^\infty {\dot M_{\rm wind}^2 
\over {v_{\rm wind}^2 r^4}} r^2 dr
\propto {\dot M_{\rm wind}^2 \over {v_{\rm ph}^2 R_{\rm ph}}}
\label{free-free-wind}
\end{equation}
for optically thin ejecta, where $F_\nu$ is the flux at the frequency $\nu$.
For Kato \& Hachisu's (1994) optically thick nova wind models,
we have $N_e \propto \rho_{\rm wind}$ and $N_i \propto \rho_{\rm wind}$,
where $\rho_{\rm wind}$ is the density of winds, $\dot M_{\rm wind}$ is the
wind mass-loss rate, $v_{\rm ph}$ and $R_{\rm ph}$ are the velocity and
radius at the pseudophotosphere.
Here, we integrate Equation (\ref{free-free-wind}) outside the photosphere
assuming that the wind velocity is $v_{\rm wind}= v_{\rm ph}~(=$constant
in space) outside the photosphere and using the relation of continuity, 
$\rho_{\rm wind} = \dot M_{\rm wind}/ 4 \pi r^2 v_{\rm wind}$.

After the wind stops, the free-free emission light curve changes
its decline shape as
\begin{equation}
F_\nu \propto \int N_{\rm e} N_{\rm i} dV \propto {{M_{\rm ej}^2} \over
{V^2}} V \propto R^{-3} \propto t^{-3}, 
\label{free-free-wind_wind_stop}
\end{equation}
where the ejected mass of $M_{\rm ej}$ is constant in time and
we assume that the ejecta volume, $V=4\pi R^3/3$, expands as
$R \propto t$.

Thus, \citet{hac06kb} calculated free-free emission model light curves
of novae for various chemical compositions, as tabulated in 
Table \ref{chemical_composition_model}, based on the optically
thick wind model of \citet{kat94h}.
Then, the free-free flux is given by
\begin{equation}
F_\nu^{\{M_{\rm WD}\}} (t) = C \left[ {{\dot M_{\rm wind}^2}
\over {v_{\rm ph}^2 R_{\rm ph}}} \right]^{\{M_{\rm WD}\}}_{(t)},
\label{free-free_calculation_original}
\end{equation}
where $C$ is the proportionality constant in Equation
(\ref{free-free-wind}).  This equation is
essentially the same as Equation (9) of \citet{hac06kb}.
Note that the flux $F_\nu$ is independent of the frequency $\nu$ 
in the case of optically thin free-free emission.  The details of 
calculations are presented in \citet{hac06kb, hac10k}.
The magnitudes of the model light curves are tabulated in 
Tables \ref{light_curves_of_novae_co3}, \ref{light_curves_of_novae_ne3},
and \ref{light_curves_of_novae_ne1} in a format of
\begin{equation}
m_{\rm ff} = -2.5 ~ \log ~ \left[ {{\dot M_{\rm wind}^2}
\over {v_{\rm ph}^2 R_{\rm ph}}} \right]^{\{M_{\rm WD}\}}_{(t)}
+ ~ G^{\{M_{\rm WD}\}},
\label{template-wind-free-free-emission}
\end{equation}
and are plotted in Figure
\ref{light_curve_combine_x45z02c15o20} for 0.55 -- 1.2 $M_\sun$ WDs
in $0.05~M_\sun$ steps (CO nova 3),
in Figure
\ref{all_mass_light_curve_model_x65z02o03ne03} for 0.7 -- 1.3 $M_\sun$ WDs
in $0.05~M_\sun$ steps (Ne nova 3),  in Figure
\ref{all_mass_light_curve_model_x35z02o20ne10} for 0.7 -- 1.3 $M_\sun$ WDs
in $0.05~M_\sun$ steps (Ne nova 1). 
The subscript $(t)$ represents the dependence on time
while the superscript $\{M_{\rm WD}\}$ corresponds to a model parameter.
The last row (15th mag) of each column  in Tables
\ref{light_curves_of_novae_co3}, \ref{light_curves_of_novae_ne3}, 
and \ref{light_curves_of_novae_ne1}
represents the magnitude at the end of the wind phase.
Here, we define the constant $G^{\{M_{\rm WD}\}}$ in Equation
(\ref{template-wind-free-free-emission}) such that
the last (lowest) point of each light curve is 15th mag.
This format helps to pack the light curve data into a short table.

\subsection{Absolute magnitudes of free-free model light curves}
\label{absolute_magnitude}
The free-free emission model light curves in Figure 
\ref{light_curve_combine_x45z02c15o20}
(also in Figures \ref{all_mass_light_curve_model_x65z02o03ne03}
and \ref{all_mass_light_curve_model_x35z02o20ne10})
have strong similarity in their shapes.
For the chemical compositions of CO nova 2, CO nova 4,
and Ne nova 2, \citet{hac06kb, hac10k, hac15k} 
showed that the model light curves are homologous among various WD masses
and almost overlap each other if they are properly squeezed/stretched
along time.  Here, we show the same properties of light curves
for other chemical compositions of CO nova 3, Ne nova 3, and Ne nova 1.
Figures \ref{all_mass_v1668_cyg_x45z02c15o20_calib_universal},
\ref{all_mass_qu_vul_x65z02o03ne03_calib_universal}, and
\ref{all_mass_v351_pup_univeral_scale_x35z02o20ne10} clearly show
that free-free emission model light curves overlap each other
after they are properly squeezed/stretched along time.
We determine the stretching factor, $f_{\rm s}$, of each UV~1455~\AA\ 
model light curve by increasing or decreasing $f_{\rm s}$
until the model light curve shape matches the observational data.
For example, we see that the evolution of the $0.95~M_\odot$ WD model
light curve is 1.05 times slower than the QU~Vul observation
($f_{\rm s} \approx 1.05$) and the $1.0~M_\odot$ WD model light curve
evolves 1.15 times faster ($f_{\rm s} \approx 0.87$).
In Figure \ref{all_mass_qu_vul_x65z02o03ne03_calib_universal},
in this way, all the UV 1455 ~\AA~ model light curves overlap
each other,  if we squeeze/stretch the time as $t' = t / f_{\rm s}$
and normalize the peak flux of UV~1455~\AA.
The values of $\log f_s$ are tabulated in Tables 
\ref{light_curves_of_novae_co3}, \ref{light_curves_of_novae_ne3},
and \ref{light_curves_of_novae_ne1}.
Using this time-stretching factor $f_s$, we can rewrite
these timescaled light curves of free-free emission as
\begin{equation}
{m'}_V^{\{M_{\rm WD}\}} (t')= 
-2.5 ~ \log ~ \left[ {{\dot M_{\rm wind}^2}
\over {v_{\rm ph}^2 R_{\rm ph}}} \right]^{\{M_{\rm WD}\}}_{(t'f_s)}
+K_V,
\label{relative_magnitude_two_free-free_convert}
\end{equation}
where $K_V$ is a constant common for all the WD masses.
Because they all overlap each other (i.e., the universal decline law),
it indicates that   
\begin{equation}
\left[ {{\dot M_{\rm wind}^2}
\over {v_{\rm ph}^2 R_{\rm ph}}} \right]^{\{M_{\rm WD}\}}_{(t'f_s)}
=
\left[ {{\dot M_{\rm wind}^2}
\over {v_{\rm ph}^2 R_{\rm ph}}} \right]^{\{0.98~M_\sun\}}_{(t')},
\label{relative_magnitude_two_free-free_equal}
\end{equation}
for all $M_{\rm WD}$ of CO nova 3.  Note that $f_s=1$ for
the $0.98~M_\sun$ WD of CO nova 3 
($f_s=1$ for the $0.96~M_\sun$ WD of Ne nova 3, and
$f_s=1$ for the $1.0~M_\sun$ WD of Ne nova 1). 
If we squeeze the timescale of a physical phenomenon
by a factor of $f_{\rm s}$ (i.e., $t' = t/f_{\rm s}$),  we covert
the frequency to $\nu ' = f_{\rm s} \nu$ and the flux of free-free
emission to $F'_{\nu '} = f_{\rm s} F_\nu$
because
\begin{equation}
{{d } \over {d t'}} = f_{\rm s} {{d} \over {d {t}} }.
\label{time-derivation-flux}
\end{equation}
Substituting $F'_{\nu'} = F'_\nu$ (independent of the frequency
in optically-thin free-free emission) into $F'_{\nu '} = f_{\rm s} F_\nu$
and integrating $F'_\nu = f_{\rm s} F_\nu$ with the $V$-filter response
function,
we have the following relation, i.e.,
\begin{eqnarray}
m'_V (t/f_s) =  m_V (t) - 2.5 \log f_{\rm s}.
\label{simple_final_scaling_flux}
\end{eqnarray}
Substituting Equation (\ref{relative_magnitude_two_free-free_equal})
into (\ref{relative_magnitude_two_free-free_convert}), and
then Equation (\ref{relative_magnitude_two_free-free_convert}) into
(\ref{simple_final_scaling_flux}), we obtain 
the apparent 
$V$ magnitudes of
\begin{equation}
m_V^{\{M_{\rm WD}\}} (t) = 2.5 \log f_{\rm s} 
-2.5 ~ \log ~ \left[ {{\dot M_{\rm wind}^2}
\over {v_{\rm ph}^2 R_{\rm ph}}} \right]^{\{0.98~M_\sun\}}_{(t/f_s)}
+ K_V,
\label{simple_scaling_flux}
\end{equation}
where note that $f_s$ is the time-scaling factor for the WD with mass
of $M_{\rm WD}$, generally not for the $0.98~M_\sun$ WD (CO nova 3).

The corresponding absolute magnitudes of the light curves can be 
readily obtained from Equation (\ref{simple_scaling_flux}) and
from the distance modulus of V1668~Cyg, i.e.,
\begin{eqnarray}
M_V^{\{M_{\rm WD}\}} (t) &=& m_V^{\{M_{\rm WD}\}} (t) 
- (m-M)_{V, \rm V1668~Cyg} \cr\cr
&=& 2.5 \log f_{\rm s} 
-2.5 ~ \log ~ \left[ {{\dot M_{\rm wind}^2}
\over {v_{\rm ph}^2 R_{\rm ph}}} \right]^{\{0.98~M_\sun\}}_{(t/f_s)} \cr\cr
& & + K_V - (m-M)_{V, \rm V1668~Cyg} \cr\cr
&=& {m'}_V^{\{M_{\rm WD}\}} (t/f_s) + 2.5 \log f_{\rm s} 
- (m-M)_{V, \rm V1668~Cyg},
\label{real_timescale_flux}
\end{eqnarray}
where $(m-M)_{V, \rm V1668~Cyg}=14.6$ is the distance modulus of V1668~Cyg
harboring a $0.98~M_\sun$ WD  (CO nova 3), and
we use Equation (\ref{relative_magnitude_two_free-free_equal}) in a
different form of
\begin{equation}
\left[ {{\dot M_{\rm wind}^2}
\over {v_{\rm ph}^2 R_{\rm ph}}} \right]^{\{M_{\rm WD}\}}_{(t)}
=
\left[ {{\dot M_{\rm wind}^2}
\over {v_{\rm ph}^2 R_{\rm ph}}} \right]^{\{0.98~M_\sun\}}_{(t/f_s)},
\label{relative_magnitude_two_free-free_equal_no2}
\end{equation}
and Equation (\ref{relative_magnitude_two_free-free_convert}) also 
in a different form of
\begin{equation}
{m'}_V^{\{M_{\rm WD}\}} (t/f_s)= 
-2.5 ~ \log ~ \left[ {{\dot M_{\rm wind}^2}
\over {v_{\rm ph}^2 R_{\rm ph}}} \right]^{\{M_{\rm WD}\}}_{(t)}
+K_V,
\label{relative_magnitude_two_free-free_convert_no2}
\end{equation}
to derive the last line of Equation (\ref{real_timescale_flux}).
The last line in Equation (\ref{real_timescale_flux})
simply means that the model light curve ${m'}_V^{\{M_{\rm WD}\}}(t')$
in Figure \ref{all_mass_v1668_cyg_x45z02c15o20_calib_universal}
is shifted horizontally by $\log f_s$ and vertically by
$2.5 \log f_s - (m-M)_{V, \rm V1668~Cyg}$ to retrieve the absolute
magnitude and real timescale.  

These retrieved absolute magnitudes 
and real timescales are plotted in Figure
\ref{all_mass_v1668_cyg_x45z02c15o20_real_scale_universal}
for CO nova 3, in Figure 
\ref{all_mass_qu_vul_x65z02o03ne03_absolute_mag}
for Ne nova 3, in Figure
\ref{all_mass_v351_pup_univeral_real_abs_x35z02o20ne10}
for Ne nova 1.
We also tabulate the absolute magnitude, $M_{\rm w}$,
at the end point of winds in Tables \ref{light_curves_of_novae_co3},
\ref{light_curves_of_novae_ne3}, and \ref{light_curves_of_novae_ne1}
and plot them in Figures \ref{bottom_ff_line_x45z02c15o20_absolute_mag},
\ref{bottom_ff_line_x65z02o03ne03_absolute_mag}, and
\ref{bottom_ff_line_x35z02o20ne10_absolute_mag}, for
CO nova 3, Ne nova 3, and Ne nova 1, respectively.
The values of Ne nova 3 are slightly lower than those for the other
chemical compositions (Figure
\ref{bottom_ff_line_x65z02o03ne03_absolute_mag}).  
Then, we retrieve the absolute magnitudes
of all model light curves in Tables \ref{light_curves_of_novae_co3},
\ref{light_curves_of_novae_ne3}, and \ref{light_curves_of_novae_ne1}, as 
\begin{equation}
M_V = m_{\rm ff} - (m_{\rm w} - M_{\rm w})
    = m_{\rm ff} - (15.0 - M_{\rm w}).
\end{equation}
Note that $K_V$ is a constant for all WD masses.
For Ne nova 3, $K_V$ is related to $G^{\{ 0.98~M_\sun \} }$ 
of V1668~Cyg as $K_V = G^{\{0.98~M_\sun \} } + 1.3$ because
$m_{\rm ff} - m'_V =m_{\rm ff} - m'_{\rm w}= G^{\{0.98 ~M_\sun \} } - K_V
= 15.0 - 16.3 = -1.3$ at the bottom of the light curve
(at the end of winds).   Here we use $m'_{\rm w}=m_{\rm w}$
because $f_{\rm s}=1$ for the $0.98~M_\sun$ WD for V1668~Cyg
and we directly read $m'_{\rm w}=m_{\rm w}=16.3$
from the $0.98~M_\sun$ WD model in Figure 
\ref{all_mass_qu_vul_m0960_x65z02o03ne03_compsite}(b).
Then, we obtain $M_{\rm w}=1.7$ for the $0.98~M_\sun$ WD model
because $M_{\rm w}=m_{\rm w} - (m-M)_V= 16.3 -14.6=1.7$.

In the same way, we have $K_V = G^{\{0.96~M_\sun \} } + 1.6$ 
for QU~Vul of Ne nova 3 
because $G^{\{0.96 ~M_\sun \} } - K_V = m_{\rm ff} - m'_{\rm w}= 
15.0 - 16.6 = -1.6$, 
and $K_V = G^{\{1.0~M_\sun \} } + 1.5$ 
for V351~Pup of Ne nova 1 
because $G^{\{1.0 ~M_\sun \} } - K_V = m_{\rm ff} - m'_{\rm w}= 
15.0 - 16.5 = -1.5$.

\subsection{MMRD relation}
\label{mmrd_relation_model_fit}

\subsubsection{CO Nova 3 based on V1668~Cyg}
We calculated free-free model light curves of various WD masses
for the chemical composition of CO nova 3.  We plot them in Figure 
\ref{light_curve_combine_x45z02c15o20}
and tabulate them in Table \ref{light_curves_of_novae_co3}.
These curves almost overlap each other if they
are properly stretched/squeezed along time, as shown in Figure
\ref{all_mass_v1668_cyg_x45z02c15o20_calib_universal}.
Using Equation (\ref{real_timescale_flux}),
we obtain the absolute magnitude of each model light curve,
as shown in Figure \ref{all_mass_v1668_cyg_x45z02c15o20_real_scale_universal}.
(Note that the ordinates of the figure is the absolute magnitude
of each light curve.)
The absolute magnitude at the end point of the wind phase, $M_{\rm w}$,
is also plotted in Figure \ref{bottom_ff_line_x45z02c15o20_absolute_mag}
(and also listed in Table \ref{light_curves_of_novae_co3}).

Figure \ref{all_mass_v1668_cyg_x45z02c15o20_real_scale_universal}
clearly shows that a more massive WD tends to have a brighter maximum
magnitude (smaller $M_{V, {\rm max}}$, denoted by a magenta filled circle)
and a faster decline rate (smaller $t_2$ or $t_3$ time).  
The apparent maximum brightness $m_{V, {\rm max}}$
of each light curve with a different WD mass is expressed as
$m_{V,{\rm max}} = m'_{V,{\rm max}} + 2.5 \log f_{\rm s}$
from Equation (\ref{simple_final_scaling_flux}).
The $t_3$ time of each model light curve with a different WD mass
is squeezed to be $t_3 = f_{\rm s} t'_3$.
Eliminating $f_{\rm s}$ from these two relations, we have
$m_{V,{\rm max}} = 2.5 \log t_3 + m'_{V,{\rm max}} - 2.5 \log t'_3$.
If we adopt the free-free model light curves, point B 
(magenta filled circle) corresponds to
the $V$ maximum of V1668~Cyg, as shown in Figure 
\ref{all_mass_v1668_cyg_x45z02c15o20_calib_universal}.
Then, we read $t'_3 = 26$ days and $m'_{V,{\rm max}} = 6.2$
along our free-free model light curves in Figures 
\ref{all_mass_v1668_cyg_x45z02c15o20_calib_universal}
and \ref{all_mass_v1668_cyg_x45z02c15o20_real_scale_universal}.
We obtained our MMRD relation as
\begin{eqnarray}
M_{V,{\rm max}} & = & m_{V,{\rm max}}  - (m-M)_V \cr
& = & 2.5 \log t_3 + m'_{V,{\rm max}} - 2.5 \log t'_3 - (m-M)_V \cr
& = & 2.5 \log t_3 -11.94.
\label{theoretical_MMRD_relation_v1668_cyg}
\end{eqnarray}
We plot this MMRD relation in Figure \ref{max_t3_point_B_scale_qu_vul}
together with the MMRD point of V1668~Cyg (based on our model
light curve, green filled square).
Note that we updated the distance modulus to 
$(m-M)_V=14.6$ for V1668~Cyg from its previous value of
$(m-M)_V=14.25$ in \citet{hac10k}.

\subsubsection{Ne Nova 3 based on QU~Vul}
We obtained the theoretical MMRD relation for Ne nova 3
based on QU~Vul.  We calculated free-free model light curves of
various WD masses for the chemical composition of Ne nova 3
(Figure \ref{all_mass_light_curve_model_x65z02o03ne03}
and Table \ref{light_curves_of_novae_ne3}).
These curves almost overlap each other if they
are properly stretched/squeezed along time
(Figure \ref{all_mass_qu_vul_x65z02o03ne03_calib_universal}).
Using Equation (\ref{real_timescale_flux}) for QU~Vul,
we obtain the absolute magnitude of each model light curve
(Figure \ref{all_mass_qu_vul_x65z02o03ne03_absolute_mag}).

In the same way as for V1668~Cyg, we obtain our MMRD relation as
\begin{eqnarray}
M_{V,{\rm max}} & = & m_{V,{\rm max}}  - (m-M)_V \cr
& = & 2.5 \log t_3 + m'_{V,{\rm max}} - 2.5 \log t'_3 - (m-M)_V \cr
& = & 2.5 \log t_3 -11.88,
\label{theoretical_MMRD_relation_qu_vul}
\end{eqnarray}
where we adopt $t'_3 = 32.7$~days, $m'_{V,{\rm max}} = 5.5$,
and $(m-M)_V=13.6$ for QU~Vul. 
Note that we read $t'_3 = 32.7$ days and $m'_{V,{\rm max}} = 5.5$
along our free-free model light curves in Figures 
\ref{all_mass_qu_vul_x65z02o03ne03_calib_universal}
and \ref{all_mass_qu_vul_x65z02o03ne03_absolute_mag}.
We plot this MMRD relation in Figure \ref{max_t3_point_B_scale_qu_vul}
together with the MMRD point of QU~Vul.  A large black open circle
corresponds to the MMRD point based on our model light curve while
a large red filled circle represents the MMRD point obtained by
\citet{dow00}.

\subsubsection{Ne Nova 1 based on V351~Pup}
We calculated free-free model light curves of various WD masses
for the chemical composition of Ne nova 1 (Figure 
\ref{all_mass_light_curve_model_x35z02o20ne10}
and Table \ref{light_curves_of_novae_ne1}).
These curves almost overlap each other if they
are properly stretched/squeezed along time (Figure
\ref{all_mass_v351_pup_univeral_scale_x35z02o20ne10}).
Using Equation (\ref{real_timescale_flux}) for V351~Pup,
we obtain the absolute magnitude of each model light curve (Figure 
\ref{all_mass_v351_pup_univeral_real_abs_x35z02o20ne10}).

In the same way as for V1668~Cyg, we obtained our MMRD relation as
\begin{eqnarray}
M_{V,{\rm max}} & = & m_{V,{\rm max}}  - (m-M)_V \cr
& = & 2.5 \log t_3 + m'_{V,{\rm max}} - 2.5 \log t'_3 - (m-M)_V \cr
& = & 2.5 \log t_3 -11.98,
\label{theoretical_MMRD_relation_v351_pup}
\end{eqnarray}
where we adopt $t'_3 = 20.4$~days, $m'_{V,{\rm max}} = 6.4$,
and $(m-M)_V=15.1$ for V351~Pup. 
Note that we read $t'_3 = 20.4$~days and $m'_{V,{\rm max}} = 6.4$
along our free-free model light curves in Figures 
\ref{all_mass_v351_pup_univeral_scale_x35z02o20ne10}
and \ref{all_mass_v351_pup_univeral_real_abs_x35z02o20ne10}.
We plot this MMRD relation in Figure \ref{max_t3_point_B_scale_v351_pup}
together with the MMRD point of QU~Vul.  A large black open circle
corresponds to the MMRD point based on our model light curve while
a large red filled circle represents the MMRD point obtained by
\citet{dow00}.

\subsection{Other empirical relations}
It is interesting to examine 
various empirical relations of classical novae on the basis of
our free-free model light curves.  Here we check the empirical formula
that the absolute magnitude 15 days after optical maximum, $M_V(15)$,
is almost common among various novae.  This relation was
proposed by \citet{bus55} with $M_V(15)= -5.2 \pm 0.1$.
Subsequently, the value of $M_V(15)= -5.60 \pm 0.43$ was reported
by \citet{coh85};
$M_V(15)= -5.23 \pm 0.39$ by \citet{van87};
$M_V(15)= -5.69 \pm 0.42$ by \citet{cap89}; and
$M_V(15)= -6.05 \pm 0.44$ by \citet{dow00}.
This relation was sometimes called the $t$15 relation 
\citep[e.g.,][]{dar06}.

We already obtained $M_V(15)= -5.95 \pm 0.25$ for 0.55 -- $1.2 ~M_\odot$ WDs
with the chemical composition of CO nova 2 \citep{hac10k}, 
$M_V(15)= -5.6 \pm 0.3$ for 0.7 -- $1.3 ~M_\sun$ WDs with Ne nova 2
\citep{hac10k}, and $M_V(15)= -5.4 \pm 0.4$ for 0.7 -- $1.05~M_\sun$ WDs
with CO nova 4 \citep{hac15k}. 
These three relations calculated from the model light curves
are consistent with the above empirical relations
obtained from the observations.

Point B (magenta filled circle)
in Figure \ref{all_mass_v1668_cyg_x45z02c15o20_calib_universal}
represents the peak of $V$ magnitude of the V1668~Cyg outburst. 
In the real timescale and absolute magnitude,
this peak is different for different WD masses as shown in Figure
\ref{all_mass_v1668_cyg_x45z02c15o20_real_scale_universal}.
The position of the absolute magnitude 15 days after the peak is
shown by green crosses.  The distribution of the green
crosses is given by 
$M_V(15)= -5.55\pm0.38$ for 0.55--$1.2~M_\sun$ WDs
with CO nova 3, as shown in Figure
\ref{all_mass_v1668_cyg_x45z02c15o20_real_scale_universal}.
Note that this relation is obtained
for the chemical composition of CO nova 3.
In the same way, we obtained $M_V(15)= -5.9\pm0.3$ 
for 0.7--$1.3~M_\sun$ WDs with Ne nova 3,
as shown in Figure \ref{all_mass_qu_vul_x65z02o03ne03_absolute_mag},
$M_V(15)= -5.8\pm0.5$ for 0.7--$1.3~M_\sun$ WDs
with Ne nova 1, as shown in Figure
\ref{all_mass_v351_pup_univeral_real_abs_x35z02o20ne10}.
These new three relations calculated from our model light curves
are all roughly consistent with the above empirical and model relations.

It should be noted that, however, 
recent extra-galactic nova surveys show no
clear evidence of the so-called $t$15 relation \citep[e.g.,][]{fer03, dar06}.
The $t$15 relation is a statistical relation like the MMRD relation.
We suppose that the maximum magnitude of a nova depends not only on
the WD mass but also on the initial hydrogen-rich envelope mass
as explained in the MMRD relation.  Therefore, the $M_V(15)$ is also
affected by the initial envelope mass, even if the WD mass is the same.
This could explain the large scatter of $M_V(15)$ distribution.

\end{document}